\documentclass[iop]{emulateapj}
\usepackage[breaklinks,colorlinks,citecolor=blue,linkcolor=magenta,urlcolor=blue]{hyperref}
\usepackage{subfigure}

\newcommand*{\mysub}[2]{\ensuremath{#1_{\mathrm{#2}}}}

\newcommand*{\Omegam}{\mysub{\Omega}{m}}

\newcommand*{\Omegal}{\mysub{\Omega}{\Lambda}}

\newcommand*{\LCDM}{$\Lambda$CDM }

\newcommand*{\amin}{\mathrm{arcmin}}

\newcommand*{\ltsim}{\ {\raise-.75ex\hbox{$\buildrel<\over\sim$}}\ }
\newcommand*{\gtsim}{\ {\raise-.75ex\hbox{$\buildrel>\over\sim$}}\ }
\newcommand*{\proptosim}{\ {\raise-.75ex\hbox{$\buildrel\propto\over\sim$}}\ }

\newcommand*{\secref}{Section}

\newcommand{\Herschel}{{\it Herschel }}

\shorttitle{The Candidate Cluster and Protocluster (CCPC) Catalog}
\shortauthors{Franck \& McGaugh}

\begin{document}
\title{The Candidate Cluster and Protocluster Catalog (CCPC) of spectroscopically identified structures spanning $2.74<\lowercase{z}<3.71$}


\author{J.R. Franck\altaffilmark{1},
	 S.S. McGaugh\altaffilmark{1}}
	 %,J.M. Schombert\altaffilmark{2}}

 \altaffiltext{1}{Case Western Reserve University, 10900 Euclid Ave., Cleveland, OH 44106}
% \altaffiltext{2}{University of Oregon, 1585 E. 13th Ave., Eugene, OR 97403}

\begin{abstract}
We have developed a search methodology to identify galaxy protoclusters at $z>2.74$, and implemented it on a sample of $\sim$14,000 galaxies with previously measured redshifts. The results of this search are recorded in the Candidate Cluster and Protocluster Catalog (CCPC). The catalog contains 12 clusters that are highly significant overdensities ($\delta_{gal}>7$), 6 of which are previously known. We also identify another 31 candidate protoclusters (including 4 previously identified structures) of lower overdensity. CCPC systems vary over a wide range of physical sizes and shapes, from small, compact groups to large, extended, and filamentary collections of galaxies. This variety persists over the range from $z=3.71$ to $z=2.74$. These structures exist as galaxy overdensities ($\delta_{gal}$) with a mean value of 2, similar to the values found for other protoclusters in the literature. The median number of galaxies for CCPC systems is 11. Virial mass estimates are large for these redshifts, with thirteen cases apparently having $M > 10^{15}\, M_{\sun}$.  If these systems are virialized, such masses would pose a challenge to \LCDM. 
\end{abstract}


\section{Introduction} \label{sec:intro}

Identifying structures over varying epochs offers the possibility of examining the assembly history of the Universe and the environmental evolution of galaxies in structures. Initial density perturbations at the earliest times, aided by gravity, give rise to present day clusters. Although still relatively rare, $z>1$ clusters are being discovered with increasing regularity based on surveys in the near-infrared (NIR) for galaxy overdensities around radio-loud sources \citep{hal98,hal01,2015arXiv150507864F}, in the field \citep{pap08,eis08}, by diffuse X-ray emission of the hot intra-cluster medium \citep{gob11,wil13}, and from the Sunyaev-Zel'Dovich (SZ) effect \citep{man14,ret14}. A larger sample of cluster candidates, especially at increasing redshift, may offer a clearer picture for structure evolution.  Protoclusters, loosely defined here as structures $z\gtrsim 2$, are much rarer, and only a handful of studies containing more than a single object at $z>2.75$ exist \citep{2007AnA...461..823V,gal12,2013ApJ...765..109D,chi14,rig14,wyl14}. With a larger catalog of clusters at various stages of assembly, the general trends of formation and galaxy evolution can be analyzed over time, instead of being dissected epoch by epoch. 

At these higher redshifts, there have been a number of amazing discoveries of groups of galaxies found at $z\sim4$ \citep{lee14,ven02}, $z>6$ \citep{2010ApJ...721.1680U,2014ApJ...792...15T}, up to $z\sim 8$ \citep{2012ApJ...746...55T}. The emergence of these protoclusters offer a test of \LCDM, as simulations provide constraints on the number of clusters and structure mass as a function of redshift and the dark energy equation of state \citep{mor11,vik09}, with greater leverage provided at larger redshifts. More fundamentally, these `pink elephants' offer a glimpse into the unexpected mysteries of this discovery driven science.

By comparing these primordial objects to their assumed, present-day manifestations as rich clusters, the physical processes that foster their evolution can be better understood. To date, a clear evolutionary path has not emerged to tie these high redshift structures to the $z=0$ clusters observed. For instance, the assembly process for the defining red sequence (RS) feature of nearby clusters is not well established \citep{fer12,fas14,fri14}. Estimating the formation epoch for the stellar population of clusters has also proven challenging. A range from $z_f=2$ to $z_f=30$ is found using a variety of methods at various wavelengths \citep{rak95,eis08,fer12,van14,wyl14}. 

To this end, we have compiled a list of 43 candidate protoclusters assembled using archival measurements in the redshift range of $2.74<z<3.71$. The majority of these candidate structures, at the time of writing, have not been previously identified to the best of our knowledge. This catalog is the largest list of high redshift protoclusters based on spectroscopic redshifts. \secref~\ref{sec:ccpc} explains the search criteria used to identify candidate structures, with the general results explored in \secref~\ref{sec:dis}. \secref~\ref{sec:sum} gives a brief summary of the findings. The Appendix contains sky position plots and $N(z)$ histograms for each protocluster.

This work assumes a \LCDM concordance cosmology, with $\Omegal = 0.7$, a matter density of $\Omegam = 0.3$, and $H_0 = 70$ km s$^{-1}$  Mpc$^{-1}$. At the redshift range of $z=2.74\rightarrow3.7$, the angular size using this cosmology is 0.47 Mpc  $\amin^{-1}$ to 0.43 Mpc  $\amin^{-1}$, 20 comoving Mpc along the line of sight has a $\Delta z\sim$ 0.019 to 0.026, whereas the corresponding age of the Universe is 2.3 to 1.7 Gyrs, respectively \citep{wri06}. %Distance Mod at z=3.7 is 47.5 

\section{The Candidate Cluster and Protocluster Catalog (CCPC)} \label{sec:ccpc}

\begin{deluxetable*}{c c c c c c c c c c c}
%\tabletypesize{\scriptsize}
\tablewidth{0pt}
\tablecolumns{10}
\tablecaption{Candidate Cluster and Protocluster Catalog (CCPC) - Strongest Candidates}
\tablehead{
\colhead{Candidate}  & \colhead{RA} & \colhead{DEC} & \colhead{Redshift} & \colhead{$\sigma_z$}  & \colhead{Number of} & \colhead{Galaxies in} & \colhead{Overdensity}& \colhead{Cluster} &  \colhead{Reference}\\
\colhead{Name}& \colhead{($deg$)}& \colhead{($deg$)}& \colhead{}& \colhead{} & \colhead{Galaxies}& \colhead{$R<10$ cMpc} & \colhead{($\delta_{gal}$)} &\colhead{Probability ($\%$)} & \colhead{}
} 
\startdata
CCPC-z27-002	&	02:21:19.92	&	-04:27:43.20	&	2.772	&	0.007	&	5	&	5	&	11.02	$\pm$	6.90\tablenotemark{a}	&	100	&	\\
CCPC-z29-001	&	09:33:35.71	&	28:44:45.89	&	2.918	&	0.005	&	9	&	8	&	11.21	$\pm$	4.76	&	100	&	\\
CCPC-z29-002	&	09:45:32.76	&	-24:29:05.28	&	2.919	&	0.009	&	26	&	26	&	12.91	$\pm$	4.55	&	100	& 1	\\
CCPC-z30-001	&	12:08:06.67	&	-30:31:05.16	&	3.035	&	0.005	&	8	&	8	&	18.78	$\pm$	10.14	&	100	& 2	\\
CCPC-z30-003	&	22:17:25.92	&	00:12:37.58	&	3.096	&	0.008	&	54*	&	25	&	12.28	$\pm$	2.42	&	100	& 3	\\
CCPC-z31-003	&	03:18:07.58	&	-25:34:55.56	&	3.133	&	0.008	&	33	&	33	&	9.80	$\pm$	2.77	&	100	& 1	\\
CCPC-z31-004	&	01:06:11.53	&	-25:46:14.52	&	3.146	&	0.006	&	6	&	6	&	7.59	$\pm$	4.65	&	85	&	\\
CCPC-z31-005	&	20:09:54.43	&	-30:41:19.68	&	3.152	&	0.007	&	12	&	12	&	17.77	$\pm$	9.19\tablenotemark{c}	&	100	& 1	\\
CCPC-z32-002	&	02:59:04.16	&	00:12:35.60	&	3.234	&	0.003	&	5	&	5	&	13.11	$\pm$	8.63	&	100	&	\\
CCPC-z33-002	&	02:03:35.13	&	11:38:06.68	&	3.372	&	0.008	&	5	&	5	&	7.44	$\pm$	4.47	&	85	& 4	\\
CCPC-z35-001	&	00:03:25.12	&	-26:04:52.68	&	3.597	&	0.003	&	4	&	4	&	10.18	$\pm$	8.05	&	100	&	\\
CCPC-z36-001	&	02:03:49.52	&	11:35:53.48	&	3.644	&	0.003	&	4	&	4	&	23.50	$\pm$	14.39	&	100	&	
\enddata
\tablenotetext{a}{No field galaxies were identified along the line of sight $\Delta z\pm 0.15$. This $\delta_{gal}$ should be treated as an upper limit. }
\tablenotetext{c}{Large numbers of diffuse field galaxies within $\Delta z\pm 0.15$ gave a $\delta_{gal}\sim 0$. By limiting the field galaxies to the same surface area (RA/DEC) as the galaxies within the protocluster, these became positive overdensities.}
\tablecomments{The names and positions for the strongest CCPC overdensities, along with the mean redshift and dispersion of galaxies. The naming scheme is explained in \secref~\ref{sec:ccpc}. For each candidate, the number of spectroscopically confirmed members is listed. If the number of galaxies is followed by a `*', this indicates that the protocluster is an extended system in which 1/3 or more members in the structure are found between $10<R<20$ comoving Mpc. The implications of this are discussed in \secref~\ref{sec:dis}. The galaxy overdensity is listed under $\delta_{gal}$, and the basis of this calculation is explained in Section 2.3. Using each candidates overdensity, we have assigned a conservative probability that the structure will collapse into a massive cluster by $z=0$ from the values in Figure 8 in \citet{chi13}, which plot the fate of overdensities within the Millennium simulation. Structures with greater than $85\%$ probability are included in this Table.  If the structure has been previously identified, its reference is included. References: (1) \citet{2007AnA...461..823V} (2) \citet{2001AnA...372L..57M} (3) \citet{1998ApJ...492..428S}  (4)  \citet{2001ApJ...549..770E} 
\label{tab:cat}}
\end{deluxetable*}

\begin{deluxetable*}{c c c c c c c c c c c}
%\tabletypesize{\scriptsize}
\tablewidth{0pt}
\tablecolumns{10}
\tablecaption{Candidate Cluster and Protocluster Catalog (CCPC) - Additional Candidates}
\tablehead{
\colhead{Candidate}  & \colhead{RA} & \colhead{DEC} & \colhead{Redshift} & \colhead{$\sigma_z$}  & \colhead{Number of} & \colhead{Galaxies in} & \colhead{Overdensity}& \colhead{Cluster} &  \colhead{Reference}\\
\colhead{Name}& \colhead{($deg$)}& \colhead{($deg$)}& \colhead{}& \colhead{} & \colhead{Galaxies}& \colhead{$R<10$ cMpc} & \colhead{($\delta_{gal}$)} &\colhead{Probability ($\%$)} & \colhead{}
} 
\startdata
CCPC-z27-001	&	17:01:03.84	&	64:11:56.40	&	2.748	&	0.006	&	8	&	8	&	2.51	$\pm$	1.37	&	18	&1	\\
CCPC-z27-003	&	03:04:42.00	&	-00:07:40.80	&	2.788	&	0.009	&	7	&	7	&	1.73	$\pm$	0.96\tablenotemark{c}	&	1	&	\\
CCPC-z27-004	&	12:36:51.84	&	62:11:56.40	&	2.803	&	0.007	&	6*	&	4	&	1.98	$\pm$	0.13	&	10	&	\\
CCPC-z27-005	&	22:32:48.96	&	-60:31:55.20	&	2.798	&	0.006	&	5	&	5	&	6.65	$\pm$	4.38	&	73	&	\\
CCPC-z28-001	&	14:24:34.08	&	22:49:04.80	&	2.814	&	0.010	&	9	&	8	&	1.97	$\pm$	0.89	&	10	&	\\
CCPC-z28-002	&	03:32:14.40	&	-27:52:37.20	&	2.818	&	0.009	&	25*	&	14	&	1.27	$\pm$	0.32	&	10	&	\\
CCPC-z28-003	&	02:59:05.76	&	00:11:27.60	&	2.825	&	0.014	&	10	&	10	&	1.01	$\pm$	0.50	&	10	&	\\
CCPC-z28-004	&	00:54:28.56	&	-23:53:49.20	&	2.857	&	0.009	&	31	&	30	&	0.75	$\pm$	0.25	&	1	& 2	\\
CCPC-z28-005	&	21:41:59.76	&	-44:13:26.40	&	2.856	&	0.005	&	23	&	23	&	3.98	$\pm$	1.72	&	48	&	\\
CCPC-z28-006	&	03:32:38.17	&	-27:47:08.16	&	2.863	&	0.009	&	14*	&	2\tablenotemark{b}	&	0.31	$\pm$	0.11	&	1	&	\\
CCPC-z28-007	&	03:04:41.13	&	-00:10:38.78	&	2.864	&	0.010	&	7	&	7	&	1.52	$\pm$	0.83	&	10	&	\\
CCPC-z29-003	&	14:18:04.13	&	52:29:54.28	&	2.924	&	0.013	&	30*	&	18	&	1.36	$\pm$	0.35	&	10	&	\\
CCPC-z29-004	&	12:36:47.71	&	62:12:55.80	&	2.931	&	0.010	&	26	&	25	&	2.93	$\pm$	0.83	&	18	&	\\
CCPC-z29-005	&	14:24:45.91	&	22:56:42.11	&	2.974	&	0.008	&	13	&	9	&	3.26	$\pm$	1.11	&	48	&	\\
CCPC-z29-006	&	02:58:53.07	&	00:09:53.60	&	2.982	&	0.013	&	6	&	6	&	0.52	$\pm$	0.28	&	1	&	\\
CCPC-z29-007	&	03:32:19.29	&	-27:43:03.14	&	2.979	&	0.012	&	14*	&	7	&	0.30	$\pm$	0.10	&	1	&	\\
CCPC-z30-002	&	14:24:30.29	&	22:52:49.91	&	3.074	&	0.012	&	15	&	11	&	1.12	$\pm$	0.41	&	10	&	\\
CCPC-z31-001	&	03:32:24.31	&	-27:41:52.44	&	3.113	&	0.009	&	27*	&	16	&	3.67	$\pm$	0.93	&	48	& 3	\\
CCPC-z31-002	&	04:22:18.53	&	-38:45:15.48	&	3.118	&	0.009	&	9	&	9	&	0.47	$\pm$	0.28	&	1	&	\\
CCPC-z31-006	&	13:49:06.70	&	-03:36:32.00	&	3.166	&	0.010	&	13	&	11	&	0.98	$\pm$	0.46	&	1	&	\\
CCPC-z31-007	&	12:36:47.35	&	62:15:12.74	&	3.179	&	0.014	&	12	&	10	&	0.55	$\pm$	0.24 \tablenotemark{c}	&	1	&	\\
CCPC-z32-001	&	12:05:19.10	&	-07:42:31.00	&	3.206	&	0.010	&	14	&	14	&	0.67	$\pm$	0.35	&	1	&	\\
CCPC-z32-003	&	03:32:34.54	&	-27:46:12.36	&	3.258	&	0.014	&	15*	&	3\tablenotemark{b}	&	1.1	$\pm$	0.43\tablenotemark{c}	&	10	&	\\
CCPC-z33-001	&	12:36:50.74	&	62:13:49.51	&	3.363	&	0.008	&	9	&	9	&	1.48	$\pm$	0.65	&	10	&	\\
CCPC-z33-003	&	03:32:34.90	&	-27:47:08.16	&	3.368	&	0.013	&	11*	&	5	&	1.70	$\pm$	0.77\tablenotemark{c}	&	10	&	\\
CCPC-z33-004	&	01:06:04.60	&	-25:46:54	&	3.388	&	0.010	&	10	&	10	&	3.63	$\pm$	1.91	&	48	& 4	\\
CCPC-z33-005	&	22:39:38.45	&	11:54:09.61	&	3.389	&	0.014	&	14	&	11	&	2.74	$\pm$	1.19	&	18	&	\\
CCPC-z34-001	&	12:36:52.42	&	62:13:37.74	&	3.423	&	0.014	&	7	&	7	&	0.61	$\pm$	0.30	&	1	&	\\
CCPC-z34-002	&	03:32:18.48	&	-27:52:06.60	&	3.476	&	0.010	&	23*	&	11	&	3.75	$\pm$	1.02	&	48	&	\\
CCPC-z36-002	&	03:32:16.90	&	-27:48:38.16	&	3.658	&	0.012	&	8*	&	5	&	0.32	$\pm$	0.14	&	1	&	\\
CCPC-z37-001	&	03:32:10.42	&	-27:40:53.04	&	3.704	&	0.013	&	11*	&	7	&	1.02	$\pm$	0.42	&	10	&	
\enddata
\tablenotetext{a}{No field galaxies were identified along the line of sight $\Delta z\pm 0.15$. This $\delta_{gal}$ should be treated as an upper limit. }
\tablenotetext{b}{Galaxies with photometric redshifts \citep{2011ApJ...736...48R} were manually removed from these candidate protoclusters, as they are uncategorized as such within NED. Without 3 or more galaxies within 2' from the RA/DEC listed, these technically violate our criteria for a candidate protocluster. From the position plots of these systems in the Appendix, it is clear that there are a significant number of galaxies within a search radius of 20 comoving Mpc, and choosing another structure position would satisfy the criteria for the CCPC, but would decrease the number of galaxy members. We have chosen to maximize the number of galaxies in these case.}
\tablenotetext{c}{Large numbers of diffuse field galaxies within $\Delta z\pm 0.15$ gave a $\delta_{gal}\sim 0$. By limiting the field galaxies to the same surface area (RA/DEC) as the galaxies within the protocluster, these became positive overdensities. In the case of CCPC-z27-004, the field $\Delta z$ range intersected CCPC-z29-004. By changing the field to $\Delta z\pm 0.05$, this system now has $\delta_{gal}> 0$.} 
\tablecomments{Identical to Table~\ref{tab:cat}, but for candidate structures with $<85\%$ probability to collapse based on their galaxy overdensity ($\delta_{gal}$). The probabilities are assigned based on overdensities analyzed by \citet{chi13} within the Millennium simulation (their Figure 8). If the structure has been previously identified, its reference is included. References: (1) \citet{2008ApJ...678L..77P} (2) \citet{2007AnA...461..823V} (3) \citet{2010ApJ...716L.200B} (4)  \citet{2003AnA...407..473F}  
\label{tab:cat2}}
\end{deluxetable*}

\begin{deluxetable*}{c c}
%\tabletypesize{\scriptsize}
\tablewidth{0pt}
\tablecolumns{2}
\tablecaption{CCPC: Member Redshift Reference List}
\tablehead{
\colhead{Candidate} & \colhead{Redshift} \\
\colhead{Name} & \colhead{References} 
} 
\startdata
CCPC-z27-001	&	\citet{2007ApJ...668...23P,2005ApJ...626..698S,2010MNRAS.405.2302H,2008PrivC.U..D...1L}	\\
CCPC-z27-002	&	\citet{2011ApJS..192....5A}	\\
CCPC-z27-003	&	\citet{2003ApJ...592..728S,1995AJ....109.1522C}	\\
CCPC-z27-004	&	\citet{2001ApJ...559..620P,2002AJ....124.1886M,2001ApJS..135...41F,2003AJ....126.1183C}	\\
...	&	\citet{2003ApJ...592..728S,2006ApJ...653.1004R}	\\
CCPC-z27-005	&	\citet{2005AnA...440..881I,2000AnA...359..489C}	\\
CCPC-z28-001	&	\citet{2003ApJ...592..728S}	\\
CCPC-z28-002	&	\citet{2010AnA...512A..12B,2008ApJ...682..985W,2001ApJ...560..127S,2009AnA...504..751S}	\\
...	&	\citet{2013ApJ...772...48P,2011MNRAS.413...80C}	\\
CCPC-z28-003	&	\citet{1998AJ....115.2184S,2003ApJ...592..728S}	\\
CCPC-z28-004	&	\citet{2007AnA...461..823V,1996ApJS..107...19M}	\\
CCPC-z28-005	&	\citet{2000ApJ...543..552S,2003AnA...407..147F}	\\
CCPC-z28-006	&	\citet{2010AnA...512A..12B,2009AnA...504..751S,2006AnA...449..951G}	\\
CCPC-z28-007	&	\citet{2003ApJ...592..728S}	\\
CCPC-z29-001	&	\citet{2003ApJ...592..728S}	\\
CCPC-z29-002	&	\citet{2007AnA...461..823V,1997AnA...326..505R,2010AnA...509A..83D}	\\
CCPC-z29-003	&	\citet{2003ApJ...592..728S,2006MNRAS.371..221G}	\\
CCPC-z29-004	&	\citet{2003ApJ...592..728S,2004AJ....127.3137C,2013ApJ...772...48P,2002AJ....124.1886M,2011ApJS..192....5A}	\\
...	&	\citet{2006ApJ...653.1004R,2001AJ....122..598D,2004ApJ...611..732C,2002AJ....124.1839B}	\\
CCPC-z29-005	&	\citet{2003ApJ...592..728S}	\\
CCPC-z29-006	&	\citet{2003ApJ...592..728S}	\\
CCPC-z29-007	&	\citet{2010AnA...512A..12B,2004AnA...428.1043L,2013ApJ...772...48P}	\\
...	&	\citet{2008ApJ...682..985W,2011ApJ...729...48B}	\\
CCPC-z30-001	&	\citet{2001AnA...372L..57M,2001AnA...374..443F}	\\
CCPC-z30-002	&	\citet{2003ApJ...592..728S,1998ApJ...494...60P}	\\
CCPC-z30-003	&	\citet{2004ApJ...606...85C,2003ApJ...592..728S,2011MNRAS.411.2336I,2000ApJ...532..170S}	\\
...	&	\citet{2004AJ....128.2073H,2004ApJ...616...71S,2011ApJ...736...18N,2012ApJ...751...29Y}	\\
...	&	\citet{2009MNRAS.400..299L,2004ApJ...614..671C,2009ApJ...691..687L}	\\
CCPC-z31-001	&	\citet{2010ApJ...716L.200B,2008ApJ...682..985W,2009AnA...504..751S,2012ApJ...744..110C}	\\
...	&	\citet{2010AnA...512A..12B,2000AnA...359..489C}	\\
CCPC-z31-002	&	\citet{2007ApJ...657..135C,1994ApJ...436..678O}	\\
CCPC-z31-003	&	\citet{2005AnA...431..793V,2012MNRAS.425..801K,2008MNRAS.389.1223M,1996ApJ...471L..11L}	\\
CCPC-z31-004	&	\citet{2004AnA...418..885N}	\\
CCPC-z31-005	&	\citet{2007AnA...461..823V}	\\
CCPC-z31-006	&	\citet{2003AnA...407..147F}	\\
CCPC-z31-007	&	\citet{2011ApJS..192....5A,2013ApJ...772...48P,2003ApJ...592..728S,2002AJ....124.1886M}	\\
...	&	\citet{2006ApJ...653.1004R,2008ApJ...689..687B,2002AJ....124.1839B}	\\
CCPC-z32-001	&	\citet{2009AnA...497..689G}	\\
CCPC-z32-002	&	\citet{2003ApJ...592..728S}	\\
CCPC-z32-003	&	\citet{2009ApJ...706..885W,2010AnA...512A..12B}	\\
...	&	\citet{2008ApJ...682..985W,2011ApJ...729...48B}	\\
CCPC-z33-001	&	\citet{1997ApJ...481..673L,2001ApJS..135...41F,2011ApJS..192....5A,1996Natur.381..759L}	\\
...	&	\citet{2003ApJ...592..728S,2013ApJ...772...48P,2006ApJ...653.1004R}	\\
CCPC-z33-002	&	\citet{2001ApJ...549..770E,2003ApJ...592..728S,1996AnA...305..450d}	\\
CCPC-z33-003	&	\citet{2010AnA...512A..12B,2012AnA...537A..16F,2011ApJ...729...48B}	\\
...	&	\citet{2008ApJ...682..985W}	\\
CCPC-z33-004	&	\citet{2004AnA...418..885N,2002AnA...393..809M}	\\
CCPC-z33-005	&	\citet{2003ApJ...592..728S}	\\
CCPC-z34-001	&	\citet{1999ApJ...513...34F,2002AJ....124.1886M,2001AJ....122..598D,2011ApJS..192....5A}	\\
...	&	\citet{2003AJ....126..632B,2003ApJ...592..728S,2006ApJ...653.1004R}	\\
CCPC-z34-002	&	\citet{2010AnA...512A..12B,2006AnA...454..423V,2004ApJS..155..271S}	\\
...	&	\citet{2008ApJ...682..985W}	\\
CCPC-z35-001	&	\citet{2003ApJ...592..728S,1997AnA...318..347S}	\\
CCPC-z36-001	&	\citet{2001ApJ...549..770E,2003ApJ...592..728S}	\\
CCPC-z36-002	&	\citet{2011AnA...528A..88G,2010AnA...512A..12B,2006AnA...454..423V,2009ApJ...706..885W}	\\
...	&	\citet{2004AnA...428.1043L,2006AnA...449..951G}	\\
CCPC-z37-001	&	\citet{2011MNRAS.413...80C,2010AnA...512A..12B,2013ApJ...772...48P}	\\
...	&	\citet{2006AnA...454..423V,2007AnA...461...39F}	
\enddata
\tablecomments{The protocluster candidates matched with references for the spectroscopic measurements of their respective members.
\label{tab:ref}}
\end{deluxetable*}

\begin{deluxetable*}{c c c c c c c }
%\tabletypesize{\scriptsize}
\tablewidth{0pt}
\tablecolumns{8}
\tablecaption{CCPC Mass Estimates}
\tablehead{
\colhead{Candidate}		&	\colhead{$\delta_m$}	&	\colhead{Overdensity}	&	\colhead{Overdensity Mass}	&	\colhead{$R_{hms}$} & \colhead{$\sigma$} & \colhead{Virial Mass} \\
\colhead{Name}		&	\colhead{($b=3$)}	&	\colhead{Volume ($cMpc^3$)}	&	\colhead{Estimate ($10^{14}$ $M_{\odot}$)}	&	\colhead{(Mpc)} & \colhead{(km/s)} & \colhead{Estimate ($10^{14}$ $M_{\odot}$)}
} 
\startdata
CCPC-z27-001	&	0.84	&	193	&	0.1	&	2250	&	486	&	2.5	\\
\textbf{CCPC-z27-002}	&	3.67	&	45	&	0.1	&	1688	&	541	&	2.3	\\
CCPC-z27-003	&	0.58	&	220	&	0.1	&	2063	&	686	&	4.5	\\
CCPC-z27-004	&	0.66\tablenotemark{c}	&	224	&	0.2	&	1125	&	582	&	1.8	\\
CCPC-z27-005	&	2.22	&	19	&	$<0.1$	&	1375	&	436	&	1.2	\\
CCPC-z28-001	&	0.66	&	587	&	0.4	&	2688	&	758	&	7.2	\\
CCPC-z28-002	&	0.42	&	11168	&	6.7	&	3938	&	735	&	9.9	\\
CCPC-z28-003	&	0.34	&	1048	&	0.6	&	4563	&	1071	&	24.4	\\
CCPC-z28-004	&	0.25	&	2642	&	1.4	&	3188	&	724	&	7.8	\\
CCPC-z28-005	&	1.33	&	828	&	0.8	&	2063	&	352	&	1.2	\\
CCPC-z28-006	&	0.10	&	6040	&	2.9	&	5313	&	672	&	11.2	\\
CCPC-z28-007	&	0.51	&	1197	&	0.8	&	3625	&	781	&	10.3	\\
\textbf{CCPC-z29-001}	&	3.74	&	601	&	1.2	&	2313	&	357	&	1.4	\\
\textbf{CCPC-z29-002}	&	4.30	&	1640	&	3.6	&	2500	&	699	&	5.7	\\
CCPC-z29-003	&	0.45	&	10147	&	6.1	&	4313	&	988	&	19.6	\\
CCPC-z29-004	&	0.98	&	3742	&	3.1	&	2375	&	736	&	6.0	\\
CCPC-z29-005	&	1.09	&	3393	&	3.1	&	3125	&	614	&	5.5	\\
CCPC-z29-006	&	0.17	&	29	&	$<0.1$	&	3500	&	995	&	16.1	\\
CCPC-z29-007	&	0.10	&	6579	&	3.1	&	4438	&	878	&	15.9	\\
\textbf{CCPC-z30-001}	&	6.26	&	71	&	0.2	&	1000	&	408	&	0.8	\\
CCPC-z30-002	&	0.37	&	3033	&	1.8	&	4063	&	917	&	15.9	\\
\textbf{CCPC-z30-003}	&	4.09	&	14192	&	30.5	&	3938	&	616	&	6.9	\\
CCPC-z31-001	&	1.22	&	13398	&	12.4	&	3313	&	629	&	6.1	\\
CCPC-z31-002	&	0.16	&	575	&	0.3	&	2438	&	645	&	4.7	\\
\textbf{CCPC-z31-003}	&	3.27	&	1740	&	3.1	&	2250	&	581	&	3.5	\\
\textbf{CCPC-z31-004}	&	2.53	&	44	&	0.1	&	1125	&	462	&	1.1	\\
\textbf{CCPC-z31-005}	&	5.92\tablenotemark{c}	&	424	&	1.3	&	2000	&	482	&	2.2	\\
CCPC-z31-006	&	0.33	&	1393	&	0.8	&	3188	&	716	&	7.6	\\
CCPC-z31-007	&	0.18\tablenotemark{c}	&	3465	&	1.9	&	4625	&	1038	&	23.2	\\
CCPC-z32-001	&	0.22	&	860	&	0.4	&	2000	&	739	&	5.1	\\
\textbf{CCPC-z32-002}	&	4.37	&	30	&	0.1	&	1063	&	218	&	0.2	\\
CCPC-z32-003	&	0.35\tablenotemark{c}	&	8715	&	5.1	&	5000	&	955	&	21.2	\\
CCPC-z33-001	&	0.49	&	154	&	0.1	&	1688	&	520	&	2.1	\\
\textbf{CCPC-z33-002}	&	2.48	&	25	&	$<0.1$	&	1063	&	514	&	1.3	\\
CCPC-z33-003	&	0.56\tablenotemark{c}	&	2348	&	1.6	&	4250	&	869	&	14.9	\\
CCPC-z33-004	&	1.21	&	245	&	0.2	&	2375	&	703	&	5.5	\\
CCPC-z33-005	&	0.91	&	2595	&	2.1	&	3125	&	974	&	13.8	\\
CCPC-z34-001	&	0.20	&	798	&	0.4	&	2563	&	976	&	11.3	\\
CCPC-z34-002	&	1.25	&	6521	&	6.2	&	3875	&	667	&	8.0	\\
\textbf{CCPC-z35-001}	&	3.39	&	4	&	$<0.1$	&	813	&	214	&	0.2	\\
\textbf{CCPC-z36-001}	&	7.83	&	50	&	0.2	&	1000	&	194	&	0.2	\\
CCPC-z36-002	&	0.11	&	1564	&	0.7	&	1688	&	743	&	4.3	\\
CCPC-z37-001	&	0.34	&	3725	&	2.1	&	3563	&	857	&	12.2	

\enddata
\tablenotetext{c}{As in Table~\ref{tab:cat}, these overdensities were found near strong, extended fields within $\Delta z\pm0.15$, which gave $\delta_m\sim 0$ values. The field galaxy numbers, when limited to the same aperture as the overdensity, show stronger mass overdensities. In the case of CCPC-z27-004, we merely had to limit $\Delta z$ to $\pm0.05$, as it intersected another structure (CCPC-z29-004). }
\tablecomments{The estimated mass overdensities ($\delta_m=\delta_{gal}/b$) are listed for the CCPC structures, and a linear bias parameter of $b=3$ is adopted from previous works \citep{1998ApJ...492..428S,2007AnA...461..823V}. Names in boldface are the strongest candidates and can be found in Table~\ref{tab:cat}. Based on the minimum and maximum RA/DEC values of galaxies within the overdense region, a rectangular volume encompassing the structure is listed in units of $cMpc^3$. If we assume that the volume containing $\delta_m$ will eventually collapse, we can calculate an estimated mass of the collapsed structure as $M = \rho_{crit,z}V(1+\delta_{m})$. We can also infer a mass using the virial equation $M_{vir} = \frac{2\sigma^2}{G} R_{hms}. $ The radius ($ R_{hms}$) used is approximated by $1.25*R_e $, where the effective radius contains $50\%$ of the galaxy members of a protocluster in physical units. The virial mass estimator is almost universally larger than the expected, collapsed mass of the system we infer from $\delta_m$. This suggests that these objects are not in virial equilibrium. Protoclusters with the highest probabilities (Tables~\ref{tab:cat},~\ref{tab:cat2}) typically have mass estimates within the expectations of simulated systems, while low probability systems ($<10\%$) have virial masses much larger than predicted at this epoch \citep{chi13}.
\label{tab:mass}}
\end{deluxetable*}

\subsection{Construction of the High Redshift Galaxy List}

To identify structures of galaxies in the high redshift universe, a large list of galaxies was first compiled. Utilizing the NASA Extragalactic Database (NED\footnote{\url{http://ned.ipac.caltech.edu/}}), we assembled a list of $\sim$14,000 spectroscopic redshifts for galaxies at $z \gtrsim 2.7$, removing sources that were flagged as gravitationally lensed objects. Occasionally NED will not flag galaxies with redshifts determined photometrically as PHOT, as in the case of \citet{2011ApJ...736...48R}, which had to be identified and removed manually. The redshift limit was chosen because it coincides with the effective onset of Lyman-Break Galaxies (LBGs), which were followed up spectroscopically \citep{2003ApJ...592..728S}. With the expectation that protoclusters would be observed as galaxy overdensities, we developed a simple algorithm that identifies candidate groups of galaxies within this high-redshift object list.

The spectroscopic galaxy list contains objects selected by a large range of criteria. Many galaxies in our list were part of spectroscopic surveys of \emph{Hubble/Chandra/Spitzer} Deep fields. One of these \citep{2010AnA...512A..12B} in the GOODS-S field was surveyed with the \emph{Very Large Telescope's} VIMOS grisms. The survey depth for galaxies with measured redshifts was to 24-25 AB magnitudes in $B$,$R$, and out to a redshift of $z<3.5$ with a redshift uncertainty of $\sigma_z \sim 0.0008$. Another large contributor to our spectroscopic galaxy list were LBGs that were followed up spectroscopically.  \citet{2003ApJ...592..728S} used rest-frame UV photometry to identify more than 2000 photometric, $z\sim 3$ candidates to a limiting magnitude of $R_{AB}= 25.5$ in fields totaling 0.38 square degrees. Almost 1000 redshifts were measured using Keck telescopes. Other sources of redshifts come from narrow band imaging around high redshift AGN that were followed up spectroscopically \citep{2007AnA...461..823V}, or targets selected based on combined narrow and broad band images to a typical limiting magnitude of $R_{AB}= 25.5$ \citep{2003AnA...407..147F,2009AnA...497..689G}.

It follows naturally that high redshift galaxies that are collapsing to form a nascent structure should exist as a large concentration of objects within a small volume. From the high redshift galaxy list,  groups of objects coincident on the sky were identified by searching within an angular search radius of 2 $\amin$ and a redshift range $\Delta z<0.03$ from a search galaxy. At the maximum redshift of our galaxy list ($z=4.05$), the volume probed is a cylinder with radius $4$ and depth of $\pm 20$ comoving Mpc from the estimated center of the distribution. This initial redshift depth was chosen in order not to miss candidate galaxy members, assuming that the protocluster dispersions could be $\sigma \gtrsim 0.01$ (600 km s$^{-1}$ at $z\sim4$).

The number of unique sources (those separated by at least $3''$) within the search volume were counted. Prospective groups with three or more galaxies within the aperture were added to an initial group list for further analysis. The `center' of a group in RA and DEC was determined iteratively by maximizing the number of galaxies within the sky aperture. Many prospective groups of galaxies that are in the final list of the Candidate Clusters and Protoclusters Catalog (CCPC) contained more than the minimum number of galaxies in this small volume. The heterogeneity of the sample (differing sky coverage and depths) does not permit us to make any estimate of group completeness. All we can state is that these candidates appear to be real, physical associations, and provide a lower limit to the number of galaxy members. Of the initial list of 14,000 galaxies, 603 sources are used in this work. Their references can be found in Table~\ref{tab:ref}.

\citet{2013ApJ...765..109D} has shown that by linking associations of only $3\le N\le 5 $ member galaxies within a few comoving Mpc in the zCOSMOS field at $z>1.8$, they could effectively identify nascent galaxy groups in the high redshift universe. Comparing the systems they discovered to mock galaxy light cones \citep{2007MNRAS.376....2K} from the Millennium simulation \citep{2005Natur.435..629S}, they could track the fate of these primordial systems. Their method, when coupled with the complete spectroscopic sampling of the zCOSMOS field, is expected to identify $\sim65\%$ of the protoclusters ($M_{z=0}>10^{14}$ $M_{\odot}$) within the survey volume \citep{2013ApJ...765..109D}. We adopt a similar, simple yet effective search criteria to identify initial regions of target protoclusters.

\subsection{Candidate Protocluster Criteria}

\begin{figure*}
\centering
\includegraphics[scale=0.5]{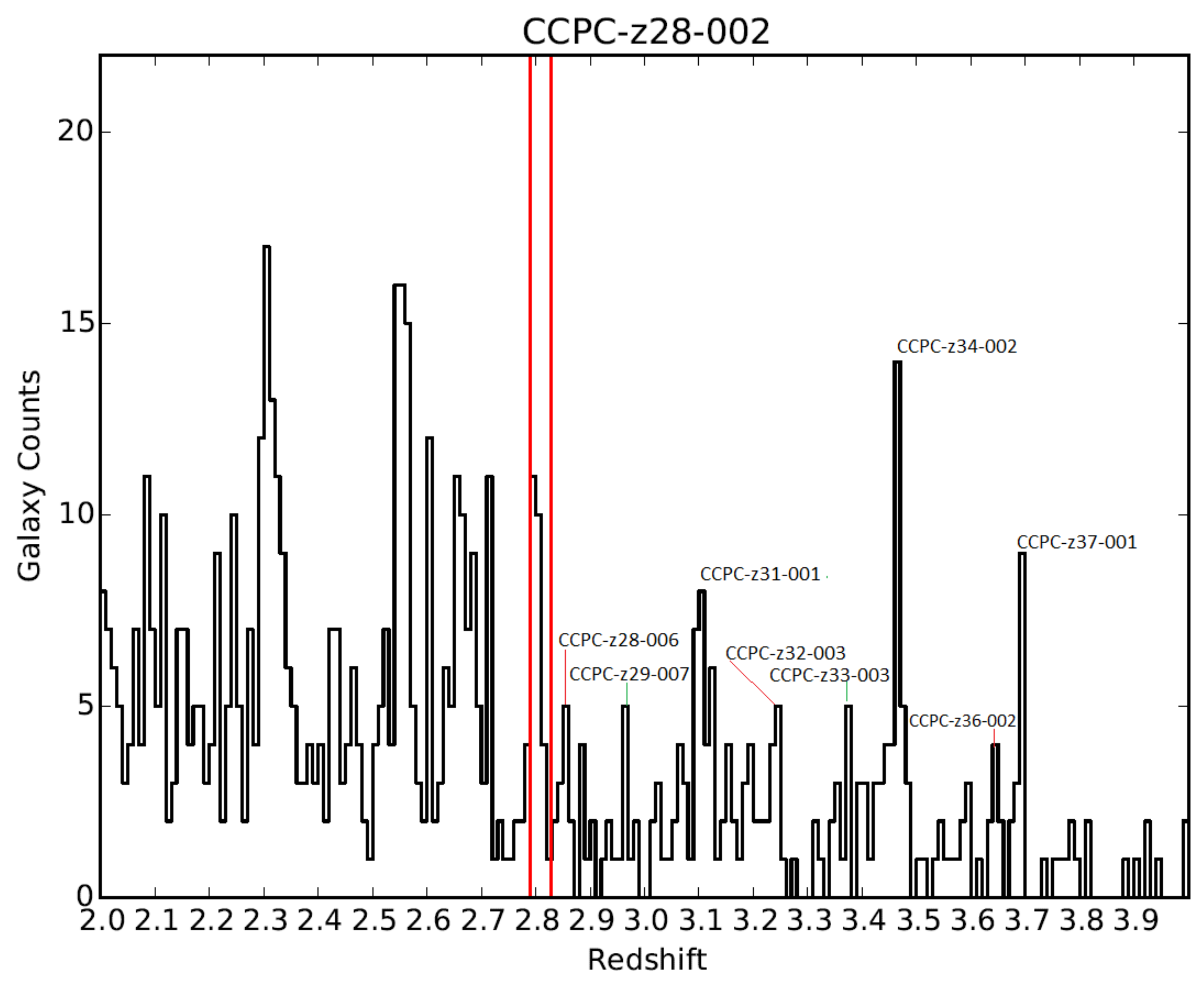}
\hfill
\caption{The number of galaxies as a function of redshift around CCPC-z28-002 at $z=2.82$. The red, vertical lines correspond to a $\Delta z$ that encompasses $\pm$20 cMpc along the line of sight from the mean redshift of the protocluster. Within this aperture of the CANDELS GOODS-S field, the partial peaks from other CCPC systems can be clearly identified. There are also strong peaks at redshifts $z\sim 2.3, 2.5,2.6$, but are beyond the scope of this paper.}
\label{fig:N_z_CANDELS}
\end{figure*}

Protoclusters are the extended, collapsing manifestations of present day clusters. As the Universe is expanding, the volume surveyed for a fixed area on the sky will be dependent on the redshift of the source. To consistently measure protoclusters iso-volumetrically, we have adopted the use of comoving Mpc (cMpc) throughout this study. This allows easy comparison to other observational studies that have also employed comoving volumes, as in \citet{rig14}. Analysis of large \LCDM simulations (e.g. Millennium Simulation) also typically utilize cMpc when tracking the growth and evolution of protoclusters at various epochs \citep{chi13,2015arXiv150608835M}. 

Generally, structures at high redshift have extended galaxy distributions of $R\gtrsim10$ comoving Mpc from the highest density region. This seemingly large volume has both theoretical and empirical bases. Analysis of the Millennium Simulation by \citet{chi13} found that the effective radii of the most massive protoclusters are typically $R_e\gtrsim 8$ cMpc at $z=3$, collapsing to $R_e\gtrsim 1$ cMpc at $z=0$. \citet{2015arXiv150608835M} also analyzed the Millennium data for protoclusters, and found that 90\% of the stellar mass of a $\gtrsim 10^{15}$ $h^{-1}\, M_{\odot}$ cluster is contained within a comoving radius of $\sim20$ $h^{-1}\, Mpc$ at $z=3$.

Observationally, \citet{rig14} investigated protocluster candidates around high redshift radio galaxies using \Herschel. They found that galaxy overdensities peak at a radius $\sim 6-7$ cMpc, with distributions that flattened out at $R\sim10$ cMpc (see their Fig 7). A protocluster identified by \citet{2008ApJ...678L..77P} at $z=2.75$ extended 20$\times$50 cMpc, while \citet{lee14} found an incredibly large structure  of three protoclusters within $72\times 72\times 25$ cMpc$^3$ at $z\sim 3.8$. These examples illustrate the importance of wide search radii in identifying these extended, perhaps filamentary structures.

To refine our initial list of galaxy groups into a catalog of protoclusters, we chose a search radius of 20 comoving Mpc from the approximate center of each initial groups' distribution. This corresponds to $11.3'$ and $9.9'$ on the sky at $z=2.75,3.7$, respectively. Many redshift surveys used in this work do not extend to $R=20$ $cMpc$, the expected size of the most massive protoclusters \citep{2015arXiv150608835M}, so this should be treated as the maximum volume probed. In redshift space, $\Delta z \sim 0.019$ corresponds to $\pm 20$ cMpc at $z=2.75$, which increases to $\Delta z \sim 0.026$ at $z=3.7$. More than 2/3 (435 out of 603 galaxies) of the redshifts have published uncertainties in NED, and the median uncertainty is $\sim$0.0008, which corresponds to an uncertainty of $\pm0.9$ comoving $Mpc$ at $z\sim3$ \citep{wri06}. This encompasses a 8\% uncertainty in the velocity dispersion, assuming a typical $\sigma_z\sim0.01$.%Corresponds to +/- 58 km/s at z=3? That is pretty good!

If within the search radius of 20 comoving Mpc on the sky (and associated length in $\Delta z$) there were 4 or more galaxies, which consist of a galaxy overdensity $\delta_{gal}>0.25$ (see Section 2.3), this candidate structure was assigned to the Candidate Cluster and Protocluster Catalog (hereafter CCPC). Although four objects does not necessarily constitute a protocluster, this is at least a group of bright galaxies within a relatively small space, and might turn out to be a richer group with a more complete redshift census within the search volume. 

\citet{chi13} illustrate that a galaxy overdensity of $\delta_{gal}\sim 1\pm1$ is representative of low mass, $\gtrsim 10^{14}$ $M_{\odot}$ clusters at $z=0$ within the Millennium simulation (see their Fig 6). This implies that even modest overdensities at high redshift may represent protoclusters.  We adopt $\delta_{gal} > 0.25$ as a working definition for candidate protoclusters, as these regions are more dense than the surrounding field while not excluding much of the overdensity distribution for the lowest mass protoclusters predicted by \citet{chi13}. Some of these objects may represent the tip of an iceberg, some may prove fictitious, while others may turn out to be filaments or void walls. At this juncture, all appear to be bona fide associations of galaxies at high redshift. Only two candidates have the minimum of 4 members. The median number of total members is 11 galaxies. Ten candidate protoclusters have over 20 members each.

Tables~\ref{tab:cat},~\ref{tab:cat2} contain the full list of candidates. The naming convention we have employed is CCPC-z, followed by the first two digits of the redshift (e.g. $z=2.9$ is indicated by 29-), and ends with a running index of objects in the respective redshift bin. Table~\ref{tab:cat} contains the sources that are at least $85\%$ likely to collapse into a cluster at $z=0$ based on the strength of their overdensity, while Table~\ref{tab:cat2} contains other overdensities that are less strong. The basis of these probabilities will be explained in the following section. The source of the individual galaxy redshift measurements for each candidate cluster can be found in Table~\ref{tab:ref}.

The search radius of $R=20$ comoving Mpc is the effective size of the most massive protoclusters, with smaller systems being much less extended \citep{chi13,2015arXiv150608835M}. In the few instances in which the survey width encompasses $R\gtrsim20$ cMpc (e.g. CANDELS GOODS-S), it is possible that we may be identifying two less massive protoclusters instead of a single, large system. However, by reducing our search radius to 10 comoving Mpc, only a few (or zero) galaxies were removed from many candidate protoclusters. Consequently, most objects appear to be more centrally concentrated than required by our search criterion, primarily because the fields-of-view (FOV) of most surveys are smaller than the maximum surface area we are probing. Upwards of 1/3 of a candidate's galaxies are found between $10<R<20$ comoving Mpc for 12 CCPC objects, which are marked by a `$*$' next to their galaxy number in Tables~\ref{tab:cat},~\ref{tab:cat2}. The implications of this are discussed in \secref~\ref{sec:dis}.

Each candidate protocluster has an associated sky position plot, showing the distribution of members in RA/DEC within 20 cMpc (black points within the outer red circle), as well as field galaxies (green $\times$'s) along the line of sight. The field galaxies serve as an illustration of the data footprint that contains the protocluster, with clear survey boundaries seen in many cases. The sky plots of the candidate structures illustrate the radial distributions of galaxies within a system. Some protoclusters are very extended (e.g. CCPC-z28-002 shown in Fig~\ref{fig:CCPC-z27-001}), while others are strongly concentrated (CCPC-z29-002) within the inner red ring (corresponding to $R=10$ $cMpc$), with the outer red circle marking the search radius of 20 comoving Mpc for comparison. The differences in these systems, as noted earlier, is primarily dependent on survey width, and has little to do with the protocluster itself. 

A $N(z)$ distribution (number of spectroscopic galaxies in redshift bins from $2<z<4$ within the search radius) for each protocluster is also located in the Appendix. The $N(z)$ plots have a large amount of variation as well, particularly with respect to differences in survey depth in our high redshift galaxy catalog. For example, CCPC-z28-005 shows an unmistakable peak of galaxies at the cluster redshift, with no other sources detected at other distances. In contrast, CCPC-z28-002, located in CANDELS GOODS-S field, contains a multitude of galaxies with peaks corresponding to 8 other CCPC members, as shown in Fig~\ref{fig:N_z_CANDELS}.

One distinct difference between this CCPC sample and the samples of \citet{2007AnA...461..823V,gal12,wyl13,rig14,wyl14} is that our protoclusters candidates were not initially targeted based on the presence of a high redshift radio galaxy (HzRG). Rather, we identify spatial coincidences among spectroscopically confirmed galaxies with $z>2.74$.  Some candidate structures in our sample contain radio galaxies, and our search methodology has recovered a number of previously identified protoclusters that were first identified using HzRGs as signposts (see Tables~\ref{tab:cat},~\ref{tab:cat2}). That we recover these previously identified structures is encouraging. 

\subsection{Candidate Protocluster Overdensities}

To statistically measure the significance of these structures, we have estimated the galaxy overdensities ($\delta_{gal}$) of each CCPC member. We computed the galaxy overdensity for each protocluster as $\delta_{gal} = (n_{proto}/n_{field})-1$, where the number density $(n_{proto})$ is the number of galaxies along the line of sight within $z\pm \sigma$  of the redshift of the structure. The $n_{field}$ density was determined by taking $z_{proto}\pm 0.15$, sans the region around the overdensity of $\pm \sigma$. A maximum field redshift range of $dz=0.15$ (excluding the overdensity region) was chosen to adequately sample the field. Increasing the protocluster overdensity redshift range to encompass the full $\pm20$ $cMpc$ (instead of $z\pm \sigma$) typically increases the value of $\delta_{gal}$, but at the expense of decreasing the field galaxies in some cases. When estimating the overdensity, we always assign galaxies to the field in cases of doubt. Some of these are likely members residing in the outskirts, so $\delta_{gal}$ is a conservative estimate of the overdensity. All listed uncertainties are Poissonian.

For the relatively modest overdensity of CCPC-z33-003, there are 6 galaxies found within the redshift space of $\Delta z=\sigma=0.012$, giving a $n_{proto} =\frac{6}{2(0.012)}$ along the length ($dz=0.024$) of the probed volume. Within the same aperture, there are 24 galaxies along the line of sight of $\Delta z=0.284$, so the field length is $\Delta-2(\sigma)=0.284-0.024=0.26$. From this, a field density of $n_{field} = \frac{24}{0.26}$ is estimated, and $\delta_{gal} = 1.70$. All values of $\delta_{gal}$ are listed in Tables~\ref{tab:cat},~\ref{tab:cat2}. The field line of sight is limited to the $\Delta z$ range where field galaxies are identified. For example, CCPC-z28-004 was previously identified using narrow-band filters by \citet{2007AnA...461..823V}, so understandably has a narrower redshift distribution (as seen in its $N(z)$ plot Fig~\ref{fig:CCPC-z28-004}) than many other systems. We identify 17 galaxies along the short length of $\Delta z$=0.028 outside of the overdensity region, which defines our field measurement. This serves as an upper limit on the field in that some of these `field' galaxies may themselves be infalling future cluster members.

In some instances, $\delta_{gal}\sim 0$ as extended field density estimates roughly matched the candidate structure densities. This is most likely the result of the relatively small number of galaxies in these structures, combined with possible intervening (sub)structure. Limiting the field galaxies to the same sky area as the structure galaxies transforms these relatively small underdensities into overdensities. In the interesting case of CCPC-z27-004, the 
edge of the field length intersects CCPC-z29-004. By limiting the field calculation to $\Delta z\pm 0.05$, the other structure is averted and a positive galaxy overdensity becomes apparent. 

It should be noted that choosing a different $\Delta z$ can also change the value of $\delta_{gal}$.  This highlights the difficulty in defining a field number density along the line of sight, as intervening (sub)structures can boost the field counts significantly. We have purposefully avoided fine-tuning the galaxy oversensities in a way to produce larger overdensities. Most importantly, we urge caution in drawing conclusions from these galaxy overdensities, as the definition of the field, coupled with the selection effects of the sample, can alter the significance of the structure.

Seven structures had 2 or fewer field counts in calculating their overdensity, with CCPC-z27-002 having none at all. In these instances, we have manually added 7 field galaxies to their number density to provide a lower limit to the actual $\delta_{gal}$. Some of these candidate protoclusters, like CCPC-z30-001 and CCPC-z31-005, have incredibly large calculated galaxy overdensities ($\delta_{gal}>$10). These are the result of an overdensity coupled with a very weak field density measurement along the line of sight as a selection effect of narrowband imaging, which targets galaxies at a specific redshift \citep{2007AnA...461..823V}. As \citet{2015arXiv150608835M} point out, this is one of the dangers in calculating galaxy overdensities using narrowband imaging, as field counts can be under-represented. %Even with the field density boost, many of these $\delta_{gal}$ are larger than those with non-artificial field values.

\citet{chi13} calculated the expected galaxy overdensities for protoclusters identified within the Millennium Simulation, combined with the semi-analytic model of \citet{2011MNRAS.413..101G}. For the lowest mass protoclusters ($M_{z=0}<3\times10^{14}$ $M_{\odot}$) in the redshift range of $2<z<5$, in cubes with sides of 25 cMpc, a $\delta_{gal}\sim 1\pm 1$ is typical for galaxies with a $SFR > 1$ $M_{\odot}$. This overdensity increases to $\delta_{gal}\sim3$ for the most massive ($M_{z=0}>10^{15}$ $M_{\odot}$) cluster progenitors. Table 5 in \citet{chi13} contains a list of more than 20 protoclusters found in the literature, with measured $\delta_{gal}$ ranging from $0.7_{-0.6}^{0.8}$ \citep{2007AnA...461..823V} to 16$\pm7$ \citep{2014ApJ...792...15T}. The median value of galaxy overdensity within the CCPC is $\delta_{gal}\sim 2$, which agrees well with values from the literature and theoretical predictions \citep{chi13}. 

In Tables~\ref{tab:cat},~\ref{tab:cat2} we provide a conservative estimate of the probability that a given candidate structure will collapse into a cluster at $z=0$ based on its $\delta_{gal}$ value. These estimates are taken from the z = 3 case in Figure 8 of Chiang et al. (2013), which plots the fraction of overdensities which will form a $M>10^{14}$ $M_{\odot}$ halo at the present day. We made the conservative assumptions that our protoclusters onservative assumptions that the galaxies identified as cluster members are the most biased tracers of mass (the brightest galaxies), the overdensity volumes are $[15$ $cMpc]^3$, and that none of the field galaxies used to calculate our overdensity values will become part of the cluster. If some of our member galaxies are less biased tracers (e.g. $M_* < 10^{10}$ $M_{\odot}$), the probability can increase by more than 20$\%$ for a given $\delta_{gal}$ \citep{chi13}. If the box volume is increased to $[25$ $cMpc]^3$, a $\delta_{gal}=0.86$ has a $50\%$ of collapsing into a cluster, while for the smaller box volume of $[15$ $cMpc]^3$, a $\delta_{gal}=3.43$ is required \citep{chi13}. Systems not fated to be clusters can end up as groups of galaxies, or may simply be false positives.

We list our candidate structures in Tables~\ref{tab:cat} and ~\ref{tab:cat2}.
Table~\ref{tab:cat} lists the most probable protoclusters, containing
12 candidates that have at least an 
$85\%$ chance of collapsing
into a cluster by $z=0$ according to \citet{chi13}.
Lower probability structures 
are listed in Table 2. Six out of the twelve protoclusters in Table~\ref{tab:cat}
have been previously identified, while only 4
of 31 candidates in Table~\ref{tab:cat2} were  found in the literature. 
Typically, overdensities of $\delta_{gal} > 7$ are expected
to be protoclusters at this redshift.

It is important to remember that these probabilities are based on analysis of the Millennium simulation by \citet{chi13}. There is no guarantee that this represents the Universe in which we reside. Therefore, these probabilities should be treated as a mere guide.

\subsection{Mass Estimates}

Using the positions of galaxies within each CCPC system and their respective redshift dispersions $\sigma$, we have provided a crude mass estimate for each structure in Table~\ref{tab:mass}. We estimate the mean harmonic separation as $R_{hms}\approx 1.25 R_e$, where $R_e$ is the radius from the approximate center of the galaxy distribution that contains half of all member galaxies in the system. The $R_{hms}$ value for each CCPC protocluster is recorded in Table~\ref{tab:mass}. Another variation of an $R_{hms}$ estimator using Equation 2 from \citet{car96} with weights set to unity %R_hms3/R_e is about 2, R_hms3/R_hms ~4
provides values roughly a factor of 2 larger, so the $R_e$ approximation is conservative. The virial mass is then estimated by

%M_hms3/M_e is about 4, M_hms3/M_hms ~9, and M_e/M_hms ~2
\begin{equation}
M = \frac{2\sigma^2}{G} R_{hms}.
\end{equation}

There is no guarantee that these systems are virialized, symmetric structures, and so we stress that these mass estimates are merely indicative. 

Although some systems have seemingly reasonable masses for a present-day cluster ($\gtrsim 10^{14}$-$10^{15}$ $M_{\odot}$), these estimates are much larger than theoretical predictions for a $z=3$ protocluster. Figure 2 in \citet{chi13} tracks the halo mass of protoclusters as a function of redshift within the Millennium Simulation. At $z=3$, the main halo of a protocluster of mass $\sim3\times10^{13}$ $M_{\odot}$ will grow into a $10^{15}$ $M_{\odot}$ cluster at $z=0$. Smaller clusters ($\gtrsim10^{14}$ $M_{\odot}$ at $z=0$) will be in the range $\sim5\times10^{12}$-$10^{13}$ $M_{\odot}$ at $z=3$. Almost all of the mass estimates for our protoclusters exceed $10^{14}$ $M_{\odot}$, with 13 structures above  $M\gtrsim10^{15}$ $M_{\odot}$. Either these CCPC systems are orders of magnitude larger than \LCDM predicts at this epoch, or more probably, the crude mass estimate we have employed is a poor representation of the physical nature of these early structures. Only 3 systems (CCPC-z32-002, CCPC-z35-001, CCPC-z36-001) have reasonable mass estimates of $<3\times10^{13}$ $M_{\odot}$, and each has 5 or fewer member galaxies. \citet{chi13} find only $2\%$ of almost 3000 clusters have $M_{z=0}>10^{15}$ $M_{\odot}$ within the Millennium Simulation's comoving box (sides of 500 $h^{-1}$ $Mpc$). With the significant number of  $M_{z\sim3}^{vir}>10^{15}$ $M_{\odot}$ protoclusters in the CCPC, this suggests that the virial masses are overestimates, or massive clusters emerge earlier than anticipated.

However, comparing the results of simulations to actual observables can be problematic. For instance, \citet{chi13} report that a $z=0$, $M>10^{15}$ $M_{\odot}$ protocluster in the Millennium Simulation would have a velocity dispersion along the line of sight of $400\pm 60$ km s$^{-1}$ at $z=3$. Many members in the CCPC are well above this velocity dispersion, with some at $\gtrsim 1000$ km s$^{-1}$. Converting the radius from \citet{chi13} for the progenitor of a $10^{15}$ $M_{\odot}$ cluster from cMpc into physical units, and setting $\sigma_v = 400$ km s$^{-1}$, Equation 2 provides a mass estimate of $4\times 10^{14}$ $M_{\odot}$. This `observable' mass is an order of magnitude larger than the mass computed by the simulation for the same protocluster ($\sim3\times10^{13}$ $M_{\odot}$).

As an illustrative exercise, doubling the velocity dispersion to $\sigma_v = 800$ km s$^{-1}$ (more typical of CCPC systems), we recover a comparable mass estimate ($\sim10^{15}$$M_{\odot}$) to those listed in Table~\ref{tab:mass}, as expected. This increased velocity dispersion is not unique to our analysis. As pointed out in \citet{2015arXiv150503877C}, other known protoclusters have line of sight velocities of $\sigma_v = 900$ km s$^{-1}$ at $z\sim3$. The larger line of sight dispersions of observed $\sigma_z$ values than the theoretical predictions boosts our structure mass estimates by orders of magnitude over their simulated sibling protoclusters. It is therefore necessary to take great care to measure apples with apples when comparing observations to simulations.

Taken at face value, the masses we compute are problematic for \LCDM. \citet{mor11} predict that no clusters should exist with $M > 6 \times 10^{14}$ $h^{-1}$ $M_{\sun}$ at $z = 2$, and they should be smaller still at higher redshift.  The CCPC contains 14 clusters that exceed this virial mass, all at $z > 2.7$. Galaxies at high redshift also appear to have a similar issue, in that the number density of massive halos is $\sim10^2$ larger than predicted by \LCDM \citep{2015arXiv150601377S}. We note in passing that clusters this large are predicted to exist at redshifts of 2-3 by MOND \citep{1998MNRAS.296.1009S,2002MNRAS.331..909N,2004ApJ...611...26M,2013ApJ...772...10K,2015CaJPh..93..250M}.

Our mass estimates assume that the objects we identify as cluster candidates are virialized systems.  This may not yet be the case at these redshifts, as they are still in the process of collapse.  We therefore urge great caution in interpreting these results.

Another estimator of protocluster mass is the use of galaxy overdensities ($\delta_{gal}$) as biased tracers of mass overdensity ($\delta_{m}$), as utilized by \citet{1998ApJ...492..428S,2007AnA...461..823V}. The equation adopted is
\begin{equation}
M = \rho_{crit,z}V(1+\delta_{m})
\end{equation}
where $\delta_{m} = \delta_{gal}/b$, with $b$ existing as a bias parameter, with values assumed to be $\sim 3-6$  \citep{1998ApJ...492..428S,2007AnA...461..823V}. Take CCPC-z33-003 as an example again, for which we computed a $\delta_{gal}$= 1.59 in Section 2.3. At $z\sim3$, the comoving critical density is $\rho_{crit}=4.2\times10^{10}$ $M_{\odot}$ $cMpc^{-3}$ within our cosmology.  Assuming a bias parameter of $b=3$ and a computed volume of approximately 3300 
 $cMpc^{3}$, the implied $z=0$ mass that this structure may collapse to is $2.2\times10^{14}$ $M_{\odot}$. The virial mass estimate for this system was $14.9\times10^{14}$ $M_{\odot}$. For the entire sample of the CCPC, the median mass estimate is $8 \times10^{13}$ $M_{\odot}$.
 
There are a number of systematic uncertainties within this calculation. The bias parameter $b$ is typically assumed to be in the range 3-6, although this value has never been directly measured. As discussed previously, the value of $\delta_{gal}$ (and therefore $\delta_{m}$) is strongly dependent on the definition of field galaxies, or lack thereof. One of the largest contributors to the mass estimate is the volume $V$ that is expected to collapse by $z=0$, which is calculated using the minimum and maximum RA/DEC values in a rectangular box along the line of sight $\Delta z=\sigma$. If the width of the field of observations is small, the calculated volume of a collapsing protocluster will be a lower limit. For instance, recomputing the overdensity volume of $\Delta z_{proto}=20$ $cMpc$ and not $\sigma$,  and a field width of $\Delta z=0.05$ instead of the $\pm$0.15 adopted in this work, the median $\delta_{gal}$ value is unchanged from this work, although with greater scatter. However, the increase in volume of the candidate structures boost the overdensity mass from $8 \times10^{13}$ $M_{\odot}$ to  $4 \times10^{14}$ $M_{\odot}$. 

There appears to be no correlation between the virial mass estimates and the masses inferred from the overdensities. The two estimates are physically distinct, as the former ostensibly measures the current system/halo mass, while the latter is an estimated collapsed mass. Yet some link should exist between them, as massive main halos generally correlate with massive $z=0$ clusters within simulations \citep{chi13,2015arXiv150608835M}. The fact that there is no connection implies that systematic uncertainties dominate, and that one or both of these methods are inherently flawed. As a result, these mass estimates should be treated with caution.

It should be noted that the systems identified as having the highest probability (85$\%$ or larger) of collapsing into a cluster in Table~\ref{tab:cat} based on the strength of their overdensity, typically have reasonable mass estimates (both virial and $\delta_{m}$) of $\sim10^{15}$ $M_{\odot}$ or less in Table~\ref{tab:mass}. A notable exception is CCPC-z30-003, but this is the richest structure in the CCPC and was previously identified by \citet{1998ApJ...492..428S}. Encouragingly, the candidates with lowest probability ($<10\%$) of collapsing into a cluster often have the most anomalously large virial mass estimates of $> \times10^{15}$ $M_{\odot}$.

\section{Discussion} \label{sec:dis}

\begin{figure*}
\centering
\begin{subfigure}
\centering
\includegraphics[scale=0.5]{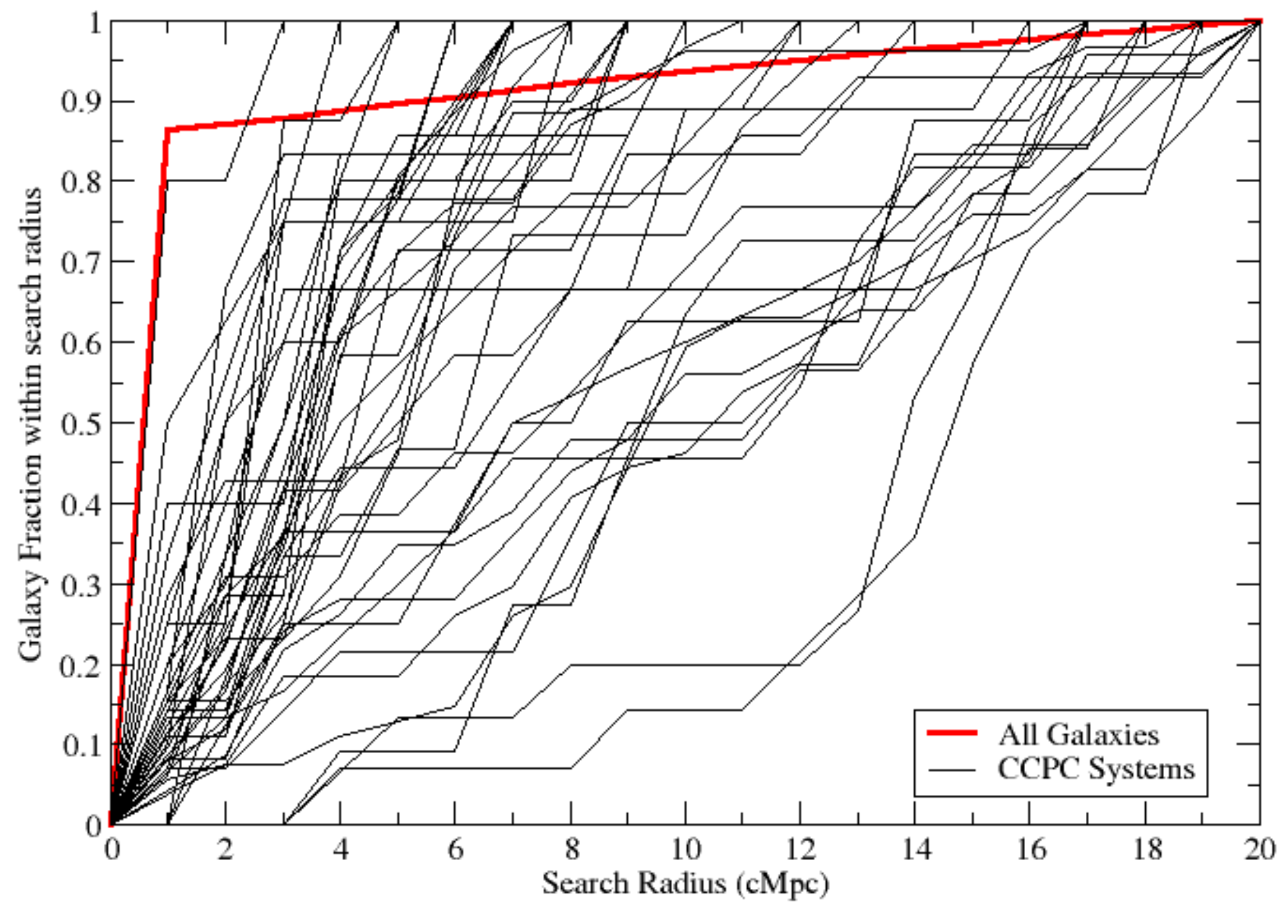}
\end{subfigure}
\hfill
\begin{subfigure}
\centering
\includegraphics[scale=0.5]{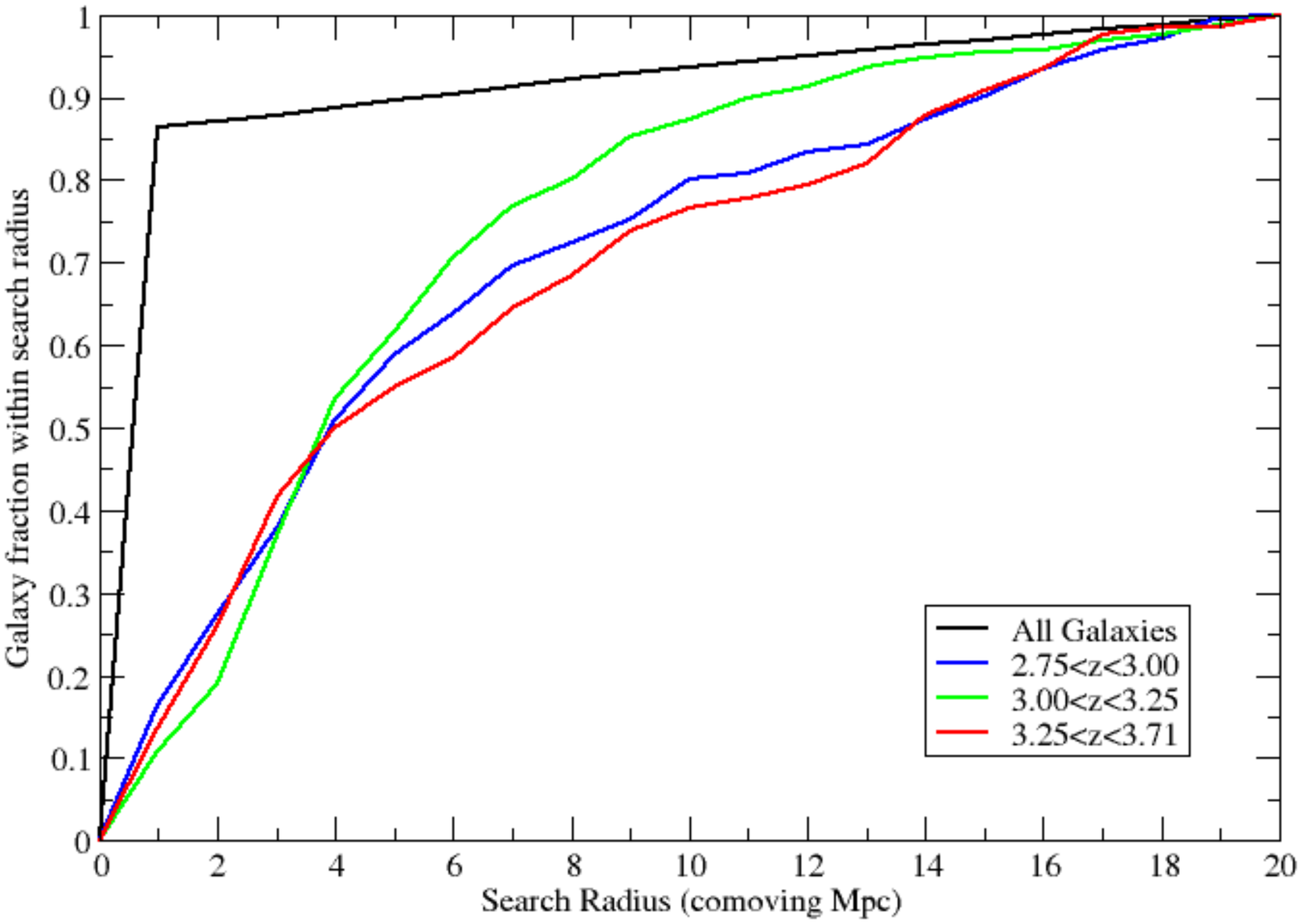}%Did not change
\end{subfigure}
\hfill
\caption{$Top:$ The fractional number of galaxies as a function of radius in individual CCPC structures. There is a  large spatial variability, as some objects are very compact while others are diffuse, extended systems. The red line represents the mean distribution for 14000 galaxies and any companions found within $R<20$ comoving Mpc (see \secref~\ref{sec:dis}). The candidate protoclusters (Black Lines) clearly have more neighbors than is typical in the field, as well as different distributions.  $Bottom:$ Mean distribution of the fraction of total members in the CCPC as a function of search radius in comoving units. The black line distribution represents the mean All Galaxies sample. The CCPC catalog is divided into three redshift bins, which show no significant difference in the mean distributions, but there is significant scatter between CCPC members within each bin (as seen in the top figure). It may appear that some evolution is taking place from $2.75<z<3.00$ (blue line) to $3.00<z<3.25$ (green line), in that the more distant systems are more centrally concentrated on average. However, the trend is reversed when looking at the mean distributions of $3.00<z<3.25$ to $3.25<z<3.71$ (red line).}
\label{fig:dist}
\end{figure*}

\begin{figure*}
\centering
\includegraphics[scale=0.5]{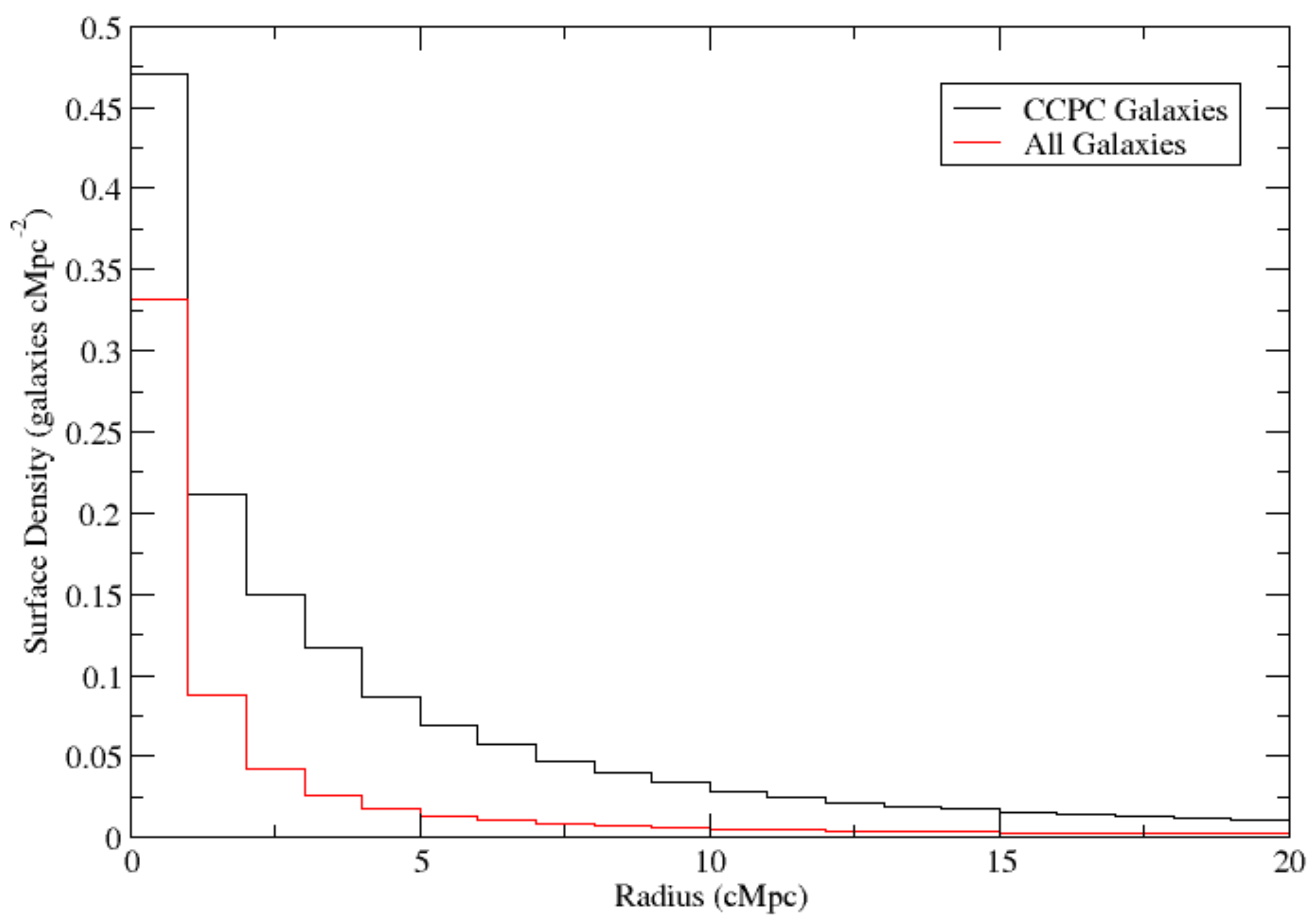}%Did not change from v6 really.
\hfill
\caption{The mean surface density profile of All Galaxies (14,000 spectroscopic sources with $z>2.75$) that were used in the initial list to identify structures is shown in red, with the mean density profile of the CCPC sources shown in black. Most galaxies in the All Galaxies list have no companions within 20 $cMpc$, although galaxies within the CCPC structures are included in All Galaxies. The CCPC surface density profile has more galaxies within the inner R=1 $cMpc$ bin, and has a more gradual decrease in density than the distribution of galaxies in general, as expected for galaxy overdensities.}
\label{fig:density}
\end{figure*}

\subsection{Confirmation tests of structure}

To confirm that these CCPC targets are indeed structures, we have used our initial list of $\sim$14,000 $z > 2.75$ objects as a benchmark sample. We searched around each galaxy (including galaxies identified within CCPC systems), and plotted any companions that were within the same comoving volume as the candidate protoclusters (search radius of 20 $cMpc$, $\Delta z$ corresponding to $\pm20$ $cMpc$). Many of these galaxies had no spectroscopic companions within the volume. We plotted the mean distribution of these galaxies as a function of comoving radius from the central source (in many cases the sole source) in Fig~\ref{fig:dist} as the red line in the top panel, and the black line in the bottom plot. 

For a simple metric of the distribution of galaxies within protoclusters at $z>2.75$, we have added the distribution of each individual CCPC system in the top panel of Fig~\ref{fig:dist} to the distribution of all galaxies. The distributions can be seen in the striking variability in concentration of these early mass overdensities. Some structures are fully contained within a radius of 5 comoving Mpc, while others do not even have 40\% of their total members by $R\gtrsim$10 cMpc. In our analysis, these variations do not appear to be dependent of the redshift of the system, as there are both extended and compact structures at the redshift extrema of the catalog. 

As a further test, we computed the mean fraction of all CCPC members within the circular 1 Mpc annuli from the central galaxy out to 20 comoving Mpc in the bottom panel of Fig~\ref{fig:dist}. 
Separating the candidates into three redshift bins $2.75<z<3.00$ (19 objects, blue line in Fig~\ref{fig:dist}), $3.00<z<3.25$ (12 objects, green line), and $3.25<z<3.7$ (12 objects, red line) shows there is no significant distribution difference between the bins, despite the $>0.5$ Gyr time difference between $z\sim3.7\rightarrow2.75$. It appears that although the mean distribution generally does not vary as a function of redshift, individual protoclusters have a wide range of spatial differences for their members. If the total sample is limited to members within $R\sim10$ comoving Mpc, the distributions still show no significant difference as a function of redshift, but have considerably less scatter, as expected. This is a possible indication that in general, protoclusters at these epochs have similar assembly histories. There also does not seem to be a significant trend in the number of galaxy members as the redshift decreases in the CCPC, with a similar median galaxy membership (13) occurring in structures between $3.00<z<3.25$ when compared to the median over all candidates $2.74<z<3.71$ (11 galaxies). Median membership in the bins $2.74<z<3.00$ and $3.25<z<3.71$ are 10 and 9 galaxies, respectively. 

Fig~\ref{fig:density} illustrates the difference in surface density of the structures in the CCPC (black line) compared with all of the galaxies in the initial list of $\sim$14,000 spectroscopic galaxies (red line). The red distribution of galaxies is consistent with a single object with no other companions found within $R_{search}=1$ $cMpc$ ($\Sigma = \frac{1 \, gal}{\pi(1 \, cMpc)^2}=0.33 \, cMpc^{-2}$ in the first bin), and rapidly decreasing as the radius increases. CCPC galaxies have a larger central concentration ($0.47 \, cMpc^{-2}$) within 1 cMpc radius, and a more gradual decrease in surface density. A two-sample Kolmogrov-Smirnov test gives a value of $KS=0.69$, suggesting that these cumulative densities are distinct distributions. In $log-log$ space, the slope of all galaxies is steeper (-1.65) compared to the -1.5 slope for CCPC galaxies. Interestingly, outside of $R\sim5$ $cMpc$, the slopes are both $\sim - 1.5$, suggesting a possible characteristic radius at which the galaxy overdensity and the field begin to merge. 

We note that Figures ~\ref{fig:dist},~\ref{fig:density} should be only used as tests of structure in comparison to the field galaxy distributions, and should not interpreted as the expected radial profile of protocluster systems. These data were taken from a number of different surveys of various widths and selection techniques. Not all protoclusters are centered within a survey's FOV, nor do they necessarily encompass the full search radius of $R=20$ $cMpc$, as can be seen in the sky plot of each CCPC member within the Appendix. Thus, the cumulative distribution will be strongly dependent on the survey characteristics and the protoclusters position within the survey's width.

A further test that a candidate protocluster's galaxies are not merely coincident on the sky, but rather constitute physically associated structures can be found by plotting the number of galaxies at each redshift along the same line of sight as the structure in a $N(z)$ plot. Protoclusters, as an overdensity of galaxies, should be visible as a peak in counts within a small redshift bin. The Appendix contains a $N(z)$ plot for each CCPC member in the right panel (Fig~\ref{fig:CCPC-z27-001}). For many of these protocluster systems, the overdensity peak is unmistakable from the background distribution (like CCPC-z30-003), or there are no other intervening sources along the line of sight, as in CCPC-z28-005. In deeper surveys, the distinction between structure and field galaxies becomes harder to identify, as peaks can be smaller relative to the continuum. Fig~\ref{fig:N_z_CANDELS} shows the $N(z)$ plot for CCPC-z28-002 in the CANDELS GOODS-S survey. The number of galaxies in each redshift bin is larger than in other pointings, presumably because this field has been surveyed more deeply.

As the CANDELS GOODS-S field is essentially a pencil-beam survey between the redshift values $2.74<z<3.71$ (roughly $50\times50\times900$ $cMpc$ in our assumed cosmology), we can compare the number and richness of our candidate structures within this pointing to the expectations of stochastic alignments of galaxies arising from Poissonian fluctuations in a smooth density field. \citet{she01} computed a toy-model of galaxy clustering with a density distribution of galaxies $n$, in which the expected number of systems $N$ with $M$ galaxy members and linking length $l$ can be expected for a sample size of $N_G$ objects along the line of sight via
\begin{equation}
N_M =N_g e^{-2nl} [1-e^{-nl}]^{M-1}. 
\end{equation}
If the Poissonian expectation of structure is lower than the recovered number of structures, this is an indication that physical structures exist in these data, are not mere chance alignments of galaxies. 

To utilize this simple model, the linking length $l$ must be computed. Using the auto-correlation function
\begin{equation}
\xi (l) = \frac{N_{DD}}{N_{DR}} \frac{n_R}{n_D} -1
\end{equation}
one measures the excess Data-Data pairs of galaxies $N_{DD}$ over the number of Data-Random pairs ($N_{DR}$) of galaxies in the interval $l\pm \Delta l$ \citep{dav83}. Within the GOODS-S field, $\xi$ was found to be $\gtrsim 1$  at $l<1.78$ $cMpc$, effectively measuring the separation $l$ at which clustering is strongest. 

A one-dimensional, Friends-of-Friends \citep{huc82,pre82} algorithm was constructed to identify structures of galaxies separated by a linking length $l <1.78$ $cMpc$ within the CANDELS GOODS-S catalog. We then compared the number of structures with $M$ members ($N_M$) with the expectation value using the toy model of \citet{she01}. Within the data set, the number of systems with $4<M<30$ galaxies identified using the Friends-of-Friends algorithm always exceeded the expected, Poissonian value in our analysis. For instance, there are 7 structures identified with $M\ge9$ galaxies along the line of sight, where only one such structure should exist by chance. As a check of the \citet{she01} model, we ran a Monte Carlo simulation of 500 randomized fields, identifying galaxy systems using our Friends-of-Friends algorithm. The Monte Carlo results provides a slightly smaller estimate of the number of structures at almost all values of $M$, but are generally similar. For example, for $M=5$, the Monte Carlo predicts 7.6 systems to be identified, where  $N_M=8.6$ in the toy model. At $M=9$, where $N_M\sim1$, the Monte Carlo predicts only 0.1 structures to exist, on average. 

It is important to emphasize that this analysis is specific to this single field, as the parameters $l,n$, and $N_G$ are unique to this survey. To apply the results of the toy model to a another field (or the CCPC as a whole) with a different selection function would not be a physically-motivated comparison.

Based on the combination of (1) positive galaxy overdensities ($\delta_{gal}$ listed in Tables~\ref{tab:cat},~\ref{tab:cat2}) within the protocluster volume, (2) radial distribution profiles of CCPC structures that look significantly different from the stochastic positions of field galaxies (Figs~\ref{fig:dist},~\ref{fig:density}), (3) redshift distributions of galaxies along the line of sight that show up as strong peaks in the $N(z)$ plots (Fig~\ref{fig:CCPC-z27-001}), (4) the excess number of galaxy groups over Poissonian expectations in the deep CANDELS GOODS-S survey, and (5) that we recover a number of previously identified, rare protoclusters in the literature suggests that the galaxy associations identified in the CCPC are strong candidate protoclusters.

\subsection{Extended Protoclusters}

As mentioned in \secref~\ref{sec:ccpc}, there appears to be two populations of structures within the CCPC. While many protocluster candidates add few members outside a 10 comoving Mpc radius, 12 CCPCs have a $1/3$ or more of member galaxies at $10<R<20$ comoving Mpc from the center of the distribution. In Tables~\ref{tab:cat},~\ref{tab:cat2}, these extended sources are indicated by a `$*$' next to their galaxy counts. By examining the galaxy position plots of each candidate protocluster in Fig~\ref{fig:CCPC-z27-001} in the Appendix of this paper, the differences between the extended and centrally concentrated structures are easily distinguishable. %These outer galaxies may not even be bound to the system! A galaxy moving at 500 km/s in 11.5 Gyr only moves about 5 Mpc

These extended objects (CCPC-z27-004, CCPC-z28-002, CCPC-z28-006, CCPC-z29-003, CCPC-z29-007, CCPC-z30-003, CCPC-z31-001, CCPC-z32-003, CCPC-z33-003, CCPC-z34-002, CCPC-z36-002, CCPC-z37-001) tend to have larger numbers of members overall, and the richest candidate (CCPC-z30-003) is among their number. They span almost the entire range of redshifts in the sample. This may be an indication that in this epoch of the universe, protoclusters are in various stages of assembly, such that some structures are centrally concentrated while others are condensing more slowly. Protoclusters identified by \citet{2015arXiv150608835M} within the Millennium Simulation are consistent with there being a variety of evolutionary stages that exist for structures at these redshifts, with a minority ($\sim10\%$) of protoclusters having their mass concentrated in a single, dominant halo at $z>2$.

\citet{chi13} mapped the growth of protoclusters from $0<z<5$ within the Millennium Simulation to estimate observable features in future surveys. In the context of this work, the effective radius ($R_e$), within which $\sim65\%$ of the mass of bound halos are found, is particularly interesting. Regardless of redshift, the greater the mass of the system at $z=0$, the larger $R_e$ was measured to be, with the expected result that at redshifts of $z\sim5$, the radius was largest (see their Fig 2). For instance, at $z\sim3$, the effective radius is $>8$ comoving Mpc for a Coma-like $M_{(z=0)} = 10^{15}$ $M_{\odot}$ halo, while it is $R_e<5$ cMpc for a halo $M_{(z=0)} <3\times 10^{14}$ $M_{\odot}$.  It is conceivable that these extended sources could also be more massive systems. As the extended sources also typically have the largest number of galaxies within our catalog, this is a plausible explanation.%Prescotts system is also quite large at 20 x 50 cMpc.

It is important to note that a significant number of these extended objects are in the extensively studied CANDELS (PI: Faber, Ferguson) $\sim$170 arcmin$^2$ GOODS-S field, and two others were previously identified as protocluster candidates \citep{2010ApJ...716L.200B,2012ApJ...750..116U}. As these fields have greater survey depth and width, these extended objects most likely constitute a selection effect and are not structurally different than the other CCPC objects. With a wider, deeper search around the non-extended sources, it is possible that their protocluster galaxy counts could similarly grow. Of the five CCPC members that were found in the Hubble Ultra Deep Field \citep{2006AJ....132.1729B}, only one of these structures have significantly extended populations despite the impressive survey depth, while all of the CANDELS protoclusters are extended. As the Ultra Deep Field has only a $3.3'\times3.3'$, the survey width is insufficient to detect such diffuse structures. The search radius corresponding to 20 $cMpc$ at $z=3.7$ is 10$'$. 

\section{Summary} \label{sec:sum}

We present a catalog of 43 candidate protoclusters in the redshift range of $2.74<z<3.71$.  These structures were identified using published position and spectroscopic measurements within a comoving search radius of 20 cMpc. Prospective structures were initially identified by flagging groups of three or more galaxies with a $\Delta z <0.03$ and within a 2 $\amin$ radius of each other. This list was later refined by requiring at least 4 spectroscopic sources within a search radius of 20 comoving Mpc, and a galaxy overdensity of $\delta_{gal}>0.25$. The median number of galaxies in each candidate protocluster is 11, while the maximum number of galaxies is 54. There appears to be little evolution in galaxy numbers as a function of redshift. As a statistical measure of the significance of these structures, we calculated a galaxy overdensity $\delta_{gal}$ for each CCPC member following the examples of \citet{1998ApJ...492..428S} and \citet{ven02}. The median $\delta_{gal}$ value is $\sim 2$, which is comparable to the overdensities of protoclusters within simulations and observationally. Twelve of these have high probability ($85\%$ or larger) of collapsing into a cluster at $z=0$ based on their overdensity compared to analysis of Millennium run data \citep{chi13}.

There are a number of tests that suggest that the CCPC systems identified by our simple algorithm are coherent structures. The protoclusters exist as overdensity peaks, both in $\delta_{gal}$ and visually above the continuum of galaxies in the N(z) plots shown in the Appendix. They follow a distinctly different radial distribution than the mean distribution of the 14000 galaxies in which the CCPC was drawn from, as shown in Figs~\ref{fig:dist},~\ref{fig:density}, many of which do not have any companion galaxies within 20 cMpc of the central source. Furthermore, we have recovered 10 previously identified candidate protoclusters from the literature. We have also found that in the CANDELS GOODS-S field, there is significantly more structure (excess of 6 systems of 9 or more galaxies) than can be expected from Poissonian fluctuations.

The mean radial distribution of members for each CCPC structure follow a similar trend, independent of redshift. However, individual CCPC structures have a wide range of distributions, with some protoclusters completely contained within $R <10$ comoving Mpc while others add more than a dozen of their constituent galaxies in the range of $10<R <20$ comoving Mpc. These variations can be primarily attributed to differences in survey depth and volume for the various regions of our galaxy list. This distribution difference could be also an indication of the various stages of assembly protoclusters are in at this epoch of the Universe, in conjunction with differences in total halo mass, as more massive protoclusters are expected to have larger radii. A combination of all three seems likely.

For each system, we computed a characteristic radius ($R_{hms}$) and a rough virial mass estimate based on its size and velocity dispersion $\sigma_z$. The mass estimates assume a virialized distribution, which may not apply at this epoch. As a result, these crude protocluster mass estimates are up to $10^2$ times larger than those predicted in analysis of the Millennium Simulation by \citet{chi13}. In addition, by assuming a linear bias parameter of $b=3$, we can calculate a mass overdensity ($\delta_{m} = \delta_{gal}/b$) within the volume of the structure. This can then be transformed into a mass estimate using $M = \rho_{crit,z}V(1+\delta_{m})$ \citep{1998ApJ...492..428S,ven02}. These are estimates of the collapsed mass of the system, and have a median value of $8\times10^{13}$ $M_{\odot}$ in the CCPC. These mass estimates are uncorrelated, and caution should be exercised in interpreting these results.

This work significantly increases the number of spectroscopically confirmed, high redshift protoclusters known. To our knowledge, it represents the largest catalog of such sources to date. 

\acknowledgements

JF would like to thank Dr. Idit Zehavi for helpful discussions on statistical structure metrics. We would like to thank the referee for suggestions that improved this manuscript.
This work has been supported in part by NASA ADAP grant NNX13AH32G. This research has made use of the NASA/IPAC Extragalactic Database (NED) which is operated by the Jet Propulsion Laboratory, California Institute of Technology, under contract with the National Aeronautics and Space Administration.

\bibliography{protos}

\begin{thebibliography}{124}
\expandafter\ifx\csname natexlab\endcsname\relax\def\natexlab#1{#1}\fi

\bibitem[{{Adams} {et~al.}(2011){Adams}, {Blanc}, {Hill}, {Gebhardt}, {Drory},
  {Hao}, {Bender}, {Byun}, {Ciardullo}, {Cornell}, {Finkelstein}, {Fry},
  {Gawiser}, {Gronwall}, {Hopp}, {Jeong}, {Kelz}, {Kelzenberg}, {Komatsu},
  {MacQueen}, {Murphy}, {Odoms}, {Roth}, {Schneider}, {Tufts}, \&
  {Wilkinson}}]{2011ApJS..192....5A}
{Adams}, J.~J., {et~al.} 2011, \apjs, 192, 5

\bibitem[{{Balestra} {et~al.}(2010){Balestra}, {Mainieri}, {Popesso},
  {Dickinson}, {Nonino}, {Rosati}, {Teimoorinia}, {Vanzella}, {Cristiani},
  {Cesarsky}, {Fosbury}, {Kuntschner}, \& {Rettura}}]{2010AnA...512A..12B}
{Balestra}, I., {et~al.} 2010, \aap, 512, A12

\bibitem[{{Barger} {et~al.}(2002){Barger}, {Cowie}, {Brandt}, {Capak},
  {Garmire}, {Hornschemeier}, {Steffen}, \& {Wehner}}]{2002AJ....124.1839B}
{Barger}, A.~J., {Cowie}, L.~L., {Brandt}, W.~N., {Capak}, P., {Garmire},
  G.~P., {Hornschemeier}, A.~E., {Steffen}, A.~T., \& {Wehner}, E.~H. 2002,
  \aj, 124, 1839

\bibitem[{{Barger} {et~al.}(2008){Barger}, {Cowie}, \&
  {Wang}}]{2008ApJ...689..687B}
{Barger}, A.~J., {Cowie}, L.~L., \& {Wang}, W.-H. 2008, \apj, 689, 687

\bibitem[{{Barger} {et~al.}(2003){Barger}, {Cowie}, {Capak}, {Alexander},
  {Bauer}, {Fernandez}, {Brandt}, {Garmire}, \&
  {Hornschemeier}}]{2003AJ....126..632B}
{Barger}, A.~J., {et~al.} 2003, \aj, 126, 632

\bibitem[{{Beckwith} {et~al.}(2006){Beckwith}, {Stiavelli}, {Koekemoer},
  {Caldwell}, {Ferguson}, {Hook}, {Lucas}, {Bergeron}, {Corbin}, {Jogee},
  {Panagia}, {Robberto}, {Royle}, {Somerville}, \&
  {Sosey}}]{2006AJ....132.1729B}
{Beckwith}, S.~V.~W., {et~al.} 2006, \aj, 132, 1729

\bibitem[{{Bond} {et~al.}(2010){Bond}, {Feldmeier}, {Matkovi{\'c}}, {Gronwall},
  {Ciardullo}, \& {Gawiser}}]{2010ApJ...716L.200B}
{Bond}, N.~A., {Feldmeier}, J.~J., {Matkovi{\'c}}, A., {Gronwall}, C.,
  {Ciardullo}, R., \& {Gawiser}, E. 2010, \apjl, 716, L200

\bibitem[{{Bond} {et~al.}(2011){Bond}, {Gawiser}, \&
  {Koekemoer}}]{2011ApJ...729...48B}
{Bond}, N.~A., {Gawiser}, E., \& {Koekemoer}, A.~M. 2011, \apj, 729, 48

\bibitem[{{Cantalupo} {et~al.}(2007){Cantalupo}, {Lilly}, \&
  {Porciani}}]{2007ApJ...657..135C}
{Cantalupo}, S., {Lilly}, S.~J., \& {Porciani}, C. 2007, \apj, 657, 135

\bibitem[{{Carlberg} {et~al.}(1996){Carlberg}, {Yee}, {Ellingson}, {Abraham},
  {Gravel}, {Morris}, \& {Pritchet}}]{car96}
{Carlberg}, R.~G., {Yee}, H.~K.~C., {Ellingson}, E., {Abraham}, R., {Gravel},
  P., {Morris}, S., \& {Pritchet}, C.~J. 1996, \apj, 462, 32

\bibitem[{{Chapman} {et~al.}(2004{\natexlab{a}}){Chapman}, {Scott},
  {Windhorst}, {Frayer}, {Borys}, {Lewis}, \& {Ivison}}]{2004ApJ...606...85C}
{Chapman}, S.~C., {Scott}, D., {Windhorst}, R.~A., {Frayer}, D.~T., {Borys},
  C., {Lewis}, G.~F., \& {Ivison}, R.~J. 2004{\natexlab{a}}, \apj, 606, 85

\bibitem[{{Chapman} {et~al.}(2004{\natexlab{b}}){Chapman}, {Smail}, {Blain}, \&
  {Ivison}}]{2004ApJ...614..671C}
{Chapman}, S.~C., {Smail}, I., {Blain}, A.~W., \& {Ivison}, R.~J.
  2004{\natexlab{b}}, \apj, 614, 671

\bibitem[{{Chapman} {et~al.}(2004{\natexlab{c}}){Chapman}, {Smail},
  {Windhorst}, {Muxlow}, \& {Ivison}}]{2004ApJ...611..732C}
{Chapman}, S.~C., {Smail}, I., {Windhorst}, R., {Muxlow}, T., \& {Ivison},
  R.~J. 2004{\natexlab{c}}, \apj, 611, 732

\bibitem[{{Chiang} {et~al.}(2013){Chiang}, {Overzier}, \& {Gebhardt}}]{chi13}
{Chiang}, Y.-K., {Overzier}, R., \& {Gebhardt}, K. 2013, \apj, 779, 127

\bibitem[{{Chiang} {et~al.}(2014){Chiang}, {Overzier}, \& {Gebhardt}}]{chi14}
---. 2014, \apjl, 782, L3

\bibitem[{{Chiang} {et~al.}(2015){Chiang}, {Overzier}, {Gebhardt},
  {Finkelstein}, {Chiang}, {Hill}, {Blanc}, {Drory}, {Chonis}, {Zeimann},
  {Hagen}, {Schneider}, {Jogee}, {Ciardullo}, \&
  {Gronwall}}]{2015arXiv150503877C}
{Chiang}, Y.-K., {et~al.} 2015, ArXiv e-prints

\bibitem[{{Ciardullo} {et~al.}(2012){Ciardullo}, {Gronwall}, {Wolf},
  {McCathran}, {Bond}, {Gawiser}, {Guaita}, {Feldmeier}, {Treister}, {Padilla},
  {Francke}, {Matkovi{\'c}}, {Altmann}, \& {Herrera}}]{2012ApJ...744..110C}
{Ciardullo}, R., {et~al.} 2012, \apj, 744, 110

\bibitem[{{Conselice} {et~al.}(2003){Conselice}, {Bershady}, {Dickinson}, \&
  {Papovich}}]{2003AJ....126.1183C}
{Conselice}, C.~J., {Bershady}, M.~A., {Dickinson}, M., \& {Papovich}, C. 2003,
  \aj, 126, 1183

\bibitem[{{Conselice} {et~al.}(2011){Conselice}, {Bluck}, {Buitrago}, {Bauer},
  {Gr{\"u}tzbauch}, {Bouwens}, {Bevan}, {Mortlock}, {Dickinson}, {Daddi},
  {Yan}, {Scott}, {Chapman}, {Chary}, {Ferguson}, {Giavalisco}, {Grogin},
  {Illingworth}, {Jogee}, {Koekemoer}, {Lucas}, {Mobasher}, {Moustakas},
  {Papovich}, {Ravindranath}, {Siana}, {Teplitz}, {Trujillo}, {Urry}, \&
  {Weinzirl}}]{2011MNRAS.413...80C}
{Conselice}, C.~J., {et~al.} 2011, \mnras, 413, 80

\bibitem[{{Cowie} {et~al.}(2004){Cowie}, {Barger}, {Hu}, {Capak}, \&
  {Songaila}}]{2004AJ....127.3137C}
{Cowie}, L.~L., {Barger}, A.~J., {Hu}, E.~M., {Capak}, P., \& {Songaila}, A.
  2004, \aj, 127, 3137

\bibitem[{{Cowie} {et~al.}(1995){Cowie}, {Songaila}, {Kim}, \&
  {Hu}}]{1995AJ....109.1522C}
{Cowie}, L.~L., {Songaila}, A., {Kim}, T.-S., \& {Hu}, E.~M. 1995, \aj, 109,
  1522

\bibitem[{{Cristiani} {et~al.}(2000){Cristiani}, {Appenzeller}, {Arnouts},
  {Nonino}, {Arag{\'o}n-Salamanca}, {Benoist}, {da Costa}, {Dennefeld},
  {Rengelink}, {Renzini}, {Szeifert}, \& {White}}]{2000AnA...359..489C}
{Cristiani}, S., {et~al.} 2000, \aap, 359, 489

\bibitem[{{Davis} \& {Peebles}(1983)}]{dav83}
{Davis}, M., \& {Peebles}, P.~J.~E. 1983, \apj, 267, 465

\bibitem[{{Dawson} {et~al.}(2001){Dawson}, {Stern}, {Bunker}, {Spinrad}, \&
  {Dey}}]{2001AJ....122..598D}
{Dawson}, S., {Stern}, D., {Bunker}, A.~J., {Spinrad}, H., \& {Dey}, A. 2001,
  \aj, 122, 598

\bibitem[{{de Bruyn} {et~al.}(1996){de Bruyn}, {O'Dea}, \&
  {Baum}}]{1996AnA...305..450d}
{de Bruyn}, A.~G., {O'Dea}, C.~P., \& {Baum}, S.~A. 1996, \aap, 305, 450

\bibitem[{{Diener} {et~al.}(2013){Diener}, {Lilly}, {Knobel}, {Zamorani},
  {Lemson}, {Kampczyk}, {Scoville}, {Carollo}, {Contini}, {Kneib}, {Le Fevre},
  {Mainieri}, {Renzini}, {Scodeggio}, {Bardelli}, {Bolzonella}, {Bongiorno},
  {Caputi}, {Cucciati}, {de la Torre}, {de Ravel}, {Franzetti}, {Garilli},
  {Iovino}, {Kova{\v c}}, {Lamareille}, {Le Borgne}, {Le Brun}, {Maier},
  {Mignoli}, {Pello}, {Peng}, {Perez Montero}, {Presotto}, {Silverman},
  {Tanaka}, {Tasca}, {Tresse}, {Vergani}, {Zucca}, {Bordoloi}, {Cappi},
  {Cimatti}, {Coppa}, {Koekemoer}, {L{\'o}pez-Sanjuan}, {McCracken}, {Moresco},
  {Nair}, {Pozzetti}, \& {Welikala}}]{2013ApJ...765..109D}
{Diener}, C., {et~al.} 2013, \apj, 765, 109

\bibitem[{{Doherty} {et~al.}(2010){Doherty}, {Tanaka}, {De Breuck}, {Ly},
  {Kodama}, {Kurk}, {Seymour}, {Vernet}, {Stern}, {Venemans}, {Kajisawa}, \&
  {Tanaka}}]{2010AnA...509A..83D}
{Doherty}, M., {et~al.} 2010, \aap, 509, A83

\bibitem[{{Eisenhardt} {et~al.}(2008){Eisenhardt}, {Brodwin}, {Gonzalez},
  {Stanford}, {Stern}, {Barmby}, {Brown}, {Dawson}, {Dey}, {Doi}, {Galametz},
  {Jannuzi}, {Kochanek}, {Meyers}, {Morokuma}, \& {Moustakas}}]{eis08}
{Eisenhardt}, P.~R.~M., {et~al.} 2008, \apj, 684, 905

\bibitem[{{Ellison} {et~al.}(2001){Ellison}, {Pettini}, {Steidel}, \&
  {Shapley}}]{2001ApJ...549..770E}
{Ellison}, S.~L., {Pettini}, M., {Steidel}, C.~C., \& {Shapley}, A.~E. 2001,
  \apj, 549, 770

\bibitem[{{Fassbender} {et~al.}(2014){Fassbender}, {Nastasi}, {Santos},
  {Lidman}, {Verdugo}, {Koyama}, {Rosati}, {Pierini}, {Padilla}, {Romeo},
  {Menci}, {Bongiorno}, {Castellano}, {Cerulo}, {Fontana}, {Galametz},
  {Grazian}, {Lamastra}, {Pentericci}, {Sommariva}, {Strazzullo}, {{\v
  S}uhada}, \& {Tozzi}}]{fas14}
{Fassbender}, R., {et~al.} 2014, \aap, 568, A5

\bibitem[{{Fern{\'a}ndez-Soto} {et~al.}(2001){Fern{\'a}ndez-Soto}, {Lanzetta},
  {Chen}, {Pascarelle}, \& {Yahata}}]{2001ApJS..135...41F}
{Fern{\'a}ndez-Soto}, A., {Lanzetta}, K.~M., {Chen}, H.-W., {Pascarelle},
  S.~M., \& {Yahata}, N. 2001, \apjs, 135, 41

\bibitem[{{Fern{\'a}ndez-Soto} {et~al.}(1999){Fern{\'a}ndez-Soto}, {Lanzetta},
  \& {Yahil}}]{1999ApJ...513...34F}
{Fern{\'a}ndez-Soto}, A., {Lanzetta}, K.~M., \& {Yahil}, A. 1999, \apj, 513, 34

\bibitem[{{Ferreras} {et~al.}(2012){Ferreras}, {Pasquali}, {Khochfar},
  {Kuntschner}, {K{\"u}mmel}, {Pirzkal}, {Windhorst}, {Malhotra}, {Rhoads},
  {O'Connell}, {Cohen}, {Hathi}, {Ryan}, \& {Yan}}]{fer12}
{Ferreras}, I., {et~al.} 2012, \aj, 144, 47

\bibitem[{{Fiore} {et~al.}(2012){Fiore}, {Puccetti}, {Grazian}, {Menci},
  {Shankar}, {Santini}, {Piconcelli}, {Koekemoer}, {Fontana}, {Boutsia},
  {Castellano}, {Lamastra}, {Malacaria}, {Feruglio}, {Mathur}, {Miller}, \&
  {Pannella}}]{2012AnA...537A..16F}
{Fiore}, F., {et~al.} 2012, \aap, 537, A16

\bibitem[{{Fontanot} {et~al.}(2007){Fontanot}, {Cristiani}, {Monaco}, {Nonino},
  {Vanzella}, {Brandt}, {Grazian}, \& {Mao}}]{2007AnA...461...39F}
{Fontanot}, F., {Cristiani}, S., {Monaco}, P., {Nonino}, M., {Vanzella}, E.,
  {Brandt}, W.~N., {Grazian}, A., \& {Mao}, J. 2007, \aap, 461, 39

\bibitem[{{Franck} {et~al.}(2015){Franck}, {McGaugh}, \&
  {Schombert}}]{2015arXiv150507864F}
{Franck}, J.~R., {McGaugh}, S.~S., \& {Schombert}, J.~M. 2015, ArXiv e-prints

\bibitem[{{Frank} {et~al.}(2003){Frank}, {Appenzeller}, {Noll}, \&
  {Stahl}}]{2003AnA...407..473F}
{Frank}, S., {Appenzeller}, I., {Noll}, S., \& {Stahl}, O. 2003, \aap, 407, 473

\bibitem[{{Fritz} {et~al.}(2014){Fritz}, {Scodeggio}, {Ilbert}, {Bolzonella},
  {Davidzon}, {Coupon}, {Garilli}, {Guzzo}, {Zamorani}, {Abbas}, {Adami},
  {Arnouts}, {Bel}, {Bottini}, {Branchini}, {Cappi}, {Cucciati}, {De Lucia},
  {de la Torre}, {Franzetti}, {Fumana}, {Granett}, {Iovino}, {Krywult}, {Le
  Brun}, {Le F{\`e}vre}, {Maccagni}, {Ma{\l}ek}, {Marulli}, {McCracken},
  {Paioro}, {Polletta}, {Pollo}, {Schlagenhaufer}, {Tasca}, {Tojeiro},
  {Vergani}, {Zanichelli}, {Burden}, {Di Porto}, {Marchetti}, {Marinoni},
  {Mellier}, {Moscardini}, {Nichol}, {Peacock}, {Percival}, {Phleps}, \&
  {Wolk}}]{fri14}
{Fritz}, A., {et~al.} 2014, \aap, 563, A92

\bibitem[{{Fynbo} {et~al.}(2003){Fynbo}, {Ledoux}, {M{\"o}ller}, {Thomsen}, \&
  {Burud}}]{2003AnA...407..147F}
{Fynbo}, J.~P.~U., {Ledoux}, C., {M{\"o}ller}, P., {Thomsen}, B., \& {Burud},
  I. 2003, \aap, 407, 147

\bibitem[{{Fynbo} {et~al.}(2001){Fynbo}, {M{\"o}ller}, \&
  {Thomsen}}]{2001AnA...374..443F}
{Fynbo}, J.~U., {M{\"o}ller}, P., \& {Thomsen}, B. 2001, \aap, 374, 443

\bibitem[{{Galametz} {et~al.}(2012){Galametz}, {Stern}, {De Breuck}, {Hatch},
  {Mayo}, {Miley}, {Rettura}, {Seymour}, {Stanford}, \& {Vernet}}]{gal12}
{Galametz}, A., {et~al.} 2012, \apj, 749, 169

\bibitem[{{Georgakakis} {et~al.}(2006){Georgakakis}, {Nandra}, {Laird}, {Gwyn},
  {Steidel}, {Sarajedini}, {Barmby}, {Faber}, {Coil}, {Cooper}, {Davis}, \&
  {Newman}}]{2006MNRAS.371..221G}
{Georgakakis}, A., {et~al.} 2006, \mnras, 371, 221

\bibitem[{{Gnerucci} {et~al.}(2011){Gnerucci}, {Marconi}, {Cresci}, {Maiolino},
  {Mannucci}, {Calura}, {Cimatti}, {Cocchia}, {Grazian}, {Matteucci}, {Nagao},
  {Pozzetti}, \& {Troncoso}}]{2011AnA...528A..88G}
{Gnerucci}, A., {et~al.} 2011, \aap, 528, A88

\bibitem[{{Gobat} {et~al.}(2011){Gobat}, {Daddi}, {Onodera}, {Finoguenov},
  {Renzini}, {Arimoto}, {Bouwens}, {Brusa}, {Chary}, {Cimatti}, {Dickinson},
  {Kong}, \& {Mignoli}}]{gob11}
{Gobat}, R., {et~al.} 2011, \aap, 526, A133

\bibitem[{{Grazian} {et~al.}(2006){Grazian}, {Fontana}, {de Santis}, {Nonino},
  {Salimbeni}, {Giallongo}, {Cristiani}, {Gallozzi}, \&
  {Vanzella}}]{2006AnA...449..951G}
{Grazian}, A., {et~al.} 2006, \aap, 449, 951

\bibitem[{{Grove} {et~al.}(2009){Grove}, {Fynbo}, {Ledoux}, {Limousin},
  {M{\o}ller}, {Nilsson}, \& {Thomsen}}]{2009AnA...497..689G}
{Grove}, L.~F., {Fynbo}, J.~P.~U., {Ledoux}, C., {Limousin}, M., {M{\o}ller},
  P., {Nilsson}, K.~K., \& {Thomsen}, B. 2009, \aap, 497, 689

\bibitem[{{Guo} {et~al.}(2011){Guo}, {White}, {Boylan-Kolchin}, {De Lucia},
  {Kauffmann}, {Lemson}, {Li}, {Springel}, \& {Weinmann}}]{2011MNRAS.413..101G}
{Guo}, Q., {et~al.} 2011, \mnras, 413, 101

\bibitem[{{Hall} \& {Green}(1998)}]{hal98}
{Hall}, P.~B., \& {Green}, R.~F. 1998, \apj, 507, 558

\bibitem[{{Hall} {et~al.}(2001){Hall}, {Sawicki}, {Martini}, {Finn},
  {Pritchet}, {Osmer}, {McCarthy}, {Evans}, {Lin}, \& {Hartwick}}]{hal01}
{Hall}, P.~B., {et~al.} 2001, \aj, 121, 1840

\bibitem[{{Hayashino} {et~al.}(2004){Hayashino}, {Matsuda}, {Tamura},
  {Yamauchi}, {Yamada}, {Ajiki}, {Fujita}, {Murayama}, {Nagao}, {Ohta},
  {Okamura}, {Ouchi}, {Shimasaku}, {Shioya}, \&
  {Taniguchi}}]{2004AJ....128.2073H}
{Hayashino}, T., {et~al.} 2004, \aj, 128, 2073

\bibitem[{{Hewett} \& {Wild}(2010)}]{2010MNRAS.405.2302H}
{Hewett}, P.~C., \& {Wild}, V. 2010, \mnras, 405, 2302

\bibitem[{{Huchra} \& {Geller}(1982)}]{huc82}
{Huchra}, J.~P., \& {Geller}, M.~J. 1982, \apj, 257, 423

\bibitem[{{Inoue} {et~al.}(2011){Inoue}, {Kousai}, {Iwata}, {Matsuda},
  {Nakamura}, {Horie}, {Hayashino}, {Tapken}, {Akiyama}, {Noll}, {Yamada},
  {Burgarella}, \& {Nakamura}}]{2011MNRAS.411.2336I}
{Inoue}, A.~K., {et~al.} 2011, \mnras, 411, 2336

\bibitem[{{Iwata} {et~al.}(2005){Iwata}, {Inoue}, \&
  {Burgarella}}]{2005AnA...440..881I}
{Iwata}, I., {Inoue}, A.~K., \& {Burgarella}, D. 2005, \aap, 440, 881

\bibitem[{{Katz} {et~al.}(2013){Katz}, {McGaugh}, {Teuben}, \&
  {Angus}}]{2013ApJ...772...10K}
{Katz}, H., {McGaugh}, S., {Teuben}, P., \& {Angus}, G.~W. 2013, \apj, 772, 10

\bibitem[{{Kitzbichler} \& {White}(2007)}]{2007MNRAS.376....2K}
{Kitzbichler}, M.~G., \& {White}, S.~D.~M. 2007, \mnras, 376, 2

\bibitem[{{Kuiper} {et~al.}(2012){Kuiper}, {Venemans}, {Hatch}, {Miley}, \&
  {R{\"o}ttgering}}]{2012MNRAS.425..801K}
{Kuiper}, E., {Venemans}, B.~P., {Hatch}, N.~A., {Miley}, G.~K., \&
  {R{\"o}ttgering}, H.~J.~A. 2012, \mnras, 425, 801

\bibitem[{{Lanzetta} {et~al.}(1996){Lanzetta}, {Yahil}, \&
  {Fern{\'a}ndez-Soto}}]{1996Natur.381..759L}
{Lanzetta}, K.~M., {Yahil}, A., \& {Fern{\'a}ndez-Soto}, A. 1996, \nat, 381,
  759

\bibitem[{{Law}(2008)}]{2008PrivC.U..D...1L}
{Law}, D.~I. 2008, NED: Public Communication, 1

\bibitem[{{Le Fevre} {et~al.}(1996){Le Fevre}, {Deltorn}, {Crampton}, \&
  {Dickinson}}]{1996ApJ...471L..11L}
{Le Fevre}, O., {Deltorn}, J.~M., {Crampton}, D., \& {Dickinson}, M. 1996,
  \apjl, 471, L11

\bibitem[{{Le F{\`e}vre} {et~al.}(2004){Le F{\`e}vre}, {Vettolani}, {Paltani},
  {Tresse}, {Zamorani}, {Le Brun}, {Moreau}, {Bottini}, {Maccagni}, {Picat},
  {Scaramella}, {Scodeggio}, {Zanichelli}, {Adami}, {Arnouts}, {Bardelli},
  {Bolzonella}, {Cappi}, {Charlot}, {Contini}, {Foucaud}, {Franzetti},
  {Garilli}, {Gavignaud}, {Guzzo}, {Ilbert}, {Iovino}, {McCracken}, {Mancini},
  {Marano}, {Marinoni}, {Mathez}, {Mazure}, {Meneux}, {Merighi}, {Pell{\`o}},
  {Pollo}, {Pozzetti}, {Radovich}, {Zucca}, {Arnaboldi}, {Bondi}, {Bongiorno},
  {Busarello}, {Ciliegi}, {Gregorini}, {Mellier}, {Merluzzi}, {Ripepi}, \&
  {Rizzo}}]{2004AnA...428.1043L}
{Le F{\`e}vre}, O., {et~al.} 2004, \aap, 428, 1043

\bibitem[{{Lee} {et~al.}(2014){Lee}, {Dey}, {Hong}, {Reddy}, {Wilson},
  {Jannuzi}, {Inami}, \& {Gonzalez}}]{lee14}
{Lee}, K.-S., {Dey}, A., {Hong}, S., {Reddy}, N., {Wilson}, C., {Jannuzi},
  B.~T., {Inami}, H., \& {Gonzalez}, A.~H. 2014, \apj, 796, 126

\bibitem[{{Lehmer} {et~al.}(2009{\natexlab{a}}){Lehmer}, {Alexander}, {Geach},
  {Smail}, {Basu-Zych}, {Bauer}, {Chapman}, {Matsuda}, {Scharf}, {Volonteri},
  \& {Yamada}}]{2009ApJ...691..687L}
{Lehmer}, B.~D., {et~al.} 2009{\natexlab{a}}, \apj, 691, 687

\bibitem[{{Lehmer} {et~al.}(2009{\natexlab{b}}){Lehmer}, {Alexander},
  {Chapman}, {Smail}, {Bauer}, {Brandt}, {Geach}, {Matsuda}, {Mullaney}, \&
  {Swinbank}}]{2009MNRAS.400..299L}
---. 2009{\natexlab{b}}, \mnras, 400, 299

\bibitem[{{Lowenthal} {et~al.}(1997){Lowenthal}, {Koo}, {Guzm{\'a}n},
  {Gallego}, {Phillips}, {Faber}, {Vogt}, {Illingworth}, \&
  {Gronwall}}]{1997ApJ...481..673L}
{Lowenthal}, J.~D., {et~al.} 1997, \apj, 481, 673

\bibitem[{{Mantz} {et~al.}(2014){Mantz}, {Abdulla}, {Carlstrom}, {Greer},
  {Leitch}, {Marrone}, {Muchovej}, {Adami}, {Birkinshaw}, {Bremer}, {Clerc},
  {Giles}, {Horellou}, {Maughan}, {Pacaud}, {Pierre}, \& {Willis}}]{man14}
{Mantz}, A.~B., {et~al.} 2014, ArXiv e-prints

\bibitem[{{Maschietto} {et~al.}(2008){Maschietto}, {Hatch}, {Venemans},
  {R{\"o}ttgering}, {Miley}, {Overzier}, {Dopita}, {Eisenhardt}, {Kurk},
  {Meurer}, {Pentericci}, {Rosati}, {Stanford}, {van Breugel}, \&
  {Zirm}}]{2008MNRAS.389.1223M}
{Maschietto}, F., {et~al.} 2008, \mnras, 389, 1223

\bibitem[{{McCarthy} {et~al.}(1996){McCarthy}, {Kapahi}, {van Breugel},
  {Persson}, {Athreya}, \& {Subrahmanya}}]{1996ApJS..107...19M}
{McCarthy}, P.~J., {Kapahi}, V.~K., {van Breugel}, W., {Persson}, S.~E.,
  {Athreya}, R., \& {Subrahmanya}, C.~R. 1996, \apjs, 107, 19

\bibitem[{{McGaugh}(2004)}]{2004ApJ...611...26M}
{McGaugh}, S.~S. 2004, \apj, 611, 26

\bibitem[{{McGaugh}(2015)}]{2015CaJPh..93..250M}
---. 2015, Canadian Journal of Physics, 93, 250

\bibitem[{{Mehlert} {et~al.}(2002){Mehlert}, {Noll}, {Appenzeller}, {Saglia},
  {Bender}, {B{\"o}hm}, {Drory}, {Fricke}, {Gabasch}, {Heidt}, {Hopp},
  {J{\"a}ger}, {M{\"o}llenhoff}, {Seitz}, {Stahl}, \&
  {Ziegler}}]{2002AnA...393..809M}
{Mehlert}, D., {et~al.} 2002, \aap, 393, 809

\bibitem[{{M{\"o}ller} \& {Fynbo}(2001)}]{2001AnA...372L..57M}
{M{\"o}ller}, P., \& {Fynbo}, J.~U. 2001, \aap, 372, L57

\bibitem[{{Mortonson} {et~al.}(2011){Mortonson}, {Hu}, \& {Huterer}}]{mor11}
{Mortonson}, M.~J., {Hu}, W., \& {Huterer}, D. 2011, \prd, 83, 023015

\bibitem[{{Moth} \& {Elston}(2002)}]{2002AJ....124.1886M}
{Moth}, P., \& {Elston}, R.~J. 2002, \aj, 124, 1886

\bibitem[{{Muldrew} {et~al.}(2015){Muldrew}, {Hatch}, \&
  {Cooke}}]{2015arXiv150608835M}
{Muldrew}, S.~I., {Hatch}, N.~A., \& {Cooke}, E.~A. 2015, ArXiv e-prints

\bibitem[{{Nestor} {et~al.}(2011){Nestor}, {Shapley}, {Steidel}, \&
  {Siana}}]{2011ApJ...736...18N}
{Nestor}, D.~B., {Shapley}, A.~E., {Steidel}, C.~C., \& {Siana}, B. 2011, \apj,
  736, 18

\bibitem[{{Noll} {et~al.}(2004){Noll}, {Mehlert}, {Appenzeller}, {Bender},
  {B{\"o}hm}, {Gabasch}, {Heidt}, {Hopp}, {J{\"a}ger}, {Seitz}, {Stahl},
  {Tapken}, \& {Ziegler}}]{2004AnA...418..885N}
{Noll}, S., {et~al.} 2004, \aap, 418, 885

\bibitem[{{Nusser}(2002)}]{2002MNRAS.331..909N}
{Nusser}, A. 2002, \mnras, 331, 909

\bibitem[{{Osmer} {et~al.}(1994){Osmer}, {Porter}, \&
  {Green}}]{1994ApJ...436..678O}
{Osmer}, P.~S., {Porter}, A.~C., \& {Green}, R.~F. 1994, \apj, 436, 678

\bibitem[{{Papovich}(2008)}]{pap08}
{Papovich}, C. 2008, \apj, 676, 206

\bibitem[{{Papovich} {et~al.}(2001){Papovich}, {Dickinson}, \&
  {Ferguson}}]{2001ApJ...559..620P}
{Papovich}, C., {Dickinson}, M., \& {Ferguson}, H.~C. 2001, \apj, 559, 620

\bibitem[{{Peter} {et~al.}(2007){Peter}, {Shapley}, {Law}, {Steidel}, {Erb},
  {Reddy}, \& {Pettini}}]{2007ApJ...668...23P}
{Peter}, A.~H.~G., {Shapley}, A.~E., {Law}, D.~R., {Steidel}, C.~C., {Erb},
  D.~K., {Reddy}, N.~A., \& {Pettini}, M. 2007, \apj, 668, 23

\bibitem[{{Petry} {et~al.}(1998){Petry}, {Impey}, \&
  {Foltz}}]{1998ApJ...494...60P}
{Petry}, C.~E., {Impey}, C.~D., \& {Foltz}, C.~B. 1998, \apj, 494, 60

\bibitem[{{Pirzkal} {et~al.}(2013){Pirzkal}, {Rothberg}, {Ly}, {Malhotra},
  {Rhoads}, {Grogin}, {Dahlen}, {Noeske}, {Meurer}, {Walsh}, {Hathi}, {Cohen},
  {Bellini}, {Holwerda}, {Straughn}, {Mechtley}, \&
  {Windhorst}}]{2013ApJ...772...48P}
{Pirzkal}, N., {et~al.} 2013, \apj, 772, 48

\bibitem[{{Prescott} {et~al.}(2008){Prescott}, {Kashikawa}, {Dey}, \&
  {Matsuda}}]{2008ApJ...678L..77P}
{Prescott}, M.~K.~M., {Kashikawa}, N., {Dey}, A., \& {Matsuda}, Y. 2008, \apjl,
  678, L77

\bibitem[{{Press} \& {Davis}(1982)}]{pre82}
{Press}, W.~H., \& {Davis}, M. 1982, \apj, 259, 449

\bibitem[{{Rafelski} {et~al.}(2011){Rafelski}, {Wolfe}, \&
  {Chen}}]{2011ApJ...736...48R}
{Rafelski}, M., {Wolfe}, A.~M., \& {Chen}, H.-W. 2011, \apj, 736, 48

\bibitem[{{Rakos} \& {Schombert}(1995)}]{rak95}
{Rakos}, K.~D., \& {Schombert}, J.~M. 1995, \apj, 439, 47

\bibitem[{{Reddy} {et~al.}(2006){Reddy}, {Steidel}, {Erb}, {Shapley}, \&
  {Pettini}}]{2006ApJ...653.1004R}
{Reddy}, N.~A., {Steidel}, C.~C., {Erb}, D.~K., {Shapley}, A.~E., \& {Pettini},
  M. 2006, \apj, 653, 1004

\bibitem[{{Rettura} {et~al.}(2014){Rettura}, {Martinez-Manso}, {Stern}, {Mei},
  {Ashby}, {Brodwin}, {Gettings}, {Gonzalez}, {Stanford}, \&
  {Bartlett}}]{ret14}
{Rettura}, A., {et~al.} 2014, \apj, 797, 109

\bibitem[{{Rigby} {et~al.}(2014){Rigby}, {Hatch}, {R{\"o}ttgering},
  {Sibthorpe}, {Chiang}, {Overzier}, {Herbonnet}, {Borgani}, {Clements},
  {Dannerbauer}, {De Breuck}, {De Lucia}, {Kurk}, {Maschietto}, {Miley},
  {Saro}, {Seymour}, \& {Venemans}}]{rig14}
{Rigby}, E.~E., {et~al.} 2014, \mnras, 437, 1882

\bibitem[{{Roettgering} {et~al.}(1997){Roettgering}, {van Ojik}, {Miley},
  {Chambers}, {van Breugel}, \& {de Koff}}]{1997AnA...326..505R}
{Roettgering}, H.~J.~A., {van Ojik}, R., {Miley}, G.~K., {Chambers}, K.~C.,
  {van Breugel}, W.~J.~M., \& {de Koff}, S. 1997, \aap, 326, 505

\bibitem[{{Sanders}(1998)}]{1998MNRAS.296.1009S}
{Sanders}, R.~H. 1998, \mnras, 296, 1009

\bibitem[{{Santini} {et~al.}(2009){Santini}, {Fontana}, {Grazian}, {Salimbeni},
  {Fiore}, {Fontanot}, {Boutsia}, {Castellano}, {Cristiani}, {de Santis},
  {Gallozzi}, {Giallongo}, {Menci}, {Nonino}, {Paris}, {Pentericci}, \&
  {Vanzella}}]{2009AnA...504..751S}
{Santini}, P., {et~al.} 2009, \aap, 504, 751

\bibitem[{{Savaglio} {et~al.}(1997){Savaglio}, {Cristiani}, {D'Odorico},
  {Fontana}, {Giallongo}, \& {Molaro}}]{1997AnA...318..347S}
{Savaglio}, S., {Cristiani}, S., {D'Odorico}, S., {Fontana}, A., {Giallongo},
  E., \& {Molaro}, P. 1997, \aap, 318, 347

\bibitem[{{Schreier} {et~al.}(2001){Schreier}, {Koekemoer}, {Grogin},
  {Giacconi}, {Gilli}, {Kewley}, {Norman}, {Hasinger}, {Rosati}, {Marconi},
  {Salvati}, \& {Tozzi}}]{2001ApJ...560..127S}
{Schreier}, E.~J., {et~al.} 2001, \apj, 560, 127

\bibitem[{{Shapley} {et~al.}(2005){Shapley}, {Steidel}, {Erb}, {Reddy},
  {Adelberger}, {Pettini}, {Barmby}, \& {Huang}}]{2005ApJ...626..698S}
{Shapley}, A.~E., {Steidel}, C.~C., {Erb}, D.~K., {Reddy}, N.~A., {Adelberger},
  K.~L., {Pettini}, M., {Barmby}, P., \& {Huang}, J. 2005, \apj, 626, 698

\bibitem[{{Sheth}(2001)}]{she01}
{Sheth}, R.~K. 2001, in Annals of the New York Academy of Sciences, Vol. 927,
  The Onset of Nonlinearity in Cosmology, ed. J.~N. {Fry}, J.~R. {Buchler}, \&
  H.~{Kandrup}, 1--12

\bibitem[{{Smail} {et~al.}(2004){Smail}, {Chapman}, {Blain}, \&
  {Ivison}}]{2004ApJ...616...71S}
{Smail}, I., {Chapman}, S.~C., {Blain}, A.~W., \& {Ivison}, R.~J. 2004, \apj,
  616, 71

\bibitem[{{Songaila}(1998)}]{1998AJ....115.2184S}
{Songaila}, A. 1998, \aj, 115, 2184

\bibitem[{{Springel} {et~al.}(2005){Springel}, {White}, {Jenkins}, {Frenk},
  {Yoshida}, {Gao}, {Navarro}, {Thacker}, {Croton}, {Helly}, {Peacock}, {Cole},
  {Thomas}, {Couchman}, {Evrard}, {Colberg}, \& {Pearce}}]{2005Natur.435..629S}
{Springel}, V., {et~al.} 2005, \nat, 435, 629

\bibitem[{{Steidel} {et~al.}(1998){Steidel}, {Adelberger}, {Dickinson},
  {Giavalisco}, {Pettini}, \& {Kellogg}}]{1998ApJ...492..428S}
{Steidel}, C.~C., {Adelberger}, K.~L., {Dickinson}, M., {Giavalisco}, M.,
  {Pettini}, M., \& {Kellogg}, M. 1998, \apj, 492, 428

\bibitem[{{Steidel} {et~al.}(2000){Steidel}, {Adelberger}, {Shapley},
  {Pettini}, {Dickinson}, \& {Giavalisco}}]{2000ApJ...532..170S}
{Steidel}, C.~C., {Adelberger}, K.~L., {Shapley}, A.~E., {Pettini}, M.,
  {Dickinson}, M., \& {Giavalisco}, M. 2000, \apj, 532, 170

\bibitem[{{Steidel} {et~al.}(2003){Steidel}, {Adelberger}, {Shapley},
  {Pettini}, {Dickinson}, \& {Giavalisco}}]{2003ApJ...592..728S}
---. 2003, \apj, 592, 728

\bibitem[{{Steinhardt} {et~al.}(2015){Steinhardt}, {Capak}, {Masters}, \&
  {Speagle}}]{2015arXiv150601377S}
{Steinhardt}, C.~L., {Capak}, P., {Masters}, D., \& {Speagle}, J.~S. 2015,
  ArXiv e-prints

\bibitem[{{Storrie-Lombardi} \& {Wolfe}(2000)}]{2000ApJ...543..552S}
{Storrie-Lombardi}, L.~J., \& {Wolfe}, A.~M. 2000, \apj, 543, 552

\bibitem[{{Szokoly} {et~al.}(2004){Szokoly}, {Bergeron}, {Hasinger}, {Lehmann},
  {Kewley}, {Mainieri}, {Nonino}, {Rosati}, {Giacconi}, {Gilli}, {Gilmozzi},
  {Norman}, {Romaniello}, {Schreier}, {Tozzi}, {Wang}, {Zheng}, \&
  {Zirm}}]{2004ApJS..155..271S}
{Szokoly}, G.~P., {et~al.} 2004, \apjs, 155, 271

\bibitem[{{Toshikawa} {et~al.}(2014){Toshikawa}, {Kashikawa}, {Overzier},
  {Shibuya}, {Ishikawa}, {Ota}, {Shimasaku}, {Tanaka}, {Hayashi}, {Niino}, \&
  {Onoue}}]{2014ApJ...792...15T}
{Toshikawa}, J., {et~al.} 2014, \apj, 792, 15

\bibitem[{{Trenti} {et~al.}(2012){Trenti}, {Bradley}, {Stiavelli}, {Shull},
  {Oesch}, {Bouwens}, {Mu{\~n}oz}, {Romano-Diaz}, {Treu}, {Shlosman}, \&
  {Carollo}}]{2012ApJ...746...55T}
{Trenti}, M., {et~al.} 2012, \apj, 746, 55

\bibitem[{{Uchimoto} {et~al.}(2012){Uchimoto}, {Yamada}, {Kajisawa}, {Kubo},
  {Ichikawa}, {Matsuda}, {Akiyama}, {Hayashino}, {Konishi}, {Nishimura},
  {Omata}, {Suzuki}, {Tanaka}, {Tokoku}, \& {Yoshikawa}}]{2012ApJ...750..116U}
{Uchimoto}, Y.~K., {et~al.} 2012, \apj, 750, 116

\bibitem[{{Utsumi} {et~al.}(2010){Utsumi}, {Goto}, {Kashikawa}, {Miyazaki},
  {Komiyama}, {Furusawa}, \& {Overzier}}]{2010ApJ...721.1680U}
{Utsumi}, Y., {Goto}, T., {Kashikawa}, N., {Miyazaki}, S., {Komiyama}, Y.,
  {Furusawa}, H., \& {Overzier}, R. 2010, \apj, 721, 1680

\bibitem[{{van de Sande} {et~al.}(2014){van de Sande}, {Kriek}, {Franx},
  {Bezanson}, \& {van Dokkum}}]{van14}
{van de Sande}, J., {Kriek}, M., {Franx}, M., {Bezanson}, R., \& {van Dokkum},
  P.~G. 2014, \apjl, 793, L31

\bibitem[{{Vanzella} {et~al.}(2006){Vanzella}, {Cristiani}, {Dickinson},
  {Kuntschner}, {Nonino}, {Rettura}, {Rosati}, {Vernet}, {Cesarsky},
  {Ferguson}, {Fosbury}, {Giavalisco}, {Grazian}, {Haase}, {Moustakas},
  {Popesso}, {Renzini}, {Stern}, \& {GOODS Team}}]{2006AnA...454..423V}
{Vanzella}, E., {et~al.} 2006, \aap, 454, 423

\bibitem[{{Venemans} {et~al.}(2002){Venemans}, {Kurk}, {Miley},
  {R{\"o}ttgering}, {van Breugel}, {Carilli}, {De Breuck}, {Ford}, {Heckman},
  {McCarthy}, \& {Pentericci}}]{ven02}
{Venemans}, B.~P., {et~al.} 2002, \apjl, 569, L11

\bibitem[{{Venemans} {et~al.}(2005){Venemans}, {R{\"o}ttgering}, {Miley},
  {Kurk}, {De Breuck}, {Overzier}, {van Breugel}, {Carilli}, {Ford}, {Heckman},
  {Pentericci}, \& {McCarthy}}]{2005AnA...431..793V}
---. 2005, \aap, 431, 793

\bibitem[{{Venemans} {et~al.}(2007){Venemans}, {R{\"o}ttgering}, {Miley}, {van
  Breugel}, {de Breuck}, {Kurk}, {Pentericci}, {Stanford}, {Overzier}, {Croft},
  \& {Ford}}]{2007AnA...461..823V}
---. 2007, \aap, 461, 823

\bibitem[{{Vikhlinin} {et~al.}(2009){Vikhlinin}, {Kravtsov}, {Burenin},
  {Ebeling}, {Forman}, {Hornstrup}, {Jones}, {Murray}, {Nagai}, {Quintana}, \&
  {Voevodkin}}]{vik09}
{Vikhlinin}, A., {et~al.} 2009, \apj, 692, 1060

\bibitem[{{Willis} {et~al.}(2013){Willis}, {Clerc}, {Bremer}, {Pierre},
  {Adami}, {Ilbert}, {Maughan}, {Maurogordato}, {Pacaud}, {Valtchanov},
  {Chiappetti}, {Thanjavur}, {Gwyn}, {Stanway}, \& {Winkworth}}]{wil13}
{Willis}, J.~P., {et~al.} 2013, \mnras, 430, 134

\bibitem[{{Wright}(2006)}]{wri06}
{Wright}, E.~L. 2006, \pasp, 118, 1711

\bibitem[{{Wuyts} {et~al.}(2008){Wuyts}, {Labb{\'e}}, {Schreiber}, {Franx},
  {Rudnick}, {Brammer}, \& {van Dokkum}}]{2008ApJ...682..985W}
{Wuyts}, S., {Labb{\'e}}, I., {Schreiber}, N.~M.~F., {Franx}, M., {Rudnick},
  G., {Brammer}, G.~B., \& {van Dokkum}, P.~G. 2008, \apj, 682, 985

\bibitem[{{Wuyts} {et~al.}(2009){Wuyts}, {van Dokkum}, {Franx}, {F{\"o}rster
  Schreiber}, {Illingworth}, {Labb{\'e}}, \& {Rudnick}}]{2009ApJ...706..885W}
{Wuyts}, S., {van Dokkum}, P.~G., {Franx}, M., {F{\"o}rster Schreiber}, N.~M.,
  {Illingworth}, G.~D., {Labb{\'e}}, I., \& {Rudnick}, G. 2009, \apj, 706, 885

\bibitem[{{Wylezalek} {et~al.}(2013){Wylezalek}, {Galametz}, {Stern}, {Vernet},
  {De Breuck}, {Seymour}, {Brodwin}, {Eisenhardt}, {Gonzalez}, {Hatch},
  {Jarvis}, {Rettura}, {Stanford}, \& {Stevens}}]{wyl13}
{Wylezalek}, D., {et~al.} 2013, \apj, 769, 79

\bibitem[{{Wylezalek} {et~al.}(2014){Wylezalek}, {Vernet}, {De Breuck},
  {Stern}, {Brodwin}, {Galametz}, {Gonzalez}, {Jarvis}, {Hatch}, {Seymour}, \&
  {Stanford}}]{wyl14}
---. 2014, \apj, 786, 17

\bibitem[{{Yamada} {et~al.}(2012){Yamada}, {Matsuda}, {Kousai}, {Hayashino},
  {Morimoto}, \& {Umemura}}]{2012ApJ...751...29Y}
{Yamada}, T., {Matsuda}, Y., {Kousai}, K., {Hayashino}, T., {Morimoto}, N., \&
  {Umemura}, M. 2012, \apj, 751, 29

\end{thebibliography}
\bibliographystyle{high_z_FINAL_arXiv}

%EXAMPLE: [protos.bib] file, style file is apj.bst

%CALL AS: 
%	PDFLATEX PROTOS_V3.TEX
%	BIBTEX PROTOS_V3.AUX
%	PDFLATEX PROTOS_V3.TEX
%	PDFLATEX PROTOS_V3.TEX
\clearpage
\appendix \label{sec:app}
%%%%%%%%%%%%%%%%%%%%%%%%%%%%%%%%%%%%%%%%%%%%%%%%%%%%%%%%%%%%%%%%%%%%%%%%%%%%%%%%%%%%%%%%%%%%%%%%%%%%%%%%%%%%%%%%%%%%%%%%%%%%%%%%%%%%%%%%%%%

Each Candidate Cluster and Protocluster entry in Tables~\ref{tab:cat},~\ref{tab:cat2} is matched with a position plot of its member galaxies in the $Left$ panel, 
and a $N(z)$ plot ($Right$ panel) showing the number of galaxies as a function of redshift within the search radius of the protocluster along 
the line of sight. The red circles in the galaxy-position plot correspond to the area covered in  10 and 20 comoving Mpc search radii calculated at the redshift
of the protocluster. Green $\times$'s mark spectroscopic field galaxies within $\Delta z\pm0.15$ found within the sky search radius, which illustrate the field of view. Note that the field of view does not always encompass a radius of 20 $cMpc$. The red vertical lines in the $N(z)$ distribution indicate the $\Delta$ from the mean redshift that correspond to $\pm20$ $cMpc$.

\begin{figure*}
\centering
\begin{subfigure}
\centering
\includegraphics[height=7.5cm,width=7.5cm]{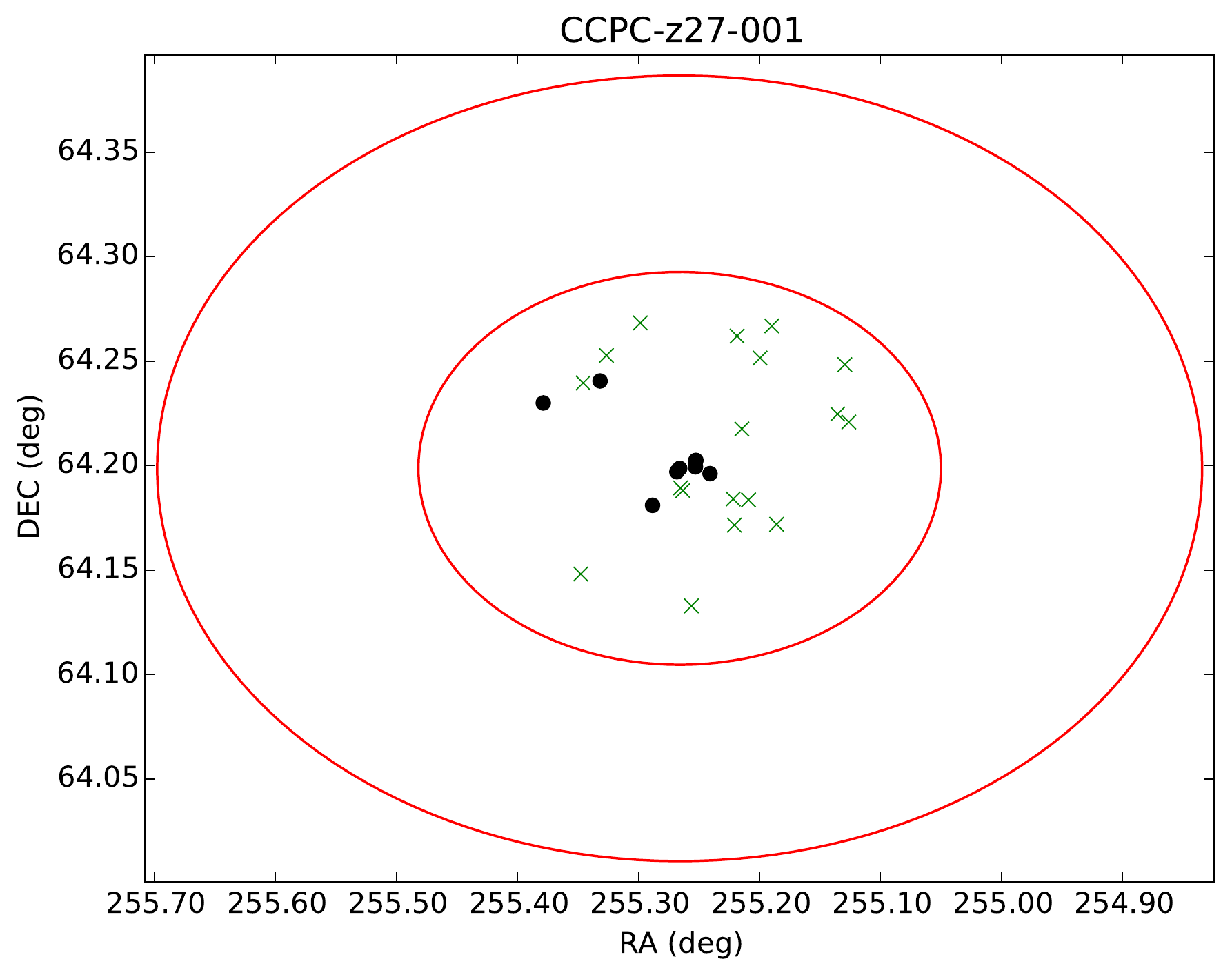}
\label{fig:CCPC-z27-001}
\end{subfigure}
\hfill
\begin{subfigure}
\centering
\includegraphics[scale=0.52]{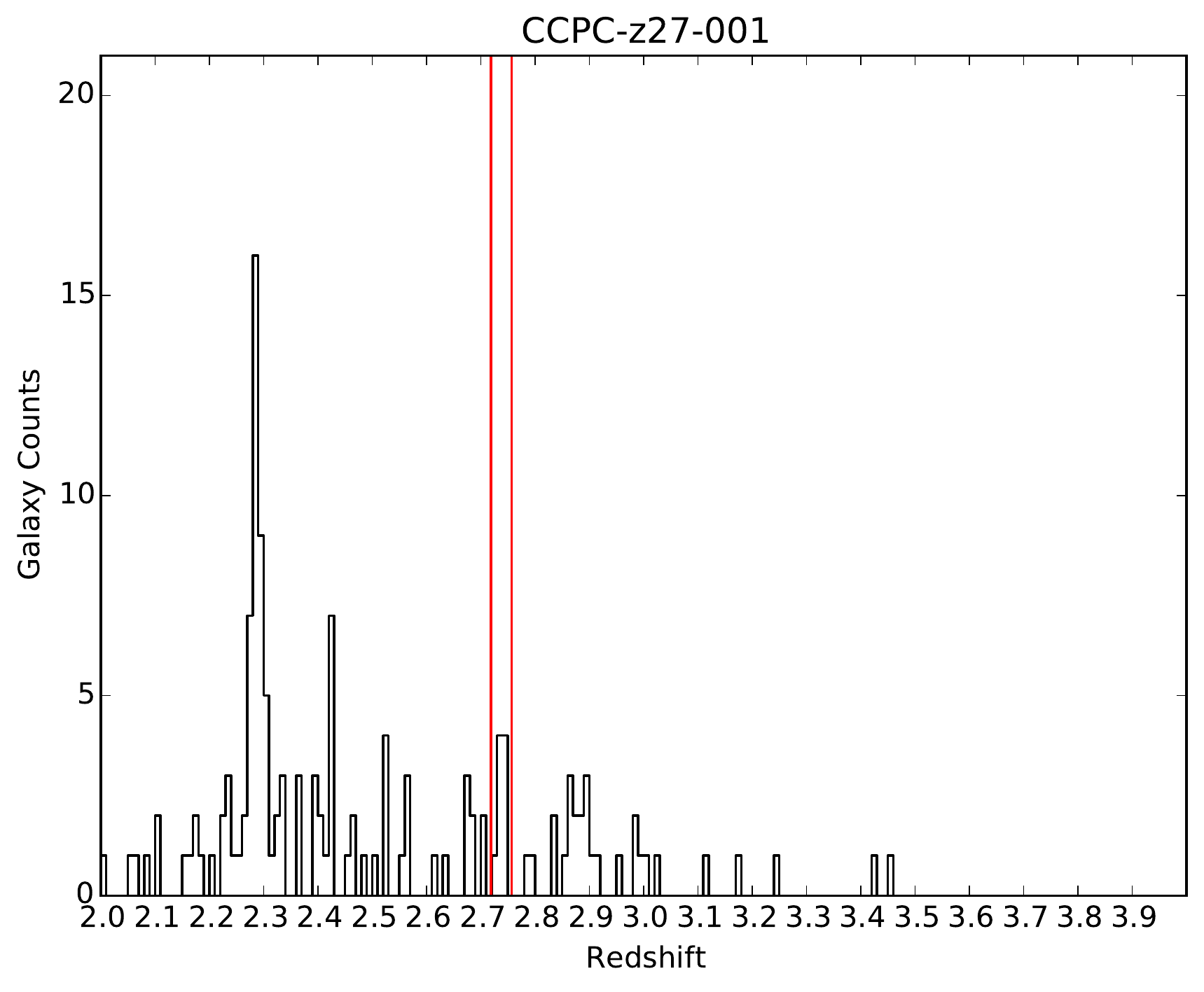}
\label{fig:CCPC-z27-001}
\end{subfigure}
\hfill
\caption{$Left:$ The outer red circle represents a radius of 20 cMpc from the center of the protocluster, while the inner circle corresponds to $R=10$ $cMpc$. The black points are the protocluster members, while the green $\times$ symbols indicate field galaxies within $\Delta z \pm 0.15$ of the redshift of the structure. The FOV of many of the surveys used in this work, as traced by the distribution of field galaxies, are not wide enough to probe out to the maximum search radius of our algorithm  ($R=20$ $cMpc$).   $Right:$ $N(z)$ plot, with the number of galaxies plotted as a function of redshift. The red vertical lines represent $\pm$20 comoving Mpc from the mean redshift of the system.}
\end{figure*}

\begin{figure*}
\centering
\begin{subfigure}
\centering
\includegraphics[height=7.5cm,width=7.5cm]{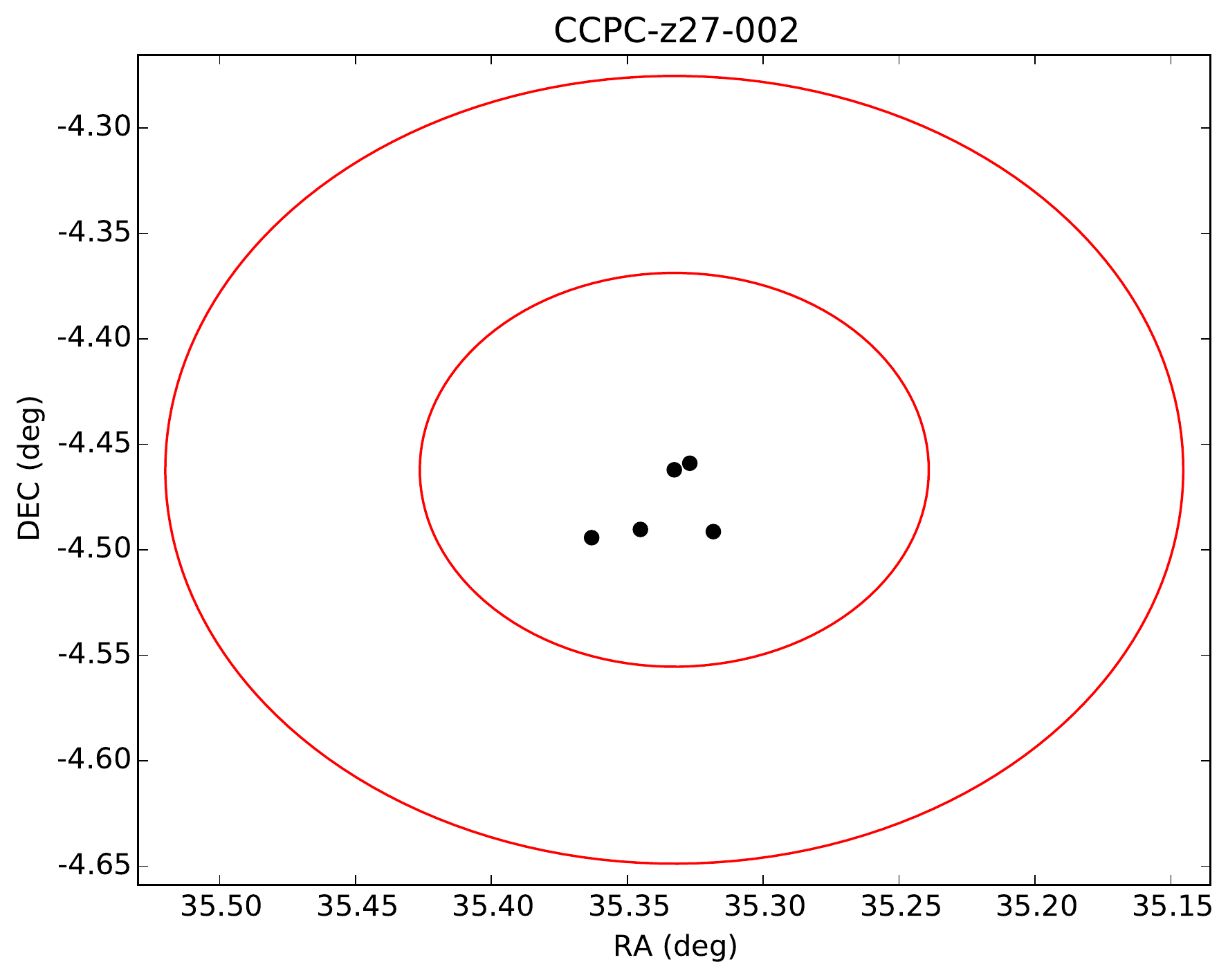}
\label{fig:CCPC-z27-002}
\end{subfigure}
\hfill
\begin{subfigure}
\centering
\includegraphics[scale=0.52]{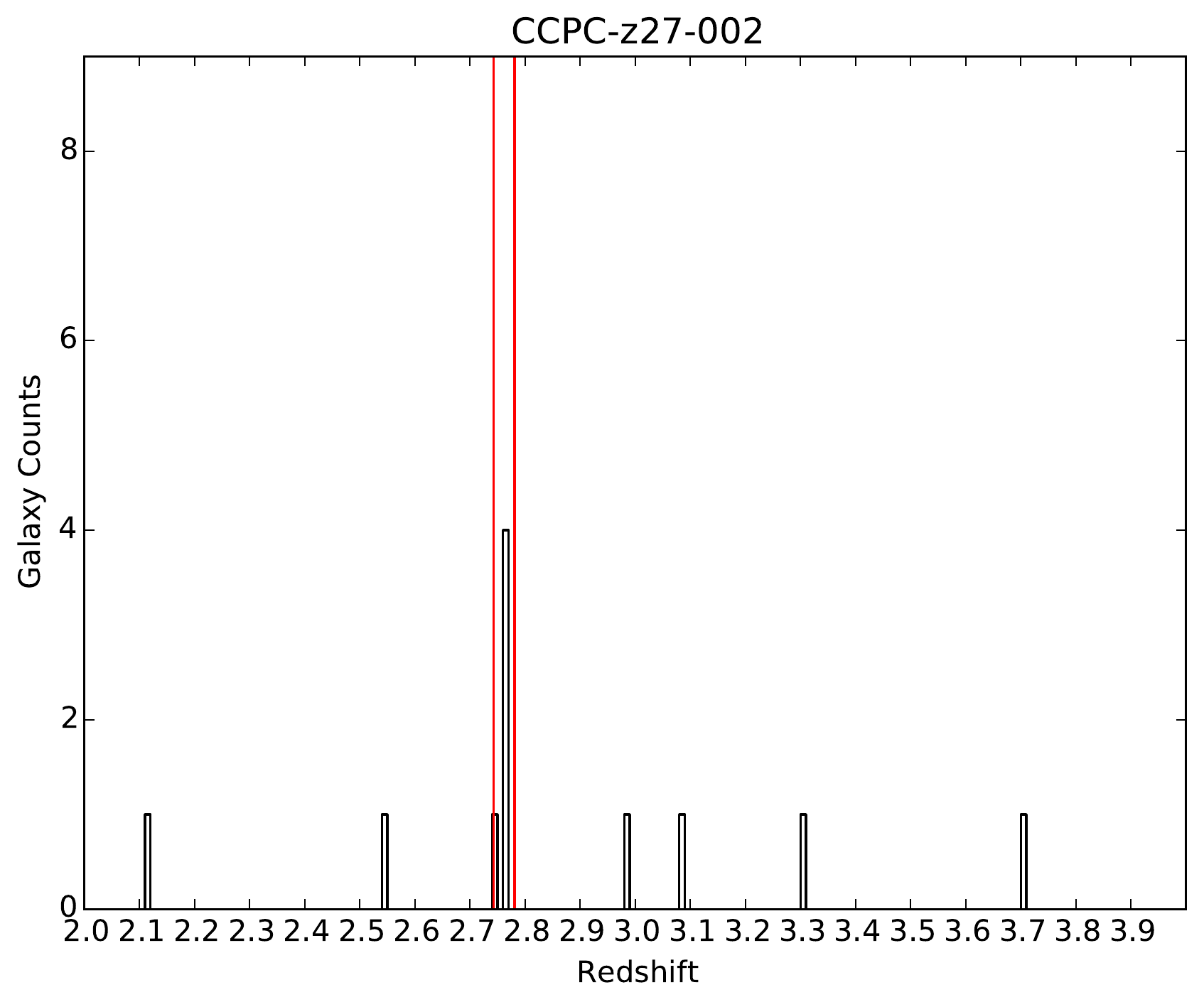}
\label{fig:CCPC-z27-002}
\end{subfigure}
\hfill
\end{figure*}

\begin{figure*}
\centering
\begin{subfigure}
\centering
\includegraphics[height=7.5cm,width=7.5cm]{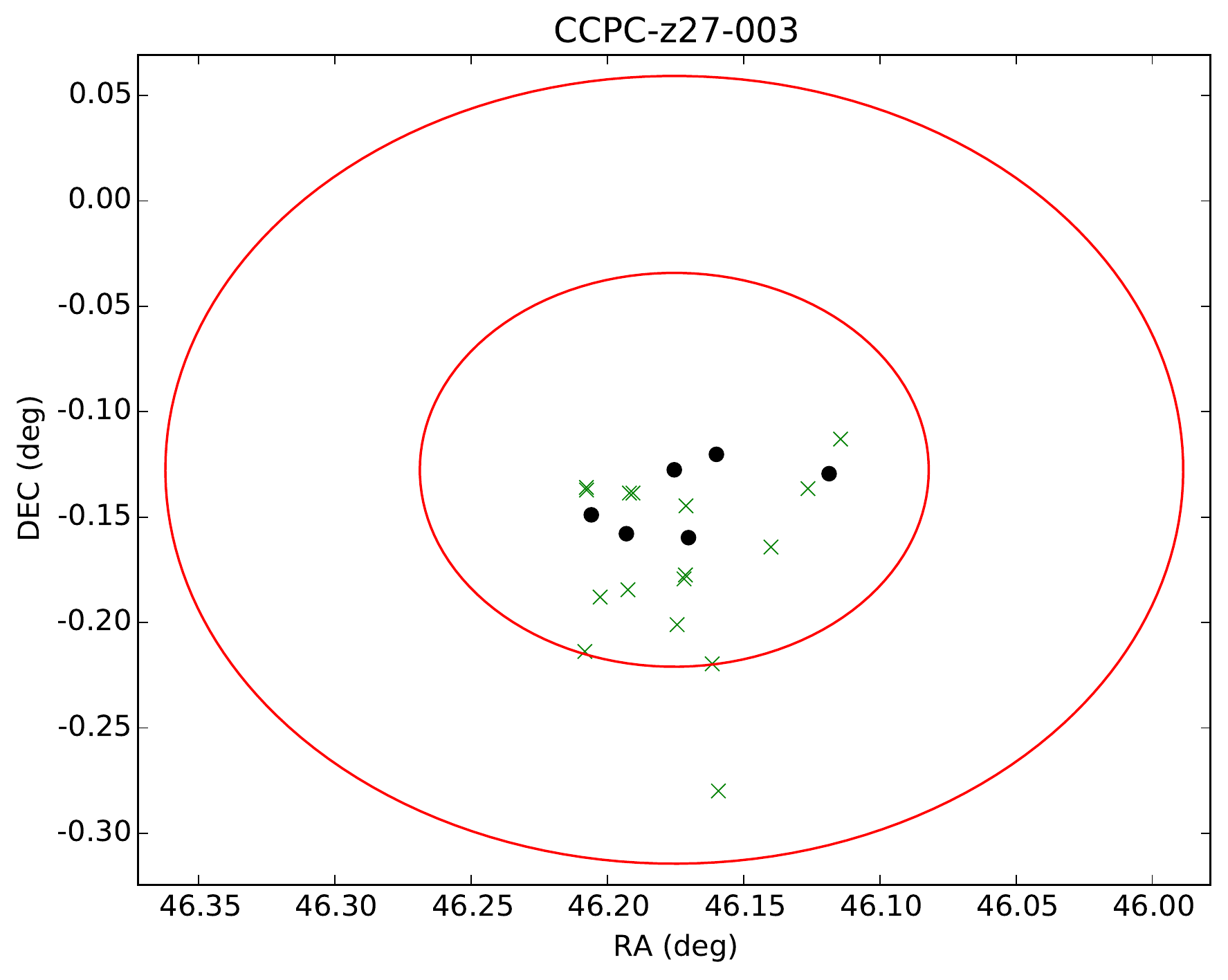}
\label{fig:CCPC-z27-003}
\end{subfigure}
\hfill
\begin{subfigure}
\centering
\includegraphics[scale=0.52]{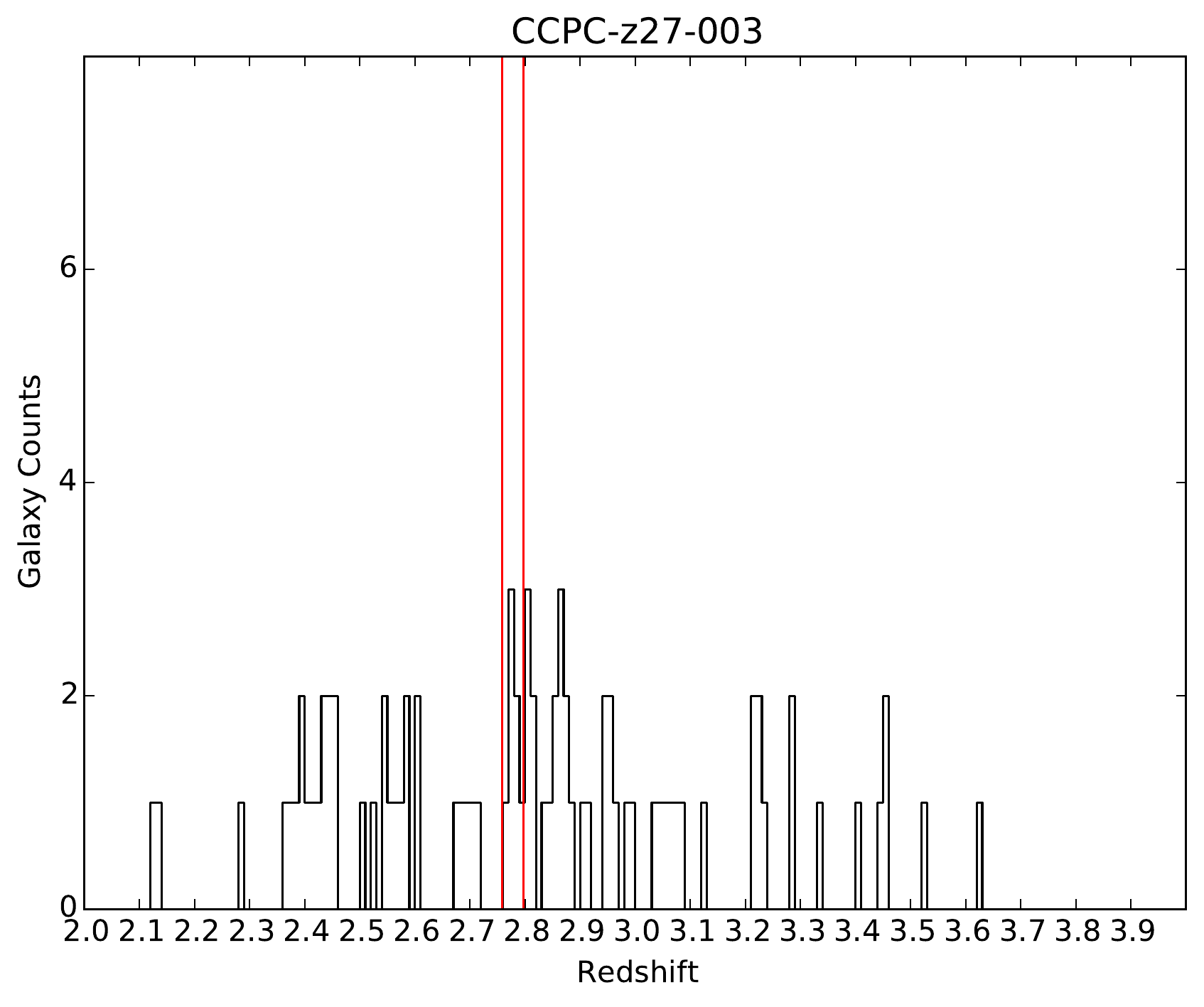}
\label{fig:CCPC-z27-003}
\end{subfigure}
\hfill
\end{figure*}

\clearpage

\begin{figure*}
\centering
\begin{subfigure}
\centering
\includegraphics[height=7.5cm,width=7.5cm]{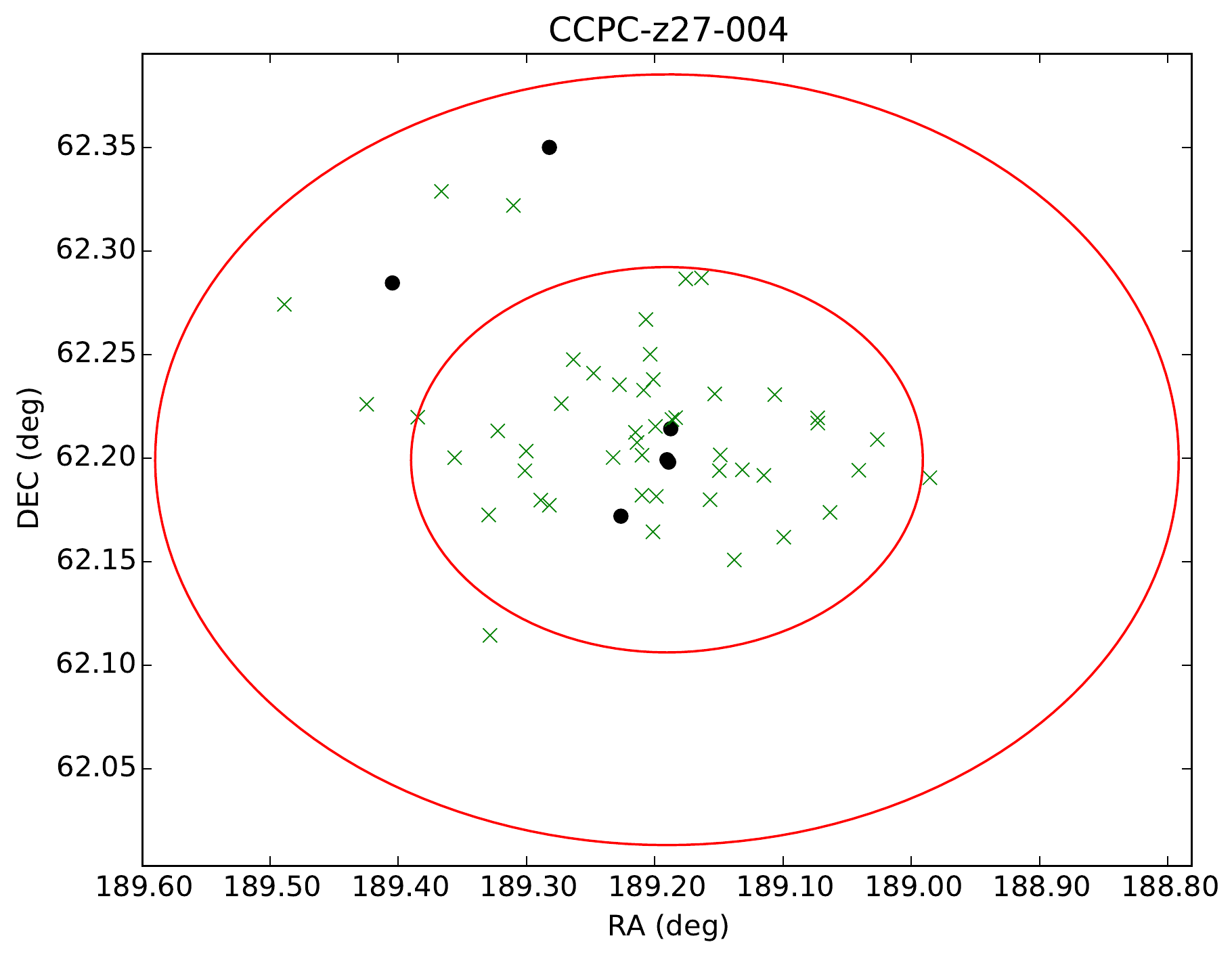}
\label{fig:CCPC-z27-004}
\end{subfigure}
\hfill
\begin{subfigure}
\centering
\includegraphics[scale=0.52]{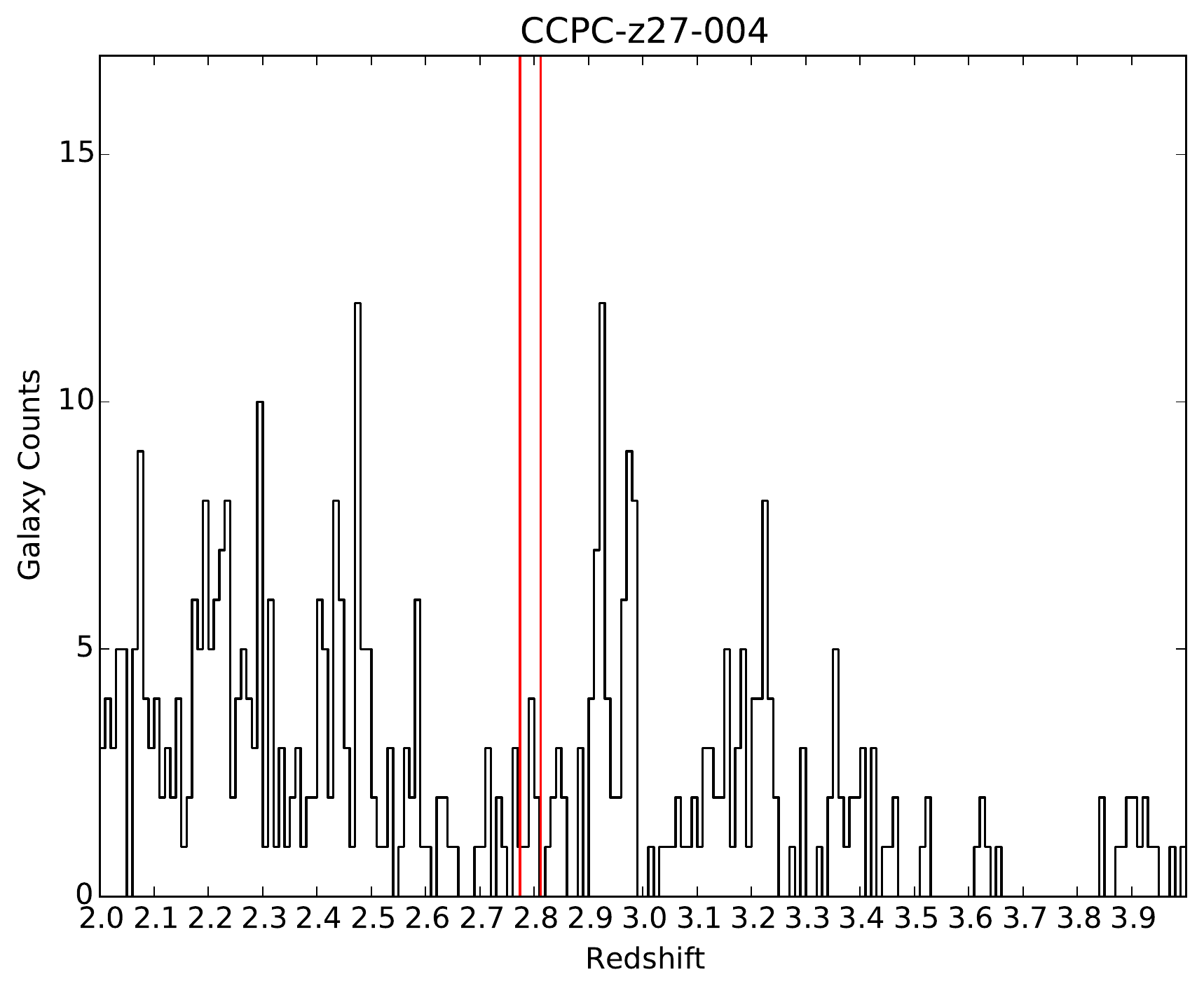}
\label{fig:CCPC-z27-004}
\end{subfigure}
\hfill
\end{figure*}

\begin{figure*}
\centering
\begin{subfigure}
\centering
\includegraphics[height=7.5cm,width=7.5cm]{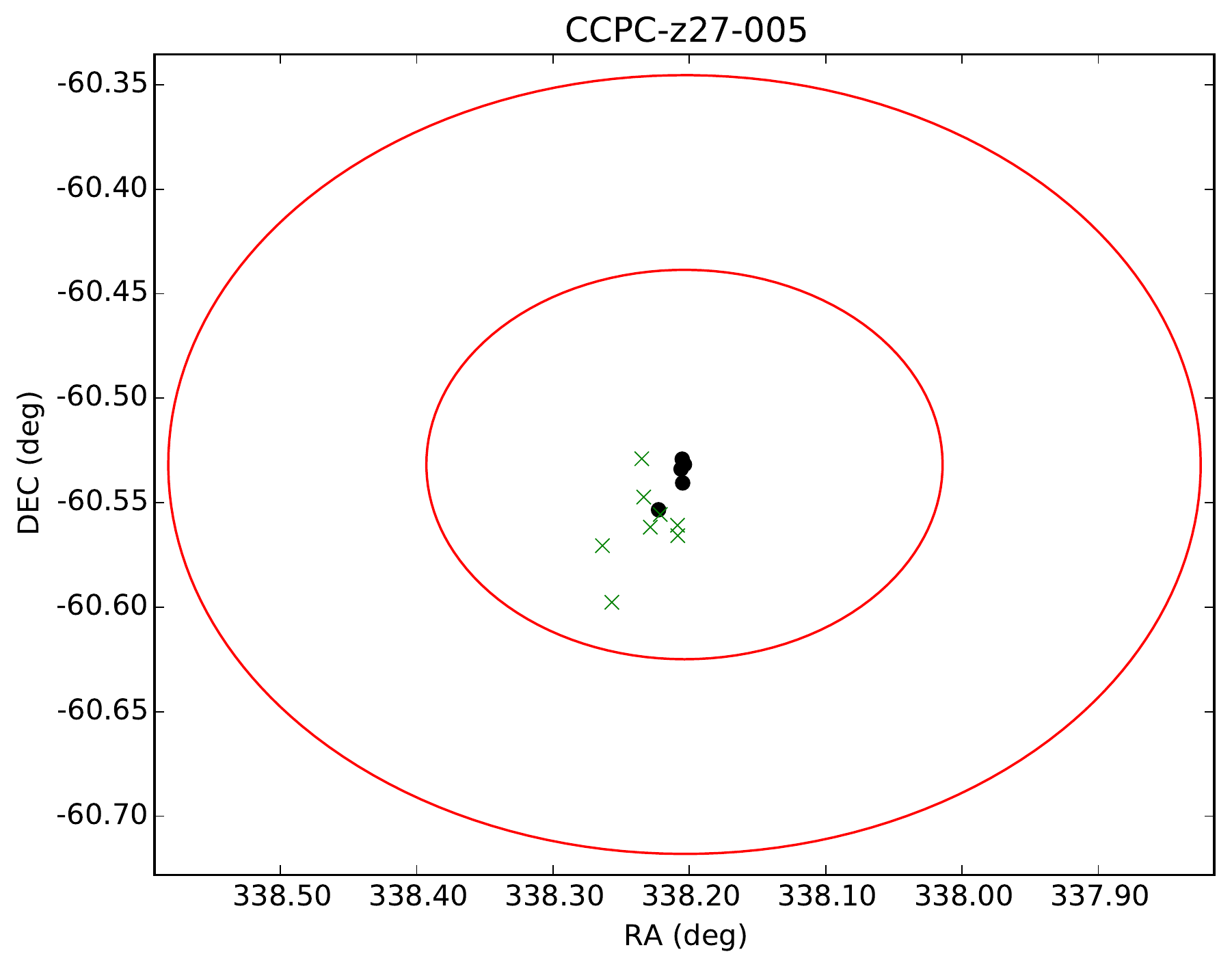}
\label{fig:CCPC-z27-005}
\end{subfigure}
\hfill
\begin{subfigure}
\centering
\includegraphics[scale=0.52]{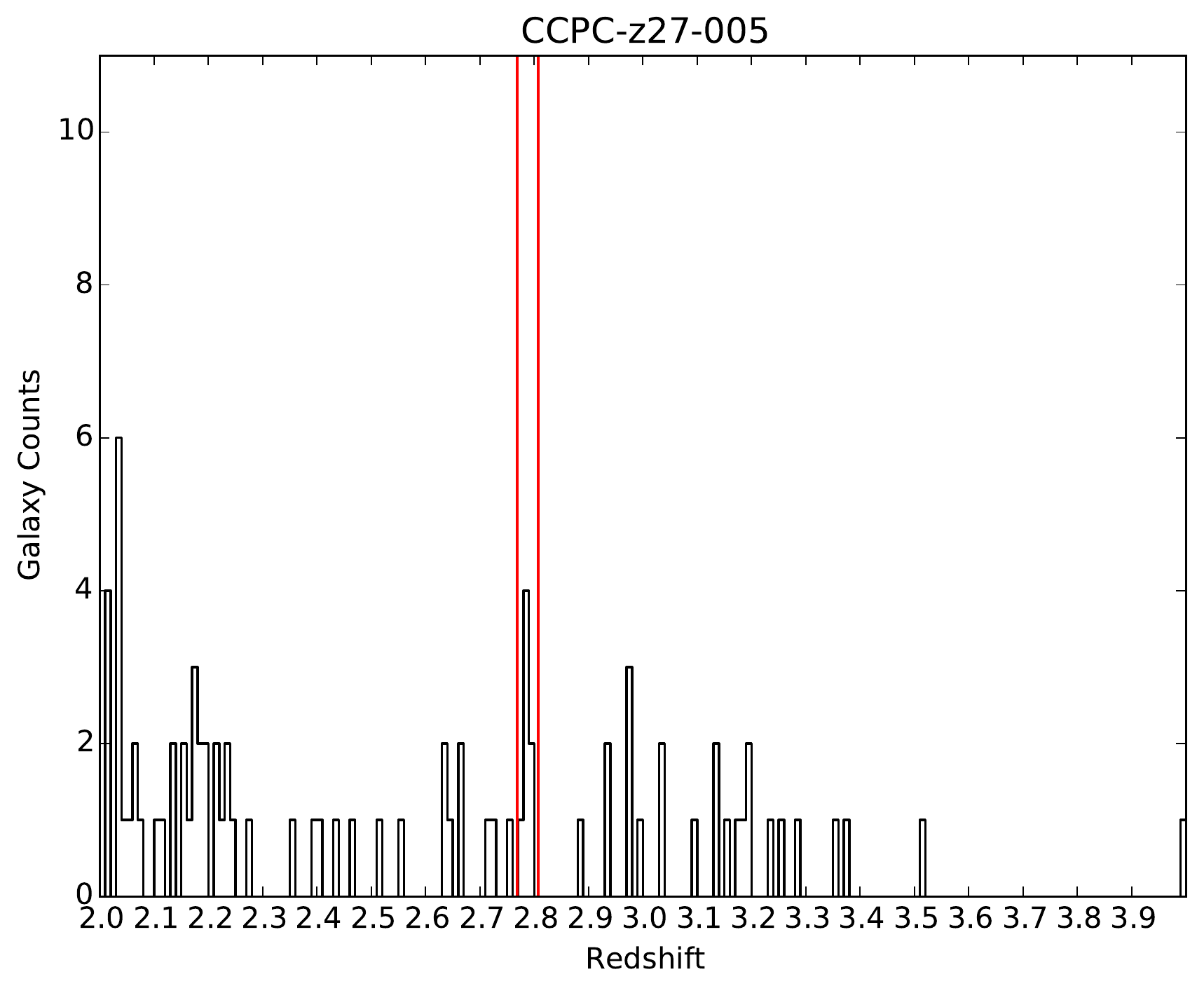}
\label{fig:CCPC-z27-005}
\end{subfigure}
\hfill
\end{figure*}

\clearpage 

\begin{figure*}
\centering
\begin{subfigure}
\centering
\includegraphics[height=7.5cm,width=7.5cm]{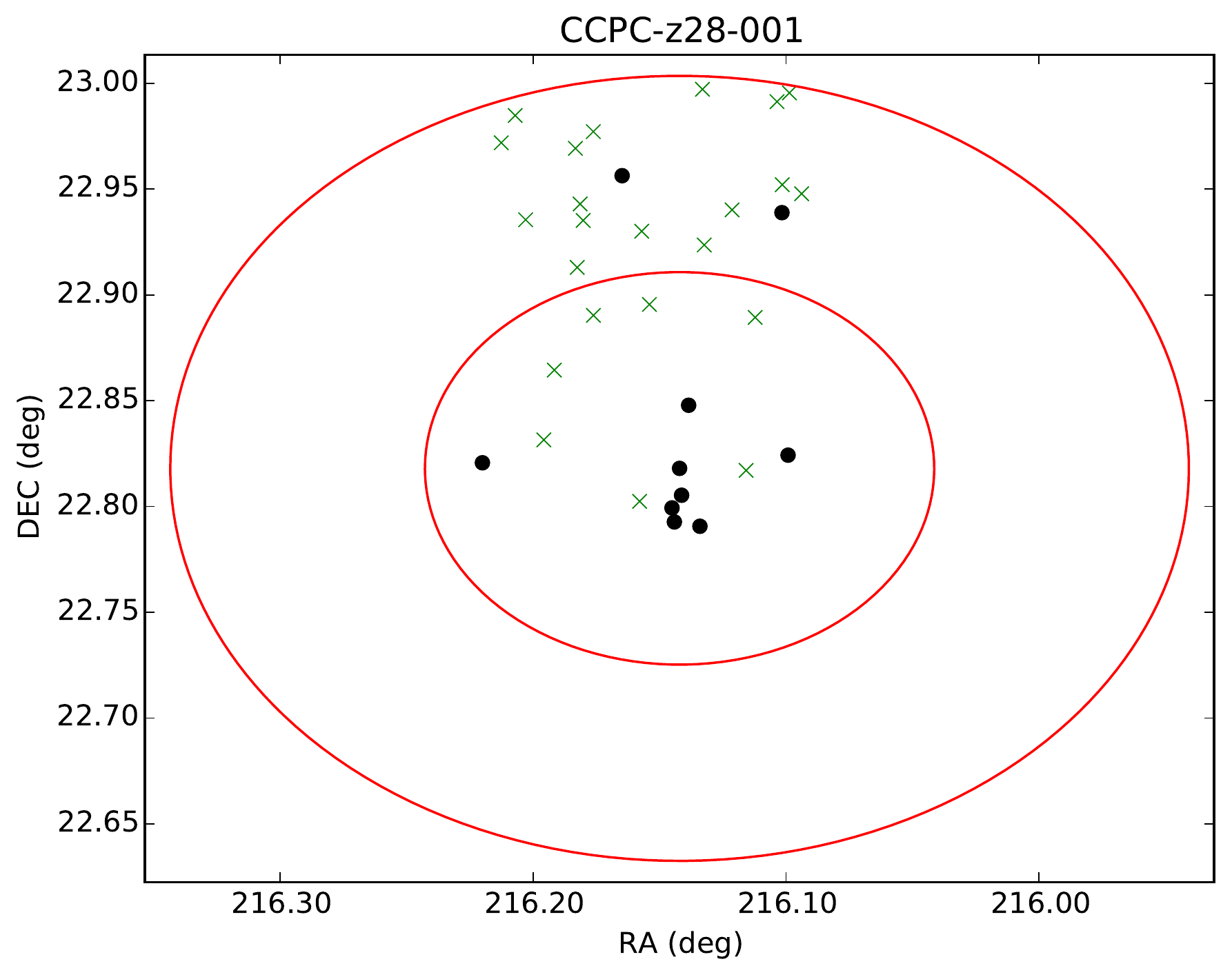}
\label{fig:CCPC-z28-001}
\end{subfigure}
\hfill
\begin{subfigure}
\centering
\includegraphics[scale=0.52]{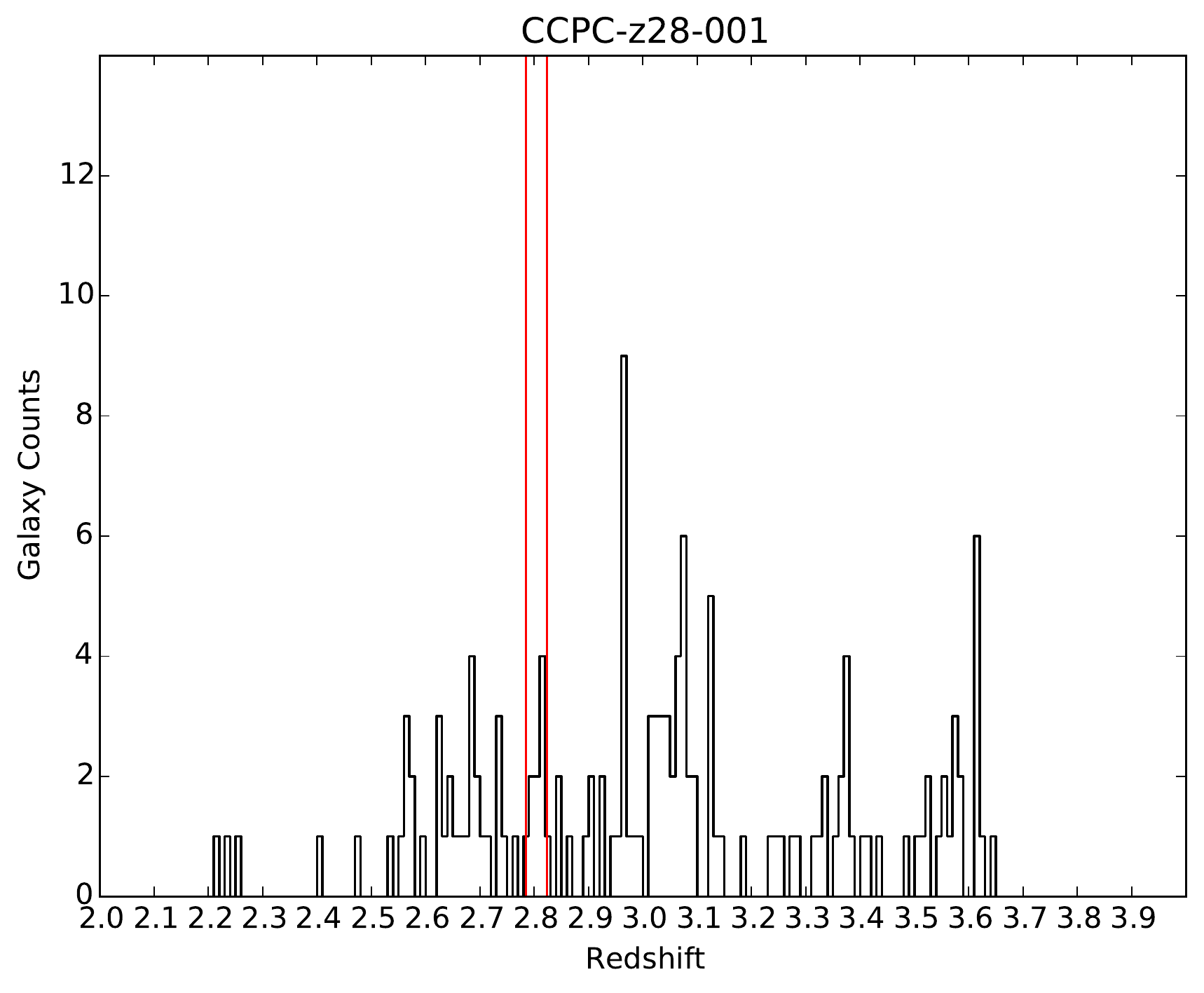}
\label{fig:CCPC-z28-001}
\end{subfigure}
\hfill
\end{figure*}

\begin{figure*}
\centering
\begin{subfigure}
\centering
\includegraphics[height=7.5cm,width=7.5cm]{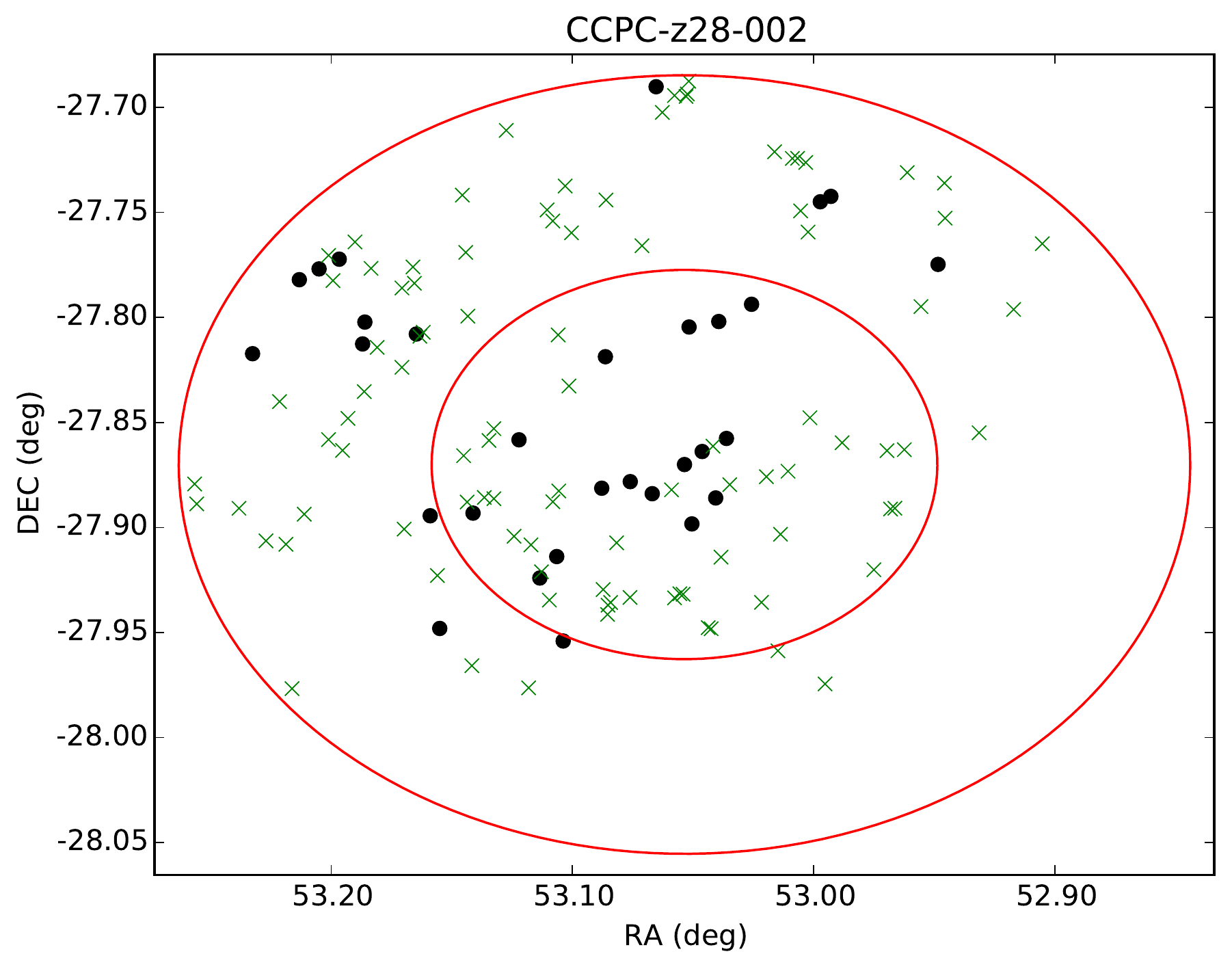}
\label{fig:CCPC-z28-002}
\end{subfigure}
\hfill
\begin{subfigure}
\centering
\includegraphics[scale=0.52]{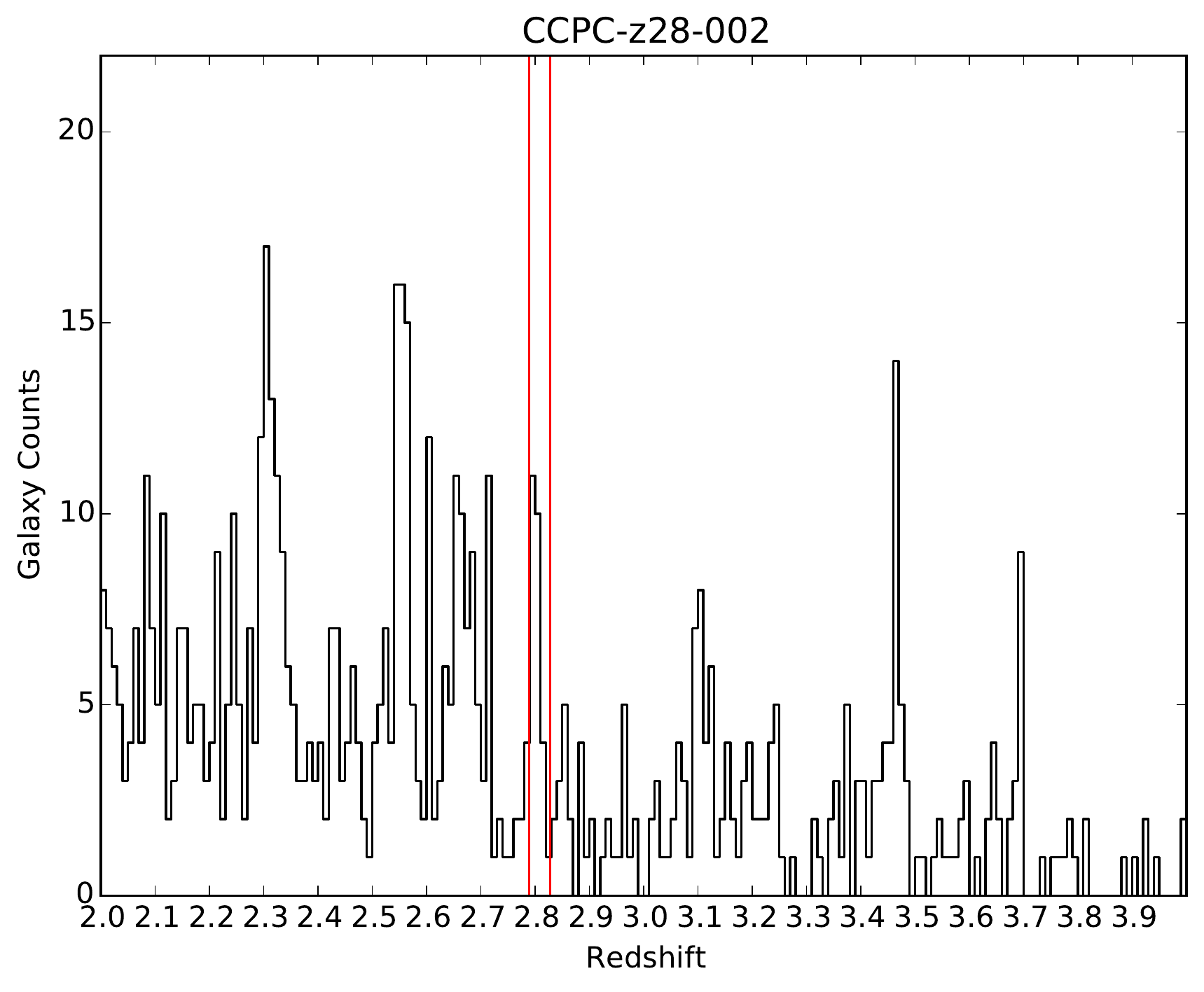}
\label{fig:CCPC-z28-002}
\end{subfigure}
\hfill
\end{figure*}

\begin{figure*}
\centering
\begin{subfigure}
\centering
\includegraphics[height=7.5cm,width=7.5cm]{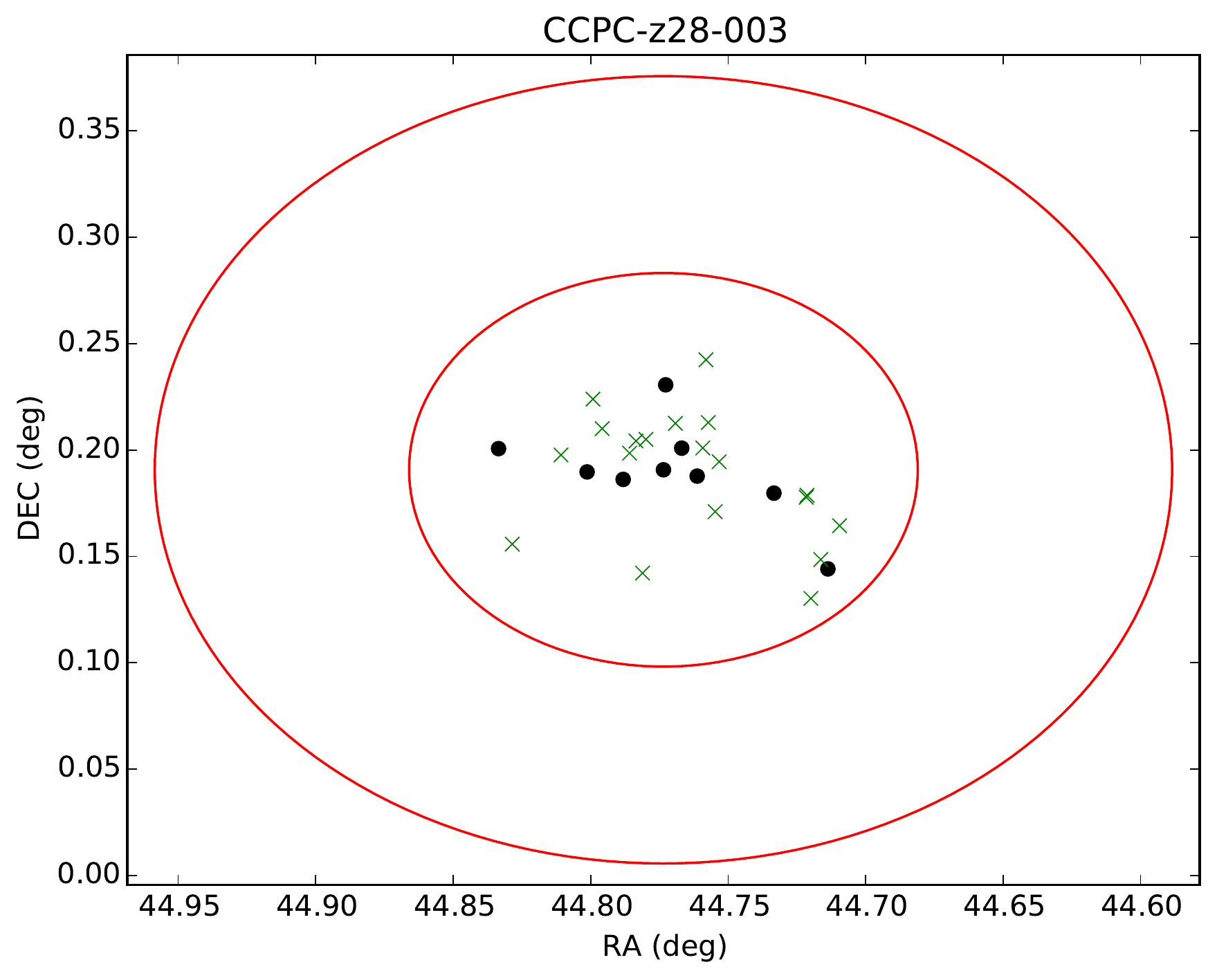}
\label{fig:CCPC-z28-003}
\end{subfigure}
\hfill
\begin{subfigure}
\centering
\includegraphics[scale=0.52]{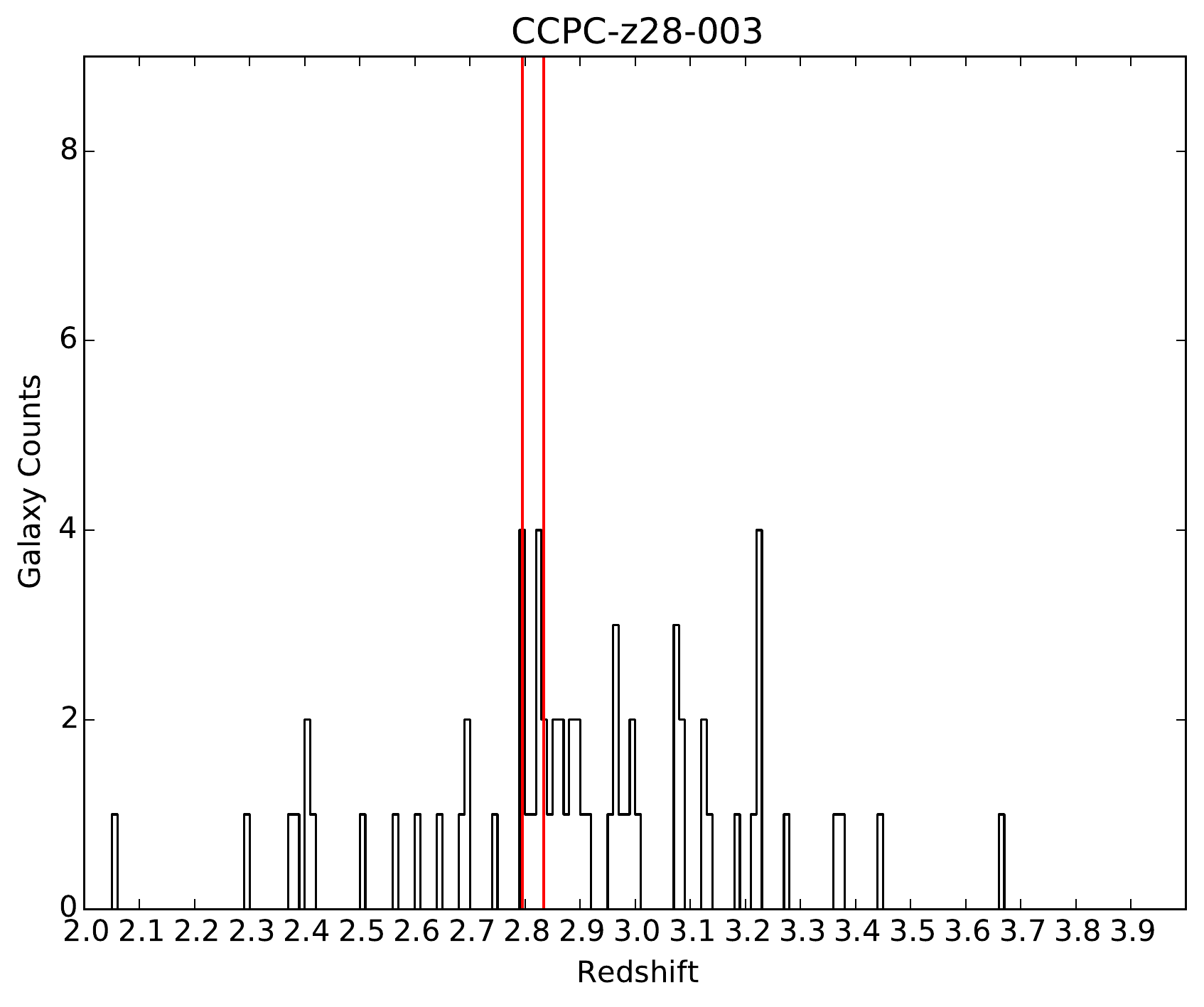}
\label{fig:CCPC-z28-003}
\end{subfigure}
\hfill
\end{figure*}
\clearpage

\begin{figure*}
\centering
\begin{subfigure}
\centering
\includegraphics[height=7.5cm,width=7.5cm]{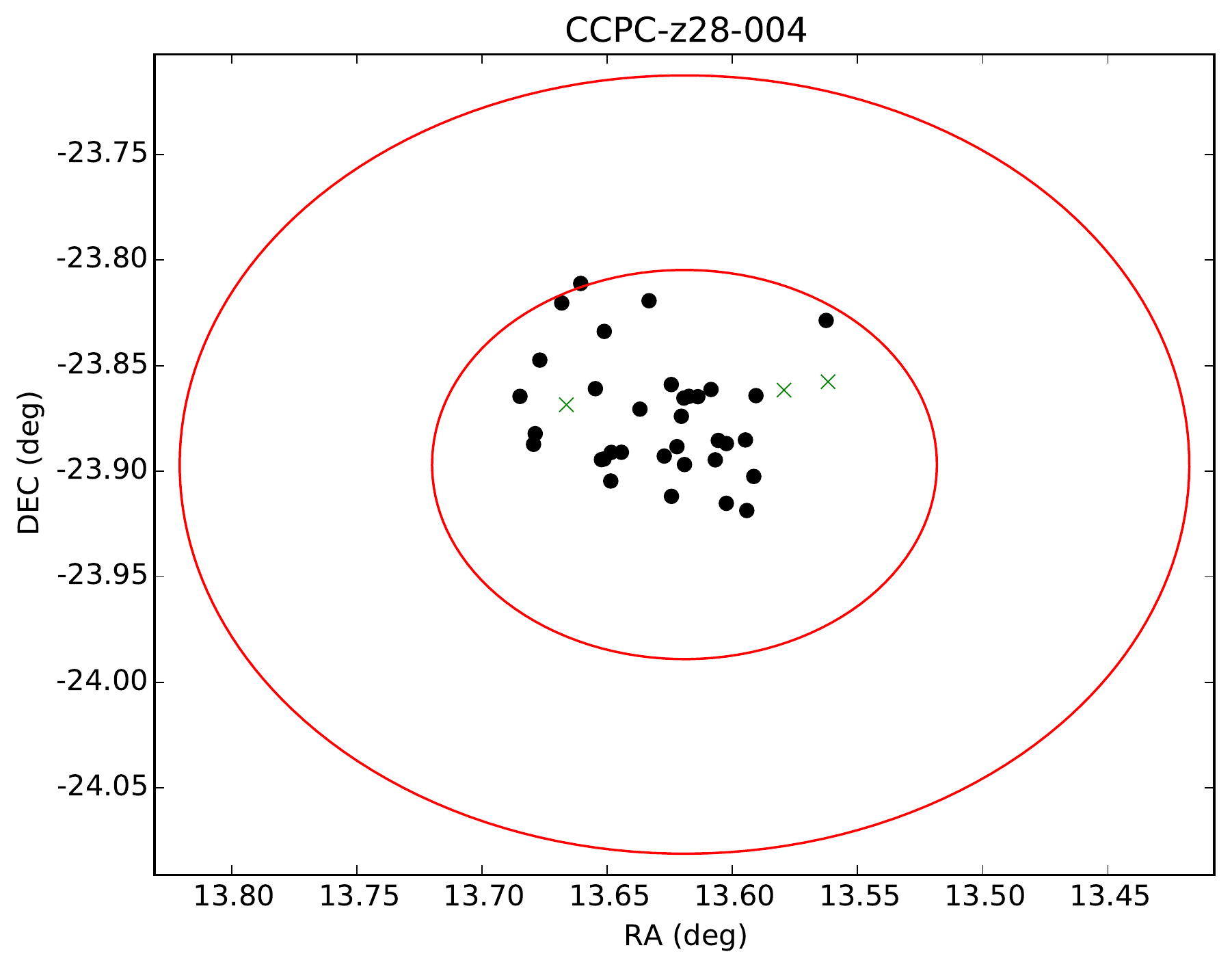}
\label{fig:CCPC-z28-004}
\end{subfigure}
\hfill
\begin{subfigure}
\centering
\includegraphics[scale=0.52]{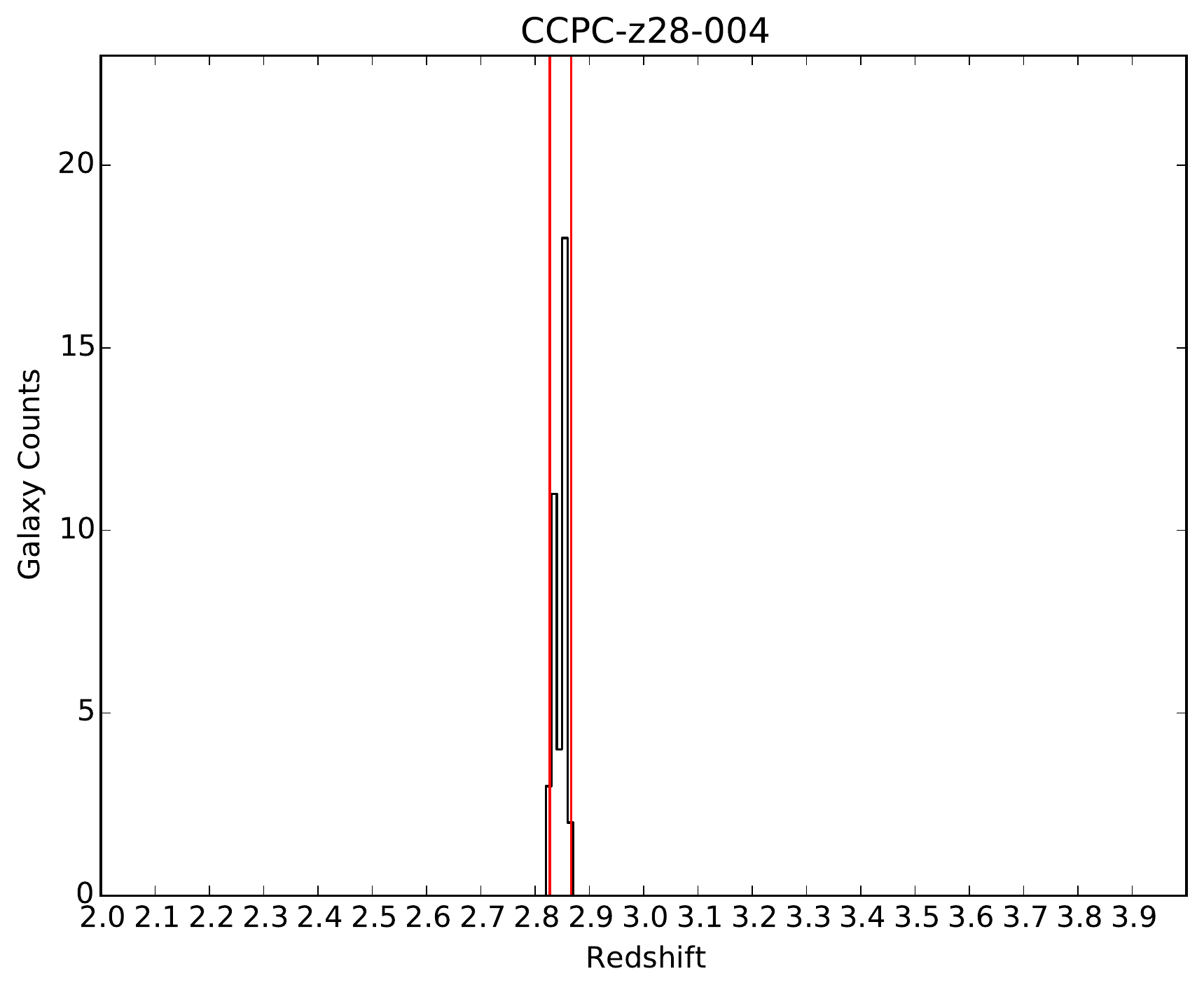}
\label{fig:CCPC-z28-004}
\end{subfigure}
\hfill
\end{figure*}

\begin{figure*}
\centering
\begin{subfigure}
\centering
\includegraphics[height=7.5cm,width=7.5cm]{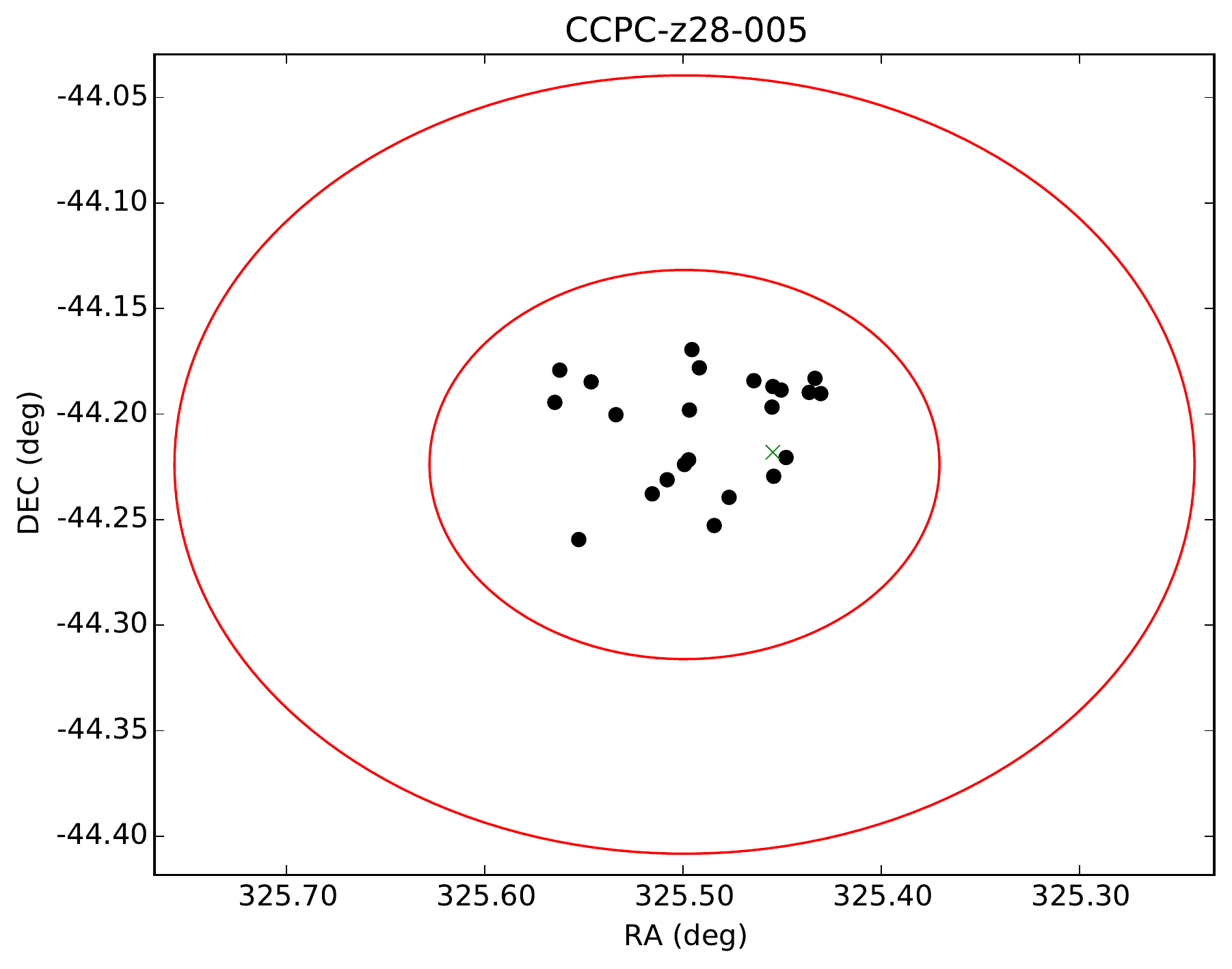}
\label{fig:CCPC-z28-005}
\end{subfigure}
\hfill
\begin{subfigure}
\centering
\includegraphics[scale=0.52]{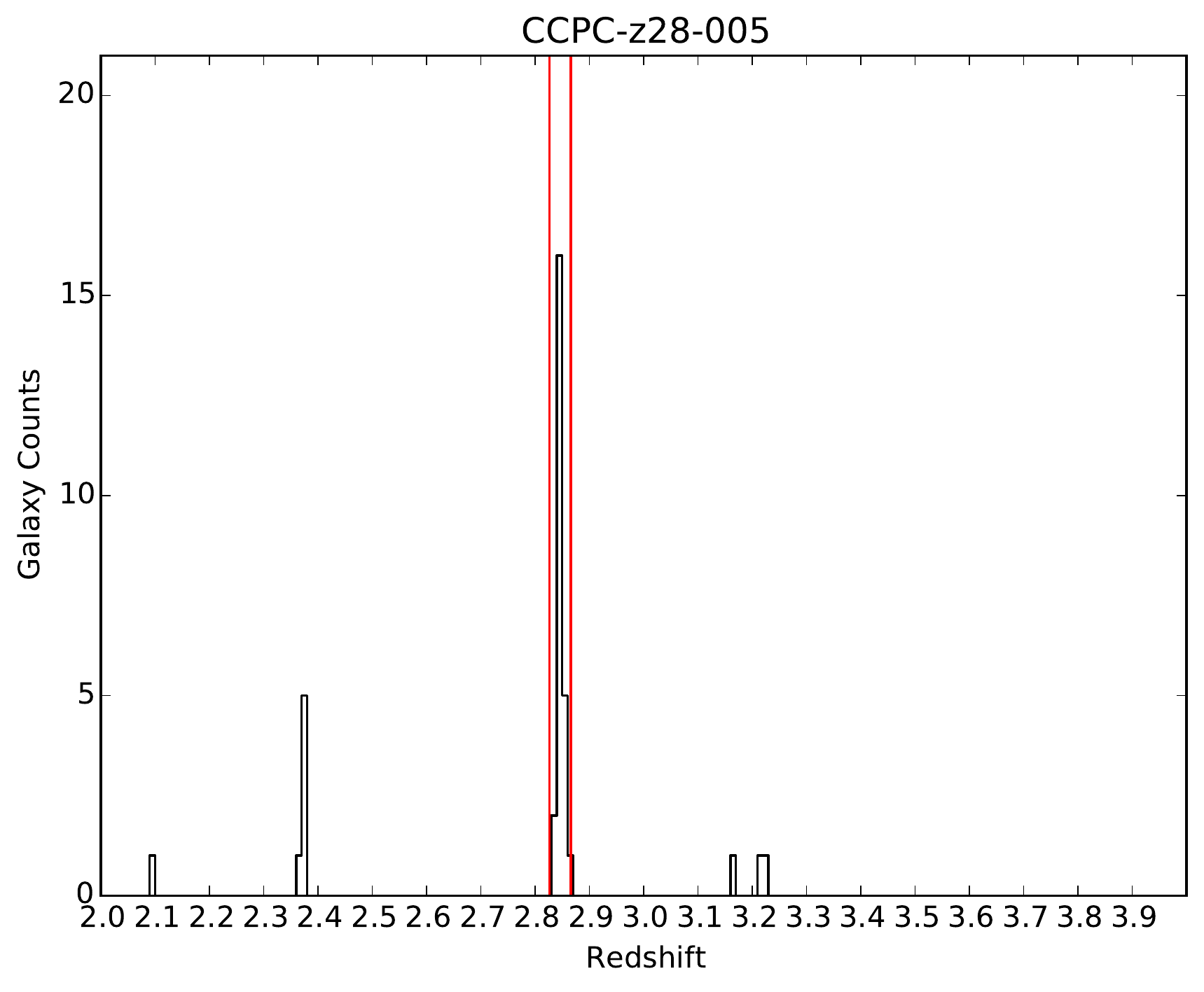}
\label{fig:CCPC-z28-005}
\end{subfigure}
\hfill
\end{figure*}

\begin{figure*}
\centering
\begin{subfigure}
\centering
\includegraphics[height=7.5cm,width=7.5cm]{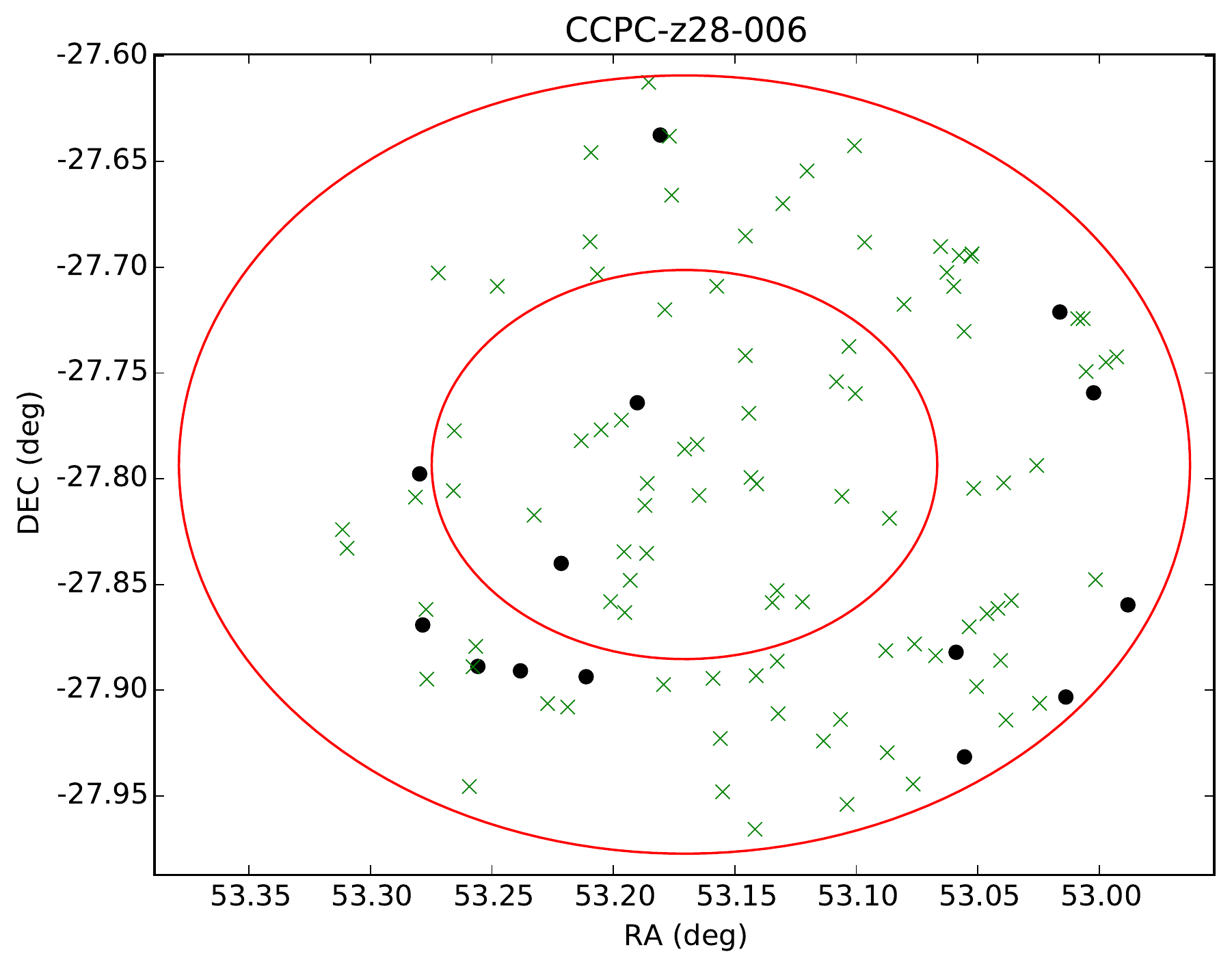}
\label{fig:CCPC-z28-006}
\end{subfigure}
\hfill
\begin{subfigure}
\centering
\includegraphics[scale=0.52]{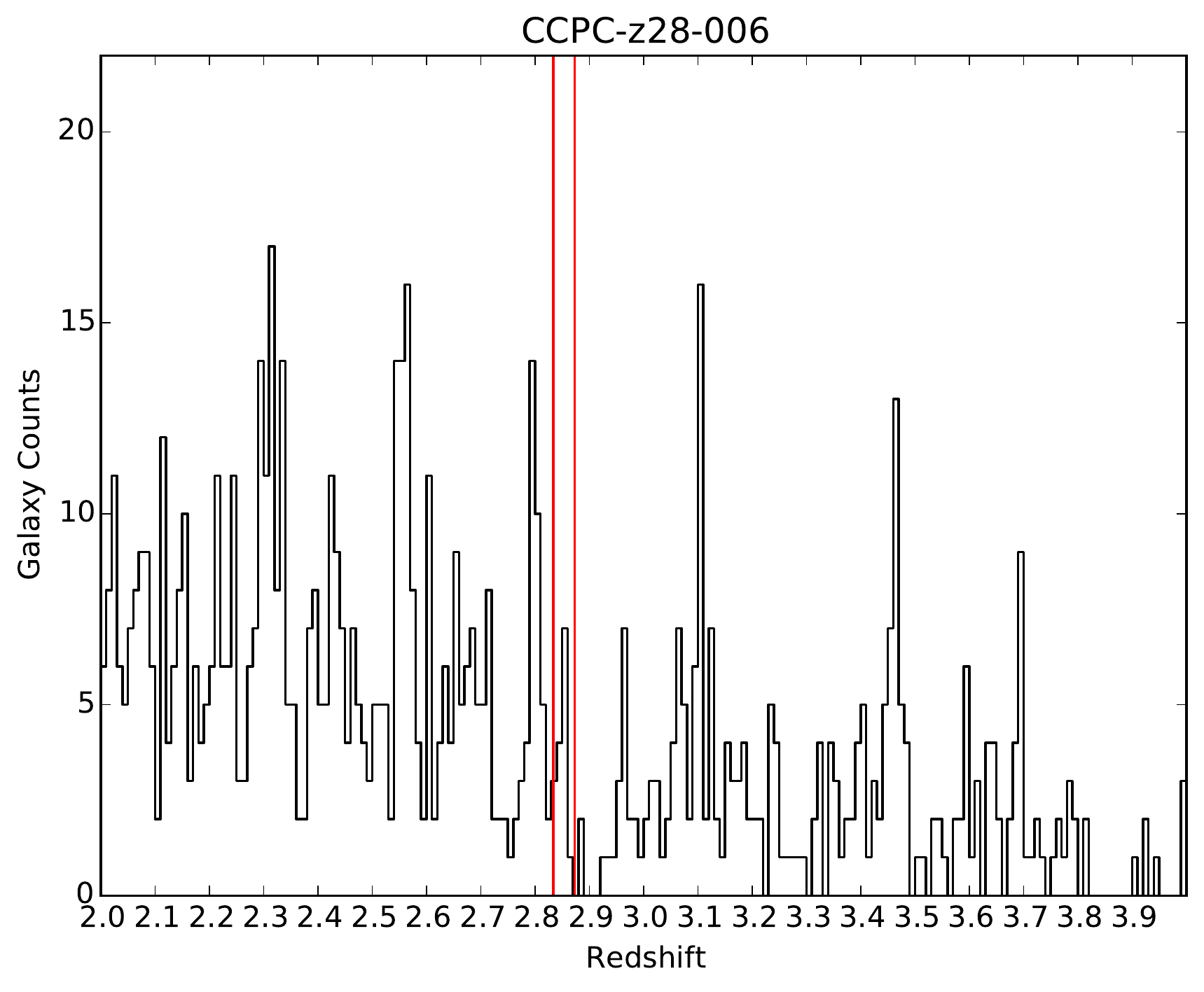}
\label{fig:CCPC-z28-006}
\end{subfigure}
\hfill
\end{figure*}
\clearpage 

\begin{figure*}
\centering
\begin{subfigure}
\centering
\includegraphics[height=7.5cm,width=7.5cm]{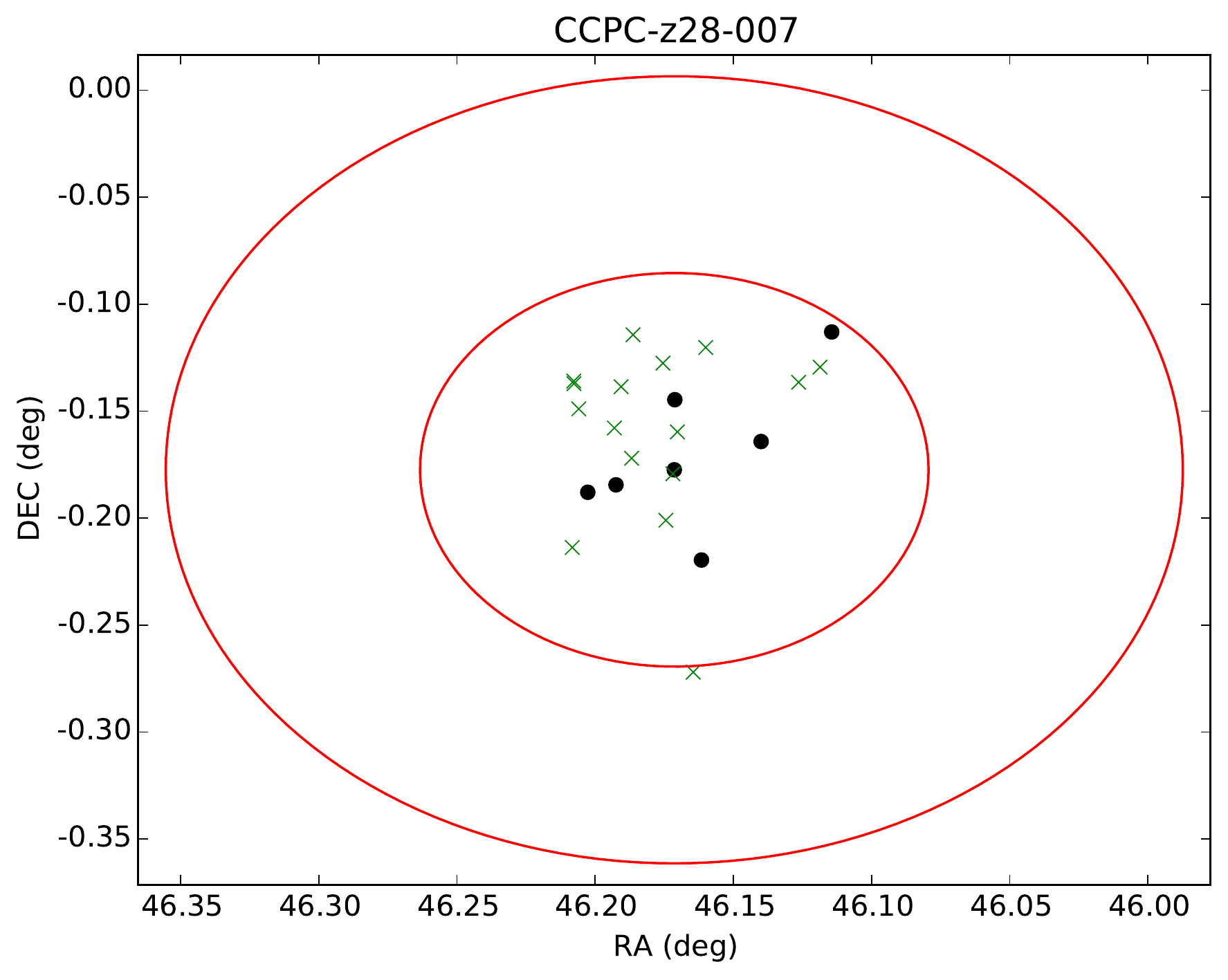}
\label{fig:CCPC-z28-007}
\end{subfigure}
\hfill
\begin{subfigure}
\centering
\includegraphics[scale=0.52]{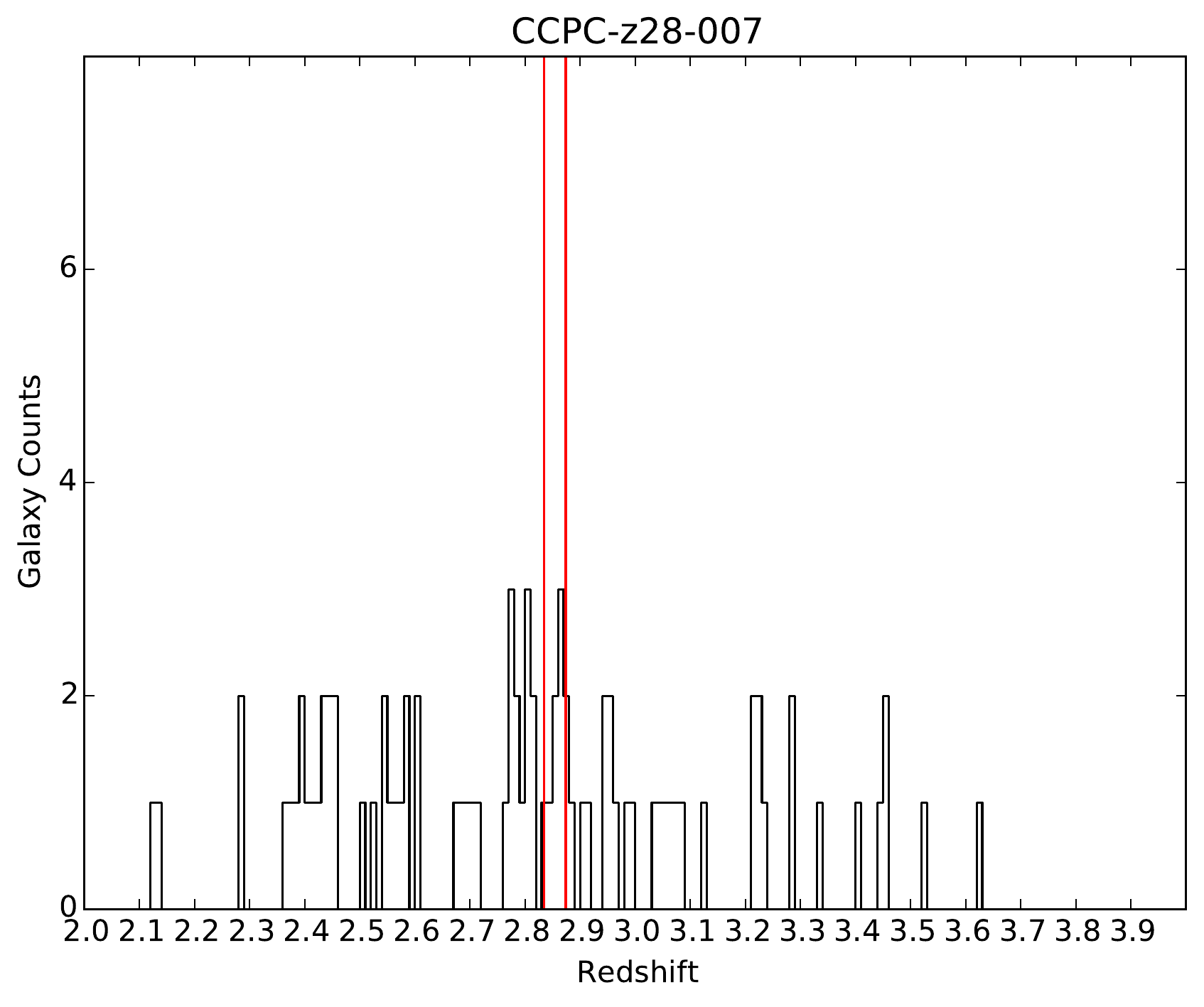}
\label{fig:CCPC-z28-007}
\end{subfigure}
\hfill
\end{figure*}

\begin{figure*}
\centering
\begin{subfigure}
\centering
\includegraphics[height=7.5cm,width=7.5cm]{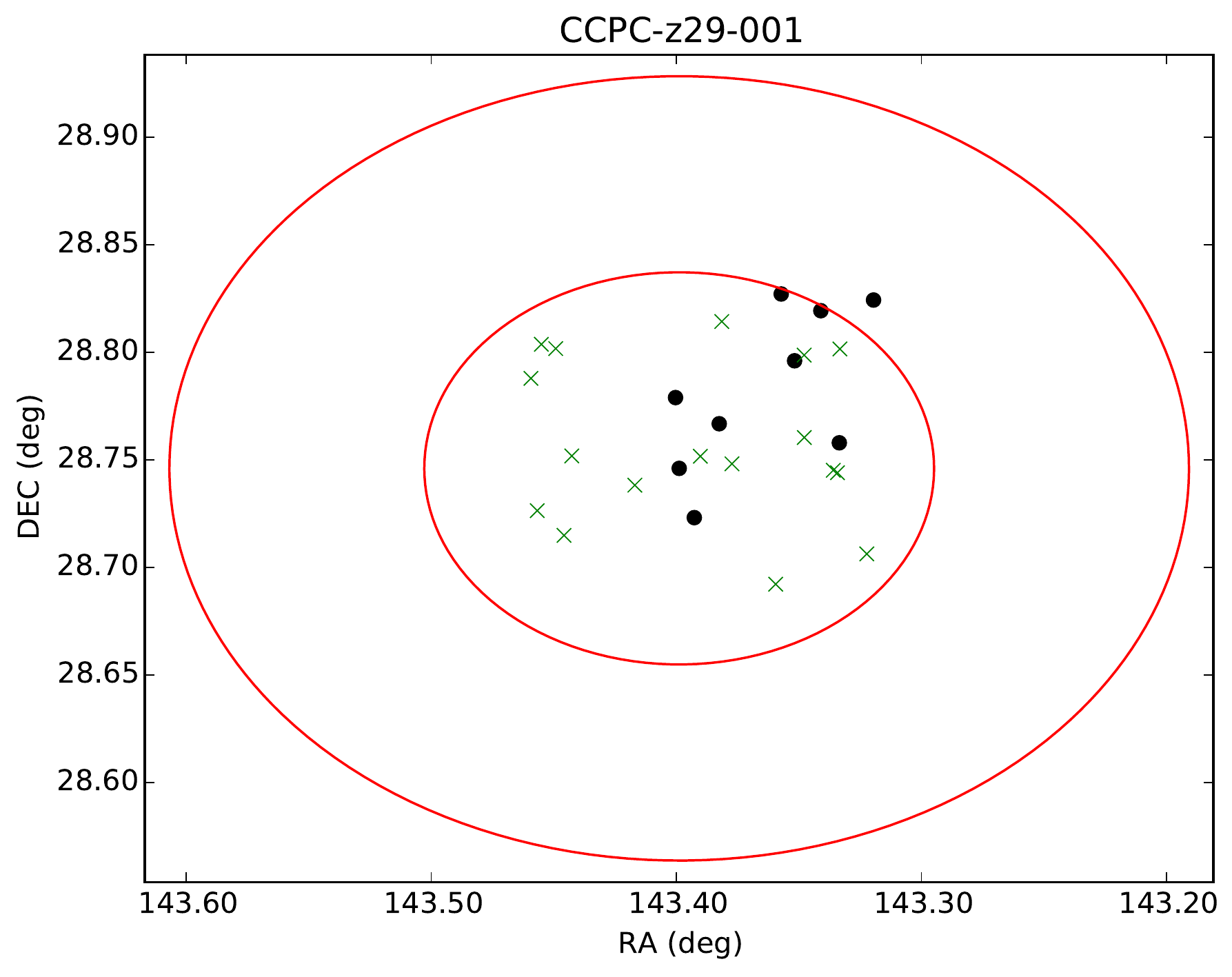}
\label{fig:CCPC-z29-001}
\end{subfigure}
\hfill
\begin{subfigure}
\centering
\includegraphics[scale=0.52]{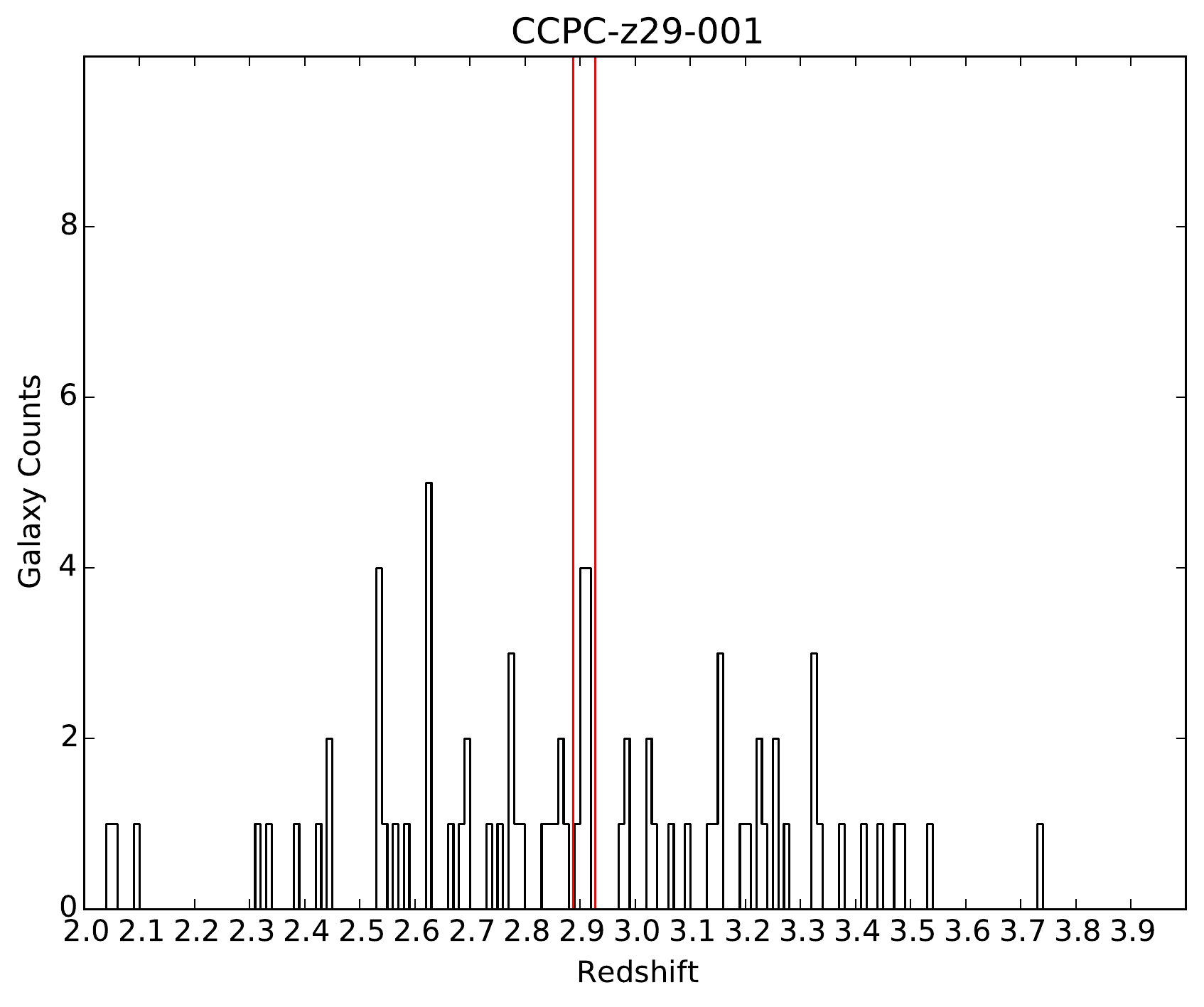}
\label{fig:CCPC-z29-001}
\end{subfigure}
\hfill
\end{figure*}

\begin{figure*}
\centering
\begin{subfigure}
\centering
\includegraphics[height=7.5cm,width=7.5cm]{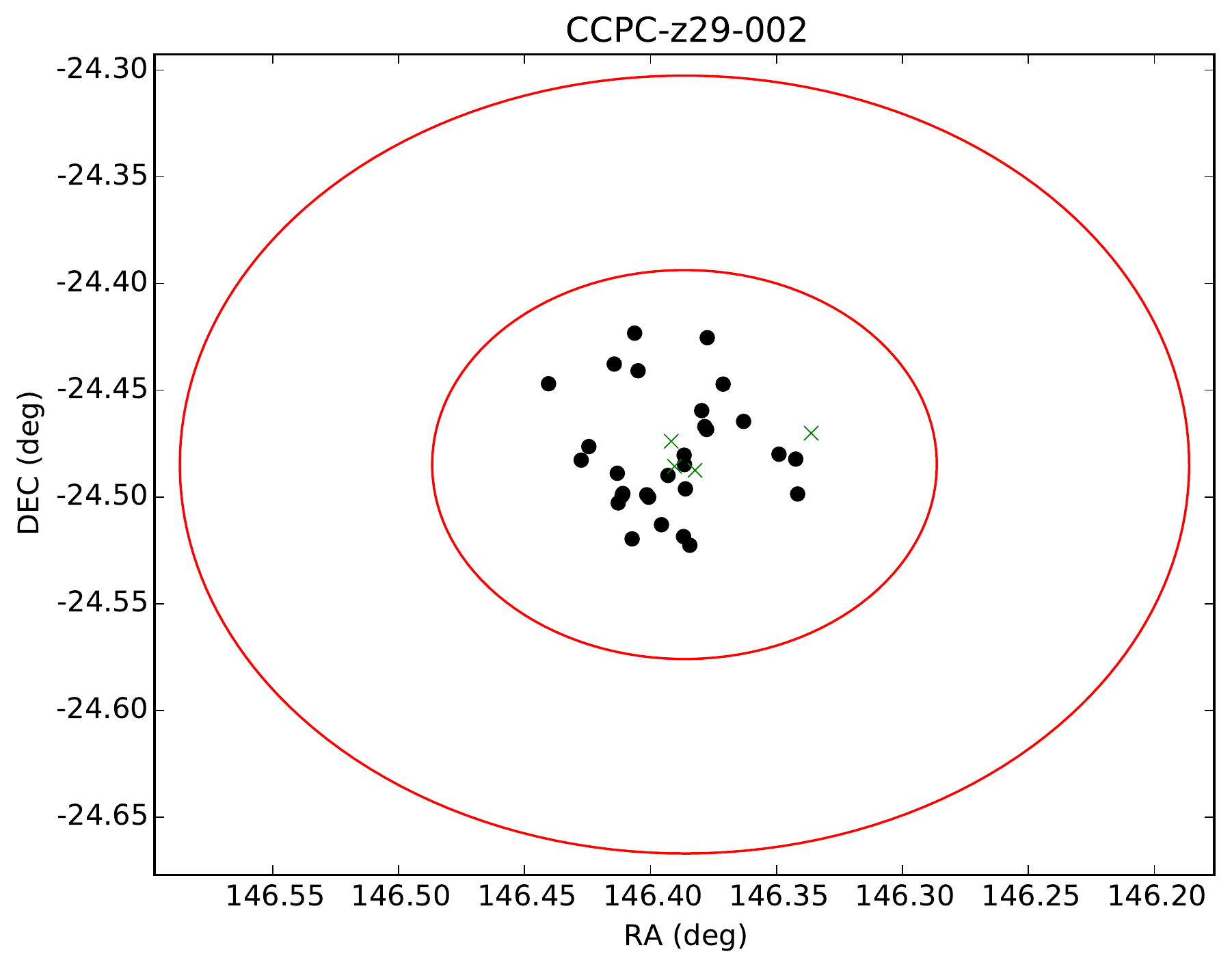}
\label{fig:CCPC-z29-002}
\end{subfigure}
\hfill
\begin{subfigure}
\centering
\includegraphics[scale=0.52]{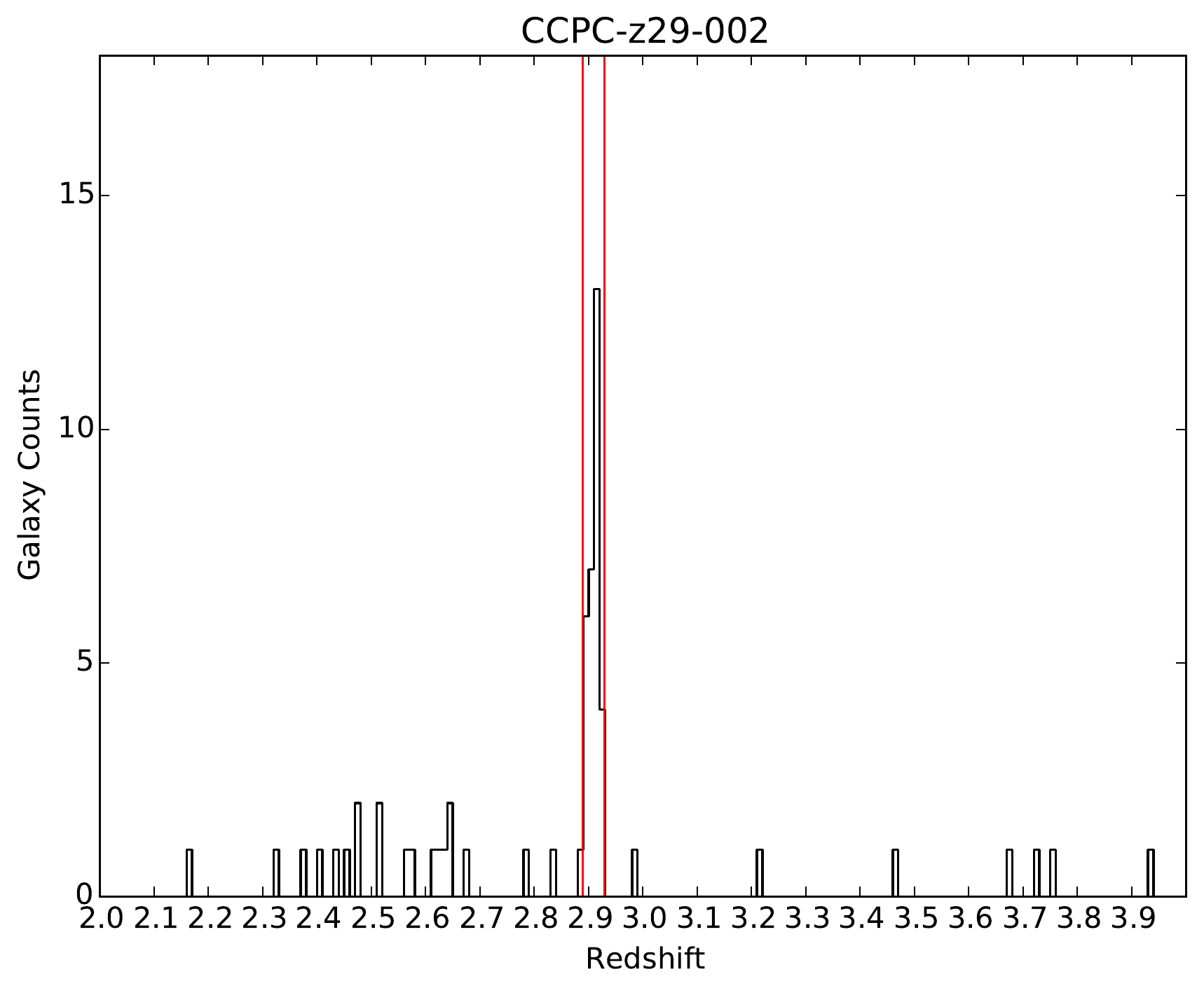}
\label{fig:CCPC-z29-002}
\end{subfigure}
\hfill
\end{figure*}
\clearpage 

\begin{figure*}
\centering
\begin{subfigure}
\centering
\includegraphics[height=7.5cm,width=7.5cm]{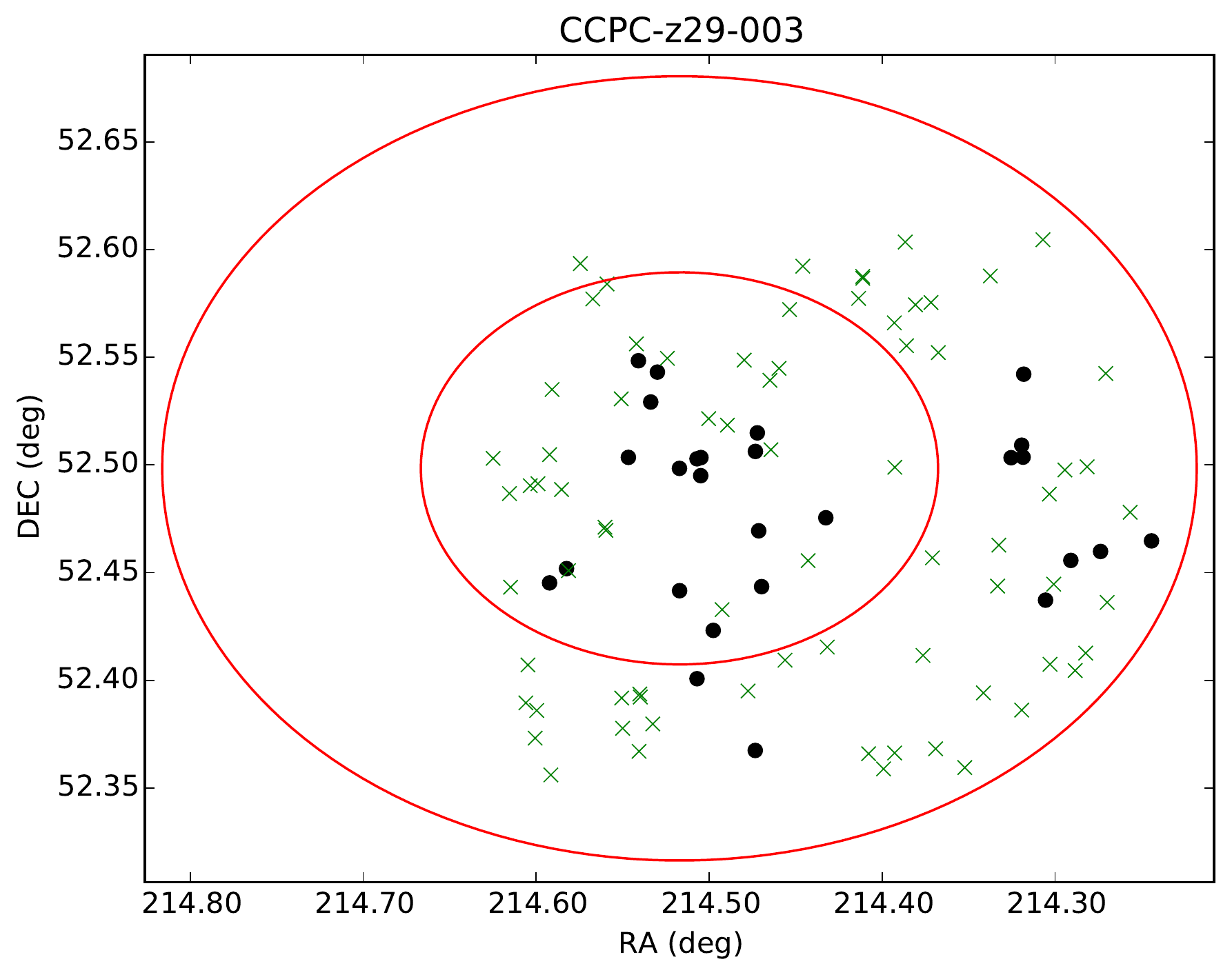}
\label{fig:CCPC-z29-003}
\end{subfigure}
\hfill
\begin{subfigure}
\centering
\includegraphics[scale=0.52]{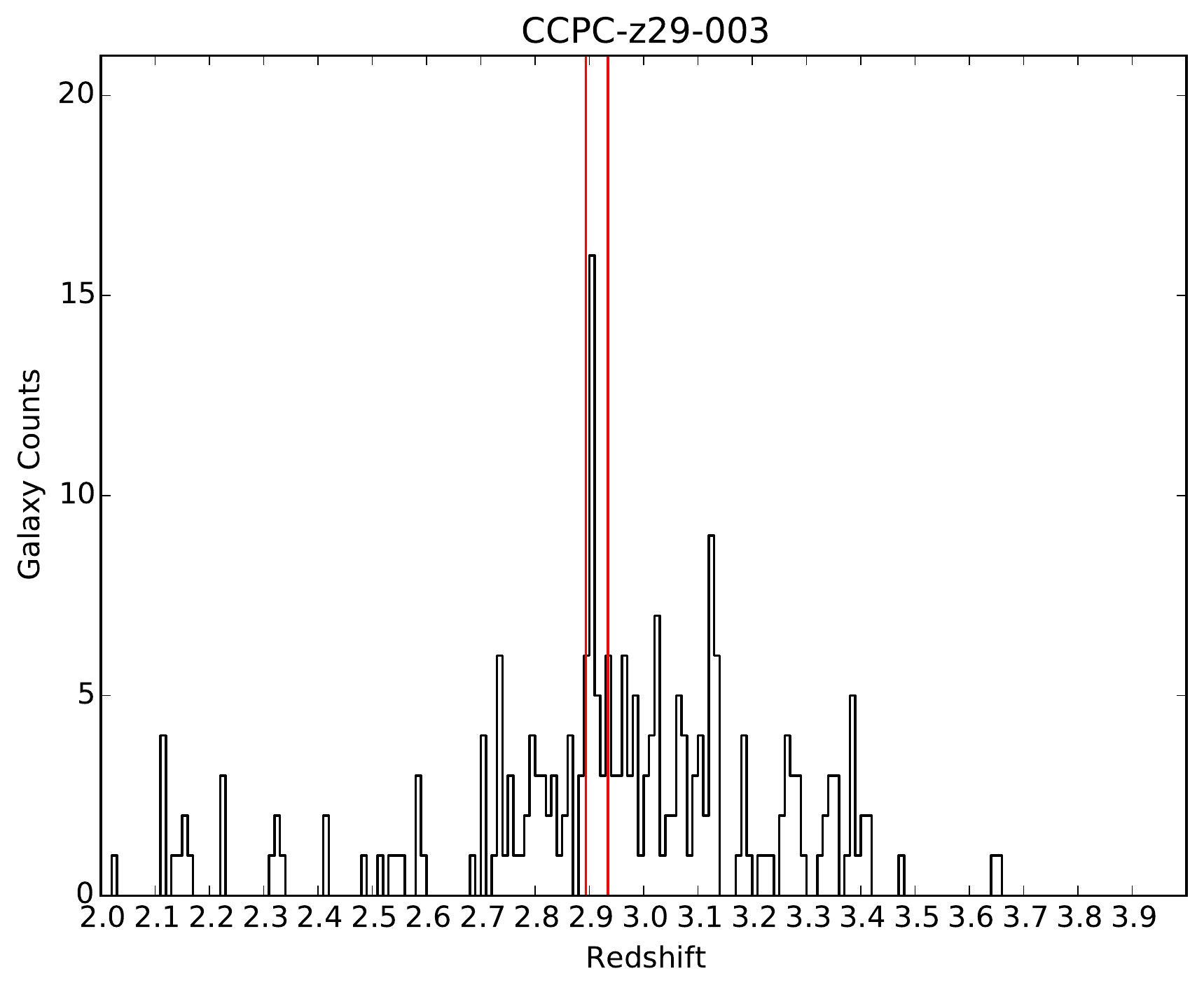}
\label{fig:CCPC-z29-003}
\end{subfigure}
\hfill
\end{figure*}

\begin{figure*}
\centering
\begin{subfigure}
\centering
\includegraphics[height=7.5cm,width=7.5cm]{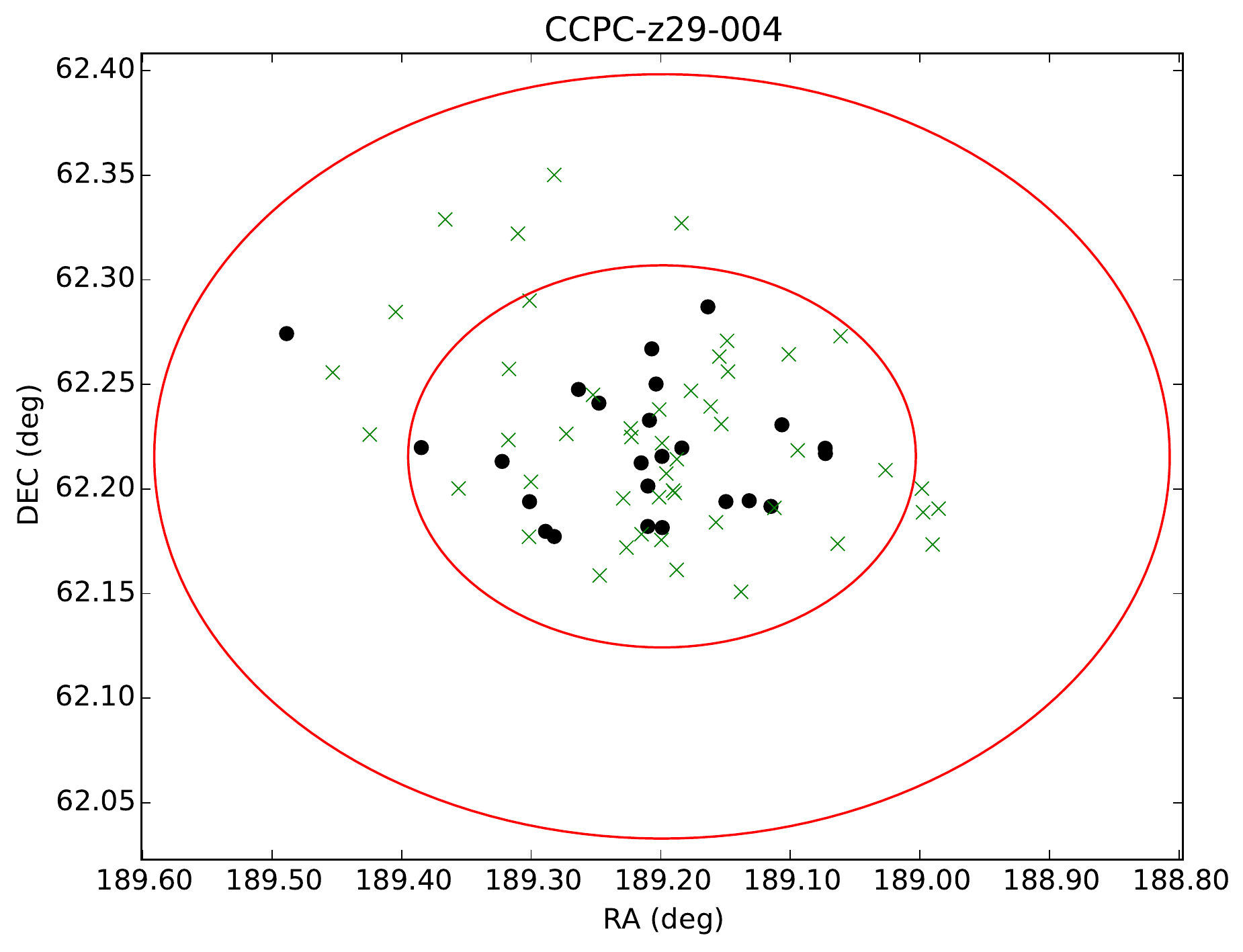}
\label{fig:CCPC-z29-004}
\end{subfigure}
\hfill
\begin{subfigure}
\centering
\includegraphics[scale=0.52]{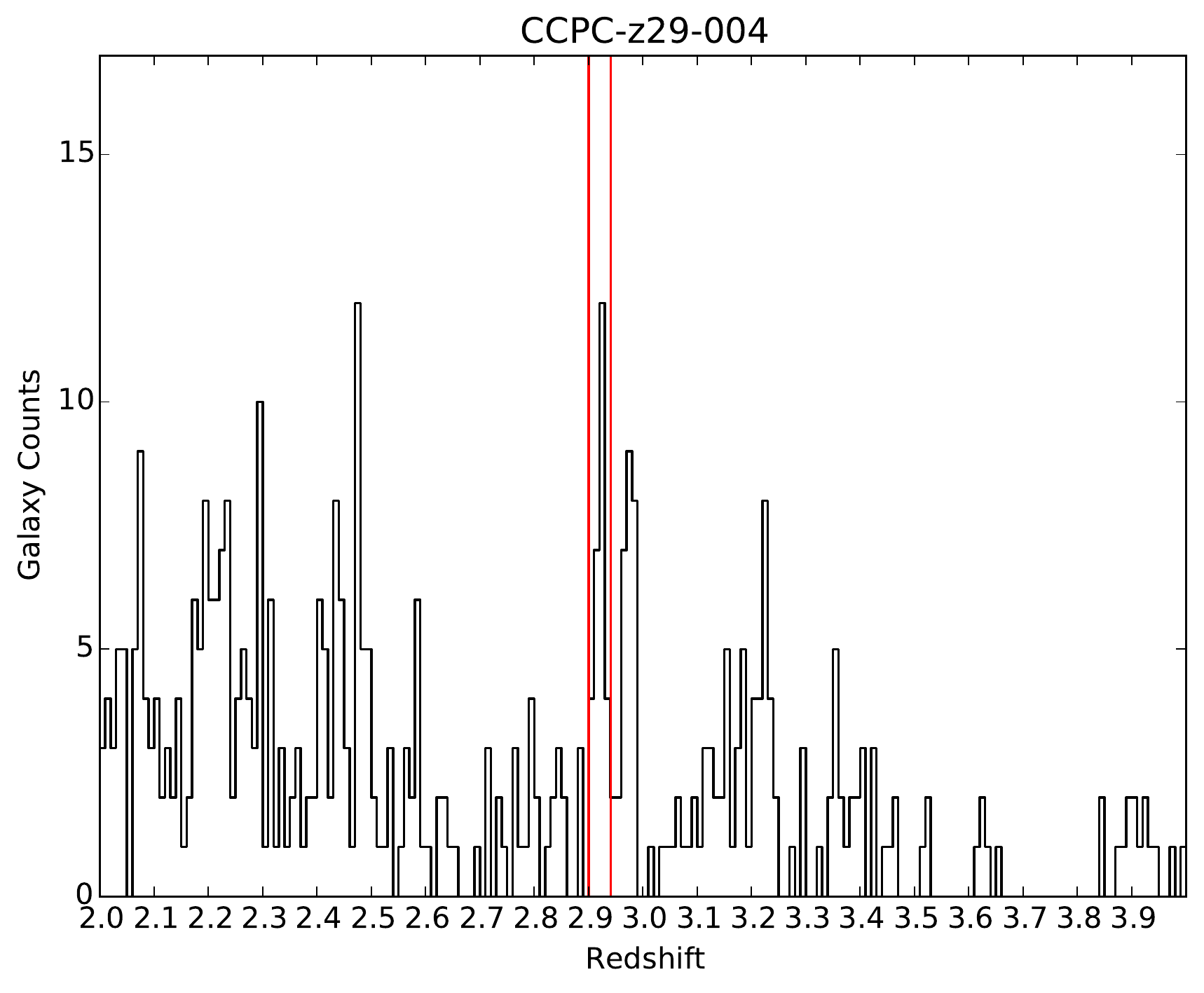}
\label{fig:CCPC-z29-004}
\end{subfigure}
\hfill
\end{figure*}

\begin{figure*}
\centering
\begin{subfigure}
\centering
\includegraphics[height=7.5cm,width=7.5cm]{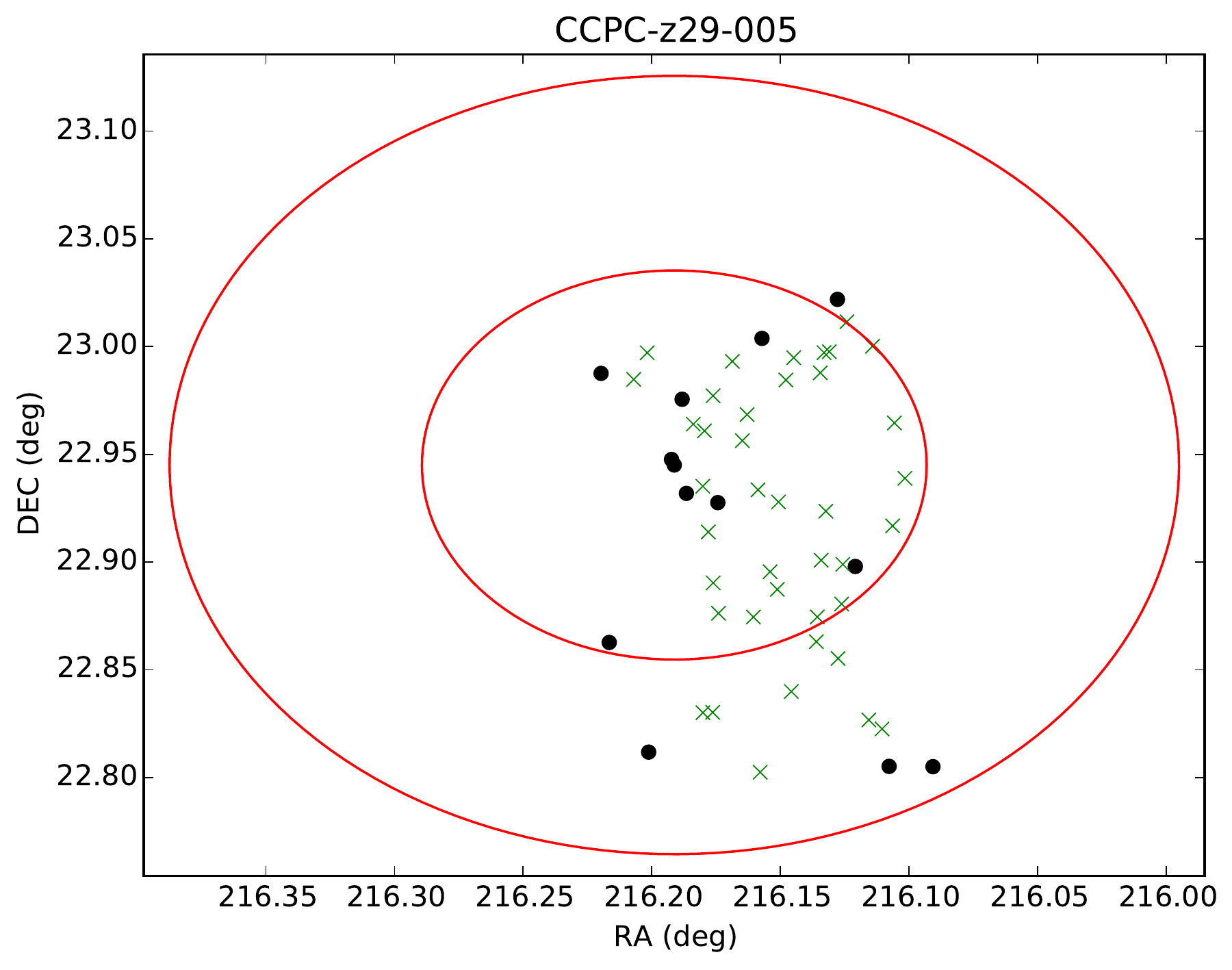}
\label{fig:CCPC-z29-005}
\end{subfigure}
\hfill
\begin{subfigure}
\centering
\includegraphics[scale=0.52]{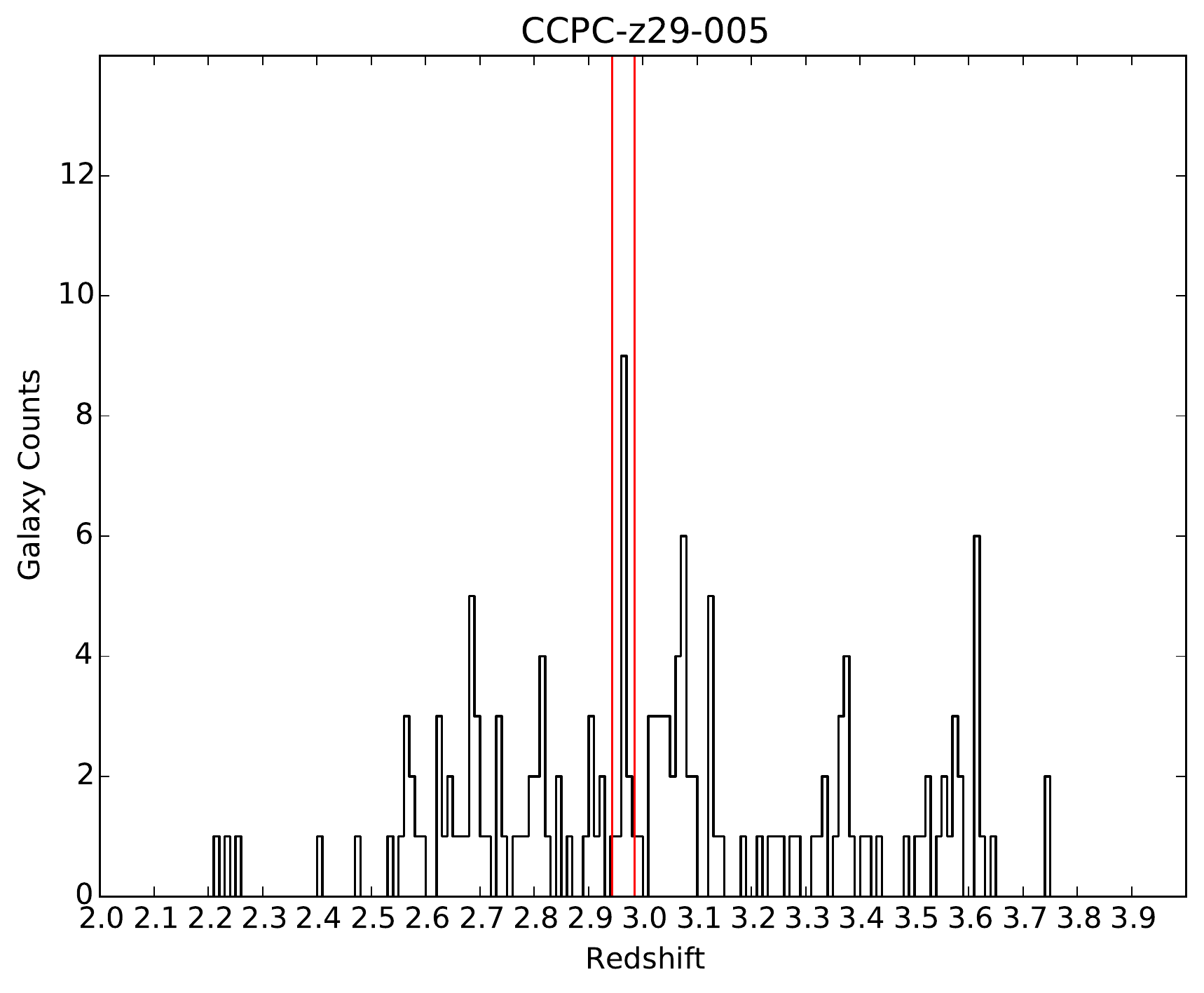}
\label{fig:CCPC-z29-005}
\end{subfigure}
\hfill
\end{figure*}
\clearpage 

\begin{figure*}
\centering
\begin{subfigure}
\centering
\includegraphics[height=7.5cm,width=7.5cm]{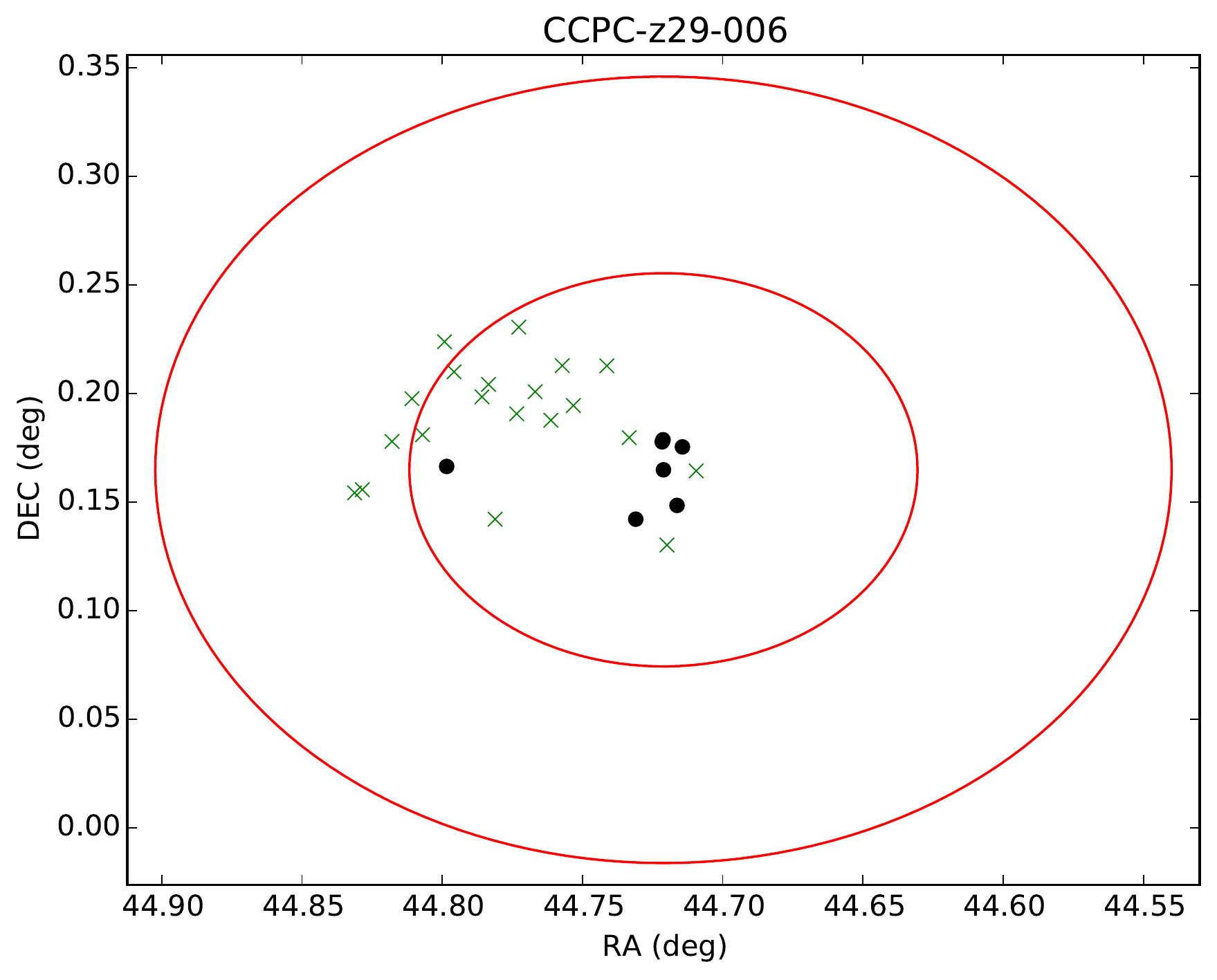}
\label{fig:CCPC-z29-006}
\end{subfigure}
\hfill
\begin{subfigure}
\centering
\includegraphics[scale=0.52]{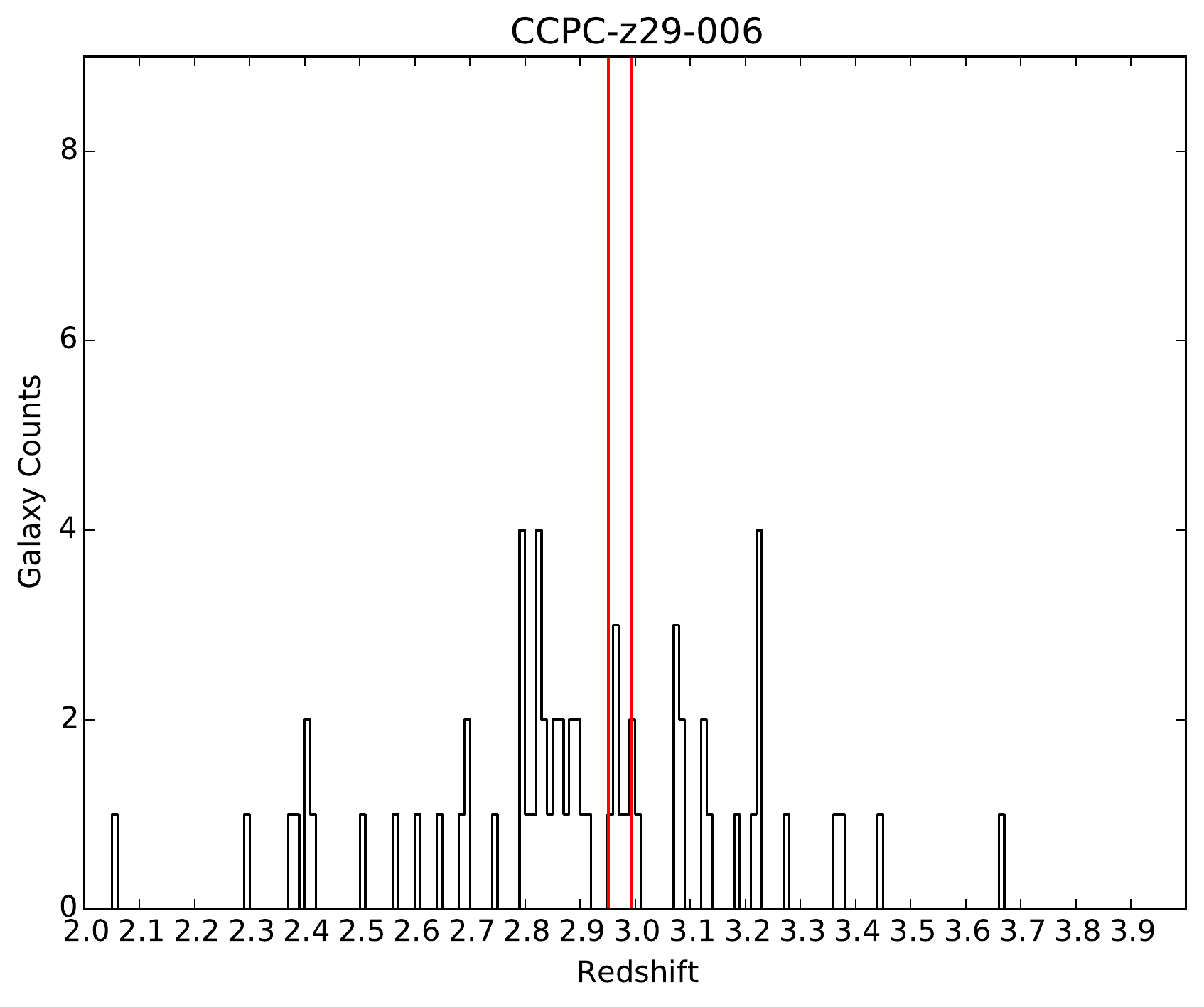}
\label{fig:CCPC-z29-006}
\end{subfigure}
\hfill
\end{figure*}

\begin{figure*}
\centering
\begin{subfigure}
\centering
\includegraphics[height=7.5cm,width=7.5cm]{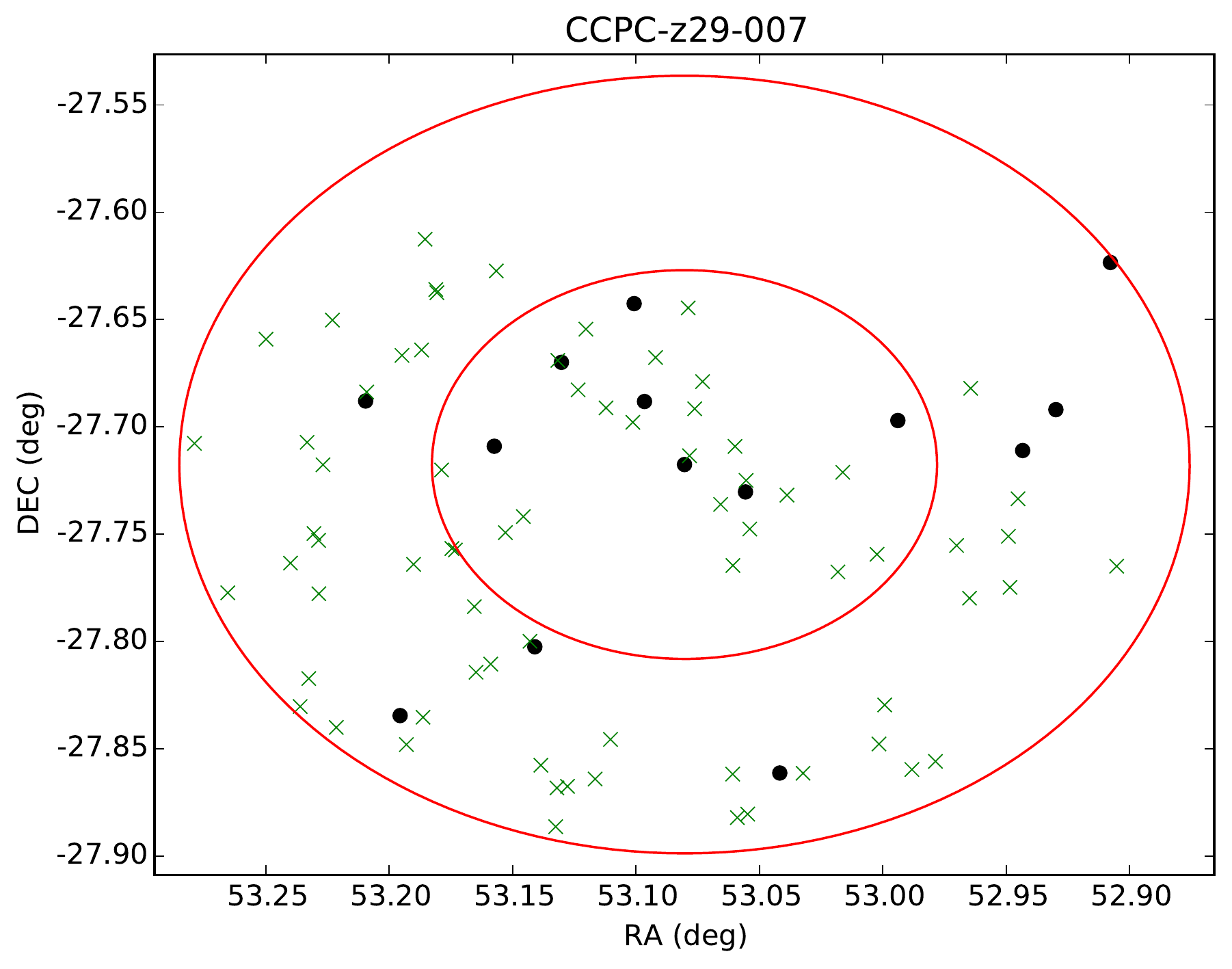}
\label{fig:CCPC-z29-007}
\end{subfigure}
\hfill
\begin{subfigure}
\centering
\includegraphics[scale=0.52]{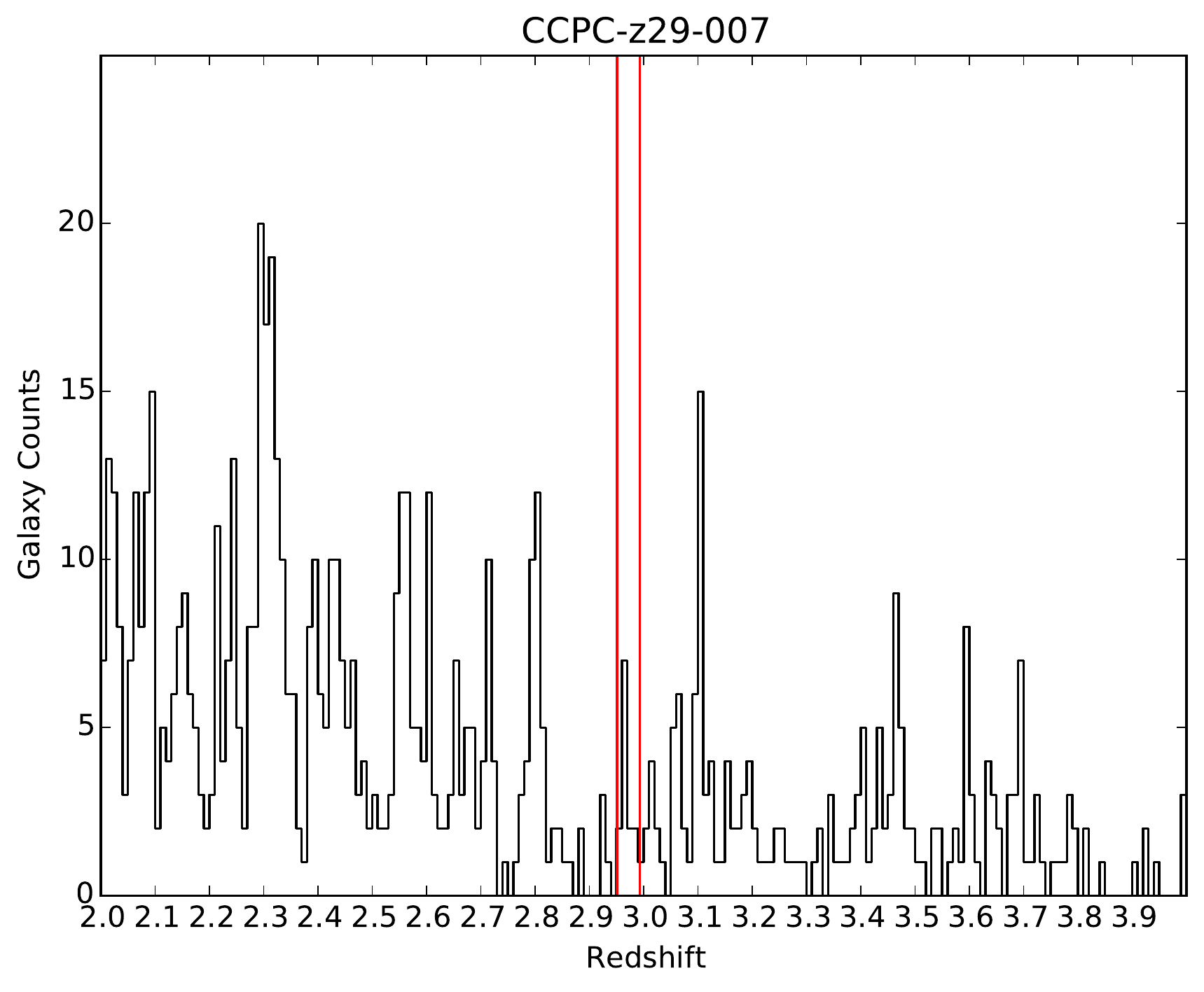}
\label{fig:CCPC-z29-007}
\end{subfigure}
\hfill
\end{figure*}

\begin{figure*}
\centering
\begin{subfigure}
\centering
\includegraphics[height=7.5cm,width=7.5cm]{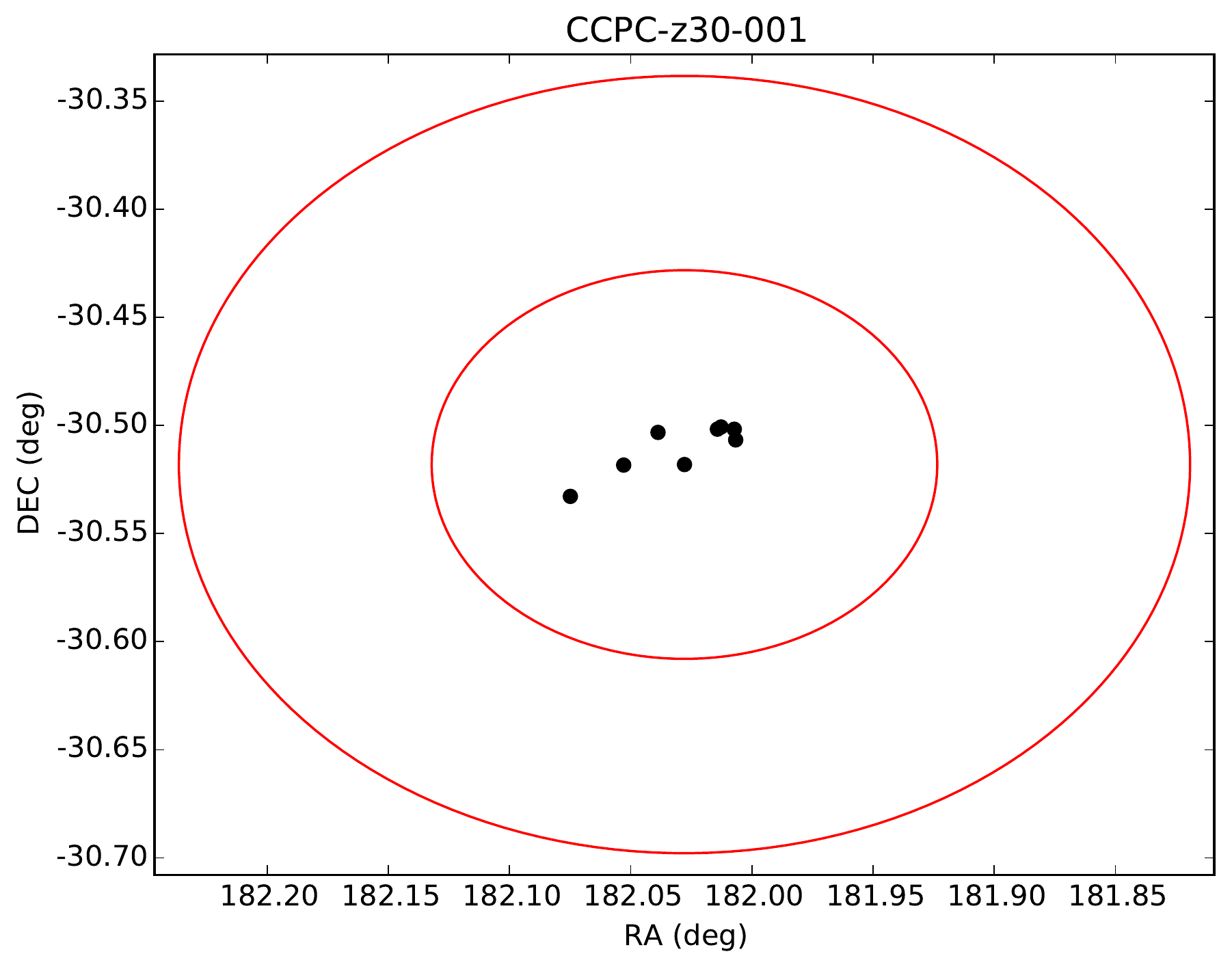}
\label{fig:CCPC-z30-001}
\end{subfigure}
\hfill
\begin{subfigure}
\centering
\includegraphics[scale=0.52]{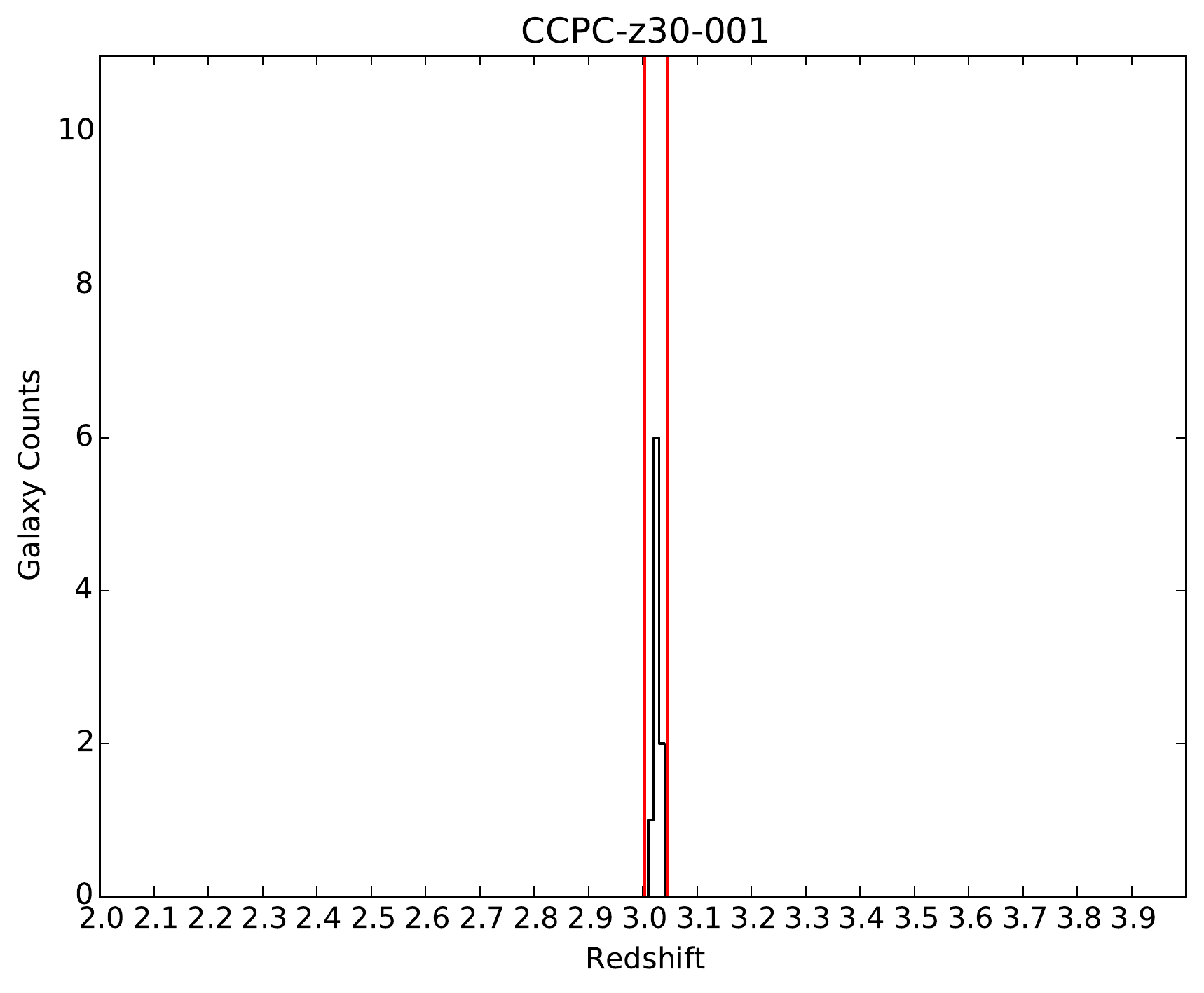}
\label{fig:CCPC-z30-001}
\end{subfigure}
\hfill
\end{figure*}
\clearpage 

\begin{figure*}
\centering
\begin{subfigure}
\centering
\includegraphics[height=7.5cm,width=7.5cm]{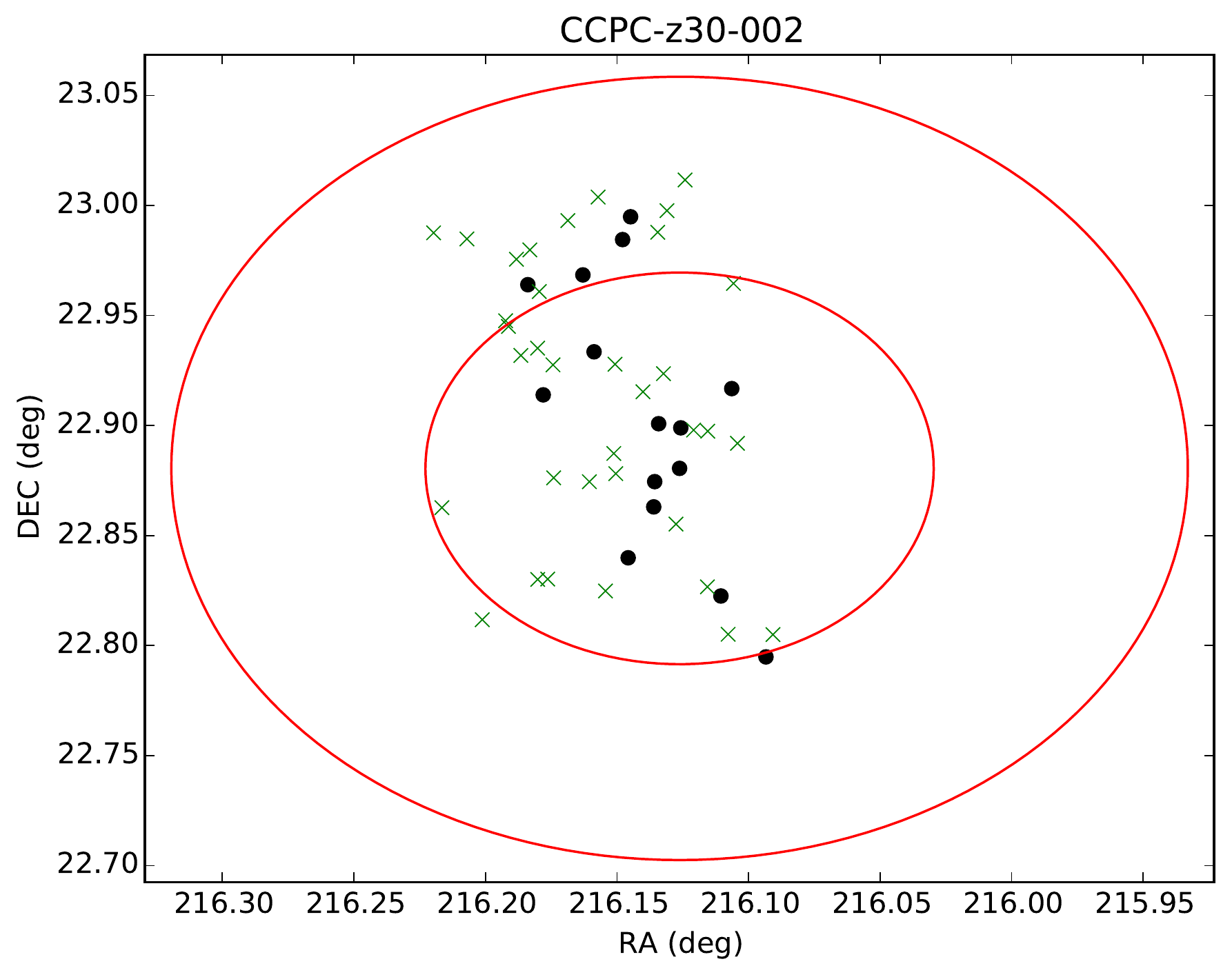}
\label{fig:CCPC-z30-002}
\end{subfigure}
\hfill
\begin{subfigure}
\centering
\includegraphics[scale=0.52]{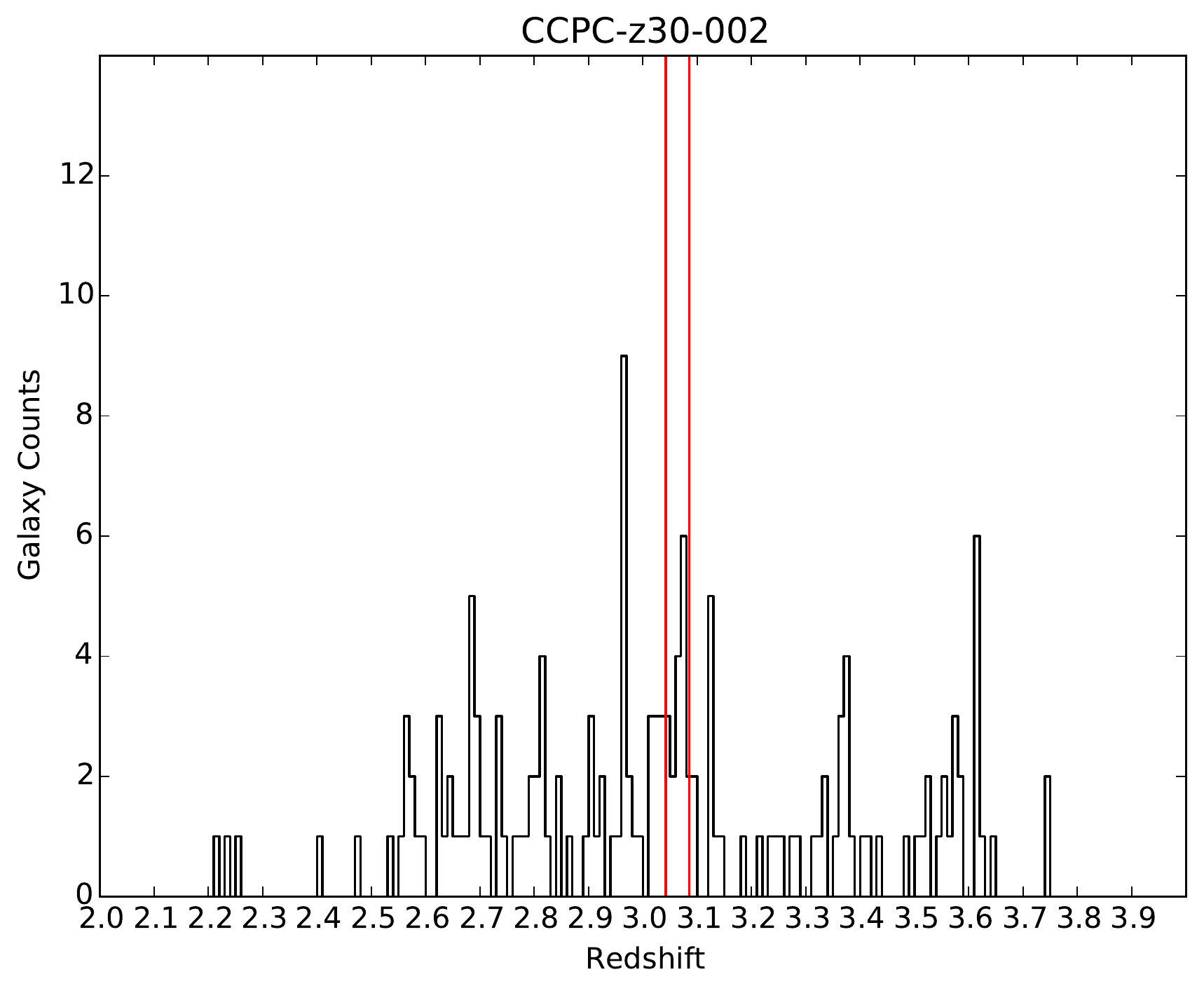}
\label{fig:CCPC-z30-002}
\end{subfigure}
\hfill
\end{figure*}

\begin{figure*}
\centering
\begin{subfigure}
\centering
\includegraphics[height=7.5cm,width=7.5cm]{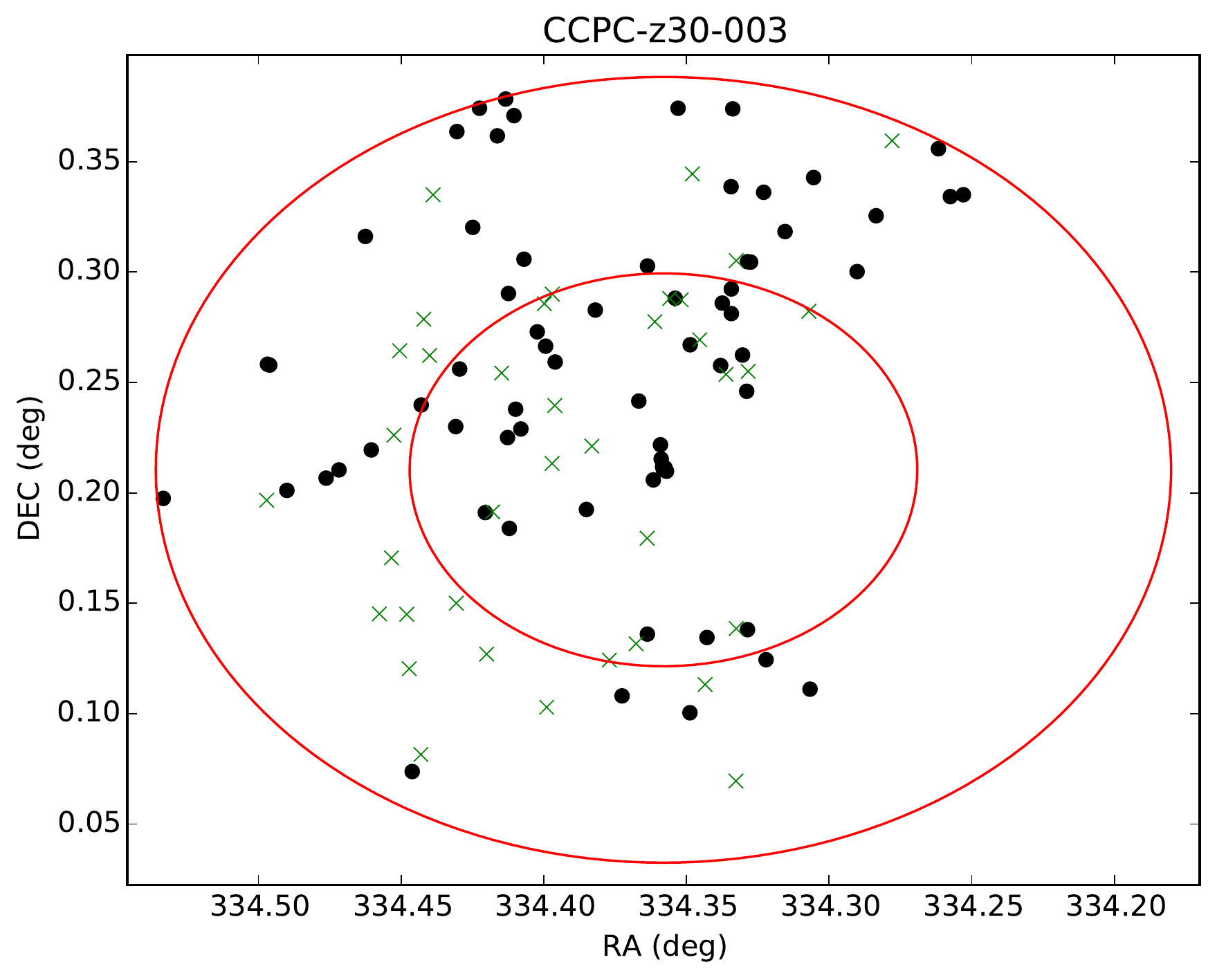}
\label{fig:CCPC-z30-003}
\end{subfigure}
\hfill
\begin{subfigure}
\centering
\includegraphics[scale=0.52]{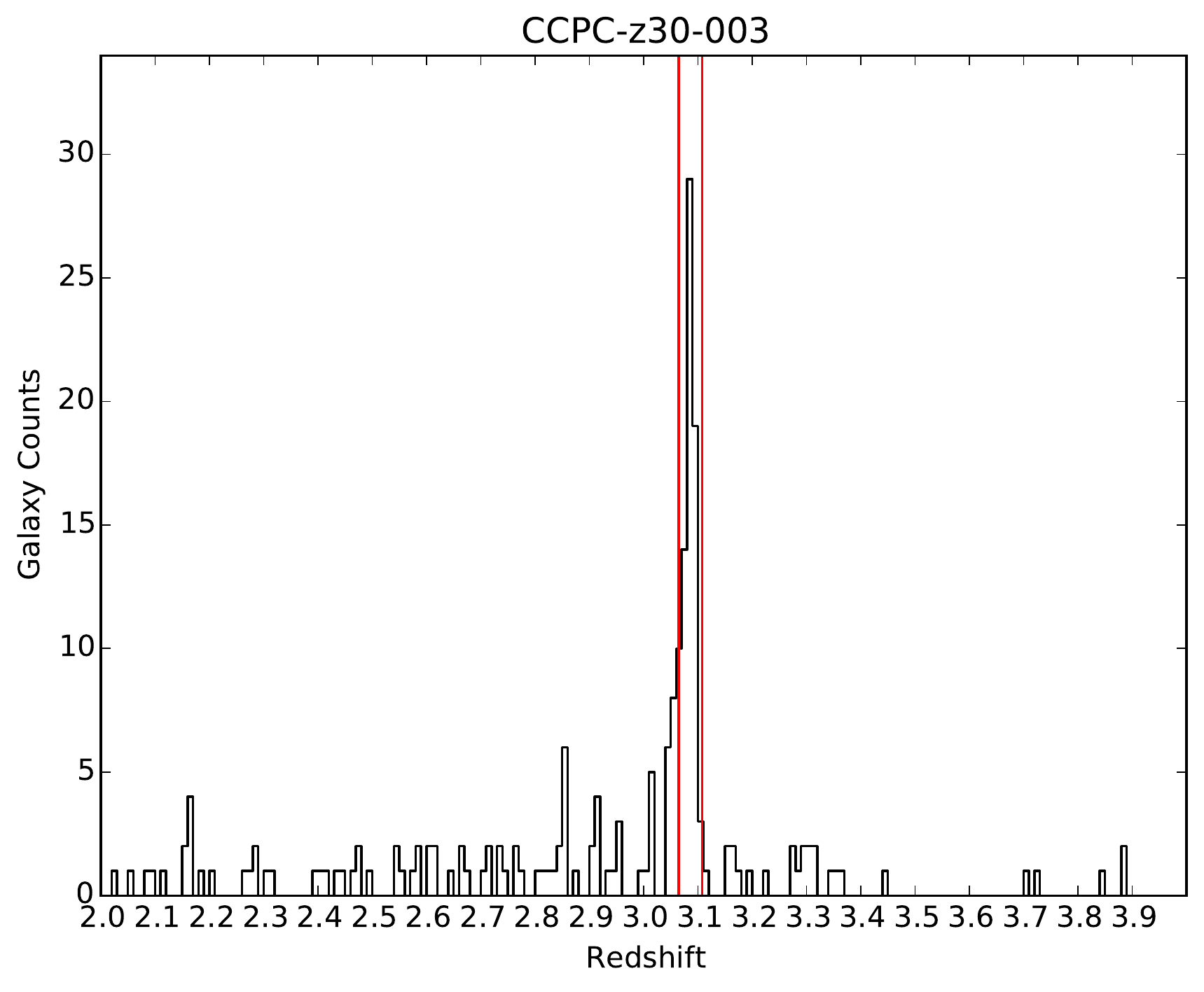}
\label{fig:CCPC-z30-003}
\end{subfigure}
\hfill
\end{figure*}

\begin{figure*}
\centering
\begin{subfigure}
\centering
\includegraphics[height=7.5cm,width=7.5cm]{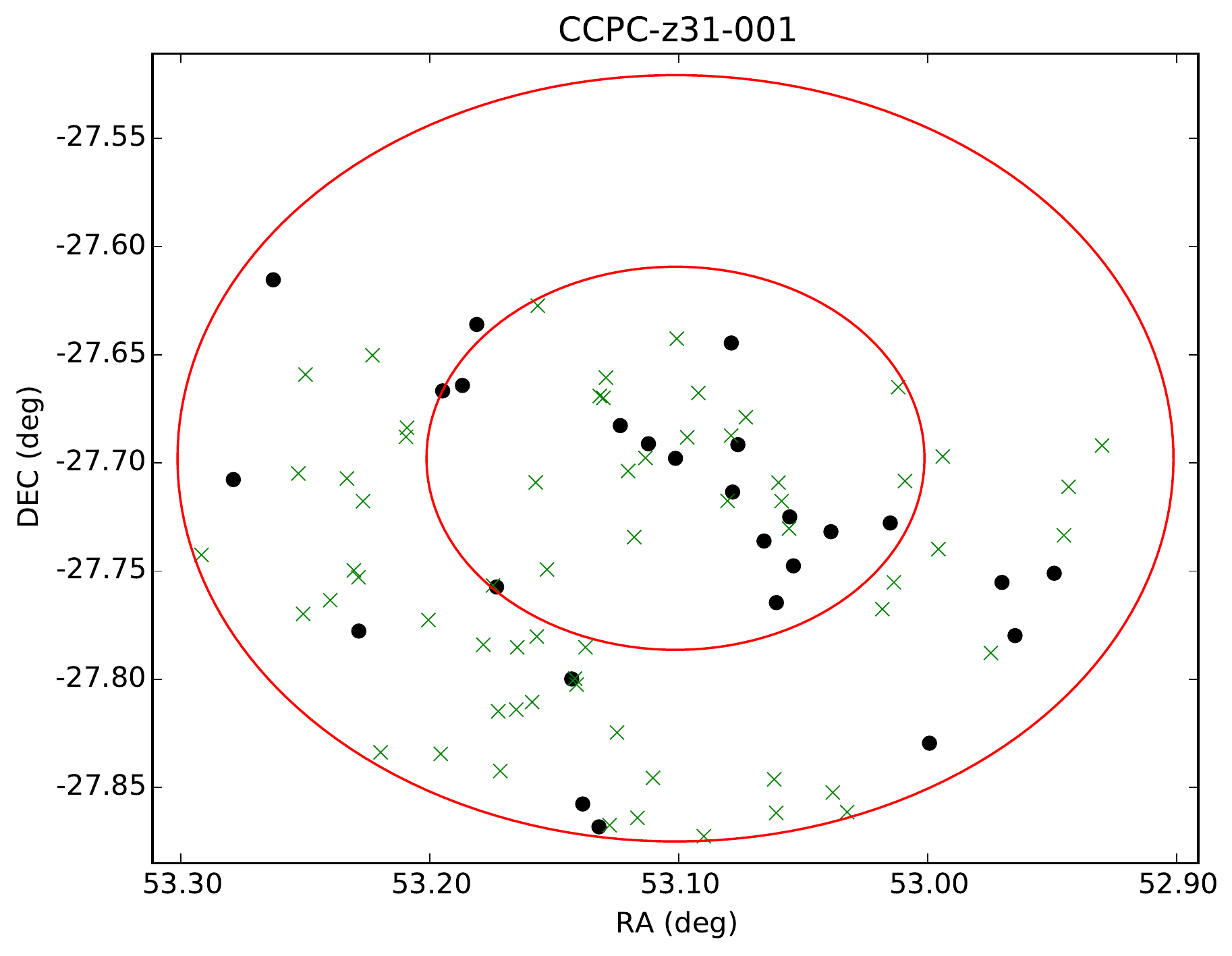}
\label{fig:CCPC-z31-001}
\end{subfigure}
\hfill
\begin{subfigure}
\centering
\includegraphics[scale=0.52]{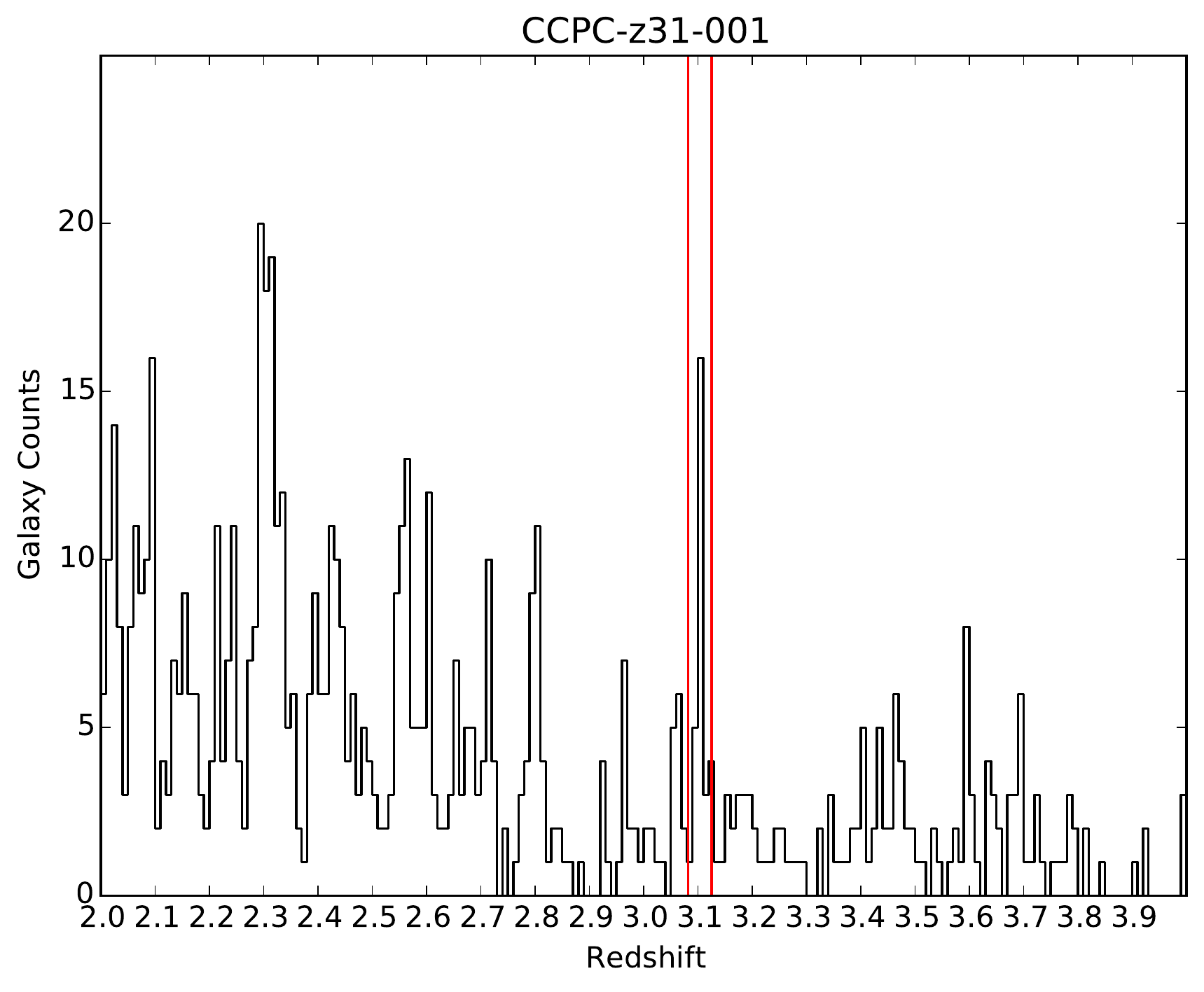}
\label{fig:CCPC-z31-001}
\end{subfigure}
\hfill
\end{figure*}
\clearpage 

\begin{figure*}
\centering
\begin{subfigure}
\centering
\includegraphics[height=7.5cm,width=7.5cm]{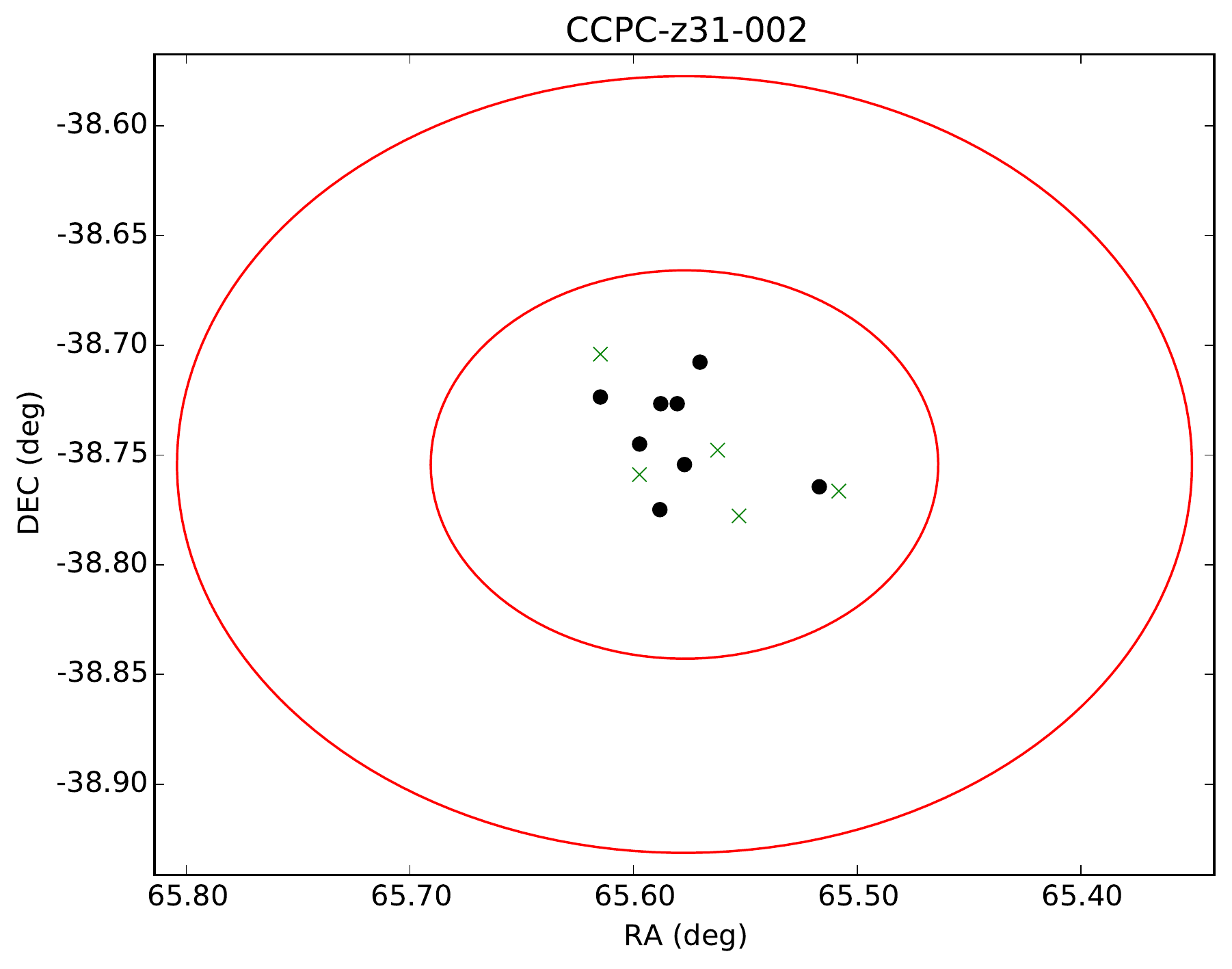}
\label{fig:CCPC-z31-002}
\end{subfigure}
\hfill
\begin{subfigure}
\centering
\includegraphics[scale=0.52]{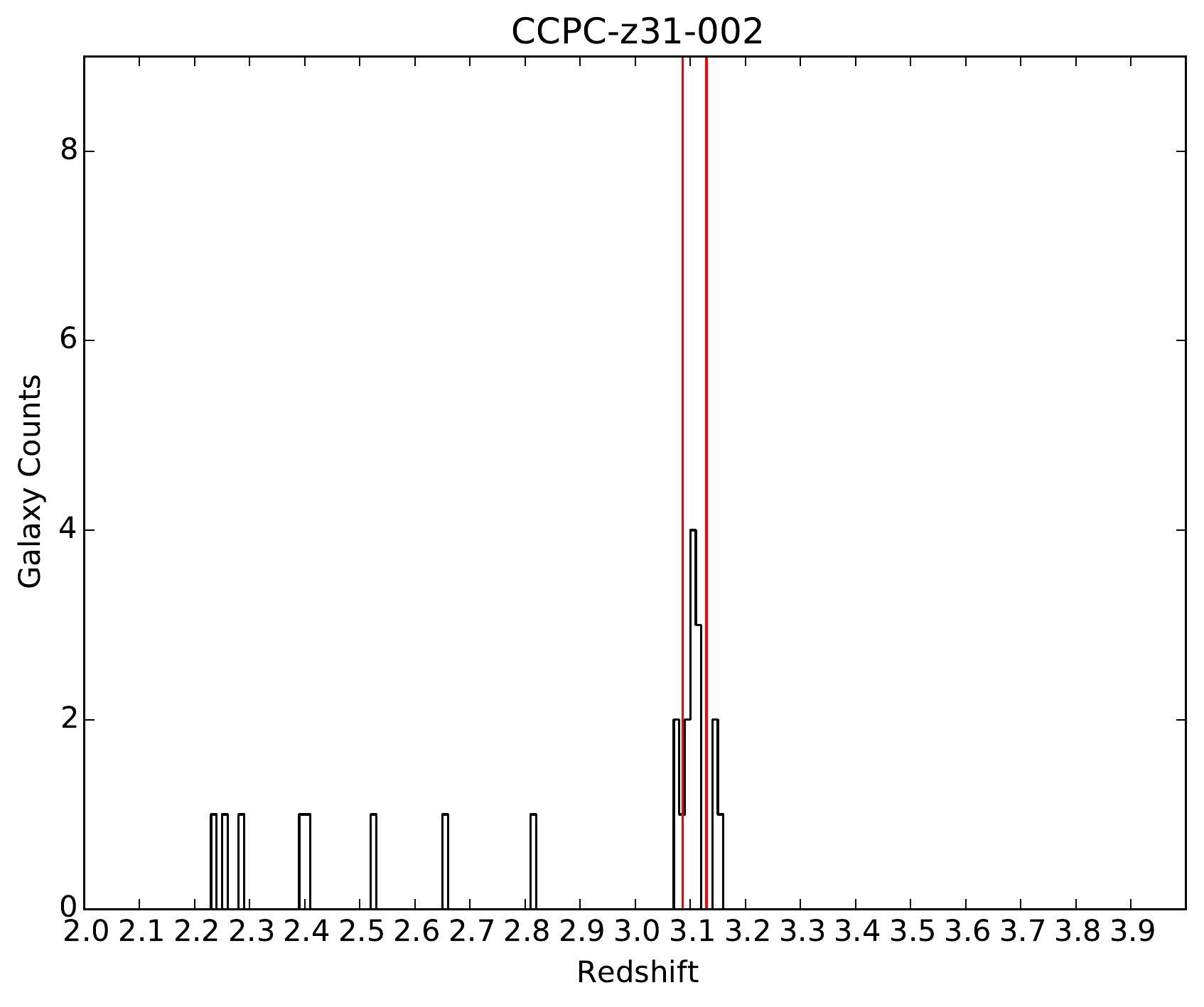}
\label{fig:CCPC-z31-002}
\end{subfigure}
\hfill
\end{figure*}

\begin{figure*}
\centering
\begin{subfigure}
\centering
\includegraphics[height=7.5cm,width=7.5cm]{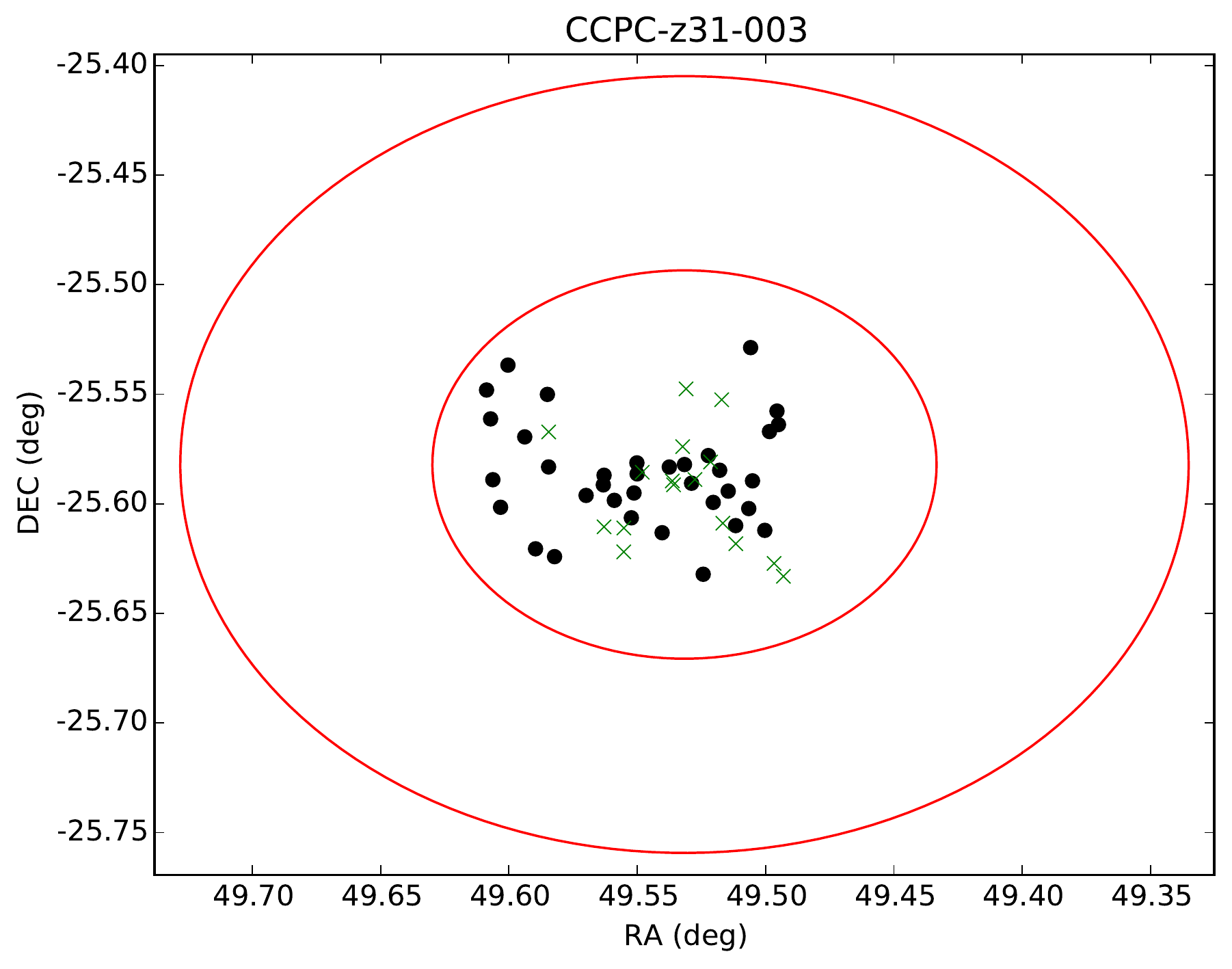}
\label{fig:CCPC-z31-003}
\end{subfigure}
\hfill
\begin{subfigure}
\centering
\includegraphics[scale=0.52]{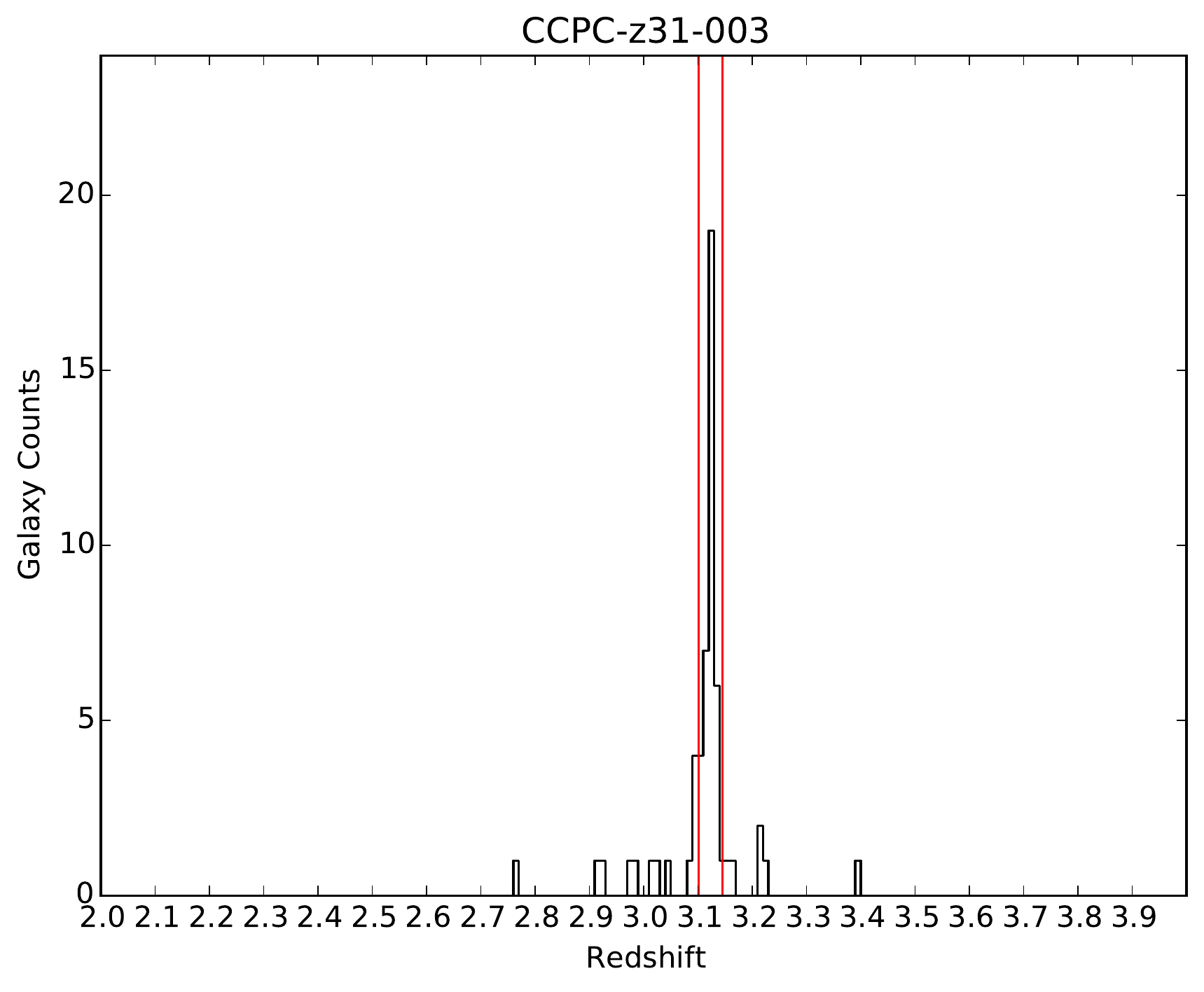}
\label{fig:CCPC-z31-003}
\end{subfigure}
\hfill
\end{figure*}

\begin{figure*}
\centering
\begin{subfigure}
\centering
\includegraphics[height=7.5cm,width=7.5cm]{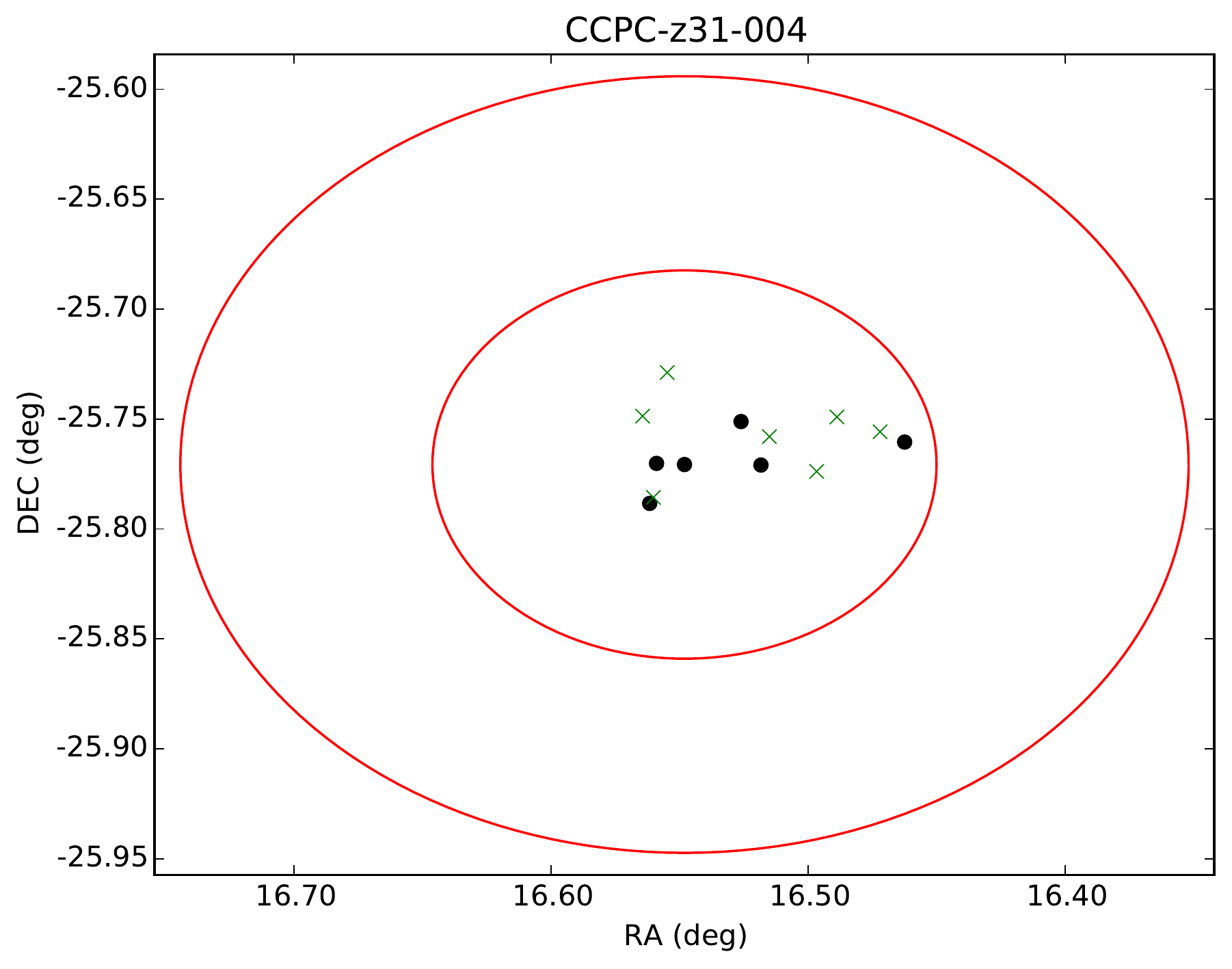}
\label{fig:CCPC-z31-004}
\end{subfigure}
\hfill
\begin{subfigure}
\centering
\includegraphics[scale=0.52]{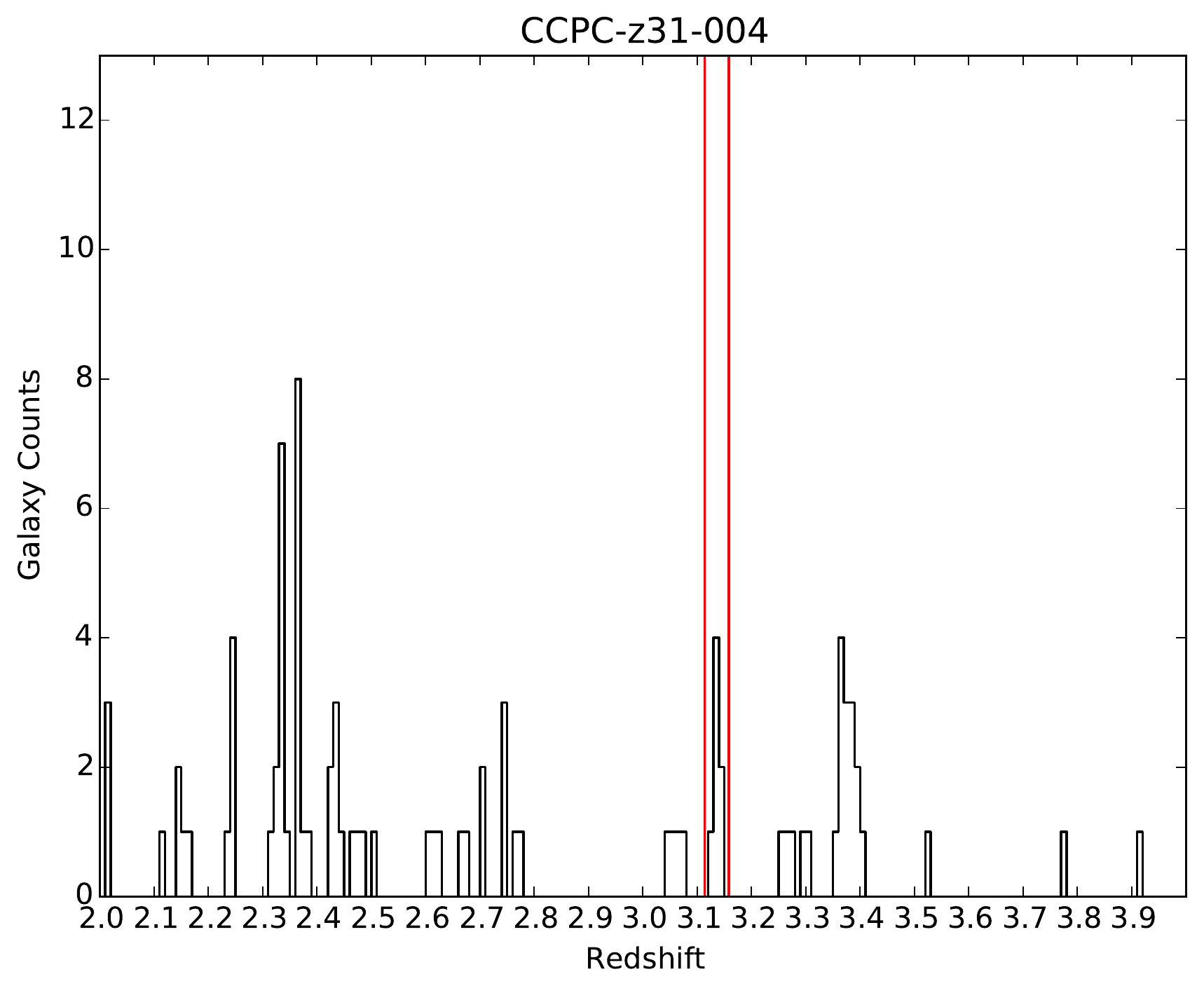}
\label{fig:CCPC-z31-004}
\end{subfigure}
\hfill
\end{figure*}
\clearpage 

\begin{figure*}
\centering
\begin{subfigure}
\centering
\includegraphics[height=7.5cm,width=7.5cm]{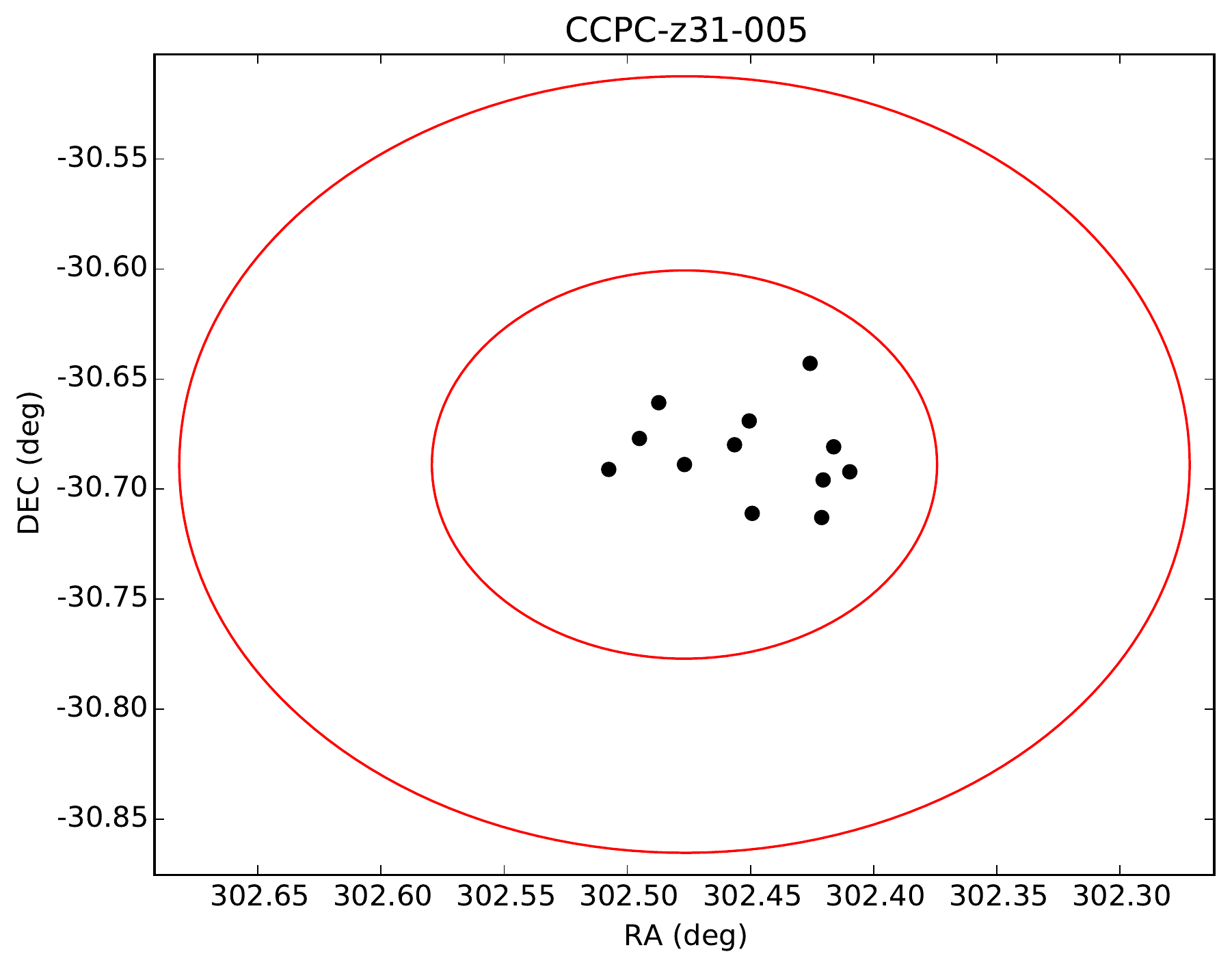}
\label{fig:CCPC-z31-005}
\end{subfigure}
\hfill
\begin{subfigure}
\centering
\includegraphics[scale=0.52]{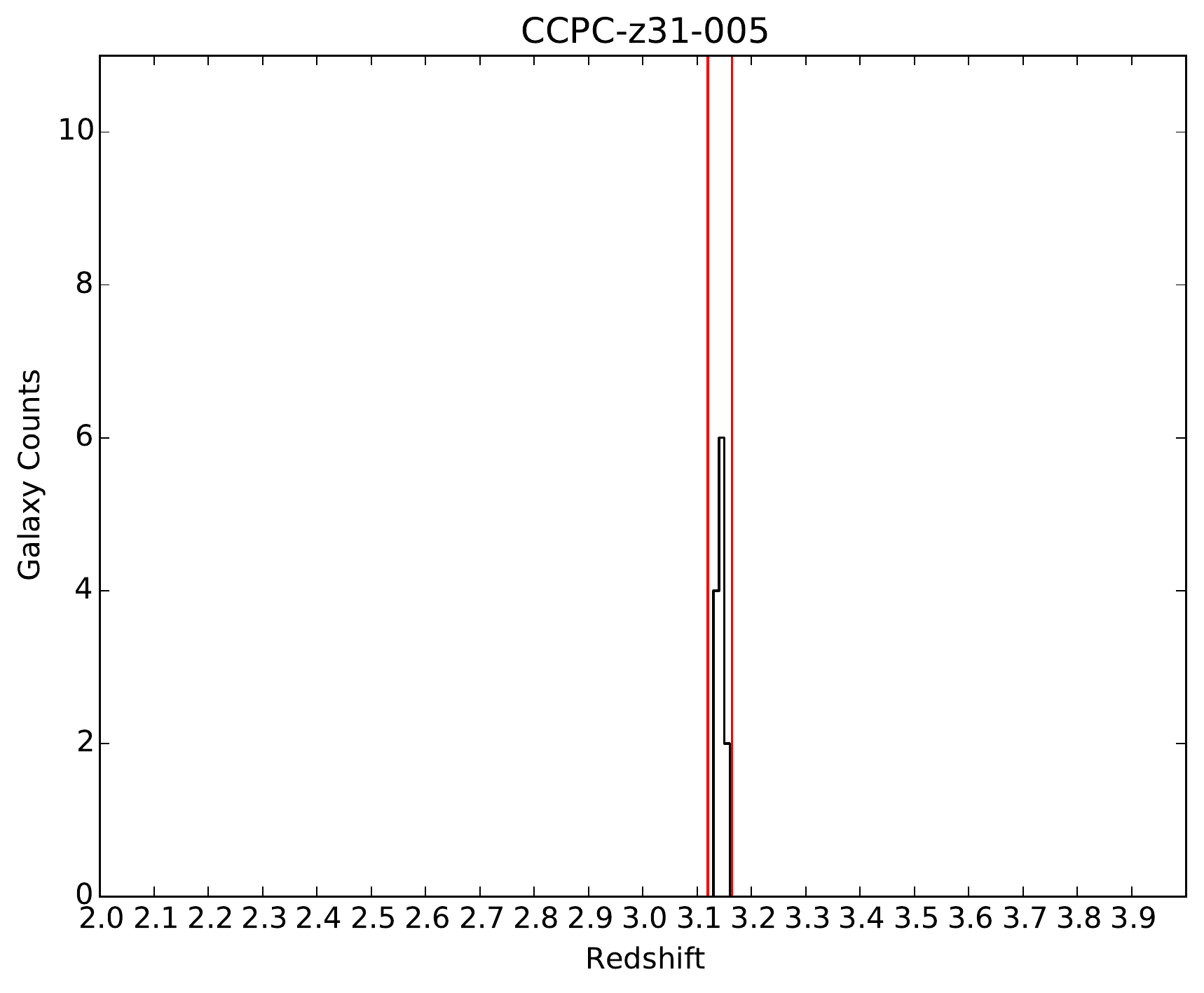}
\label{fig:CCPC-z31-005}
\end{subfigure}
\hfill
\end{figure*}

\begin{figure*}
\centering
\begin{subfigure}
\centering
\includegraphics[height=7.5cm,width=7.5cm]{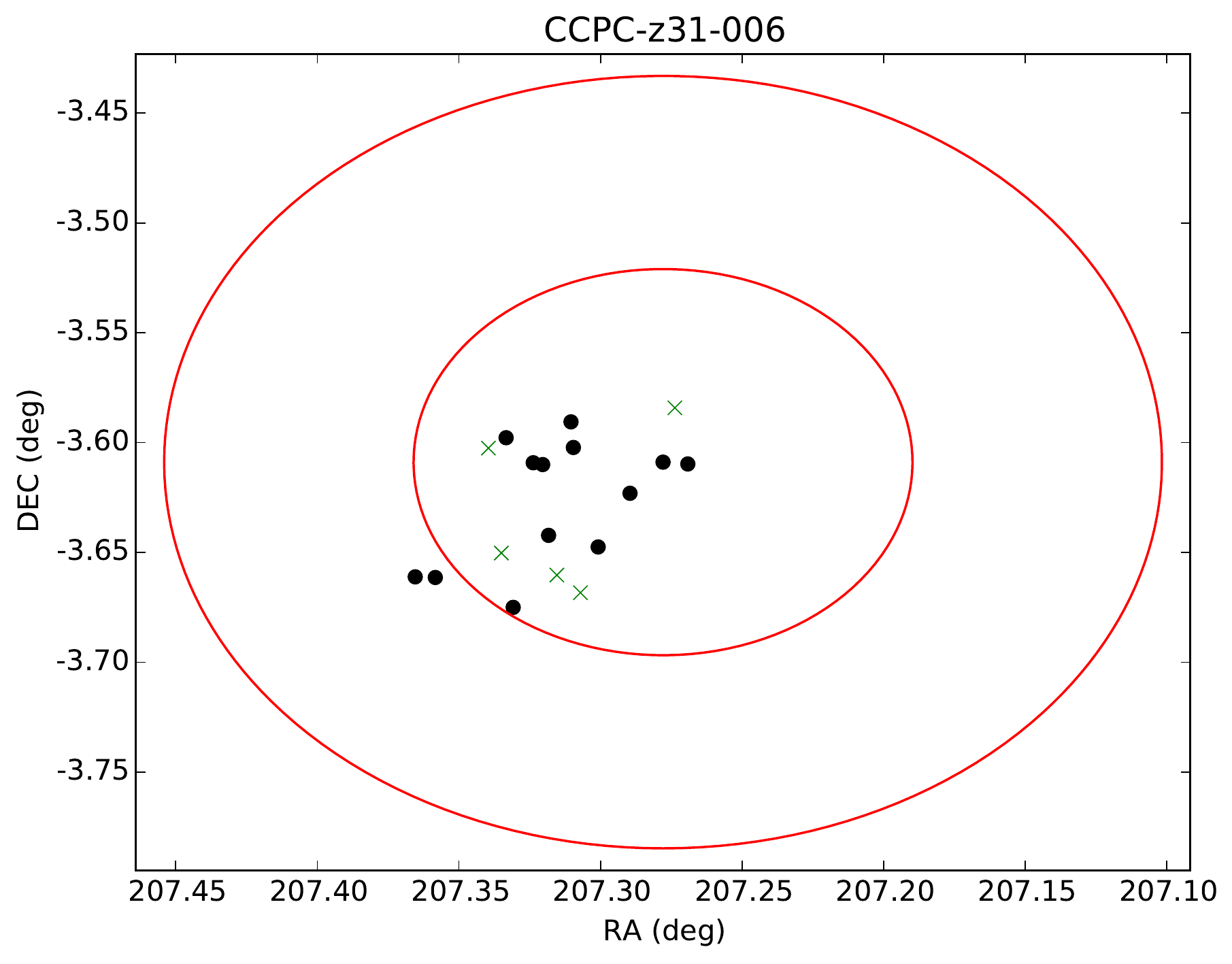}
\label{fig:CCPC-z31-006}
\end{subfigure}
\hfill
\begin{subfigure}
\centering
\includegraphics[scale=0.52]{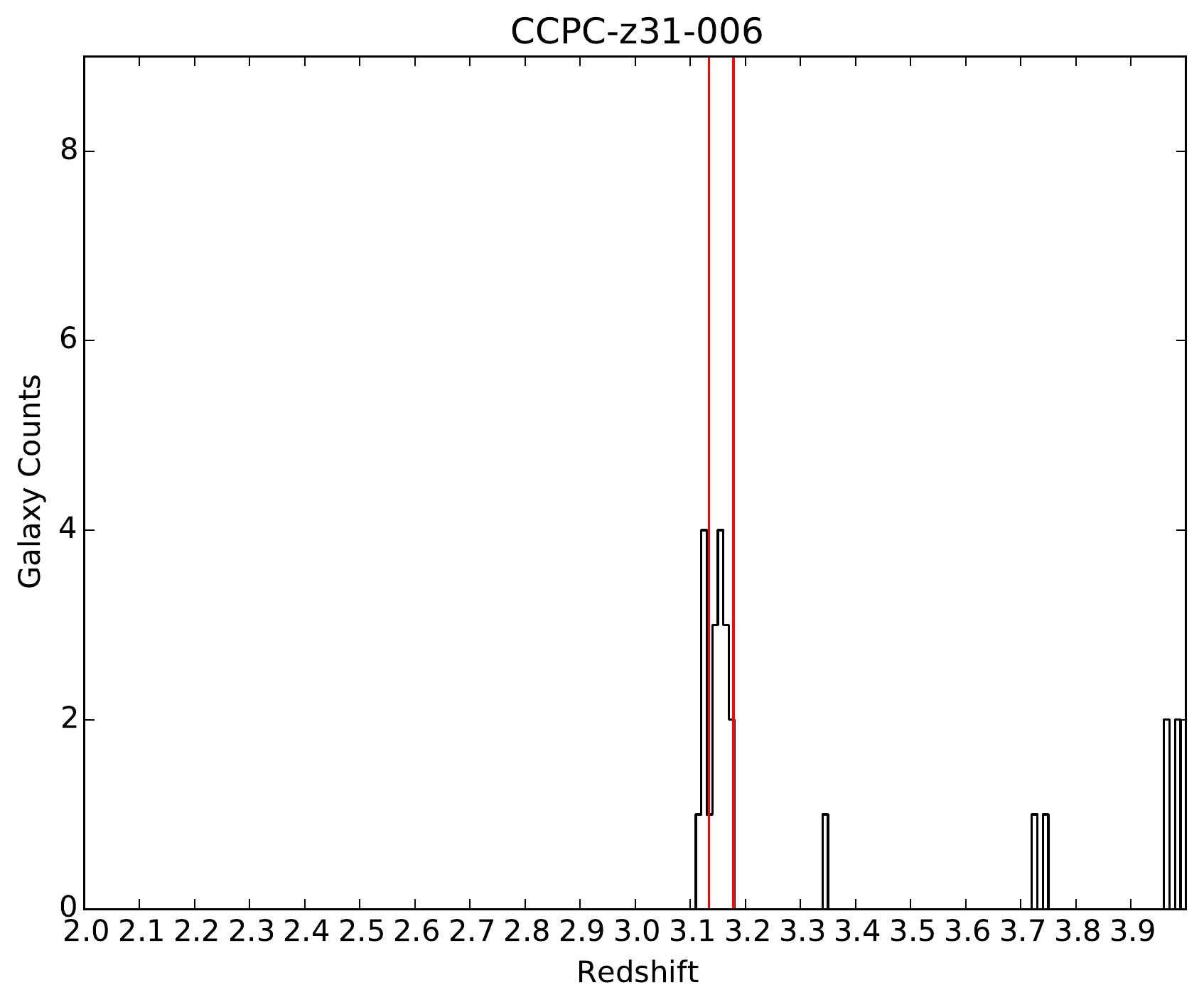}
\label{fig:CCPC-z31-006}
\end{subfigure}
\hfill
\end{figure*}

\begin{figure*}
\centering
\begin{subfigure}
\centering
\includegraphics[height=7.5cm,width=7.5cm]{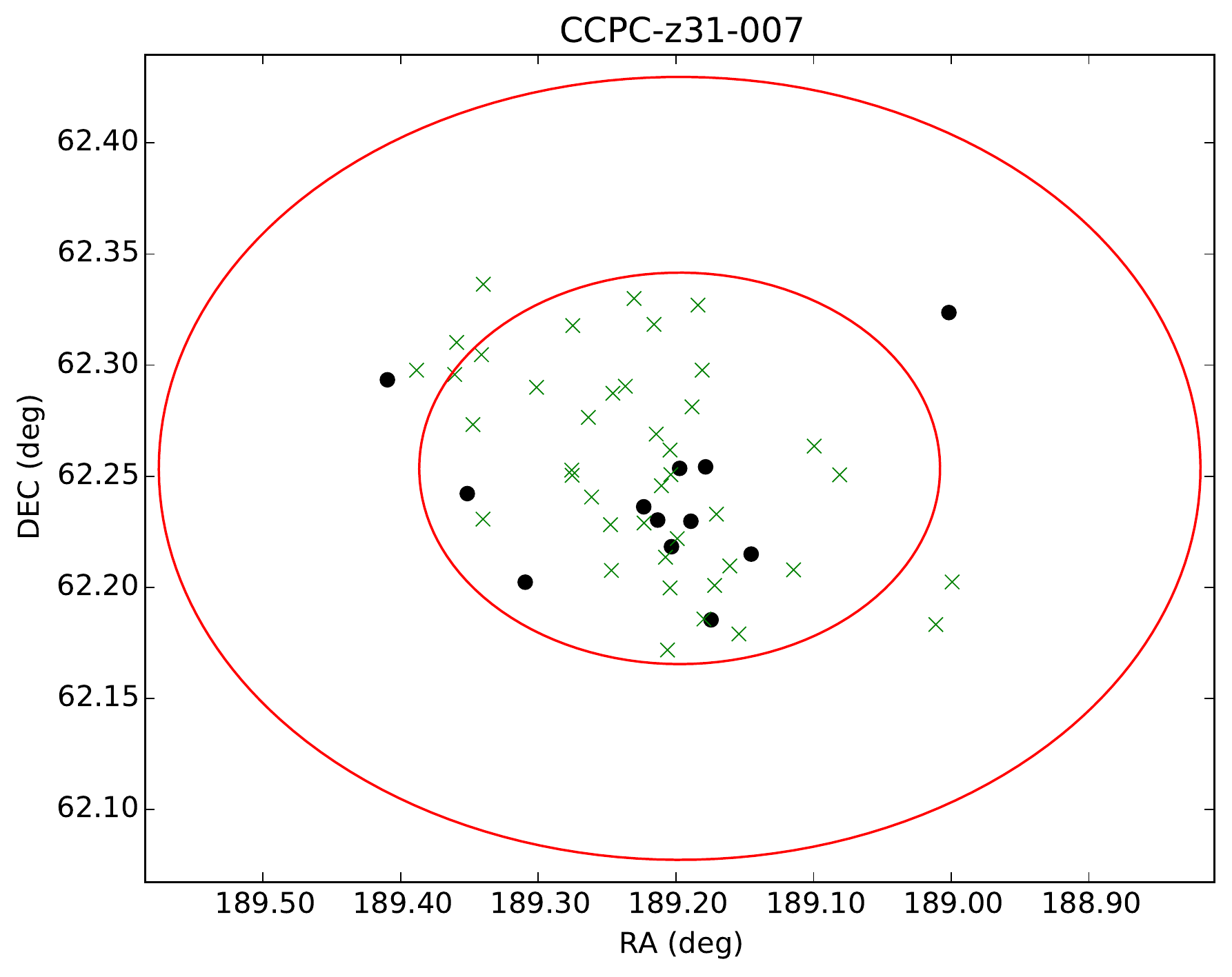}
\label{fig:CCPC-z31-007}
\end{subfigure}
\hfill
\begin{subfigure}
\centering
\includegraphics[scale=0.52]{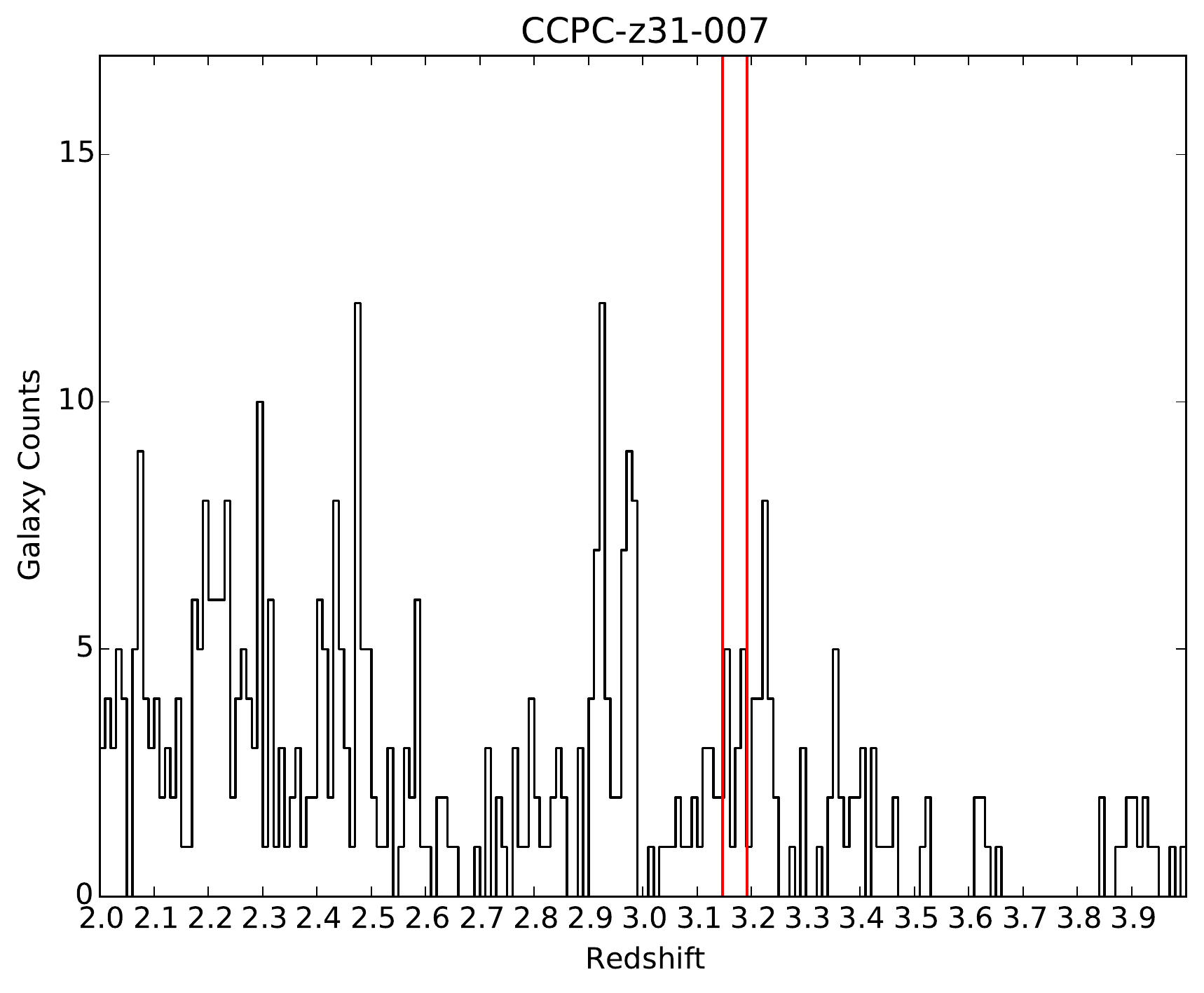}
\label{fig:CCPC-z31-007}
\end{subfigure}
\hfill
\end{figure*}
\clearpage

\begin{figure*}
\centering
\begin{subfigure}
\centering
\includegraphics[height=7.5cm,width=7.5cm]{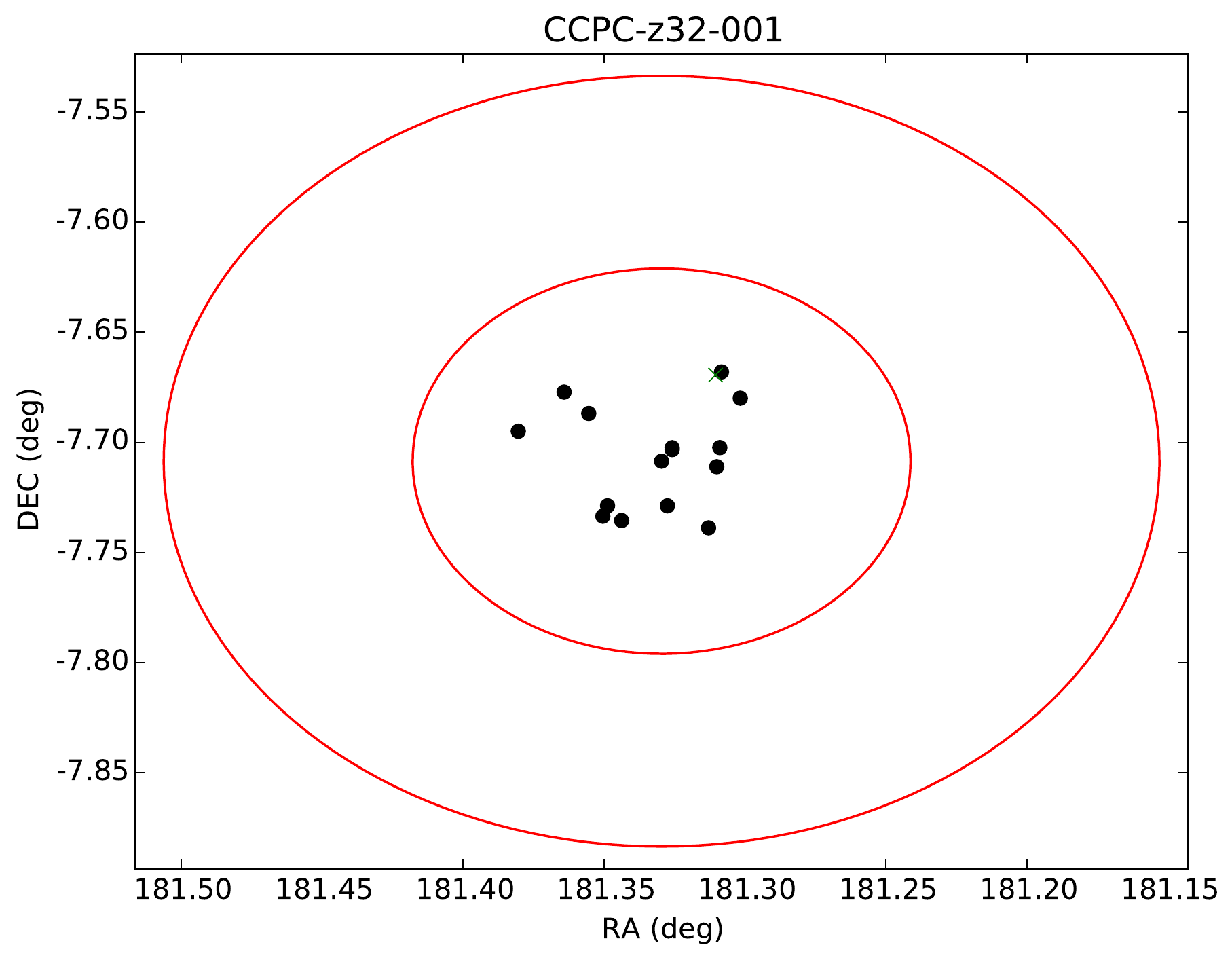}
\label{fig:CCPC-z32-001}
\end{subfigure}
\hfill
\begin{subfigure}
\centering
\includegraphics[scale=0.52]{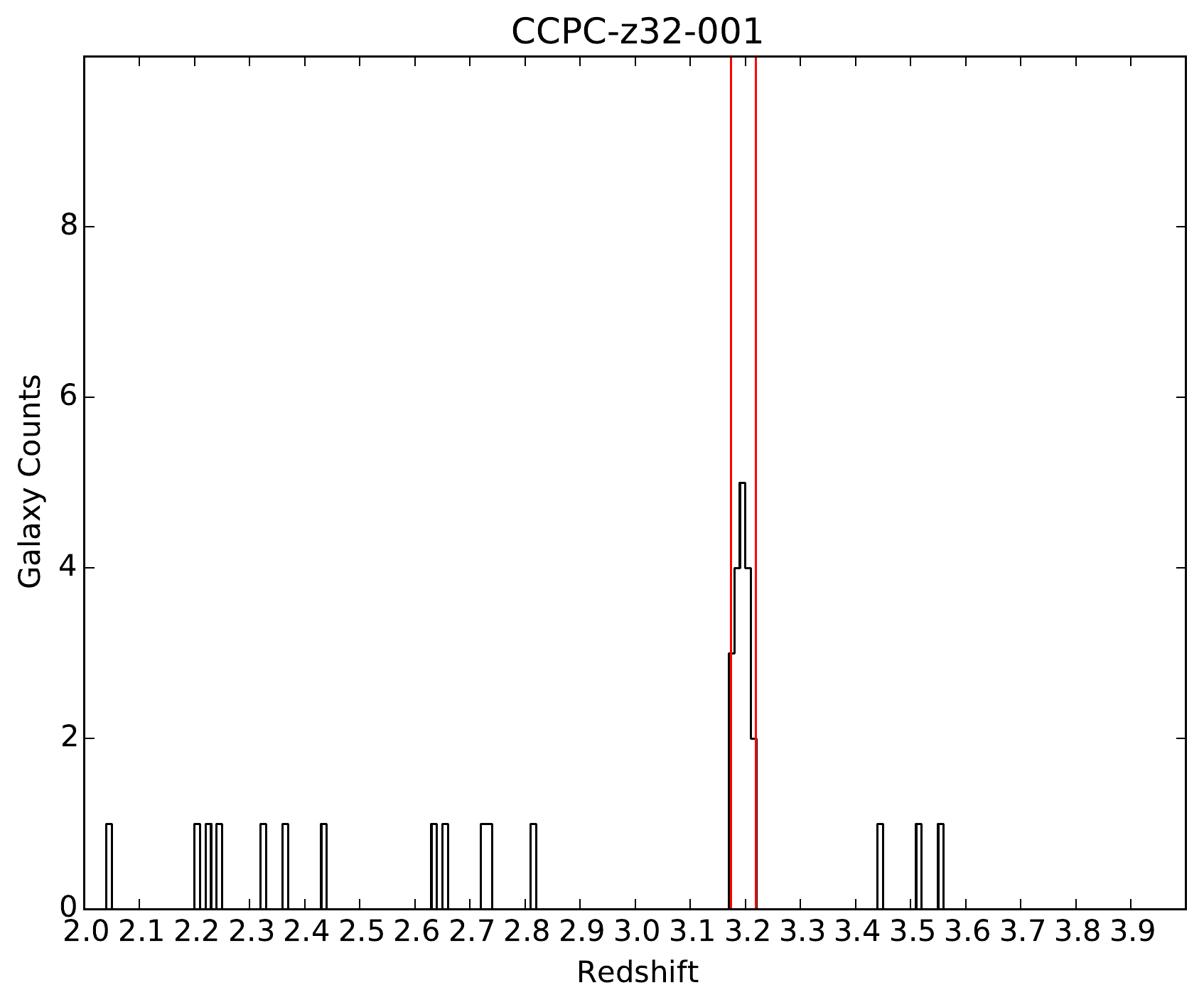}
\label{fig:CCPC-z32-001}
\end{subfigure}
\hfill
\end{figure*}

\begin{figure*}
\centering
\begin{subfigure}
\centering
\includegraphics[height=7.5cm,width=7.5cm]{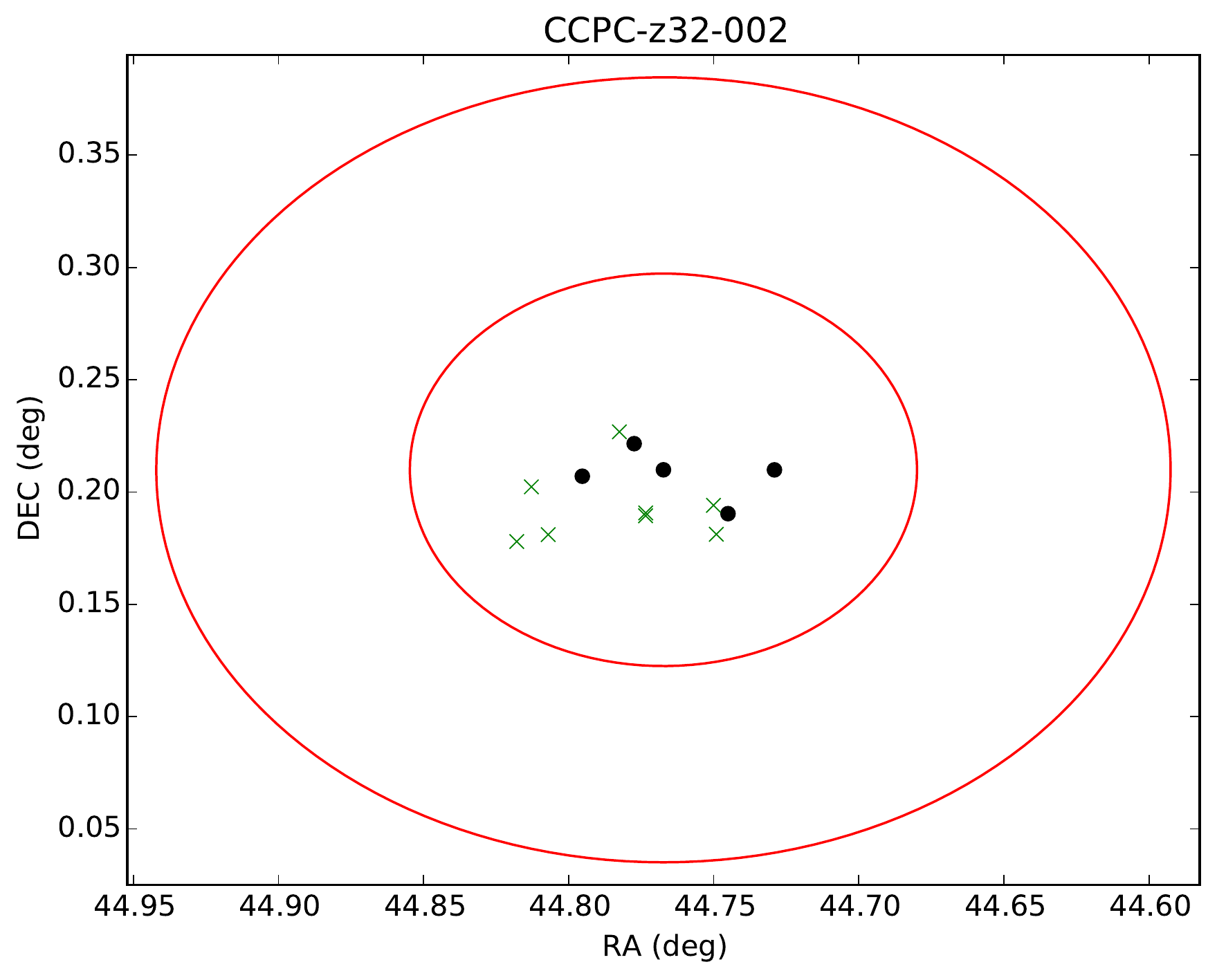}
\label{fig:CCPC-z32-002}
\end{subfigure}
\hfill
\begin{subfigure}
\centering
\includegraphics[scale=0.52]{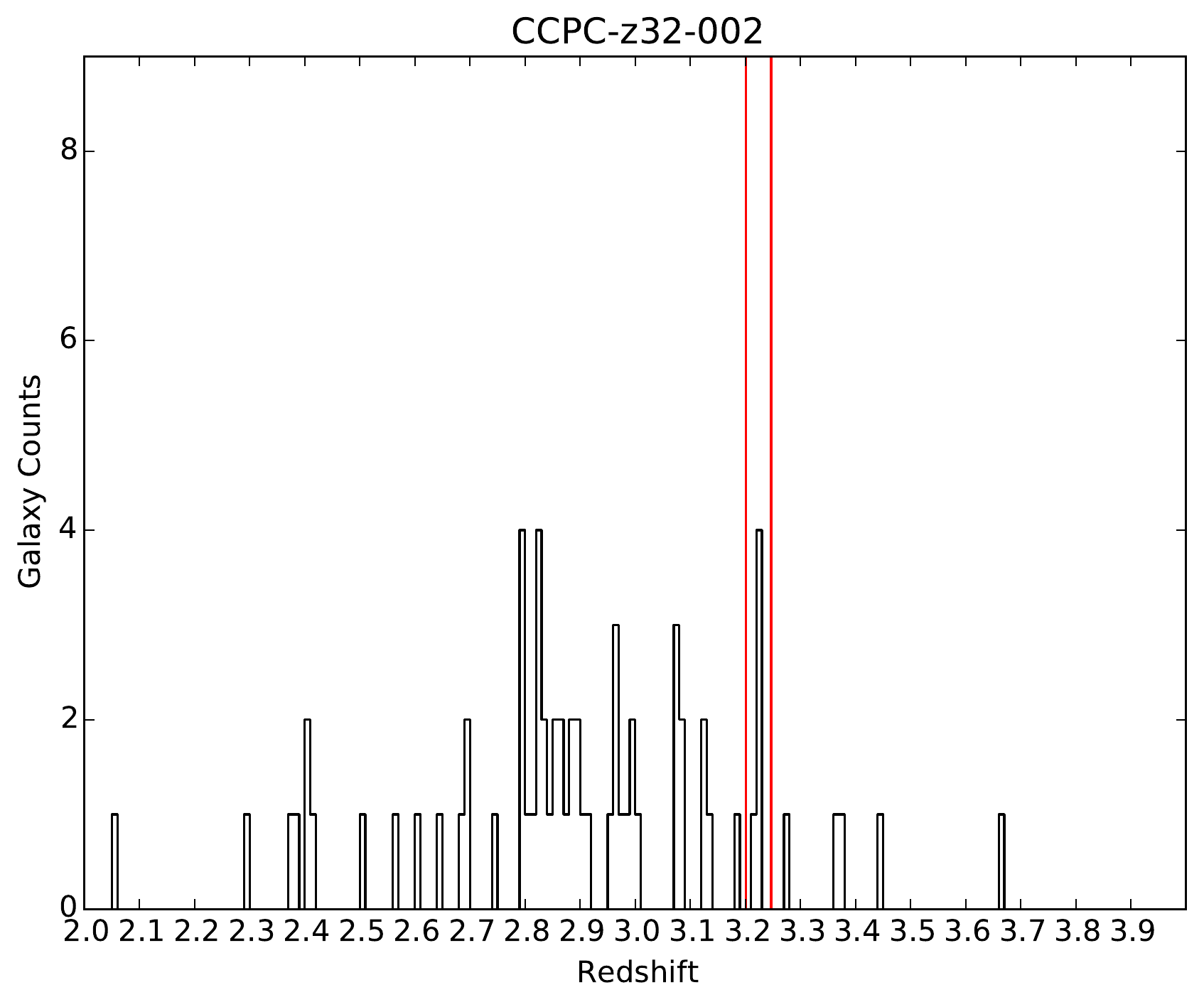}
\label{fig:CCPC-z32-002}
\end{subfigure}
\hfill
\end{figure*}

\begin{figure*}
\centering
\begin{subfigure}
\centering
\includegraphics[height=7.5cm,width=7.5cm]{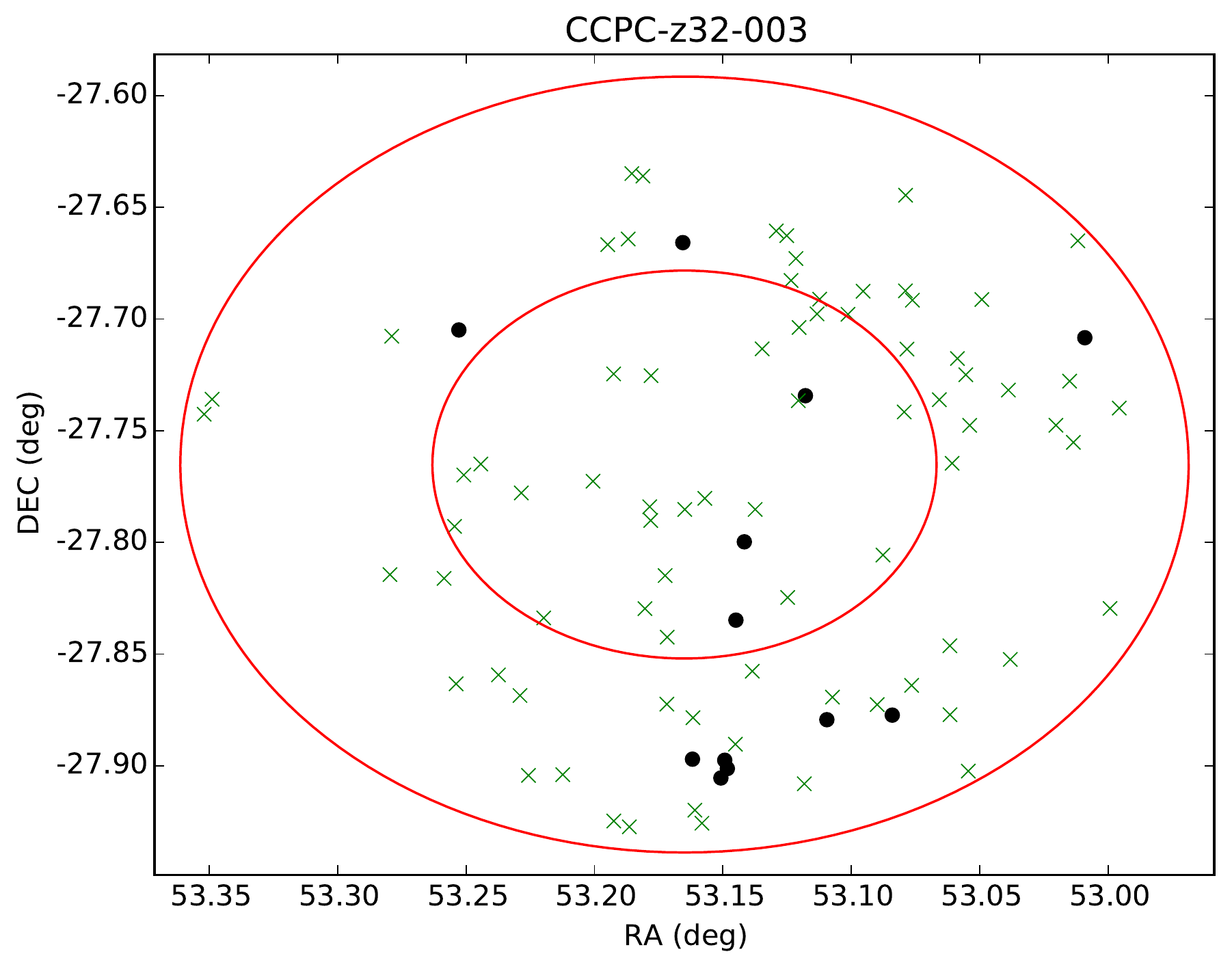}
\label{fig:CCPC-z32-003}
\end{subfigure}
\hfill
\begin{subfigure}
\centering
\includegraphics[scale=0.52]{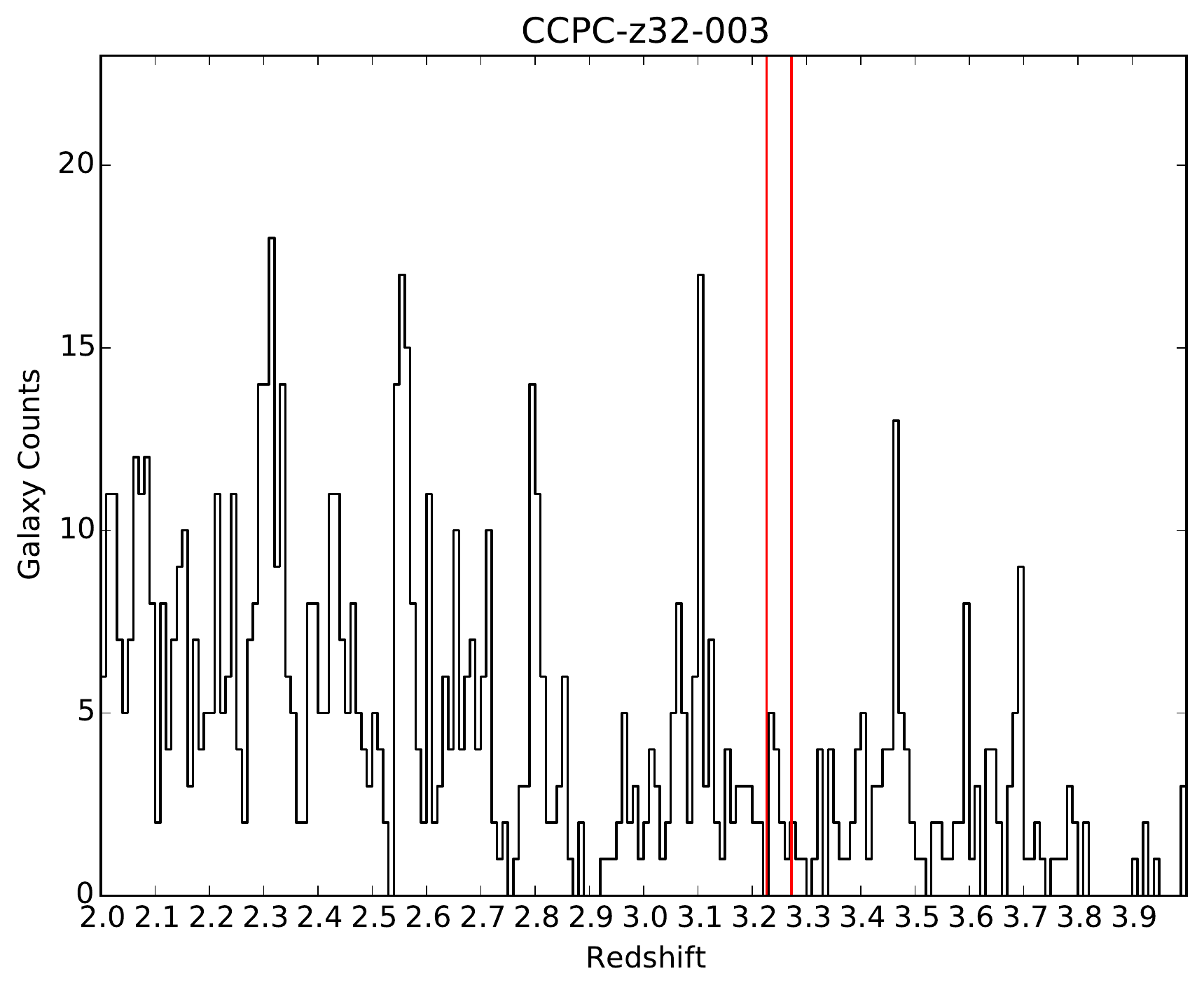}
\label{fig:CCPC-z32-003}
\end{subfigure}
\hfill
\end{figure*}
\clearpage 

\begin{figure*}
\centering
\begin{subfigure}
\centering
\includegraphics[height=7.5cm,width=7.5cm]{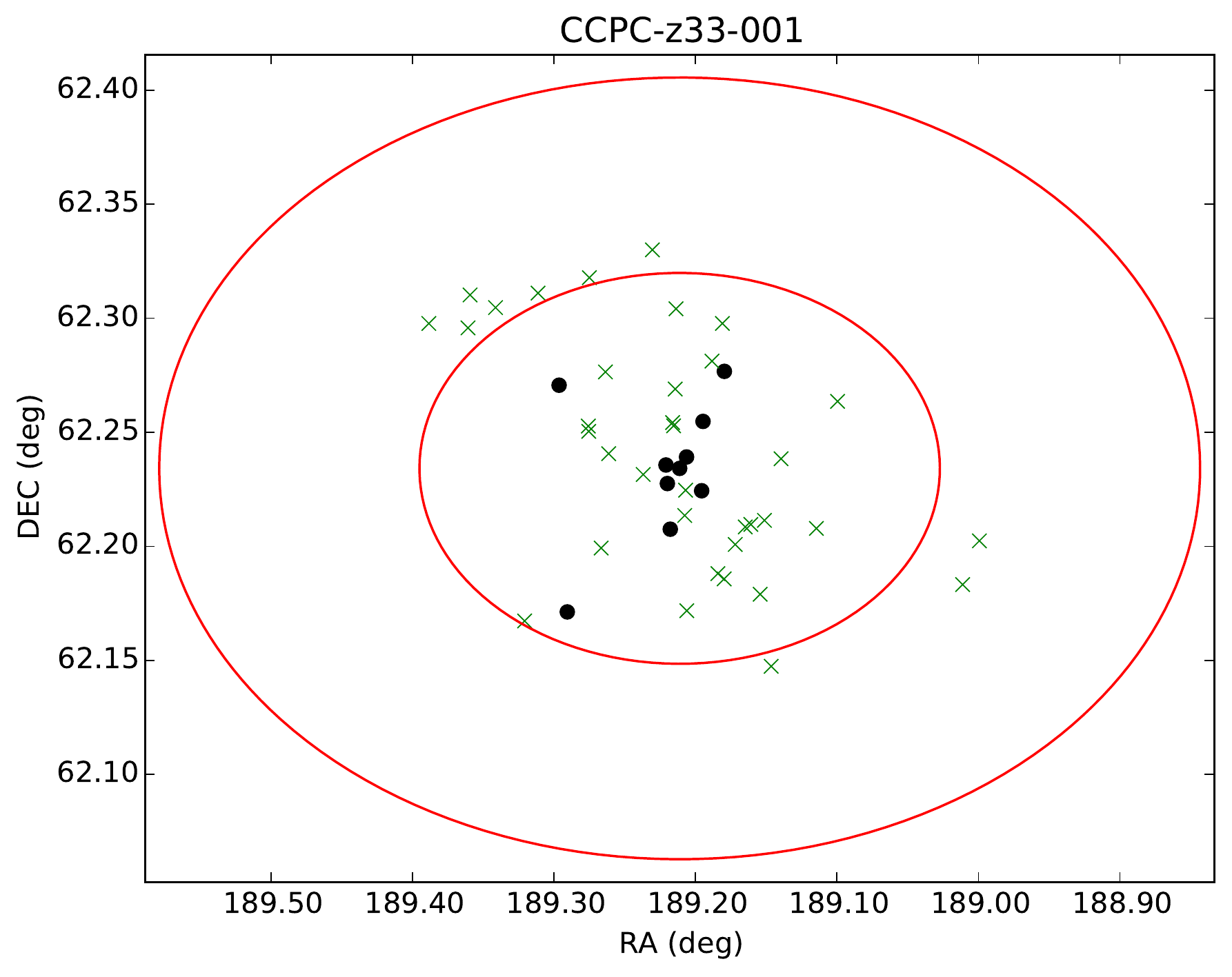}
\label{fig:CCPC-z33-001}
\end{subfigure}
\hfill
\begin{subfigure}
\centering
\includegraphics[scale=0.52]{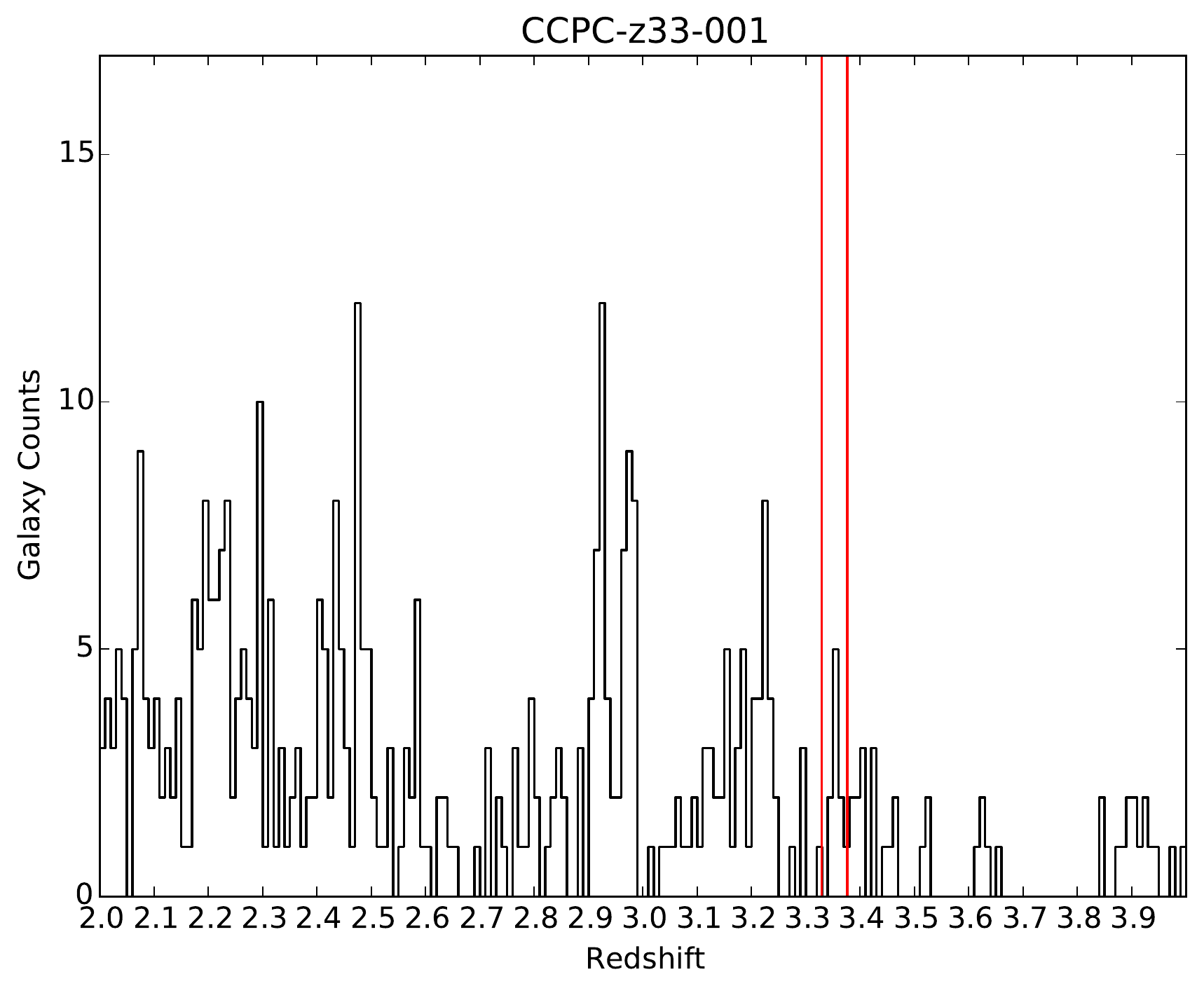}
\label{fig:CCPC-z33-001}
\end{subfigure}
\hfill
\end{figure*}

\begin{figure*}
\centering
\begin{subfigure}
\centering
\includegraphics[height=7.5cm,width=7.5cm]{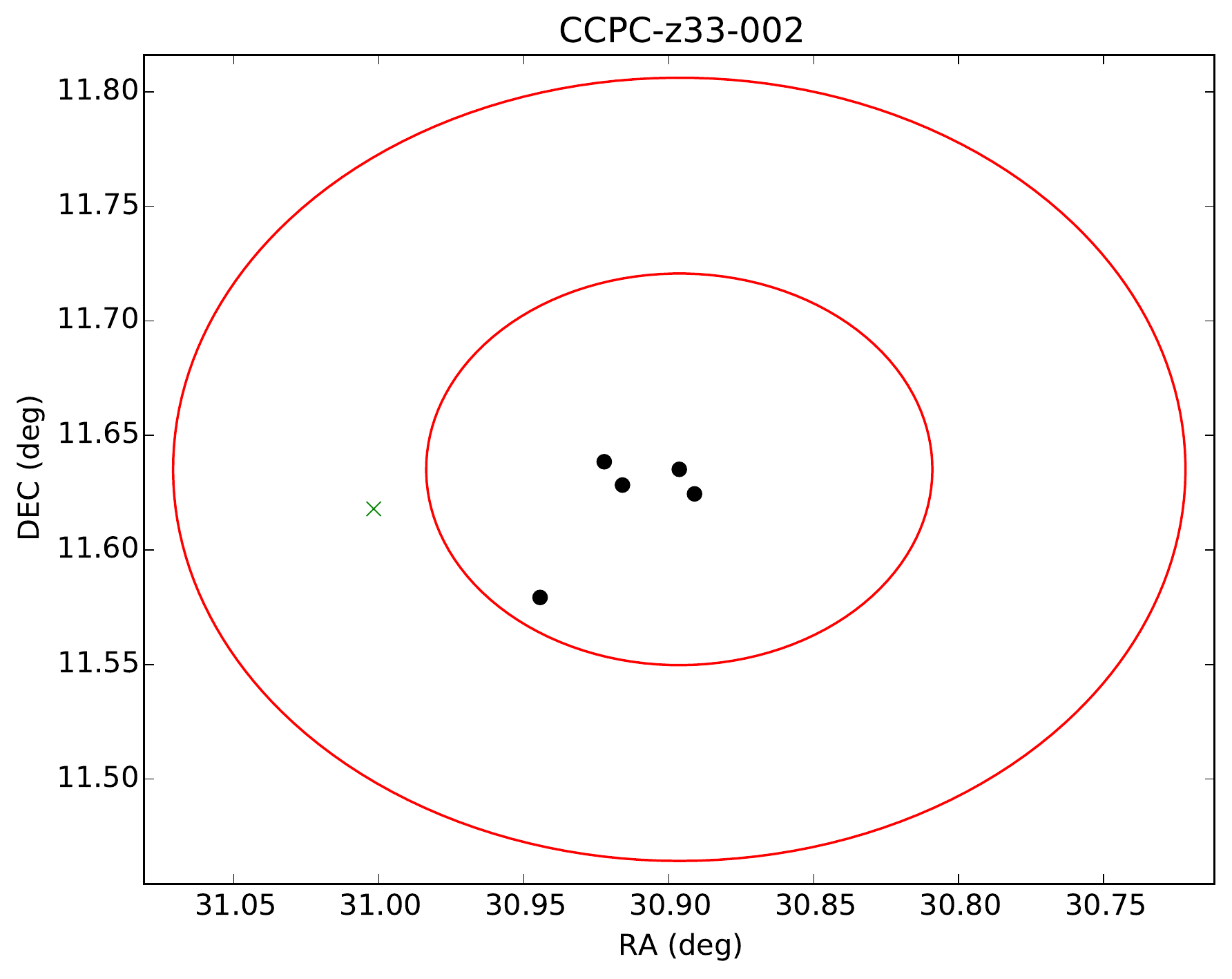}
\label{fig:CCPC-z33-002}
\end{subfigure}
\hfill
\begin{subfigure}
\centering
\includegraphics[scale=0.52]{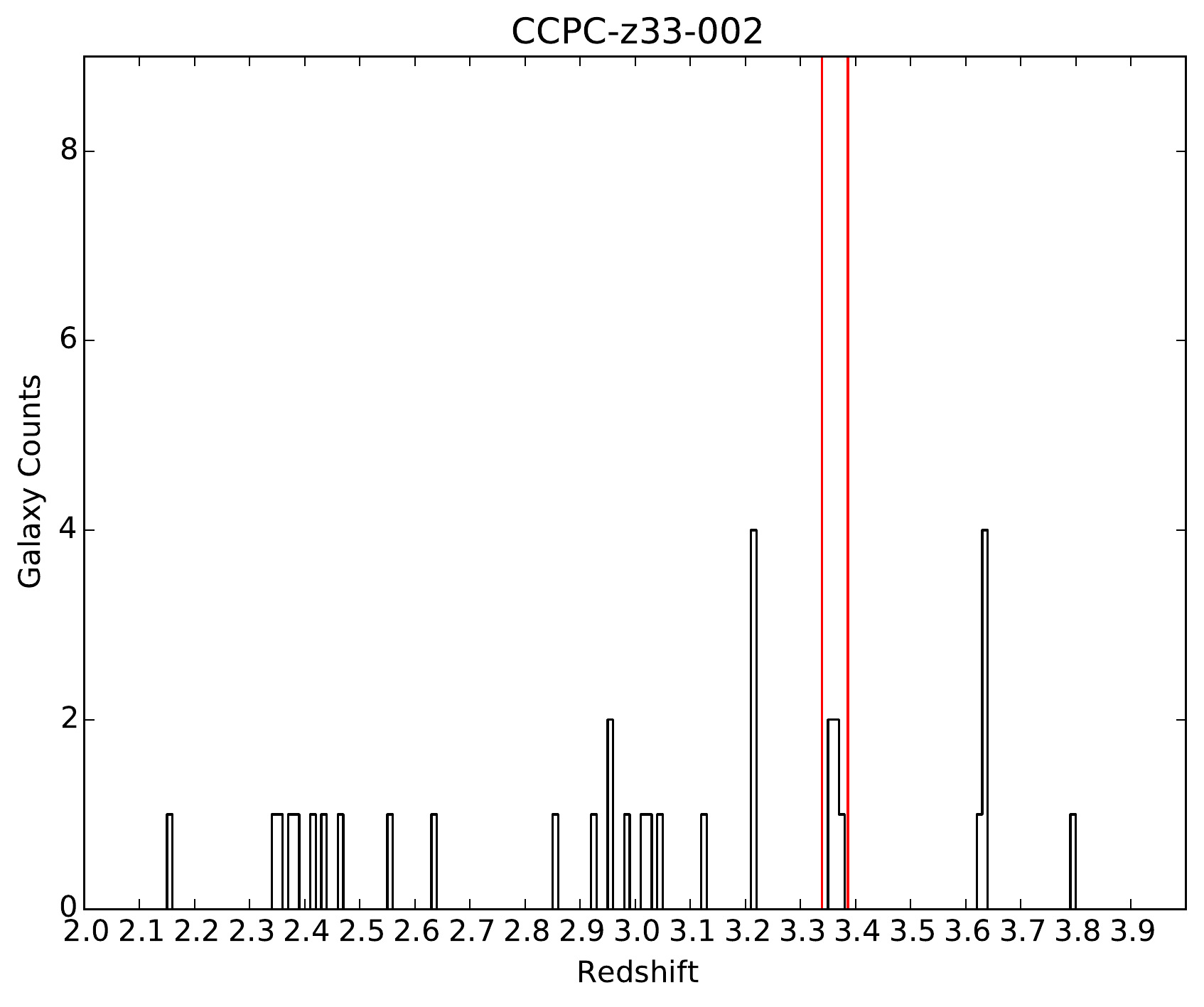}
\label{fig:CCPC-z33-002}
\end{subfigure}
\hfill
\end{figure*}

\begin{figure*}
\centering
\begin{subfigure}
\centering
\includegraphics[height=7.5cm,width=7.5cm]{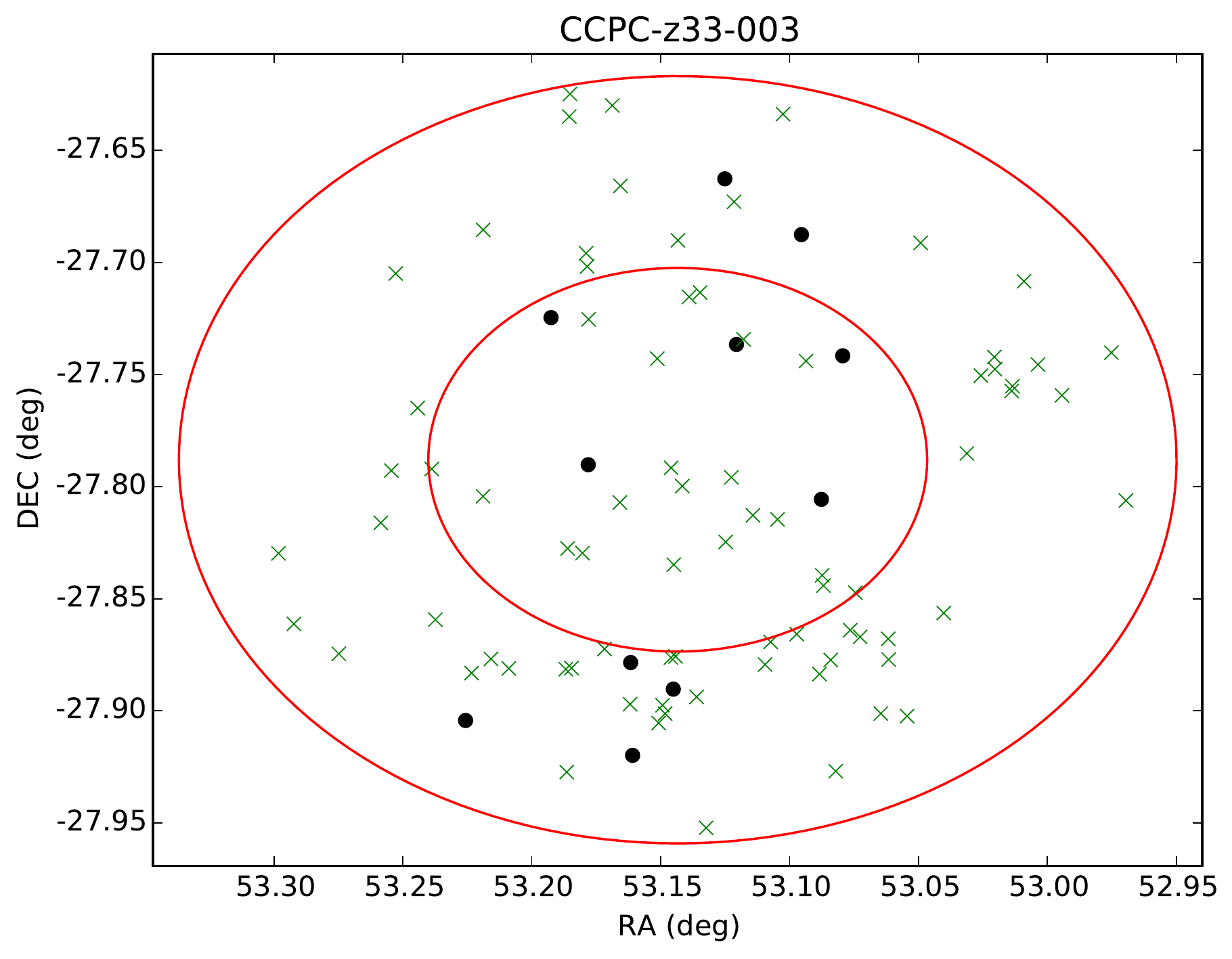}
\label{fig:CCPC-z33-003}
\end{subfigure}
\hfill
\begin{subfigure}
\centering
\includegraphics[scale=0.52]{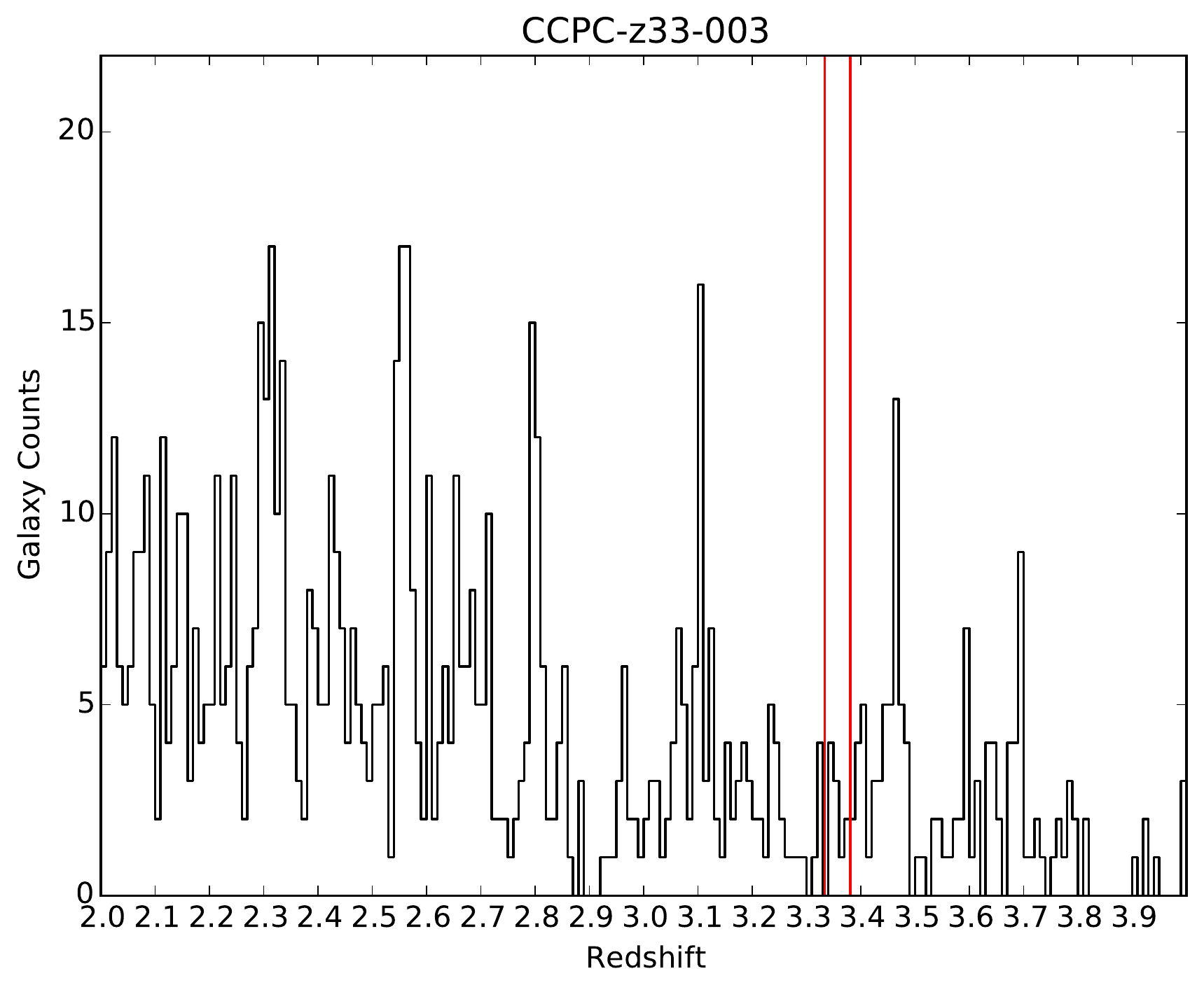}
\label{fig:CCPC-z33-003}
\end{subfigure}
\hfill
\end{figure*}
\clearpage 

\begin{figure*}
\centering
\begin{subfigure}
\centering
\includegraphics[height=7.5cm,width=7.5cm]{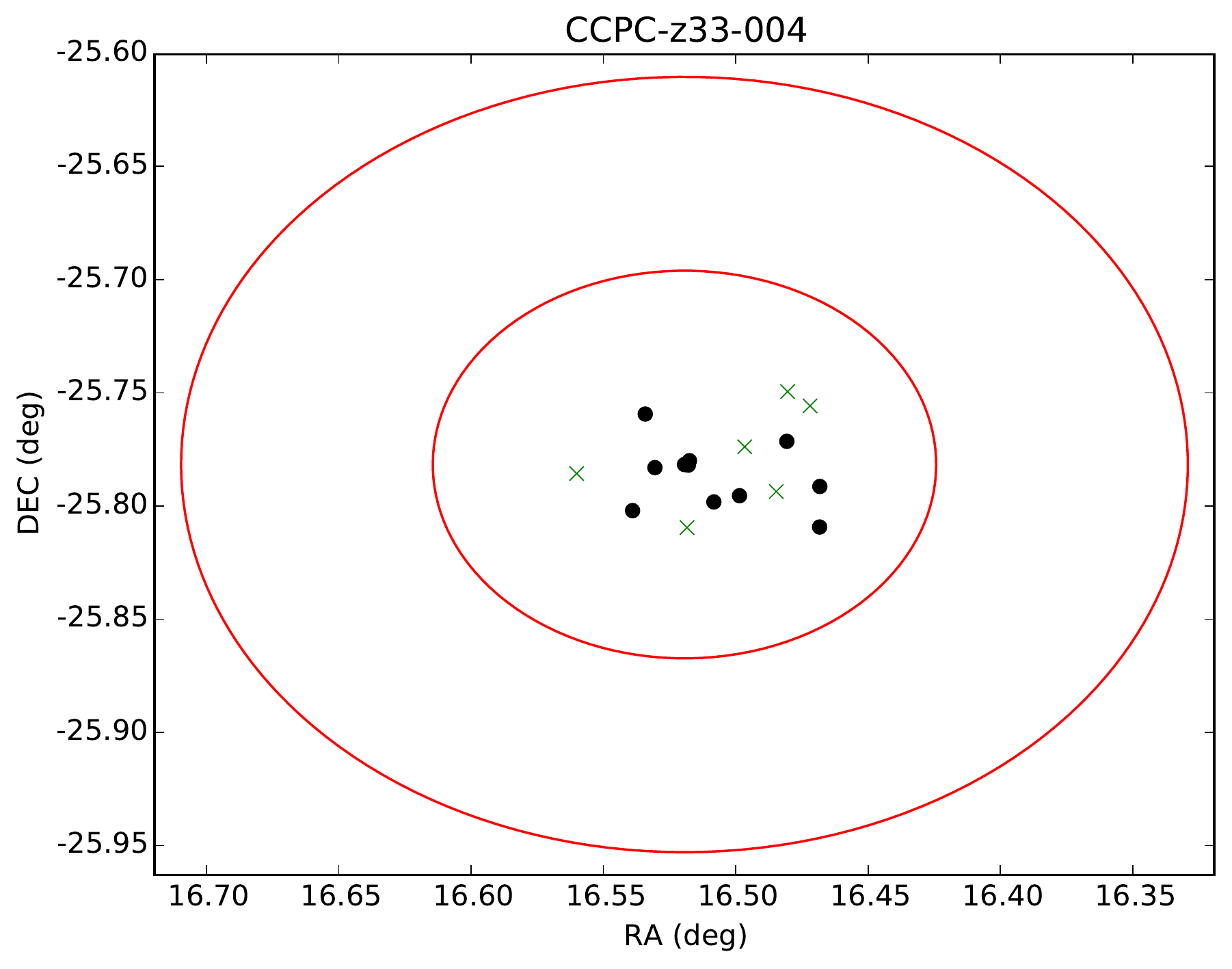}
\label{fig:CCPC-z33-004}
\end{subfigure}
\hfill
\begin{subfigure}
\centering
\includegraphics[scale=0.52]{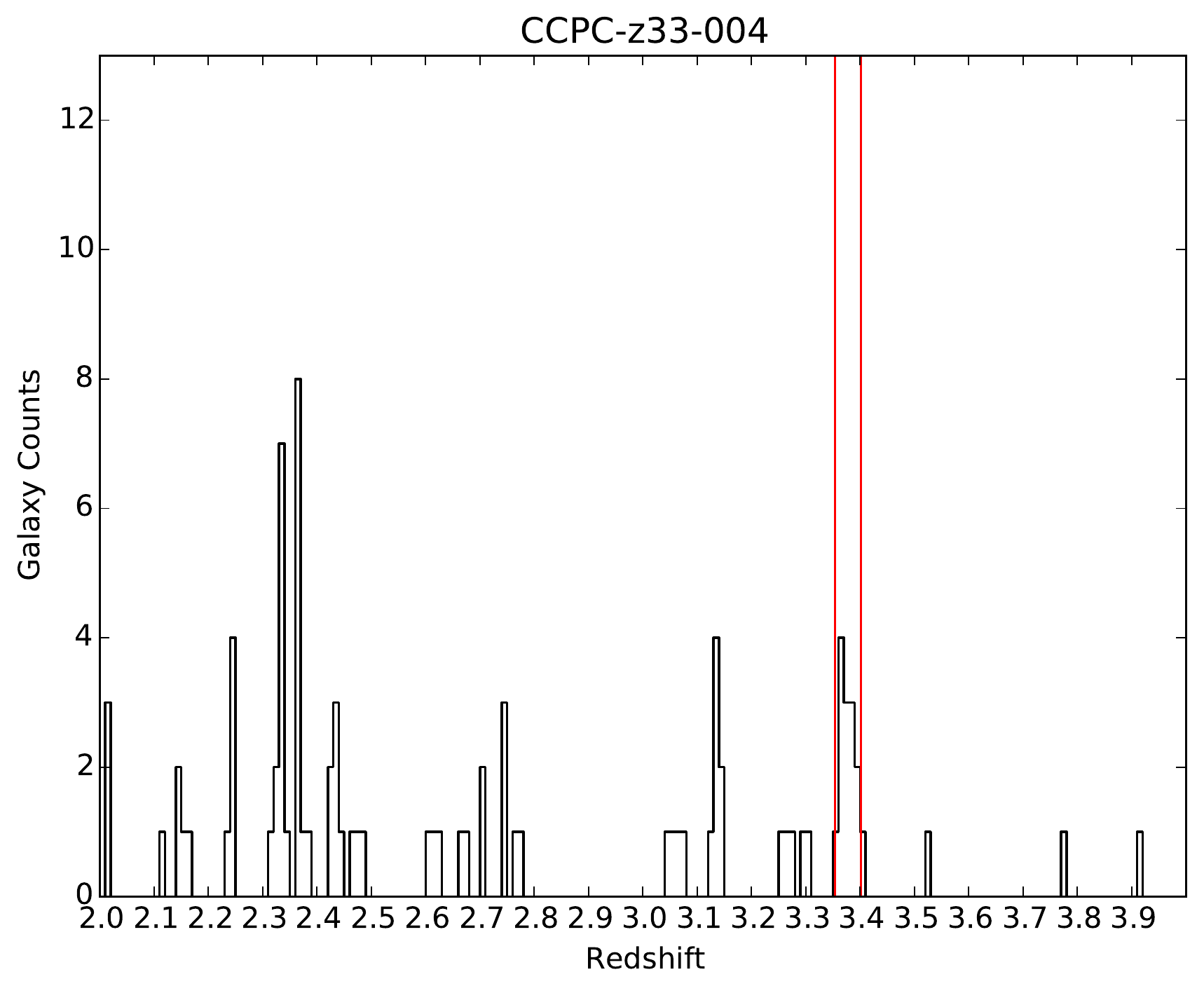}
\label{fig:CCPC-z33-004}
\end{subfigure}
\hfill
\end{figure*}

\begin{figure*}
\centering
\begin{subfigure}
\centering
\includegraphics[height=7.5cm,width=7.5cm]{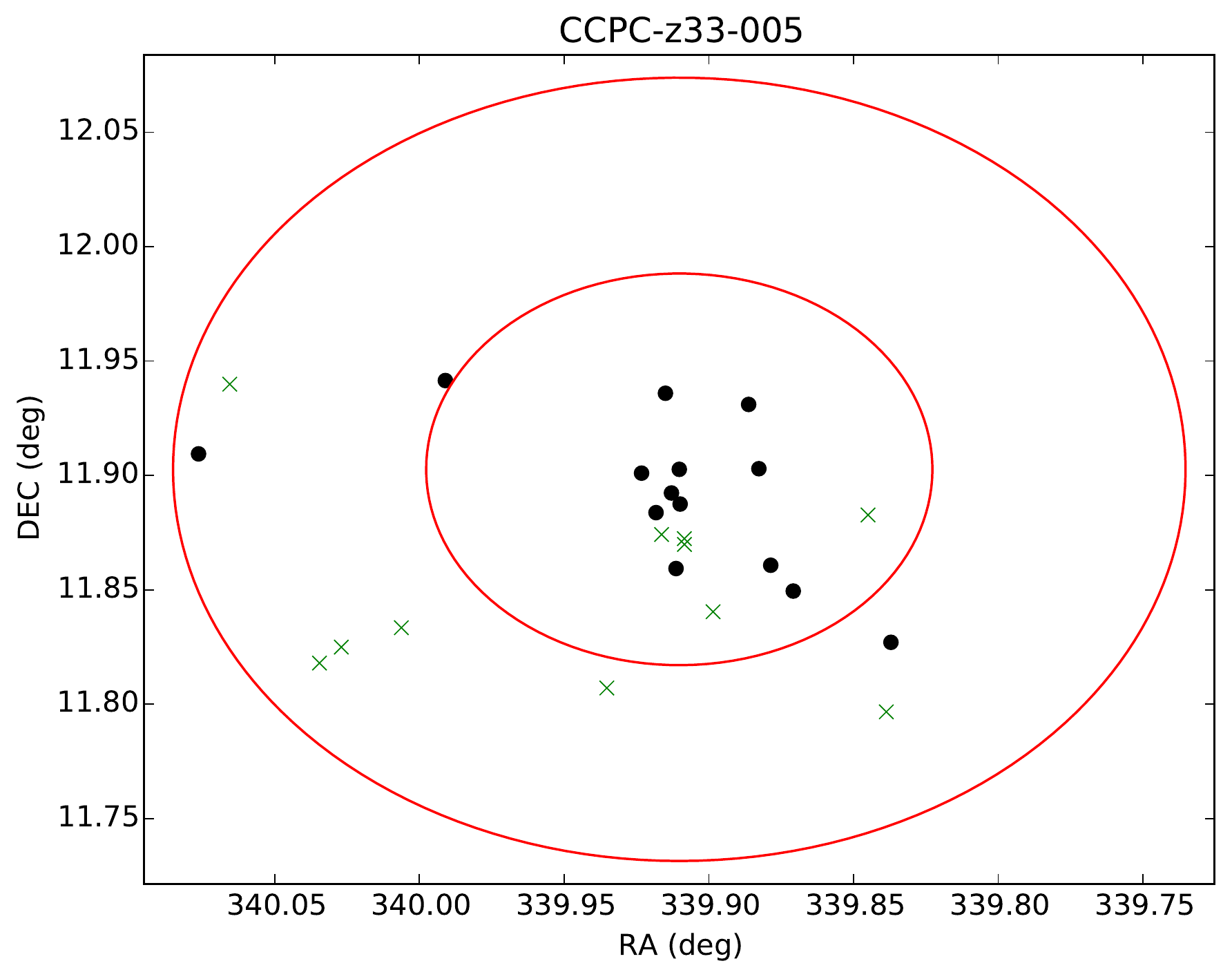}
\label{fig:CCPC-z33-005}
\end{subfigure}
\hfill
\begin{subfigure}
\centering
\includegraphics[scale=0.52]{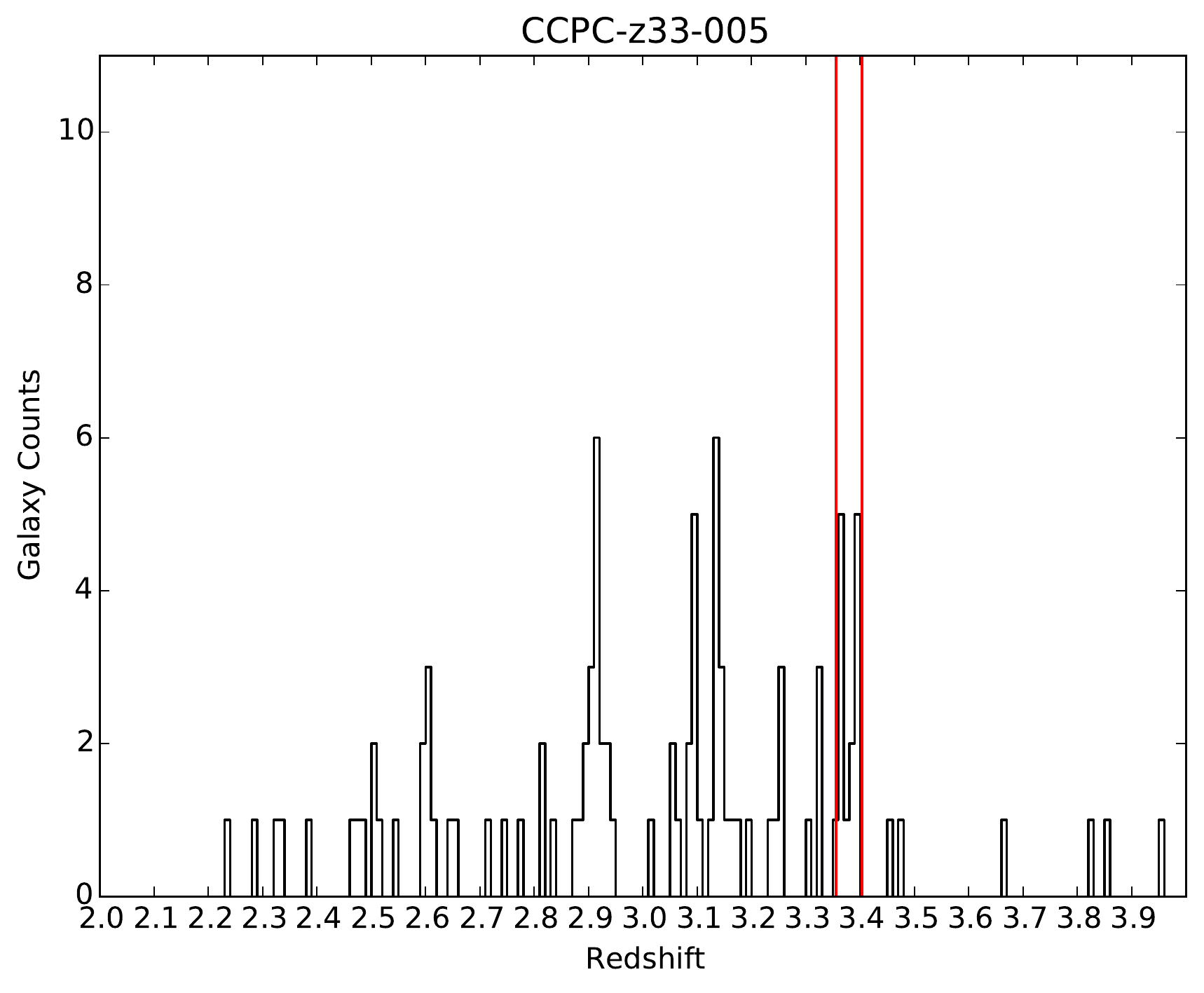}
\label{fig:CCPC-z33-005}
\end{subfigure}
\hfill
\end{figure*}

\begin{figure*}
\centering
\begin{subfigure}
\centering
\includegraphics[height=7.5cm,width=7.5cm]{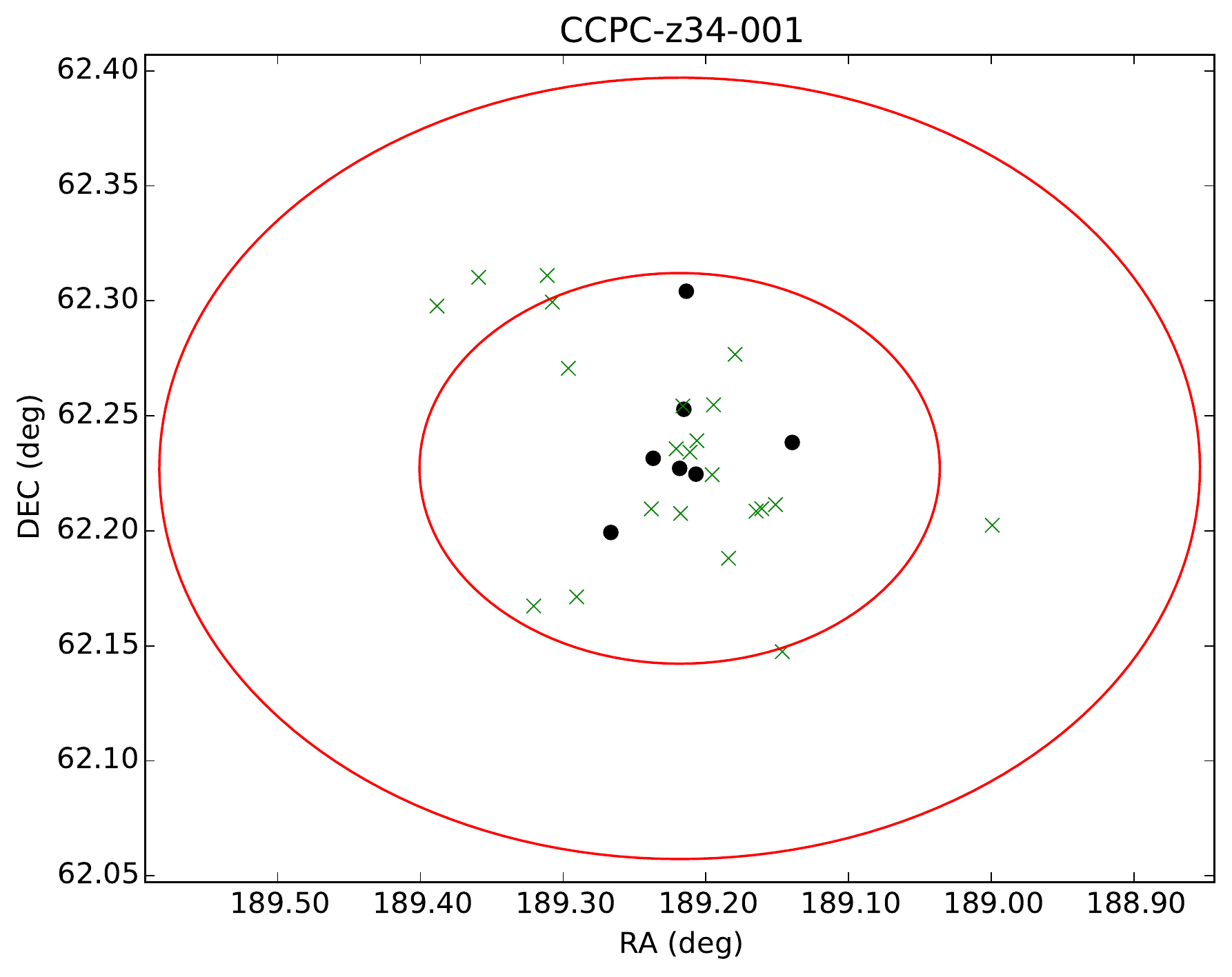}
\label{fig:CCPC-z34-001}
\end{subfigure}
\hfill
\begin{subfigure}
\centering
\includegraphics[scale=0.52]{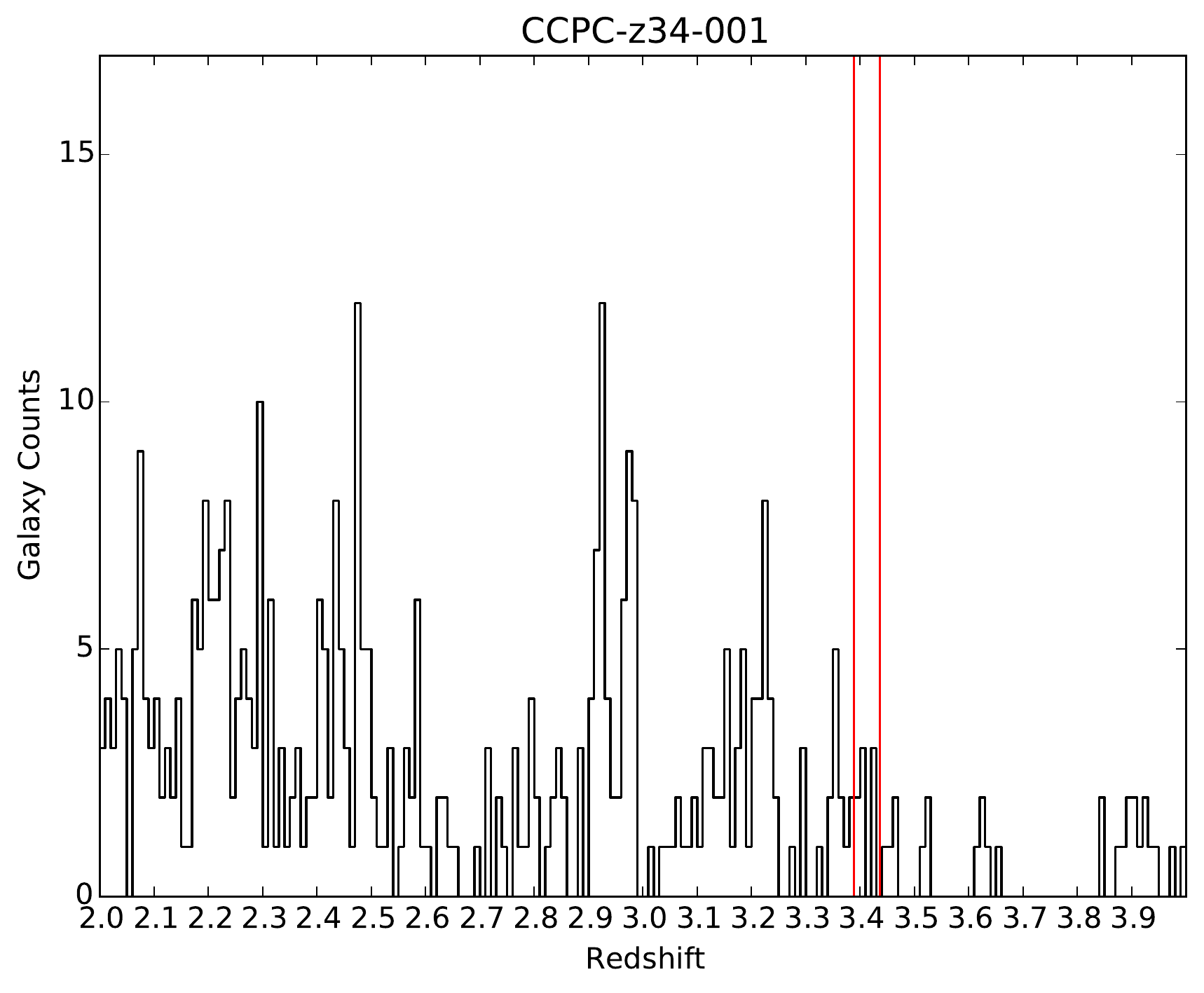}
\label{fig:CCPC-z34-001}
\end{subfigure}
\hfill
\end{figure*}
\clearpage 

\begin{figure*}
\centering
\begin{subfigure}
\centering
\includegraphics[height=7.5cm,width=7.5cm]{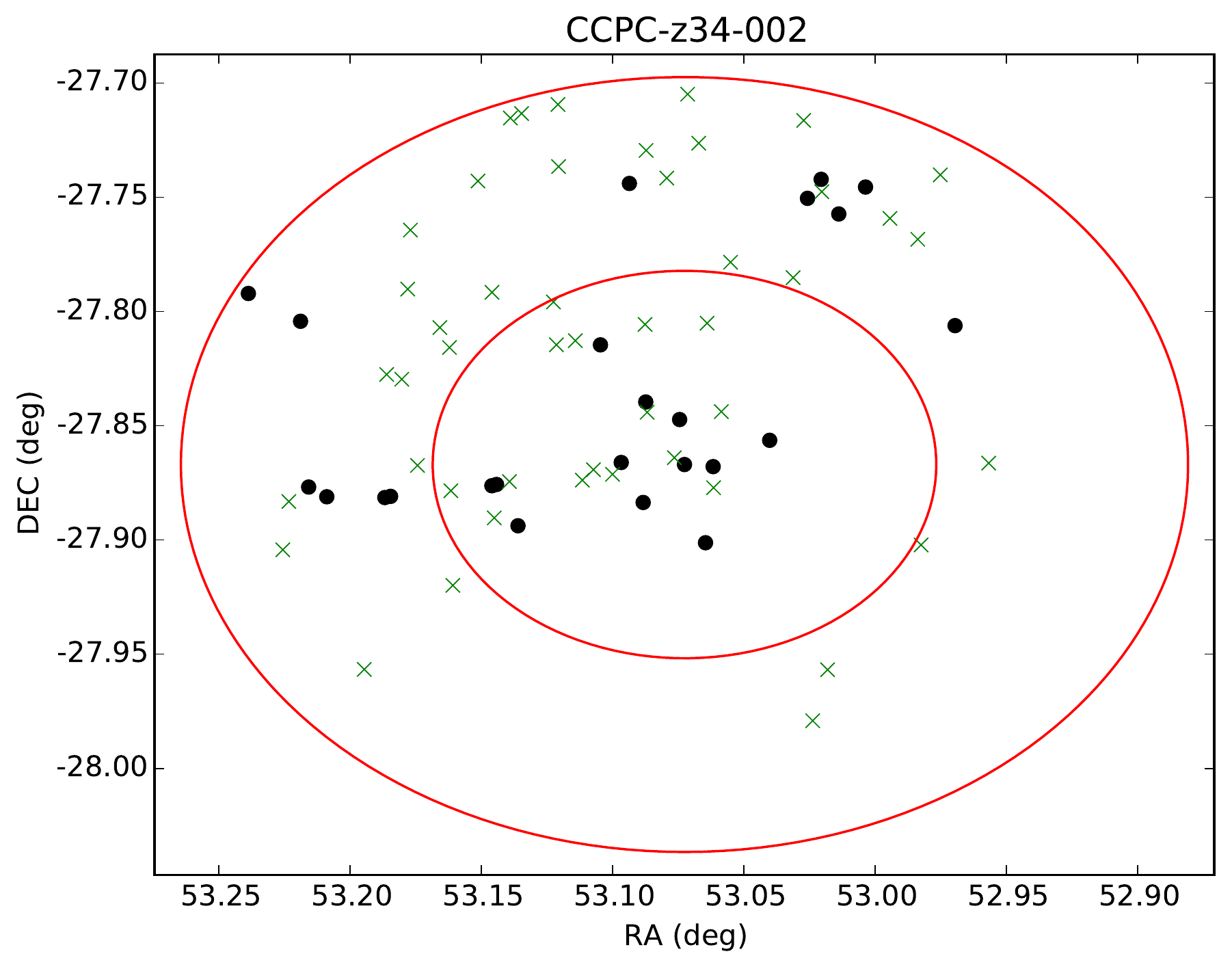}
\label{fig:CCPC-z34-002}
\end{subfigure}
\hfill
\begin{subfigure}
\centering
\includegraphics[scale=0.52]{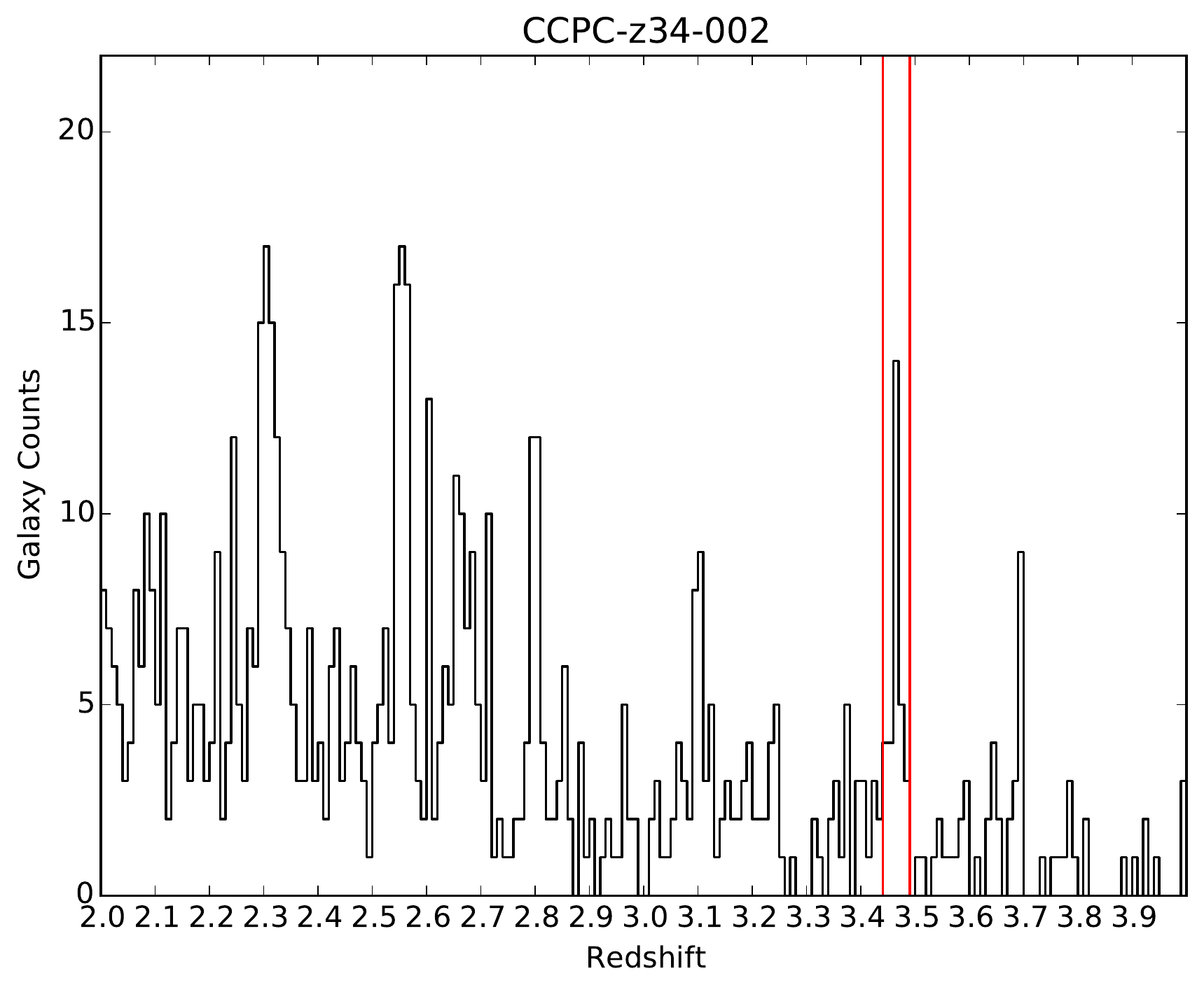}
\label{fig:CCPC-z34-002}
\end{subfigure}
\hfill
\end{figure*}

\begin{figure*}
\centering
\begin{subfigure}
\centering
\includegraphics[height=7.5cm,width=7.5cm]{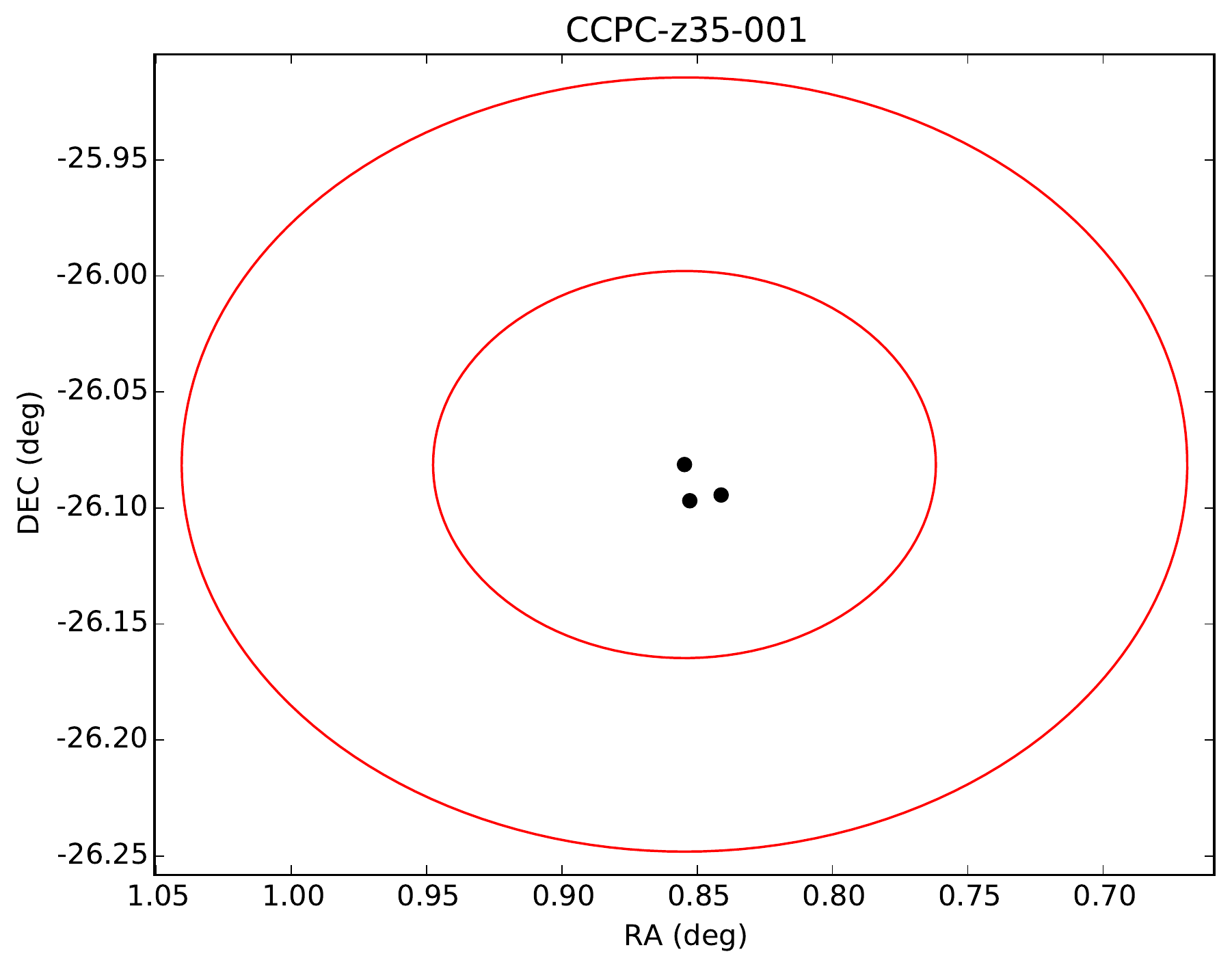}
\label{fig:CCPC-z35-001}
\end{subfigure}
\hfill
\begin{subfigure}
\centering
\includegraphics[scale=0.52]{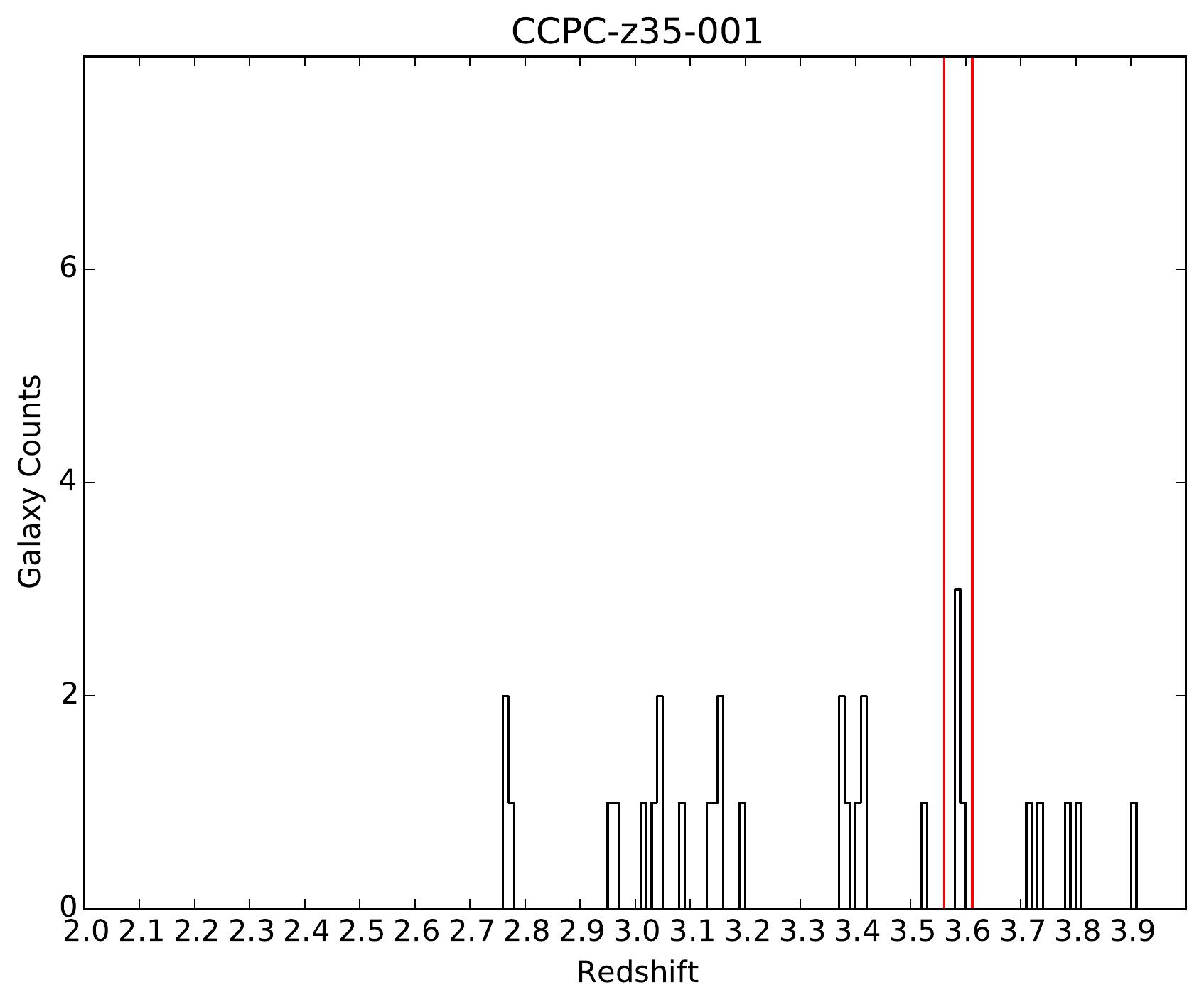}
\label{fig:CCPC-z35-001}
\end{subfigure}
\hfill
\end{figure*}

\begin{figure*}
\centering
\begin{subfigure}
\centering
\includegraphics[height=7.5cm,width=7.5cm]{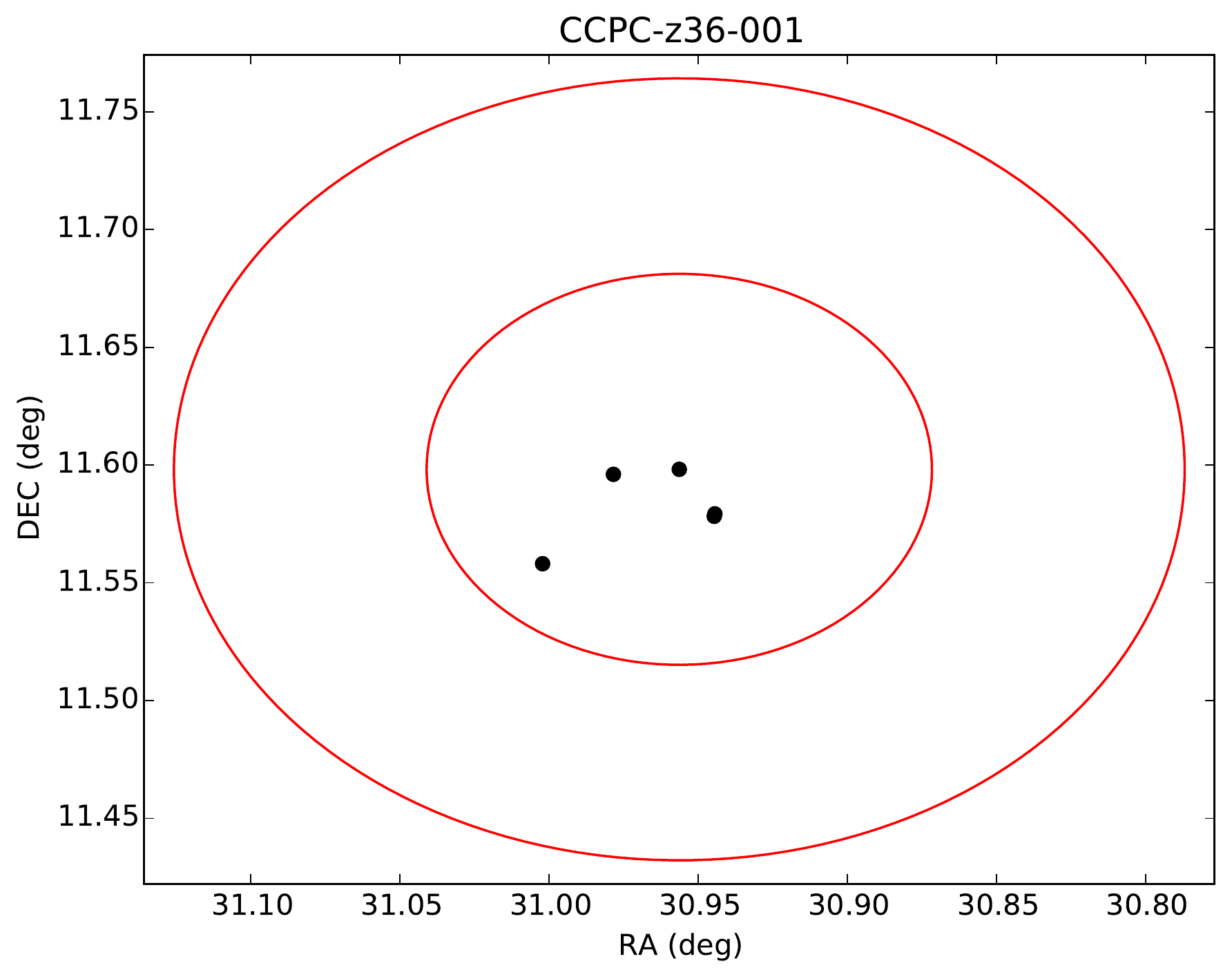}
\label{fig:CCPC-z36-001}
\end{subfigure}
\hfill
\begin{subfigure}
\centering
\includegraphics[scale=0.52]{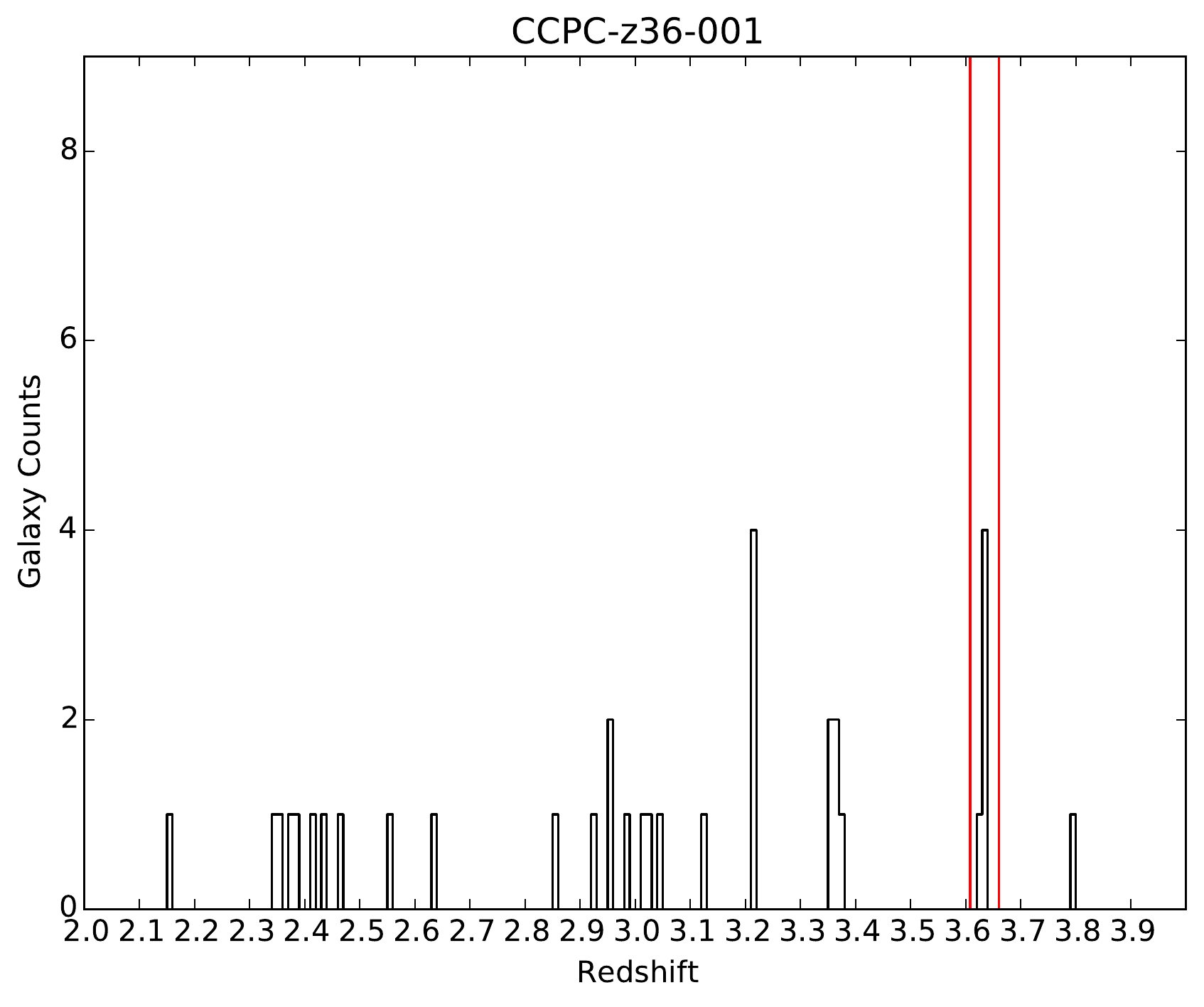}
\label{fig:CCPC-z36-001}
\end{subfigure}
\hfill
\end{figure*}
\clearpage 

\begin{figure*}
\centering
\begin{subfigure}
\centering
\includegraphics[height=7.5cm,width=7.5cm]{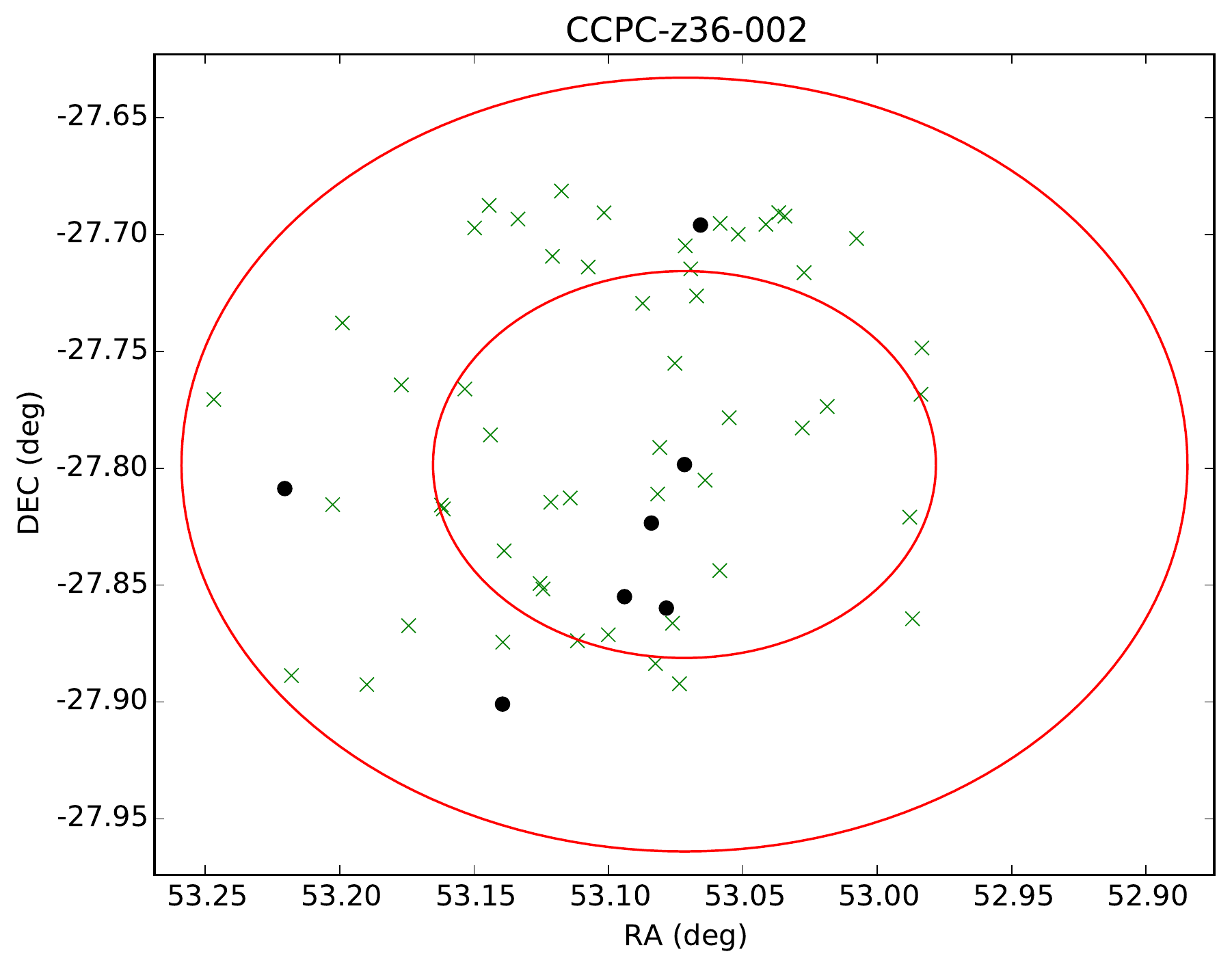}
\label{fig:CCPC-z36-002}
\end{subfigure}
\hfill
\begin{subfigure}
\centering
\includegraphics[scale=0.52]{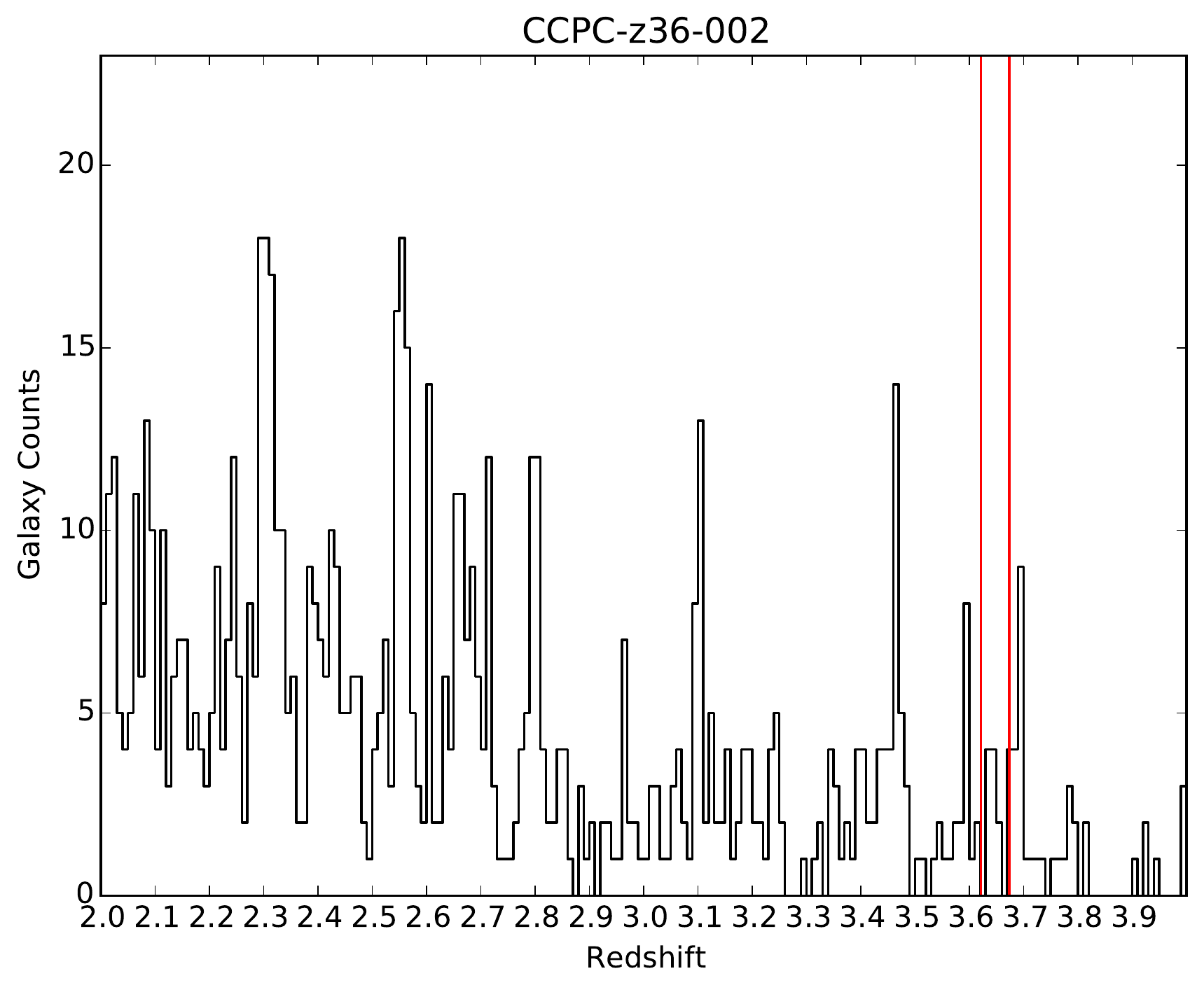}
\label{fig:CCPC-z36-002}
\end{subfigure}
\hfill
\end{figure*}

\begin{figure*}
\centering
\begin{subfigure}
\centering
\includegraphics[height=7.5cm,width=7.5cm]{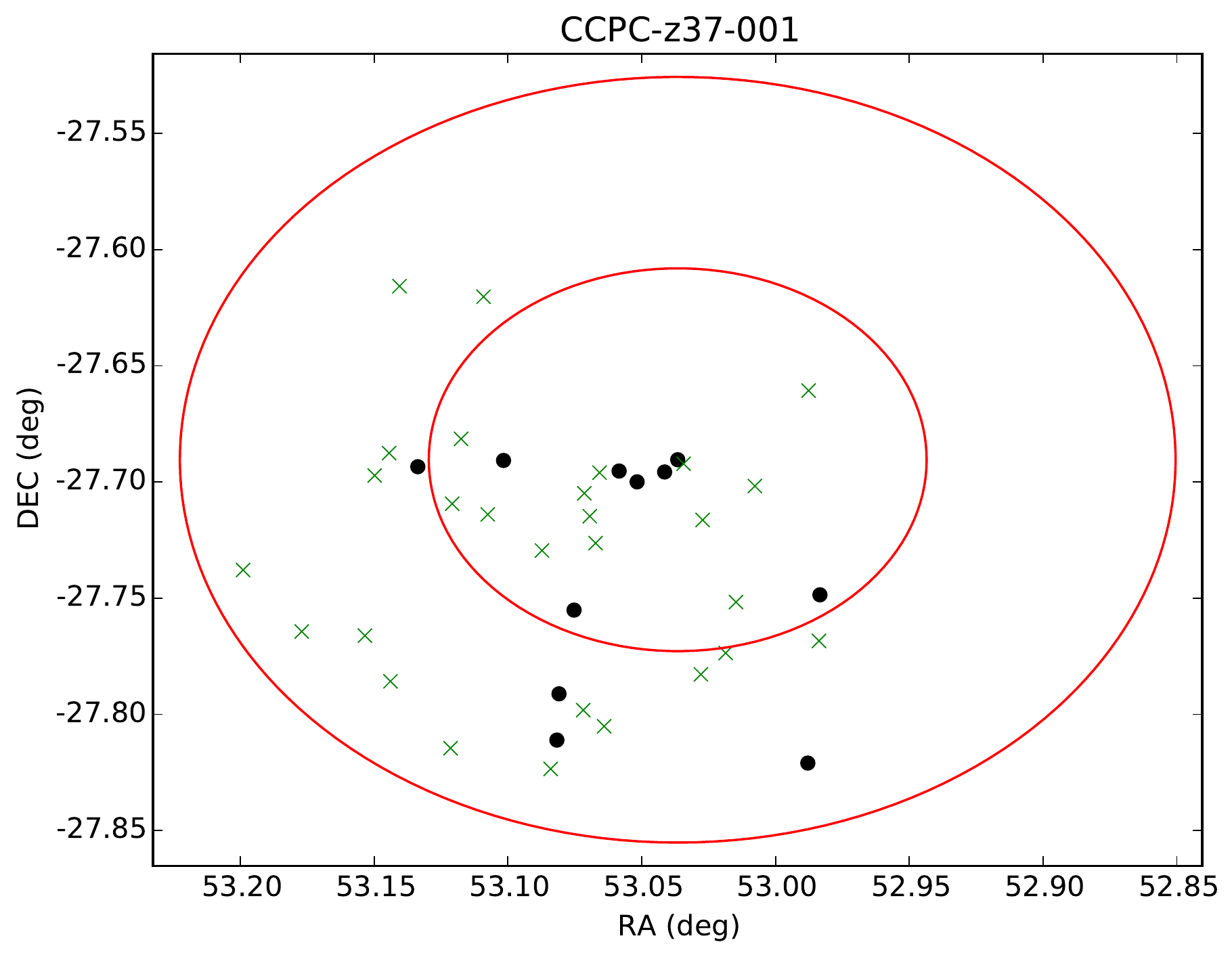}
\label{fig:CCPC-z37-001}
\end{subfigure}
\hfill
\begin{subfigure}
\centering
\includegraphics[scale=0.52]{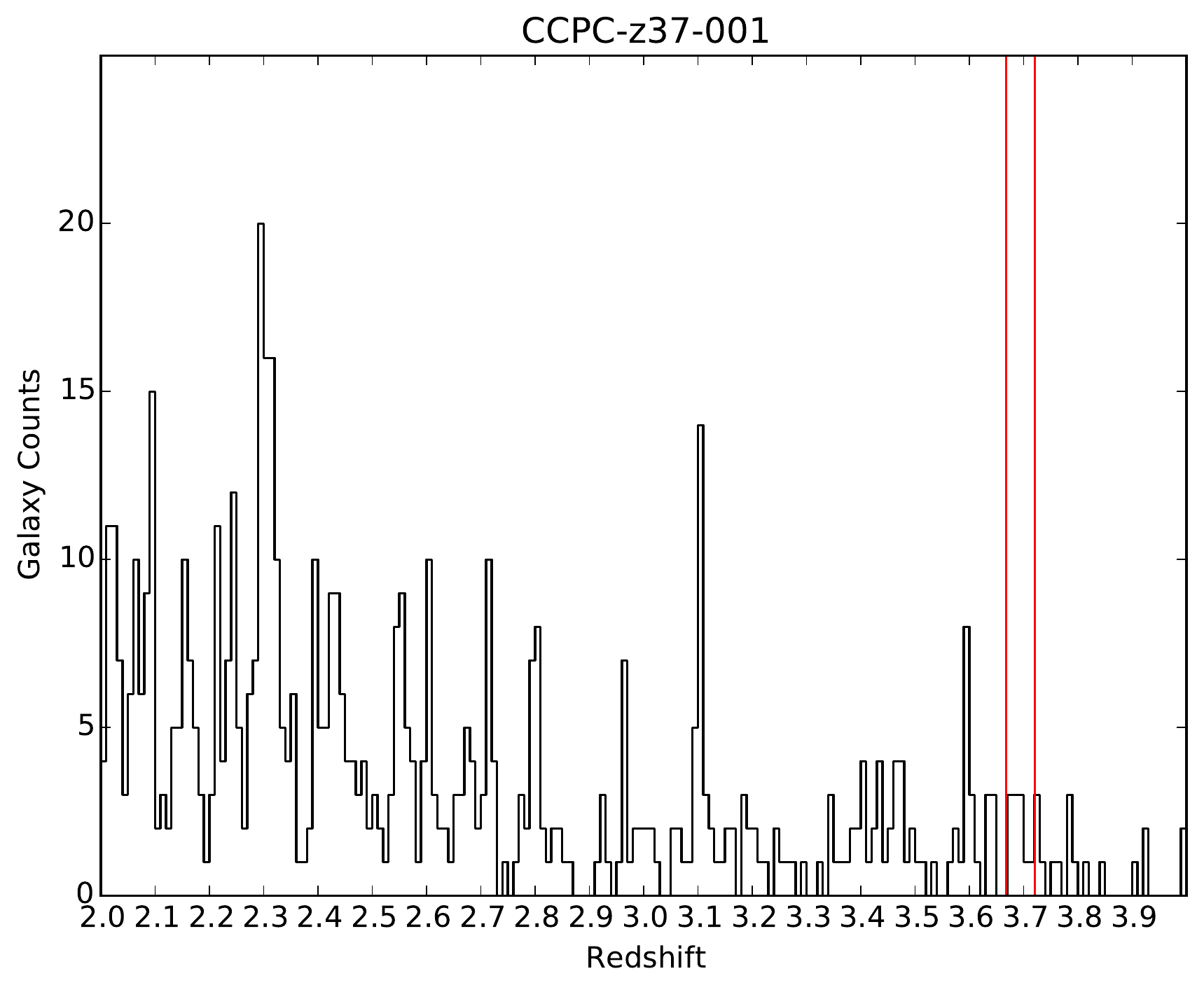}
\label{fig:CCPC-z37-001}
\end{subfigure}
\hfill
\end{figure*}

\clearpage

\end{document}